\documentclass[12pt]{article}
\usepackage{amsmath,amsfonts,amssymb}

\begin{document}
\thispagestyle{empty}

\def\theequation{\arabic{section}.\arabic{equation}}
\def\a{\alpha}
\def\b{\beta}
\def\g{\gamma}
\def\d{\delta}
\def\dd{\rm d}
\def\e{\epsilon}
\def\ve{\varepsilon}
\def\z{\zeta}
\def\B{\mbox{\bf B}}
\def\B{\mbox{\bf B}}\def\cp{\mathbb {CP}^3}
\newcommand{\h}{\hspace{0.5cm}}

\newcommand{\vf}{\varphi}
\newcommand{\ul}{\underline}
\newcommand{\p}{\partial}
\newcommand{\s}{\sigma}
\newcommand{\uz}{\underline z}
\newcommand{\us}{\underline\sigma}
\newcommand{\da}{\delta^p(\us_1 - \us_2)}
\newcommand{\la}{\lambda}
\newcommand{\La}{\Lambda}
\newcommand{\Di}{\left(\p_0-\la^{i}\p_i\right)}
\newcommand{\Dj}{\left(\p_0-\la^{j}\p_j\right)}
\newcommand{\ct}{\cal{t}}
\newcommand{\cf}{\cal{\varphi}}

\begin{titlepage}
\vspace*{1.cm}
\renewcommand{\thefootnote}{\fnsymbol{footnote}}
\begin{center}
\LARGE{BULGARIAN ACADEMY OF SCIENCES \\ Institute for Nuclear
Research and Nuclear Energy}
\vspace*{4cm}

{\Huge \bf $\mathbf{P}$-Branes Dynamics, AdS/CFT and Correlation Functions}
\end{center}

 \vskip 20mm

\baselineskip 18pt

\begin{center}

\centerline{{\huge \bf Plamen Bozhilov}}
 \vskip 0.6cm
{\huge {\bf Thesis}}

\Large{\bf presented for obtaining the scientific degree\\ "Doctor
of sciences"}

\vspace*{5cm}

\Large{\bf SOFIA, 2015}
\end{center}

\end{titlepage}

\newpage

\def\nn{\nonumber}
\def\tr{{\rm tr}\,}
\def\p{\partial}
\newcommand{\bea}{\begin{eqnarray}}
\newcommand{\eea}{\end{eqnarray}}
\renewcommand{\thefootnote}{\fnsymbol{footnote}}
\newcommand{\be}{\begin{equation}}
\newcommand{\ee}{\end{equation}}

\renewcommand{\thefootnote}{\arabic{footnote}}
\setcounter{footnote}{0}

\tableofcontents

\newpage

\setcounter{equation}{0}
\section{Introduction}
\hspace{1cm} In the attempt to unify all the forces present in
Nature, which entails having a consistent quantum theory of
gravity, superstring theories seem promising candidates. The
evidence that string theories could be unified theories is
provided by the presence in their massless spectrum of enough
particles to account for those present at low energies, including
the graviton \cite{GSW87}.

String theory is by now a vast subject with more than four decades
of active research contributing to its development. During the
last years, our understanding of string theory  has undergone a
dramatic change. One of the keys to this development is the
discovery of duality symmetries, which relate the strong and weak
coupling limits of different string theories ($S$-duality). This
led to the appearance of $M$-theory supposed to be the strong
coupling limit of all superstring theories.

Now, we will give a brief introduction to the related notions in
string theory following mainly \cite{V97}.

\subsection{\bf Brief Overview of String Theory}
\hspace{1cm} String theory is a description of dynamics of objects
with one spatial direction, which we parameterize by $\sigma$,
propagating in a space parameterized by $x^\mu$. The world-sheet
of the string is parameterized by coordinates ($\tau,\sigma $)
where each $\tau ={\rm constant}$ denotes the string at a given
time. The amplitude for propagation of a string from an initial
configuration to a final one is given by sum over world-sheets
which interpolate between the two string configurations weighed by
${\exp} (i S)$, where \bea\label{a1} S\propto\int d\tau d\sigma \
\partial_J x^\mu \partial^J x^{\nu}g_{\mu \nu}(x) \eea where
$g_{\mu \nu}$ is the metric on space-time and $J$ runs over the
$\tau$ and $\sigma$ directions. Note that by slicing the
world-sheet we will get configurations where a single string
splits to a pair or vice versa, and combinations thereof.

If we consider propagation in flat space-time where $g_{\mu
\nu}=\eta_{\mu \nu}$ the fields $x^{\mu}$ on the world-sheet,
which describe the position in space-time of each bit of string,
are free fields and satisfy the 2 dimensional equation \bea\nn
\partial _J\partial^J x^\mu
=(\partial^2_\tau -\partial^2_\sigma )x^\mu =0 . \eea The solution
of which is given by \bea\nn x^\mu(\tau,\sigma)=
x^\mu_L(\tau+\sigma )+x^\mu_R(\tau -\sigma ). \eea In particular
notice that the left- and right-moving degrees of freedom are
essentially independent. There are two basic types of strings:
{\it Closed} strings and {\it Open} strings depending on whether
the string is a closed circle or an open interval respectively. If
we are dealing with closed strings the left- and right-moving
degrees of freedom remain essentially independent but if are
dealing with open strings the left-moving modes reflecting off the
left boundary become the right-moving modes--thus the left- and
right-moving modes are essentially identical in this case.  In
this sense an open string has `half' the degrees of freedom of a
closed string and can be viewed as a `folding' of a closed string
so that it looks like an interval.

There are two basic types of string theories, bosonic and
fermionic. What distinguishes bosonic and fermionic strings is the
existence of supersymmetry on the world-sheet.  This means that in
addition to the coordinates $x^\mu$ we also have anti-commuting
fermionic coordinates $\psi^{\mu}_{L,R}$ which are space-time
vectors but fermionic spinors on the worldsheet whose chirality is
denoted by subscript $L,R$.  The action for superstrings takes the
form \bea\nn S=\int
\partial_Lx^\mu
\partial_R x^\mu +\psi_R^\mu \partial_L \psi_R^\mu +\psi_L^\mu
\partial_R \psi _R^\mu.
\eea There are two consistent boundary conditions on each of the
fermions, periodic ({\bf R}amond sector) or anti-periodic ({\bf
N}eveu-{\bf S}chwarz sector) (note that the coordinate $\sigma$ is
periodic).

A natural question arises as to what metric we should put on the
world-sheet.  In the above we have taken it to be flat.  However
in principle there is one degree of freedom that a metric can have
in two dimensions. This is because it is a $2\times 2$ symmetric
matrix (3 degrees of freedom) which is defined up to arbitrary
reparametrization of 2 dimensional space-time (2 degrees of
freedom) leaving us with one function. Locally we can take the 2
dimensional metric $g_{JK}$ to be conformally flat \bea\nn
g_{JK}={\rm exp}(\phi) \eta_{JK}. \eea Classically the action $S$
does not depend on $\phi$.  This is easily seen by noting that the
properly coordinate invariant action density goes as
 ${\sqrt{|g|}}g^{JK}\partial_J x^{\mu} \partial_K x^{\nu}\eta_{\mu\nu}$ and is
independent of $\phi$ only in $D=2$.  This is rather nice and
means that we can ignore all the local dynamics associated with
gravity on the world-sheet.  This case is what is known as the
critical string case which is the case of most interest.  It turns
out that this independence from the local dynamics of the
world-sheet metric survives quantum corrections only when the
dimension of space is 26 in the case of {\it bosonic strings} and
10 for {\it fermionic} or {\it superstrings}.  Each string can be
in a specific vibrational mode which gives rise to a particle.  To
describe the totality of such particles it is convenient to go to
`light-cone' gauge.  Roughly speaking this means that we take into
account that string vibration along their world-sheet is not
physical.  In particular for bosonic string the vibrational modes
exist only in 24 transverse directions and for superstrings they
exist in $8$ transverse directions.

Solving the free field equations for $x,\psi $ we have \bea\nn
\partial_L x^\mu = \sum_{n}\alpha_{-n}^\mu {\rm
e}^{-in(\tau+\sigma)}\\ \psi_L^\mu =\sum_n \psi_{-n}^\mu {\rm
e}^{-in(\tau+\sigma)} \eea and similarly for right-moving
oscillator modes $\tilde \alpha_{-n}^\mu$ and $\tilde
\psi_{-n}^\mu$. The sum over $n$ in the above runs over integers
for the $\alpha_{-n}$.  For fermions depending on whether we are
in the {\bf{R}} sector or {\bf {NS}} sector it runs over integers
or integers shifted by ${1/2}$ respectively. Many things decouple
between the left- and right-movers in the construction of a single
string Hilbert space and we sometimes talk only about one of them.
For the open string Fock space the left- and right-movers mix as
mentioned before, and we simply get one copy of the above
oscillators.

A special role is played by the {\it zero modes} of the
oscillators. For the $x$-fields they correspond to the center of
mass motion and thus $\alpha_0$ gets identified with the
left-moving momentum of the center of mass. In particular we have
for the center of mass \bea\nn x=\alpha_0 (\tau+\sigma) +\tilde
\alpha_0 (\tau -\sigma ), \eea where we identify \bea\nn
(\alpha_0,\tilde \alpha_0)=(P_L,P_R). \eea Note that for closed
string, periodicity of $x$ in $\sigma$ requires that $P_L=P_R=P$
which we identify with the center of mass momentum of the string.

In quantizing the fields on the strings we use the usual
(anti)commutation relations \bea\nn
[\alpha^{\mu}_{n},\alpha^{\nu}_{m}]=n\delta_{m+n,0}\eta^{\mu \nu}
\\ \nn
 \{\psi^{\mu}_{n},\psi^{\nu}_{m}\}=\eta^{\mu \nu}\delta_{m+n,0}.
\eea We choose the negative moded oscillators as creation
operators. In constructing the Fock space we have to pay special
attention to the zero modes. The zero modes of $\alpha$ should be
diagonal in the Fock space and we identify their eigenvalue with
momentum.  For $\psi$ in the $NS$ sector there is no zero mode so
there is no subtlety in construction of the Hilbert space. For the
$R$ sector, we have zero modes.  In this case the zero modes form
a Clifford algebra \bea\nn \{\psi^\mu_0,\psi^\nu_0\}=\eta^{\mu
\nu}. \eea This implies that in these cases the ground state is a
spinor representation of the Lorentz group.  Thus a typical
element in the Fock space looks like \bea\nn
\alpha_{-n_1}^{L\mu_1}....\psi_{-n_k}^{L\mu_k}...|P_L,a\rangle\otimes
\alpha_{-m_1}^{R\mu_1}....\psi_{-m_r}^{R\mu_k}...|P_R,b\rangle ,
\eea where $a,b$ label spinor states  for R sectors and are absent
in the NS case; moreover for the bosonic string we only have the
left and right bosonic oscillators.

It is convenient to define the total oscillator number as sum of
the negative oscillator numbers, for left- and right- movers
separately. $N_L=n_1+...+n_k+...$, $N_R=m_1+...+m_r+...$. The
condition that the two dimensional gravity decouple implies that
the energy momentum tensor annihilate the physical states.  The
trace of the energy momentum tensor is zero here (and in all
compactifications of string theory) and so we have two independent
components which can be identified with the left- and right-
moving hamiltonians $H_{L,R}$ and the physical states condition
requires that \bea\label{onsh} H_L=N_L+(1/2)P_L^2-\delta_L=0=
H_R=N_R+(1/2)P_R^2-\delta_R ,\eea where $\delta_{L,R}$ are normal
ordering constants which depend on which string theory and which
sector we are dealing with.  For bosonic string $\delta =1$, for
superstrings we have two cases:  For $NS$ sector $\delta ={1/2}$
and for the $R$ sector $\delta =0$. The equations (\ref{onsh})
give the spectrum of particles in the string perturbation theory.
Note that $P_L^2=P_R^2=-m^2$ and so we see that $m^2$ grows
linearly with the oscillator number $N$, up to a shift:
\bea\label{spec} (1/2)m^2=N_L-\delta_L=N_R-\delta_R. \eea

\subsubsection{Massless States of Bosonic Strings}
\hspace{1cm} Let us consider the left-mover excitations.  Since
$\delta =1$ for bosonic string, (\ref{spec}) implies that if we do
not use any string oscillations, the ground state is tachyonic
${1/2}m^2=-1$.  This clearly implies that bosonic string by itself
is not a good starting point for perturbation theory. Nevertheless
in anticipation of a modified appearance of bosonic strings in the
context of heterotic strings, let us continue to the next state.

If we consider oscillator number $N_L=1$, from (\ref{spec}) we
learn that excitation is massless.  Putting the right-movers
together with it, we find that it is given by \bea\nn
\alpha^\mu_{-1}\tilde \alpha^\nu_{-1}|P\rangle . \eea What is the
physical interpretation of these massless states? The most
reliable method is to find how they transform under the little
group for massless states which in this case is $SO(24)$.  If we
go to the light cone gauge, and count the physical states, which
roughly speaking means taking the indices $\mu$ to go over spatial
directions transverse to a null vector, we can easily deduce the
content of states.   By decomposing the above massless state under
the little group of $SO(24)$, we find that we have symmetric
traceless tensor, anti-symmetric 2-tensor, and the trace, which we
identify as arising from 26 dimensional fields \bea\label{grm}
{g_{\mu \nu},B_{\mu\nu},\phi} \eea the metric, the anti-symmetric
field $B$ and the {\it dilaton}. This triple of fields should be
viewed as the stringy multiplet for gravity.  The quantity ${\rm
\exp}[-\phi]$ is identified with the string coupling constant.
What this means is that a world-sheet configuration of a string
which sweeps a genus $g$ curve, which should be viewed as $g$-th
loop correction for string theory, will be weighed by ${\rm
exp}(-2(g-1)\phi)$. The existence of the field $B$ can also be
understood (and in some sense predicted) rather easily. If we have
a point particle it is natural to have it charged under a gauge
field, which introduces a term ${\rm exp}(i\int A)$ along the
world-line. For strings the natural generalization of this
requires an anti-symmetric 2-form to integrate over the
world-sheet, and so we say that the strings are {\it charged}
under $B_{\mu \nu}$ and that the amplitude for a world-sheet
configuration will have an extra factor of ${\rm exp}(i\int B)$.

Since bosonic string has tachyons we do not know how to make sense
of that theory by itself.

\subsubsection{Massless States of Type II Superstrings}
\hspace{1cm} Let us now consider the light particle states for
superstrings. We recall from the above discussion that there are
two sectors to consider, NS and R, separately for the left- and
the right-movers. As usual we will first treat the left- and
right-moving sectors separately and then combine them at the end.
Let us  consider the NS sector for left-movers. Then the formula
for masses (\ref{onsh}) implies that the ground state is tachyonic
with ${1/2}m^2={-1/2}$. The first excited states from the
left-movers are massless and corresponds to $\psi_{-1/2}^\mu
|0\rangle$, and so is a vector in space-time.  How do we deal with
the tachyons? It turns out that summing over the boundary
conditions of fermions on the world-sheet amounts to keeping the
states with a fixed fermion number $(-1)^F$ on the world-sheet.
Since in the NS sector the number of fermionic oscillator
correlates with the integrality/half-integrality of $N$, it turns
out that the consistent choice involves keeping only the
$N=$half-integral states. This is known as the GSO projection.
Thus the tachyon is projected out and the lightest left-moving
state is a massless vector.

For the R-sector using (\ref{onsh}) we see that the ground states
are massless.  As discussed above, quantizing the zero modes of
fermions implies that they are spinors.  Moreover GSO projection,
which is projection on a definite $(-1)^F$ state, amounts to
projecting to spinors of a given chirality. So after GSO
projection we get a massless spinor of a definite chirality. Let
us denote the spinor of one chirality by $s$ and the other one by
$s'$.

Now let us combine the left- and right-moving sectors together.
Here we run into two distinct possibilities: A) The GSO
projections on the left- and right-movers are different and lead
in the R sector to ground states with different chirality. B) The
GSO projections on the left- and right-movers are the same and
lead in the R sector to ground states with the same chirality. The
first case is known as type IIA superstring and the second one as
type IIB.  Let us see what kind of massless modes we get for
either of them.  From NS$\otimes$NS we find for both type IIA,B
\bea\nn NS\otimes NS \rightarrow v\otimes v \rightarrow (g_{\mu
\nu},B_{\mu \nu},\phi ) \eea From the $NS \otimes R$ and $R
\otimes NS$ we get the fermions of the theory (including the
gravitinos). However the IIA and IIB differ in that the gravitinos
of IIB are of the same chirality, whereas for IIA they are of the
opposite chirality.  This implies that IIB is a chiral theory
whereas IIA is non-chiral. Let us move to the R$\otimes$ R sector.
We find \bea\nn IIA: R\otimes R =s\otimes s' \rightarrow (A_\mu,
C_{\mu \nu \rho})\\ \label{cont}IIB: R\otimes R =s\otimes
s\rightarrow (\chi,B'_{\mu \nu}, D_{\mu\nu\rho \lambda}), \eea
where all the tensors appearing above are fully antisymmetric.
Moreover $D_{\mu \nu \rho \lambda}$ has a self-dual field strength
$F=dD=*F$. It turns out that to write the equations of motion in a
unified way it is convenient to consider a generalized gauge
fields ${\cal A}$ and ${\cal B}$ in the IIA and IIB case
respectively by adding all the fields in the RR sector together
with the following properties: i) ${\cal A} ({\cal B})$ involve
all the odd (even) dimensional antisymmetric fields. ii) the
equation of motion is $d{\cal A}=* d{\cal A}$.  In the case of all
fields (except $D_{\mu\nu\lambda \rho}$) this equation allows us
to solve for the forms with degrees bigger than 4 in terms of the
lower ones and moreover it implies the field equation $d*dA=0$
which is the familiar field equation for the gauge fields. In the
case of the $D$-field it simply gives that its field strength is
self-dual.

\subsubsection{Open Superstring: Type I String}
\hspace{1cm} In the case of type IIB theory in 10 dimensions, we
note that the left- and right-moving degrees of freedom on the
worldsheet are the same.  In this case we can `mod out' by a
reflection symmetry on the string; this means keeping only the
states in the full Hilbert space which are invariant under the
left-/right-moving exchange of quantum numbers.  This is simply
projecting the Hilbert space onto the invariant subspace of the
projection operator $P={1\over 2}(1+\Omega)$ where $\Omega$
exchanges left- and right-movers. $\Omega$ is known as the
orientifold operation as it reverses the orientation on the
world-sheet.
 Note that this symmetry
only exists for IIB and not for IIA theory (unless we accompany it
with a parity reflection in spacetime).  Let us see which bosonic
states we will be left with after this projection. {}From the
NS-NS sector $B_{\mu\nu}$ is odd and projected out and thus we are
left with the symmetric parts of the tensor product \bea\nn
NS-NS\rightarrow (v\otimes v)_{symm.}=(g_{\mu \nu},\phi ). \eea
From the R-R sector since the degrees of freedom are fermionic
from each sector we get, when exchanging left- and right-movers an
extra minus sign which thus means we have to keep anti-symmetric
parts of the tensor product \bea\nn R-R\rightarrow (s\otimes
s)_{anti-symm.}={\tilde B}_{\mu \nu}. \eea This is not the end of
the story, however.  In order to make the theory consistent we
need to introduce a new sector in this theory involving open
strings.  This comes about from the fact that in the R-R sector
there actually is a 10 form gauge potential which has no
propagating degree of freedom, but acquires a tadpole.
Introduction of a suitable open string sector cancels this
tadpole.

As noted before the construction of open string sector Hilbert
space proceeds as in the closed string case, but now, the
left-moving and right-moving modes become indistinguishable due to
reflection off the boundaries of open string.  We thus get only
one copy of the oscillators. Moreover we can associate
`Chan-Paton' factors to the boundaries of open string . To cancel
the tadpole it turns out that we need 32 Chan-Paton labels on each
end. We still have two sectors corresponding to the NS and R
sectors.  The NS sector gives a vector field $A_\mu$ and the R
sector gives the gaugino.  The gauge field $A_{\mu}$ has two
additional labels coming from the end points of the open string
and it turns out that the left-right exchange projection of the
type IIB theory translates to keeping the antisymmetric component
of $A_\mu =-A_\mu^T$, which means we have an adjoint of $SO(32)$.
Thus all put together, the bosonic degrees of freedom are \bea\nn
(g_{\mu \nu},{\tilde B}_{\mu\nu},\phi)+(A_\mu)_{SO(32)}. \eea We
should keep in mind here that $\tilde B$ came not from the NS-NS
sector, but from the R-R sector.

\subsubsection{Heterotic Strings}
\hspace{1cm} Heterotic string is a combination of bosonic string
and superstring, where roughly speaking the left-moving degrees of
freedom are as in the bosonic string and the right-moving degrees
of freedom are as in the superstring. It is clear that this makes
sense for the construction of the states because the left- and
right-moving sectors hardly talk with each other.  This is almost
true, however they are linked together by the zero modes of the
bosonic oscillators which give rise to momenta $(P_L,P_R)$.
Previously we had $P_L=P_R$ but now this cannot be the case
because $P_L$ is 26 dimensional but $P_R$ is 10 dimensional.  It
is natural to decompose $P_L$ to a 10+16 dimensional vectors,
where we identify the 10 dimensional part of it with $P_R$. It
turns out that for the consistency of the theory the extra 16
dimensional component should belong to the root lattice of
$E_8\times E_8$ or a $Z_2$ sublattice of $SO(32)$ weight lattice.
In either of these two cases the vectors in the lattice with
$(length)^2=2$ are in one to one correspondence with non-zero
weights in the adjoint of $E_8\times E_8$ and $SO(32)$
respectively.  These can also be conveniently represented (through
bosonization) by 32 fermions:  In the case of $E_8\times E_8$ we
group them to two groups of 16 and consider independent NS, R
sectors for each group.
 In the case of $SO(32)$ we only
have one group of $32$ fermions with either NS or R boundary
conditions.

Let us tabulate the massless modes using (\ref{spec}). The
right-movers can be either NS or R. The left-moving degrees of
freedom start out with a tachyonic mode.  But (\ref{spec}) implies
that this is not satisfying the level-matching condition because
the right-moving ground state is at zero energy.  Thus we should
search on the left-moving side for states with $L_0=0$ which means
from (\ref{spec}) that we have either $N_L=1$ or $(1/2)P_L^2=1$,
where $P_L$ is an internal 16 dimensional vector in one of the two
lattices noted above. The states with $N_L=1$ are \bea\nn 16\oplus
v , \eea where 16 corresponds to the oscillation direction in the
extra 16 dimensions and $v$ corresponds to vector in 10
dimensional spacetime. States with $(1/2)P_L^2=1$ correspond to
the non-zero weights of the adjoint of $E_8\times E_8$ or $SO(32)$
which altogether correspond to $480$ states in both cases.  The
extra 16 $N_L=1$ modes combine with these $480$ states to form the
adjoints of $E_8\times E_8$ or $SO(32)$ respectively. The
right-movers give, as before, a $v\oplus s$ from the NS and R
sectors respectively.  So putting the left- and right-movers
together we finally get for the massless modes \bea\nn (v\oplus
Adj)\otimes (v\oplus s). \eea Thus the bosonic states are
$(v\oplus Adj)\otimes v$ which gives \bea\nn (g_{\mu \nu},B_{\mu
\nu},\phi ; A_{\mu}), \eea where the $A_\mu$ is in the adjoint of
$E_8\times E_8$ or $SO(32)$. Note that in the $SO(32)$ case this
is an {\it identical} spectrum to that of type I strings.

\subsubsection{Summary}
\hspace{1cm} To summarize, we have found 5 consistent strings in
10 dimensions: Type IIA with $N=2$ non-chiral supersymmetry, type
IIB with $N=2$ chiral supersymmetry, type I with N=1 supersymmetry
and gauge symmetry $SO(32)$ and heterotic strings with N=1
supersymmetry with $SO(32)$ or $E_8\times E_8$ gauge symmetry.
Note that as far as the massless modes are concerned we only have
four inequivalent theories, because heterotic $SO(32)$ theory and
Type I theory have the same light degrees of freedom. In
discussing compactifications it is sometimes natural to divide the
discussion between two cases depending on how many supersymmetries
we start with.  In this context we will refer to the type IIA and
B as $N=2$ {\it theories} and Type I and heterotic strings as
$N=1$ {\it theories}.

\subsection{\bf String Compactifications}
\hspace{1cm} So far we have only talked about superstrings
propagating in 10 dimensional Minkowski spacetime.  If we wish to
connect string theory to the observed four dimensional spacetime,
somehow we have to get rid of the extra 6 directions. One way to
do this is by assuming that the extra 6 dimensions are tiny and
thus unobservable in the present day experiments. In such
scenarios we have to understand strings propagating not on ten
dimensional Minkowski spacetime but on four dimensional Minkowski
spacetime times a compact 6 dimensional manifold $K$.  In order to
gain more insight it is convenient to consider compactifications
not just to 4 dimensions but to arbitrary dimensional spacetimes,
in which case the dimension of $K$ is variable.

The choice of $K$ and the string theory we choose to start in 10
dimensions will lead to a large number of
 theories in diverse dimensions,
which have different number of supersymmetries and different low
energy effective degrees of freedom. In order to get a handle on
such compactifications it is useful to first classify them
according to how much supersymmetry they preserve.  This is useful
because the higher the number of supersymmetry the less the
quantum corrections there are.

If we consider a general manifold $K$ we find that the
supersymmetry is completely broken.  This is the case we would
really like to understand, but it turns out that string
perturbation theory always breaks down in such a situation;  this
is intimately connected with the fact that typically cosmological
constant is generated by perturbation theory and this destabilizes
the Minkowski solution. For this reason we do not even have a
single example of such a class whose dynamics we understand.
Instead if we choose $K$ to be of a special type we can preserve a
number of supersymmetries.

For this to be the case, we need $K$ to admit some number of
covariantly constant spinors.  This is the case because the number
of supercharges which are `unbroken' by compactification is
related to how many covariantly constant spinors we have. To see
this note that if we wish to define a {\it constant} supersymmetry
transformation, since a space-time spinor, is also a spinor of
internal space, we need in addition a constant spinor in the
internal compact directions. The basic choices are manifolds with
trivial holonomy (flat tori are the only example), $SU(n)$
holonomy (Calabi-Yau n-folds), $Sp(n)$ holonomy (4n dimensional
manifolds),
 7-manifolds of $G_2$ holonomy
and 8-manifolds of $Spin(7)$ holonomy.

\subsubsection{Toroidal Compactifications}
\hspace{1cm} The space with maximal number of covariantly constant
spinors is the flat torus $T^d$.  This is also the easiest to
describe the string propagation in. The main modification to the
construction of the Hilbert space from flat non-compact space in
this case involves relaxing the condition $P_L=P_R$ because the
string can wrap around the internal space and so $X$ does not need
to come back to itself as we go around $\sigma$. In particular if
we consider compactification on a circle of radius $R$ we can have
\bea\nn (P_L,P_R)=({n\over 2R}+mR,{n\over 2R}-mR). \eea Here $n$
labels the center of mass momentum of the string along the circle
and $m$ labels how many times the string is winding around the
circle.  Note that the spectrum of allowed $(P_L,P_R)$ is
invariant under $R\rightarrow 1/2R$.  All that we have to do is to
exchange the momentum and winding modes ($n\leftrightarrow m$).
This symmetry is a consequence of what is known as $T$-duality.

If we compactify on a d-dimensional torus $T^d$ it can be shown
that $(P_L,P_R)$ belong to a 2d dimensional lattice with signature
$(d,d)$.  Moreover this lattice is integral, self-dual and even.
Evenness means, $P_L^2-P_R^2$ is even for each lattice vector.
Self-duality means that any vector which has integral product with
all the vectors in the lattice sits in the lattice as well.  It is
an easy exercise to check these condition in the one dimensional
circle example given above.  Note that we can change the radii of
the torus and this will clearly affect the $(P_L,P_R)$.  Given any
choice of a d-dimensional torus compactifications, all the other
ones can be obtained by doing an $SO(d,d)$ Lorentz boost on
$(P_L,P_R)$ vectors. Of course rotating $(P_L,P_R)$ by an
$O(d)\times O(d)$ transformation does not change the spectrum of
the string states, so the totality of such vectors is given by
\bea\nn SO(d,d)\over SO(d)\times SO(d). \eea Some Lorentz boosts
will not change the lattice and amount to relabeling the states.
These are the boosts that sit in $O(d,d;Z)$ (i.e. boosts with
integer coefficients), because they can be undone by choosing a
new basis for the lattice by taking an integral linear combination
of lattice vectors.  So the space of inequivalent choices are
actually given by \bea\nn SO(d,d)\over SO(d)\times SO(d)\times
O(d,d;Z). \eea The $O(d,d;Z)$ generalizes the T-duality considered
in the 1-dimensional case.

\subsubsection{Compactifications on $K3$}
\hspace{1cm} The four dimensional manifold $K3$ is the only
compact four dimensional manifold, besides $T^4$, which admits
covariantly constant spinors.  In fact it has exactly half the
number of covariantly constant spinors as on $T^4$ and thus
preserves half of the supersymmetry that would have been preserved
upon toroidal compactification. More precisely the holonomy of a
generic four manifold is $SO(4)$.  If the holonomy resides in an
$SU(2)$ subgroup of $SO(4)$ which leaves an $SU(2)$ part of
$SO(4)$ untouched, we end up with one chirality of $SO(4)$ spinor
being unaffected by the curvature of $K3$, which allows us to
define supersymmetry transformations as if $K3$ were flat (note a
spinor of $SO(4)$ decomposes as $({\bf 2}, {\bf 1})\oplus ({\bf
1},{\bf 2})$ of $SU(2)\times SU(2)$).

There are a number of realizations of $K3$, which are useful
depending on which question one is interested in. Perhaps the
simplest description of it is in terms of {\it orbifolds}. This
description of $K3$ is very close to toroidal compactification and
differs from it by certain discrete isometries of the $T^4$ which
are used to (generically) identify points which are in the same
{\it orbit} of the discrete group. Another description is as a 19
complex parameter family of $K3$ defined by an algebraic equation.

Consider a $T^4$ which for simplicity we take to be parametrized
by four real coordinates $x_i$ with $i=1,...,4$, subject to the
identifications $x_i\sim x_i+1$. It is sometimes convenient to
think of this as two complex coordinates $z_1=x_1+ix_2$ and
$z_2=x_3+ix_4$ with the obvious identifications.  Now we identify
the points on the torus which are mapped to each other under the
$Z_2$ action (involution) given by reflection in the coordinates
$x_i\rightarrow -x_i$, which is equivalent to \bea\nn
z_i\rightarrow -z_i . \eea Note that this action has $2^4=16$
fixed points given by the choice of midpoints or the origin in any
of the four $x_i$. The resulting space is \ul{singular} at any of
these 16 fixed points because the angular degree of freedom around
each of these points is cut by half. Put differently, if we
consider any primitive loop going `around' any of these 16 fixed
point, it corresponds to an open curve on $T^4$ which connects
pairs of points related by the $Z_2$ involution. Moreover the
parallel transport of vectors along this path, after using the
$Z_2$ identification, results in a flip of the sign of the vector.
This is true no matter how small the curve is.  This shows that we
cannot have a smooth manifold at the fixed points.

When we move away from the orbifold points of $K3$ the description
of the geometry of $K3$ in terms of the properties of the $T^4$
and the $Z_2$ twist become less relevant, and it is natural to ask
about other ways to think about $K3$.  In general a simple way to
define complex manifolds is by imposing complex equations in a
compact space known as the projective $n$-space ${\bf CP}^n$. This
is the space of complex variables $(z_1,...,z_{n+1})$ excluding
the origin and subject to the identification \bea\nn
(z_1,...,z_{n+1})\sim \lambda (z_1,...,z_{n+1}) \qquad \lambda
\not=0 . \eea One then considers the vanishing locus of a
homogeneous polynomial of degree $d$, $W_d(z_i)=0$ to obtain an
$n-1$ dimensional subspace of ${\bf CP}^n$. An interesting special
case is when the degree is $d=n+1$. In this case one obtains an
$n-1$ complex dimensional manifold which admits a Ricci-flat
metric.  This is the case known as Calabi-Yau. For example, if we
take the case $n=2$, by considering cubics in it \bea\nn
z_1^3+z_2^3+z_3^3+az_1z_2z_3=0 \eea we obtain an elliptic curve,
i.e. a torus of complex dimension 1 or real dimension 2.  The next
case would be $n=3$ in which case, if we consider a quartic
polynomial in ${\bf CP}^3$ we obtain the 2 complex dimensional
$K3$ manifold: \bea\nn W=z_1^4+z_2^4+z_3^4+z_4^4 +{\rm
deformations}=0 . \eea There are 19 inequivalent quartic terms we
can add.  This gives us a 19 dimensional complex subspace of 20
dimensional complex moduli of the $K3$ manifold.  Clearly this way
of representing $K3$ makes the complex structure description of it
very manifest, and makes the Kahler structure description
implicit.

Note that for a generic quartic polynomial the $K3$ we obtain is
non-singular.  This is in sharp contrast with the orbifold
construction which led us to 16 singular points.  It is possible
to choose parameters of deformation which lead to singular points
for $K3$. For example if we consider \bea\nn
z_1^4+z_2^4+z_3^4+z_4^4+4 z_1z_2z_3z_4=0 \eea it is easy to see
that the resulting $K3$ will have a singularity (one simply looks
for non-trivial solutions to $dW=0$).

There are other ways to construct Calabi-Yau manifolds and in
particular $K3$'s.  One natural generalization to the above
construction is to consider weighted projective spaces where the
$z_i$ are identified under different rescalings. In this case one
considers quasi-homogeneous polynomials to construct submanifolds.

\subsubsection{Calabi-Yau Threefolds}
\hspace{1cm} Calabi-Yau threefolds are manifolds with $SU(3)$
holonomy. The compactification on manifolds of $SU(3)$ holonomy
preserves 1/4 of the supersymmetry.  In particular if we
compactify $N=2$  theories on Calabi-Yau threefolds we obtain
$N=2$ theories in $d=4$, whereas if we consider $N=1$ theories we
obtain $N=1$ theories in $d=4$.

If we wish to construct the Calabi-Yau threefolds as toroidal
orbifolds we need to consider six dimensional tori, three complex
dimensional, which have discrete isometries residing in $SU(3)$
subgroup of the $O(6)=SU(4)$ holonomy group.  A simple example is
if we consider the product of three copies of $T^2$ corresponding
to the Hexagonal lattice and mod out by a simultaneous ${\bf Z}_3$
rotation on each torus (this is known as the `Z-orbifold'). This
$Z_3$ transformation has $27$ fixed points which can be blown up
to give rise to a smooth Calabi-Yau.

We can also consider description of Calabi-Yau threefolds in
algebraic geometry terms for which the complex deformations of the
manifold can be typically realized as changes of coefficients of
defining equations, as in the $K3$ case. For instance we can
consider the projective 4-space ${\bf CP}^4$ defined by 5 complex
not all vanishing coordinates $z_i$ up to overall rescaling, and
consider the vanishing locus of a homogeneous degree 5 polynomial
\bea\nn P_5(z_1,...,z_5)=0. \eea This defines a Calabi-Yau
threefold, known as the quintic three-fold. This can be
generalized to the case of product of several projective spaces
with more equations. Or it can be generalized by taking the
coordinates to have different homogeneity weights. This will give
a huge number of Calabi-Yau manifolds.

\subsection{\bf Solitons in String Theory}
\hspace{1cm} Solitons arise in field theories when the vacuum
configuration of the field has a non-trivial topology which allows
non-trivial wrapping of the field configuration at spatial
infinity around the vacuum manifold.  These will carry certain
topological charge related to the `winding' of the field
configuration around the vacuum configuration. Examples of
solitons include magnetic monopoles in four dimensional
non-abelian gauge theories with unbroken $U(1)$, cosmic strings
and domain walls. The solitons naively play a less fundamental
role than the fundamental fields which describe the quantum field
theory. In some sense we can think of the solitons as `composites'
of more fundamental elementary excitations. However as is well
known, at least in certain cases, this is just an illusion. In
certain cases it turns out that we can reverse the role of what is
fundamental and what is composite by considering a different
regime of parameter.  In such regimes the soliton may be viewed as
the elementary excitation and the previously viewed elementary
excitation can be viewed as a soliton. A well known example of
this phenomenon happens in 2 dimensional field theories. Most
notably the boson/fermion equivalence in the two dimensional
sine-Gordon model, where the fermions may be viewed as solitons of
the sine-Gordon model and the boson can be viewed as a composite
of fermion-anti-fermion excitation.  Another example is the
T-duality we have already discussed in the context of 2
dimensional world sheet of strings which exchanges the radius of
the target space with its inverse.  In this case the winding modes
may be viewed as the solitons of the more elementary excitations
corresponding to the momentum modes.  As discussed before
$R\rightarrow 1/R$ exchanges momentum and winding modes. In
anticipation of generalization of such dualities to string theory,
it is thus important to study various types of solitons that may
appear in string theory.

As already mentioned solitons typically carry some conserved
topological charge.  However in string theory every conserved
charge is a gauge symmetry.  In fact this is to be expected from a
theory which includes quantum gravity.  This is because the global
charges of a black hole will have no influence on the outside and
by the time the black hole disappears  due to Hawking radiation,
so does the global charges it may carry. So the process of
formation and evaporation of black hole leads to a
non-conservation of global charges.  Thus for any soliton, its
conserved topological charge must be a gauge charge.  This may
appear to be somewhat puzzling in view of the fact that solitons
may be point-like as well as string-like, sheet-like etc.  We can
understand how to put a charge on a point-like object and gauge
it.  But how about the higher dimensional extended solitonic
states? Note that if we view the higher dimensional solitons as
made of point-like structures the soliton has no stability
criterion as the charge can disintegrate into little bits.

Let us review how it works for point particles (or point
solitons): We have a 1-form gauge potential $A_{\mu}$ and the
coupling of the particle to the gauge potential involves weighing
the world-line propagating in the space-time with background
$A_{\mu}$ by \bea\nn Z\rightarrow Z {\rm exp}(i \int_\gamma A),
\eea where $\gamma$ is the world line of the particle.  The gauge
principle follows from defining an action in terms of $F=dA$:
\bea\label{acti} S=\int F\wedge *F ,\eea where $*F$ is the dual of
the $F$, where we note that shifting $A\rightarrow d\epsilon$ for
arbitrary function $\epsilon$ will not modify the action.

Suppose we now consider instead of a point particle a
$p$-dimensional extended object.  In this convention $p=0$
corresponds to the case of point particles and $p=1$ corresponds
to strings and $p=2$ corresponds to membranes, etc.  We shall
refer to $p$-dimensional extended objects as $p$-branes
(generalizing `membrane').  Note that the world-volume of a
$p$-brane is a $p+1$ dimensional subspace $\gamma_{p+1}$ of
space-time.  To generalize what we did for the case of point
particles we introduce a gauge potential which is a $p+1$ form
$A_{p+1}$ and couple it to the charged $p+1$ dimensional state by
\bea\nn Z\rightarrow Z {\rm exp}(i \int_{\gamma _{p+1}} A_{p+1}).
\eea Just as for the case of the point particles we introduce the
field strength $F=dA$ which is now a totally antisymmetric $p+2$
tensor. Moreover we define the action as in (\ref{acti}), which
possesses the gauge symmetry $A\rightarrow d \epsilon$ where
$\epsilon$ is a totally antisymmetric tensor of rank $p$.

\subsubsection{Magnetically Charged States}
\hspace{1cm} The above charge defines the generalization of
electrical charges for extended objects. Can we generalize the
notion of magnetic charge? Suppose we have an electrically charged
particle in a theory with space-time dimension $D$. Then we
measure the electrical charge by surrounding the point by an
$S^{D-2}$ sphere and integrating $*F$ (which is a $D-2$ form) on
it, i.e. \bea\nn Q_E=\int_{S^{D-2}} *F. \eea Similarly it is
natural to define the magnetic charge.  In the case of $D=4$, i.e.
four dimensional space-time, the magnetically charged point
particle can be surrounded also by a sphere and the magnetic
charge is simply given by \bea\nn Q_M=\int_{S^2} F. \eea Now let
us generalize the notion of magnetic charged states for arbitrary
dimensions $D$ of space-time and arbitrary electrically charged
$p$-branes.  From the above description it is clear that the role
that $*F$ plays in measuring the electric charge is played by $F$
in measuring the magnetic charge.  Note that for a $p$-brane $F$
is $p+2$ dimensional, and $*F$ is $D-p-2$ dimensional.  Moreover,
note that a sphere surrounding a $p$-brane is a sphere of
dimension $D-p-2$. Note also that for $p=0$ this is the usual
situation.  For higher $p$, a $p$-dimensional subspace of the
space-time is occupied by the extended object and so the position
of the object is denoted by a point in the transverse $(D-1)-p$
dimensional space which is surrounded by an $S^{D-p-2}$
dimensional sphere.

Now for the magnetic states the role of $F$ and $*F$ are
exchanged: \bea\nn F\leftrightarrow *F. \eea To be perfectly
democratic we can also define a magnetic gauge potential $\tilde
A$ with the property that \bea\nn d\tilde A=*F=*dA. \eea In
particular noting that $F$ is a $p+2$ form, we learn that $*F$ is
an $D-p-2$ form and thus $\tilde A$ is an $D-p-3$ form. We thus
deduce that the magnetic state will be an $D-p-4$-brane (i.e. one
dimension lower than the degree of the magnetic gauge potential
${\tilde A}$). Note that this means that if we have an
electrically charged $p$-brane, with a magnetically charged dual
$q$-brane then we have \bea\label{mage} p+q=D-4. \eea This is an
easy sum rule to remember. Note in particular that for a
4-dimensional space-time an electric point charge ($p=0$) will
have a dual magnetic point charge ($q=0$). Moreover this is the
only space-time dimension where both the electric and magnetic
dual can be point-like.

Note that a p-brane wrapped around an r-dimensional compact object
will appear as a $p-r$-brane for the non-compact space-time.  This
is in accord with the fact that if we decompose the $p+1$ gauge
potential into an $(p+1-r)+r$ form consisting of an $r$-form in
the compact direction we will end up with an $p+1-r$ form in the
non-compact directions.  Thus the resulting state is charged under
the left-over part of the gauge potential.  A particular case of
this is when $r=p$ in which case we are wrapping a p-dimensional
extended object about a p-dimensional closed cycle in the compact
directions. This will leave us with point particles in the
non-compact directions carrying ordinary electric charge under the
reduced gauge potential which now is a 1-form.

\subsubsection{String Solitons}
\hspace{1cm} From the above discussion it follows that the charged
states will in principle exist if there are suitable gauge
potentials given by $p+1$-forms.  Let us first consider type II
strings. Recall that from the NS-NS sector we obtained an
anti-symmetric 2-form $B_{\mu\nu}$.  This suggests that there is a
1-dimensional extended object which couples to it by \bea\nn {\rm
exp}(i\int B). \eea But that is precisely how $B$ couples to the
world-sheet of the fundamental string.  We thus conclude that {\it
the fundamental string carries electric charge under the
antisymmetric field} $B$. What about the magnetic dual to the
fundamental string? According to (\ref{mage}) and setting $d=10$
and $p=1$ we learn that the dual magnetic state will be a 5-brane.
Note that as in the field theories, we expect that in the
perturbative regime for the fundamental fields, the solitons be
very massive. This is indeed the case and the 5-brane magnetic
dual can be constructed as a solitonic state of type II strings
with a mass per unit 5-volume going as $1/g_{s}^2$ where $g_{s}$
is the string coupling. Conversely, in the strong coupling regime
these 5-branes are light and at infinite coupling they become
massless, i.e. \ul{tensionless} 5-branes \cite{DKL95}.

Let us also recall that type II strings also have anti-symmetric
fields coming from the R-R sector. In particular for type IIA
strings we have 1-form $A_{\mu}$ and 3-form $C_{\mu\nu \rho}$
gauge potentials. Note that the corresponding magnetic dual gauge
fields will be 7-forms and 5-forms respectively (which are not
independent degrees of freedom).  We can also include a 9-form
potential which will have trivial dynamics in 10  dimensions. Thus
it is natural to define a generalized gauge field ${\cal A}$ by
taking the sum over all odd forms and consider the equation ${\cal
F}=*{\cal F}$ where ${\cal F}=d{\cal A}$.  A similar statement
applies to the type IIB strings where from the R-R sector we
obtain all the even-degree gauge potentials (the case with degree
zero can couple to a ``$-1$-brane'' which can be identified with
an instanton, i.e. a point in space-time). We are thus led to look
for p-branes with even $p$ for type IIA and odd $p$ for type IIB
which carry charge under the corresponding RR gauge field.  It
turns out that surprisingly enough the states in the elementary
excitations of string all are neutral under the RR fields.  We are
thus led to look for solitonic states which carry RR charge.
Indeed there are such p-branes and they are known as $D$-branes ,
as we will now review.

\subsubsection{D-Branes}
\hspace{1cm} In the context of field theories constructing
solitons is equivalent to solving classical field equations with
appropriate boundary conditions. For string theory the condition
that we have a classical solution is equivalent to the statement
that propagation of strings in the corresponding background would
still lead to a conformal theory on the worldsheet of strings, as
is the case for free theories.

In search of such stringy p-branes, we are thus led to consider
how could a p-brane modify the string propagation.  Consider an
$p+1$ dimensional plane, to be identified with the world-volume of
the $p$-brane. Consider string propagating in this background.
How could we modify the rules of closed string propagation given
this $p+1$ dimensional sheet?  The simplest way turns out to allow
closed strings to open up and end on the $p+1$ dimensional
world-volume.  In other words we allow to have a new sector in the
theory corresponding to open string with ends lying on this $p+1$
dimensional subspace. This will put Dirichlet boundary conditions
on $10-p-1$ coordinates of string endpoints.  Such $p$-branes are
called $D$-branes, with D reminding us of Dirichlet boundary
conditions. In the context of type IIA,B we also have to specify
what boundary conditions are satisfied by fermions. This turns out
to lead to consistent boundary conditions only for $p$ even for
type IIA string and $p$ odd for type IIB. This is a consequence of
the fact that for type IIA(B), left-right exchange is a symmetry
only when accompanied by a $Z_2$ spatial reflection with
determinant -1(+1). Moreover, it turns out that they do carry the
corresponding RR charge \cite{P95}.

Quantizing the new sector of type II strings in the presence of
D-branes is rather straightforward. We simply consider the set of
oscillators as before, but now remember that due to the Dirichlet
boundary conditions on some of the components of string
coordinates, the momentum of the open string lies on the $p+1$
dimensional world-volume of the D-brane. It is thus
straightforward to deduce that the massless excitations
propagating on the D-brane will lead to the dimensional reduction
of $N=1$, $U(1)$ Yang-Mills from $d=10$ to $p+1$ dimensions. In
particular the $10-(p+1)$ scalar fields living on the D-brane,
signify the D-brane excitations in the $10-(p+1)$ transverse
dimensions. This tells us that the significance of the new open
string subsector is to quantize the D-brane excitations.

An important property of D-branes is that when $N$ of them
coincide we get a $U(N)$ gauge theory on their world-volume.  This
follows because we have $N^2$ open string subsectors going from
one D-brane to another and in the limit they are on top of each
other all will have massless modes and we thus obtain the
reduction of $N=1$ $U(N)$ Yang-Mills from $d=10$ to $d=p+1$.

Another important property of D-branes is that they are BPS
states. A BPS state is a state which preserves a certain number of
supersymmetries and as a consequence of which one can show that
their mass (per unit volume) and charge are equal. This in
particular guarantees their absolute stability against decay.

If we consider the tension of D-branes, it is proportional to
$1/g_{s}$, where $g_{s}$ is the string coupling constant.  Note
that as expected at weak coupling they have a huge tension. At
strong coupling their tension goes to zero and they become
\ul{tensionless}.

We have already discussed that in $K3$ compactification of string
theory we end up with singular limits of manifolds when some
cycles shrink to zero size. What is the physical interpretation of
this singularity?

Suppose we consider for concreteness an $n$-dimensional sphere
$S^n$ with volume $\epsilon\rightarrow 0$. Then the string
perturbation theory breaks down when $\epsilon << g_{s}$, where
$g_{s}$ is the string coupling constant.  If we have  $n$-brane
solitonic states such as D-branes then we can consider a
particular solitonic state corresponding to wrapping the $n$-brane
on the vanishing $S^n$.  The mass of this state is proportional to
$\epsilon$, which implies that in the limit $\epsilon \rightarrow
0$ we obtain a massless soliton. An example of this is when we
consider type IIA compactification on $K3$ where we develop a
singularity. Then by wrapping D2-branes around vanishing $S^2$'s
of the singularity we obtain massless states, which are vectors.
This in fact implies that in this limit we obtain enhanced gauge
symmetry. Had we been considering type IIB on $K3$ near the
singularity, the lightest mode would be obtained by wrapping a
D3-brane around vanishing $S^2$'s, which leaves us with a string
state with tension of the order of $\epsilon$. This kind of regime
which exists in other examples of compactifications as well is
called the phase with \ul{tensionless} strings.

\vspace*{2.cm}

This thesis is based on \cite{1}-\cite{30}.

Our notations for the special functions we use are summarized in
Appendix A.

\setcounter{equation}{0}
\section{$\mathbf{P}$-branes dynamics in general backgrounds}
The {\it probe} branes approach for studying issues in the
string/M-theory uses an approximation, in which one neglects the
back-reaction of the branes on the background. In this sense, the
probe branes are multidimensional dynamical systems, evolving in
given, variable in general, external fields.

The probe branes method is widely used in the string/M-theory to
investigate many different problems at a classical, semiclassical
and quantum levels. The literature in this field of research can
be {\it conditionally} divided into several parts. One of them is
devoted to the properties of the probe branes themselves, e.g.,
\cite{CM98}. The subject of another part of the papers is to probe
the geometries of the string/M-theory backgrounds, e.g.,
\cite{DKL94}. One another part can be described as connected with
the investigation of the correspondence between the
string/M-theory geometries and their field theory duals, e.g.,
\cite{AdS/CFT}, \cite{K99_1}. Let us also mention the application
of the probe branes technique in the 'Mirage cosmology'- an
approach to the brane world scenario, e.g., \cite{KK99}.

In view of the wide implementation of the probe branes as a tool
for investigation of different problems in the string/M-theory, it
will be useful to have a method describing their dynamics, which
is general enough to include as many cases of interest as
possible, and on the other hand, to give the possibility for
obtaining {\it explicit exact} solutions.

Here, we propose such an approach, which is appropriate for
$p$-branes and D$p$-branes, for arbitrary worldvolume and
space-time dimensions, for tensile and tensionless branes, for
different variable background fields with minimal restrictions on
them, and finally, for different space-time and worldvolume gauges
(embeddings).

Now, we are going to consider probe $p$-branes and D$p$-branes
dynamics in $D$-dimensional string/M-theory backgrounds of general
type. Unified description for the tensile and tensionless branes
is used. We obtain exact solutions of their equations of motion
and constraints in static gauge as well as in more general gauges.
Their dynamics in the whole space-time is also analyzed and exact
solutions are found \cite{4}.

Before considering the problem for obtaining exact brane solutions in
general string theory backgrounds, it will be useful first to choose
appropriate actions, which  will facilitate our task. Generally speaking,
there are two types of brane actions - with and without square roots
\footnote{Examples of these two type of actions are the Nambu-Goto and
Polyakov actions for the string.}. The former ones are not well suited to
our purposes, because the square root introduces additional nonlinearities
in the equations of motion. Nevertheless, they have been used when
searching for exact brane solutions in fixed backgrounds, because there
are no constraints in the Lagrangian description and one has to solve only
the equations of motion. The other type of actions contain additional
worldvolume fields (Lagrange multipliers). Varying with respect to them,
one obtains constraints, which, in general, are not independent. Starting
with an action without square root, one escapes the nonlinearities
connected with the square root, but has to solve the equations of motion
and the (dependent) constraints.

Independently of their type, all actions proportional to the brane tension
cannot describe the tensionless branes. The latter appear in many important
cases in the string theory, and it is preferable to have a unified
description for tensile and tensionless branes.

Our aim now is to find brane actions, which do not
contain square roots, generate only independent constraints and give a unified
description for tensile and tensionless branes.

\subsection{$P$-brane actions}

The Polyakov type action for the bosonic $p$-brane in a $D$-dimensional
curved space-time with metric tensor $g_{MN}(x)$, interacting with a
background (p+1)-form gauge field $b_{p+1}$ via Wess-Zumino term, can be
written as \bea\label{pa} S_p^{P}=&-&\int
d^{p+1}\xi\Bigl\{\frac{T_p}{2}\sqrt{-\gamma}\left[\gamma^{mn} \p_m X^M\p_n
X^N g_{MN}(X)-(p-1)\right]\\ \nn &-&Q_p\frac{ \varepsilon^{m_1\ldots
m_{p+1}}}{(p+1)!} \p_{m_1}X^{M_1}\ldots\p_{m_{p+1}}X^{M_{p+1}}
b_{M_1\ldots M_{p+1}}(X)\Bigr\},\\ \nn && \p_m=\p/\p\xi^m,\h m,n =
0,1,\ldots,p;\h M,N = 0,1,\ldots,D-1,\eea where $\gamma$ is the determinant
of the auxiliary worldvolume metric $\gamma_{mn}$, and  $\gamma^{mn}$ is
its inverse. The position of the brane in the background space-time is
given by $x^M=X^M(\xi^m)$, and $T_p$, $Q_p$ are the $p$-brane tension and
charge, respectively. If we consider the action (\ref{pa}) as a bosonic
part of a supersymmetric one, we have to set $Q_p=\pm T_p$. In what
follows, $Q_p = T_p$.

The requirement that the variation of the action (\ref{pa}) with respect
to $\gamma_{mn}$ vanishes, leads to \bea\label{dc}
\left(\gamma^{kl}\gamma^{mn}-2\gamma^{km}\gamma^{ln}\right)G_{mn}
=(p-1)\gamma^{kl},\eea where $G_{mn}=\p_m X^M\p_n X^N g_{MN}(X)$ is the
metric induced on the $p$-brane worldvolume. Taking the trace of the above
equality, one obtains \bea\nn \gamma^{mn}G_{mn}=p+1,\eea i.e., $\gamma^{mn}$
is the inverse of $G_{mn}$: $\gamma^{mn}=G^{mn}$. If one inserts this back
into (\ref{pa}), the result will be the corresponding Nambu-Goto type
action $(G\equiv\det(G_{mn}))$: \bea\label{nga} S_p^{NG}&=& \int
d^{p+1}\xi\mathcal{L}^{NG}\\ \nn &=& - T_p\int d^{p+1}\xi\left[
\sqrt{-G}-\frac{ \varepsilon^{m_1\ldots m_{p+1}}}{(p+1)!}
\p_{m_1}X^{M_1}\ldots\p_{m_{p+1}}X^{M_{p+1}} b_{M_1\ldots
M_{p+1}}(X)\right].\eea This means that the two actions, (\ref{pa}) and
(\ref{nga}), are classically equivalent.

As already discussed, the action (\ref{nga}) contains a square root,
the constraints (\ref{dc}), following from (\ref{pa}), are not independent
and none of these actions is appropriate for description of the {\it
tensionless} branes. To find an action of the type we are looking for,
we first compute the
explicit expressions for the generalized momenta, following from
(\ref{nga}): \bea\nn P_M(\xi)=-T_p\left(\sqrt{-G}G^{0n}\p_nX^Ng_{MN}
-\p_1X^{M_1}\ldots\p_pX^{M_p}b_{MM_1\ldots M_p}\right).\eea It can be
checked that $P_M(\xi)$ satisfy the constraints \bea\nn
&&\mathcal{C}_0\equiv g^{MN}P_MP_N -2T_pg^{MN}D_{M1\ldots p}P_N
+T^2_p\left[GG^{00}+(-1)^pD_{1\ldots p M}g^{MN}D_{N1\ldots p}\right]=0, \\
\nn &&\mathcal{C}_i\equiv P_M\p_iX^M=0,\h (i=1,\ldots,p),\eea where we have
introduced the notation \bea\nn D_{M1\ldots p}\equiv b_{MM_1\ldots
M_p}\p_1X^{M_1}\ldots\p_pX^{M_p}.\eea

Let us now find the canonical Hamiltonian for this dynamical system. The
result is: \bea\nn H_{canon} = \int d^p\xi\left(P_M\p_0X^M
-\mathcal{L}^{NG}\right)= 0.\eea Therefore, according to Dirac \cite{D64},
we have to take as a Hamiltonian the linear combination of the first class
primary constraints $\mathcal{C}_n$: \footnote{In the case under
consideration, secondary constraints do not appear. The first class
property of $\mathcal{C}_n$ follows from their Poisson bracket algebra.}
\bea\nn H = \int d^p\xi\mathcal{H} = \int d^p\xi
\left(\lambda^0\mathcal{C}_0 + \lambda^i\mathcal{C}_i\right).\eea The
corresponding Hamiltonian equations of motion for $X^M$ are \bea\nn
\left(\p_0-\lambda^i\p_i\right)X^M =
2\lambda^0g^{MN}\left(P_N-T_pD_{N1\ldots p}\right),\eea from where one
obtains the explicit expressions for $P_M$ \bea\label{nm} P_M
=\frac{1}{2\lambda^0}g_{MN}\left(\p_0-\lambda^j\p_j\right)X^N
+T_pD_{M1\ldots p}.\eea With the help of (\ref{nm}), one arrives at the
following configuration space action \bea\label{oa} S_p&=&\int
d^{p+1}\xi\mathcal{L}_p=\int d^{p+1}\xi\left(P_M\p_0X^M
-\mathcal{H}\right)\\ \nn &=& \int
d^{p+1}\xi\Bigl\{\frac{1}{4\lambda^0}\Bigl[g_{MN}\left(X\right)
\left(\p_0-\lambda^i\p_i\right) X^M\left(\p_0-\lambda^j\p_j\right)X^N
-\left(2\lambda^0T_p\right)^2GG^{00}\Bigr]\\ \nn &+&T_p b_{M_0\ldots
M_p}(X)\p_0X^{M_0}\ldots\p_pX^{M_p} \Bigr\}\\ \nn &=&\int
d^{p+1}\xi\Bigl\{\frac{1}{4\lambda^0}\Bigl[
G_{00}-2\lambda^{j}G_{0j}+\lambda^{i}\lambda^{j}G_{ij}
-\left(2\lambda^0T_p\right)^2GG^{00}\Bigr]\\ \nn &+&T_p b_{M_0\ldots
M_p}(X)\p_0X^{M_0}\ldots\p_pX^{M_p} \Bigr\},\eea which does not contain
square root, generates the independent $(p+1)$ constraints, as we will
show below, and in which the limit $T_p\to 0$ may be taken.

It can be proven that this action is classically equivalent to the previous
two actions. It is enough to show that (\ref{nga}) and (\ref{oa}) are
equivalent, because we already saw that this is true for (\ref{pa}) and
(\ref{nga}).

Varying the action $S_p$ with respect to Lagrange multipliers $\lambda^m$
and requiring these variations to vanish, one obtains the constraints
\bea\label{pbic0} &&G_{00}-2\lambda^{j}G_{0j}+\lambda^{i}\lambda^{j}G_{ij}
+\left(2\lambda^0T_p\right)^2GG^{00}=0,\\
\label{pbicj} &&G_{0j}-\lambda^{i}G_{ij}=0.\eea
By using them, the Lagrangian density $\mathcal{L}_p$ from (\ref{oa}) can be
rewritten in the form \bea\label{il} \mathcal{L}_p=-T_p
\sqrt{-GG^{00}\left[G_{00}-G_{0i}\left(G^{-1}\right)^{ij}G_{j0}\right]}
+T_p b_{M_0\ldots M_p}(X)\p_0X^{M_0}\ldots\p_pX^{M_p}.\eea Now, applying
the equalities \bea\label{sg} GG^{00}=\det\left(G_{ij}\right)\equiv\mathbf{G},
\h G=\left[G_{00}-G_{0i}\left(G^{-1}\right)^{ij}G_{j0}\right]\mathbf{G},\eea
one finds that \bea\nn
G^{00}\left[G_{00}-G_{0i}\left(G^{-1}\right)^{ij}G_{j0}\right]=1.\eea
Inserting this in (\ref{il}), one obtains the Nambu-Goto type Lagrangian
density $\mathcal{L}^{NG}$ from (\ref{nga}). Thus, the classical
equivalence of the actions (\ref{nga}) and (\ref{oa}) is established.

We will work further in the gauge $\lambda^m=constants$, in which the
equations of motion for $X^M$, following from (\ref{oa}), are given by
\bea\label{pbem} &&g_{LN}\left[\Di\Dj X^N - \left(2\lambda^0T_p\right)^2
\p_i\left(\mathbf{G}G^{ij}\p_j X^N\right)\right]\\ \nn
&&+\Gamma_{L,MN}\left[\Di X^M \Dj X^N - \left(2\lambda^0T_p\right)^2
\mathbf{G}G^{ij}\p_i X^M \p_j X^N\right]\\ \nn
&&=2\la^0 T_p H^{\mathbf{b}}_{LM_0\ldots M_{p}}\p_0 X^{M_0}\ldots
\p_p X^{M_p},\eea
where $\mathbf{G}$ is defined in (\ref{sg}),
\bea\nn \Gamma_{L,MN}=g_{LK}\Gamma^K_{MN}=\frac{1}{2}\left(\p_Mg_{NL}
+\p_Ng_{ML}-\p_Lg_{MN}\right)\eea
are the components of the symmetric connection compatible with the metric
$g_{MN}$ and $H^{\mathbf{b}}_{p+2}=db_{p+1}$ is the field strength of the
$(p+1)$-form gauge potential $b_{p+1}$.

\subsection{D$p$-brane actions}

The Dirac-Born-Infeld type action for the bosonic part of the super-
D$p$-brane in a $D$-dimensional space-time with metric tensor $g_{MN}(x)$,
interacting with a background (p+1)-form Ramond-Ramond gauge field
$c_{p+1}$ via Wess-Zumino term, can be written as \bea\label{dbia}
S^{DBI}&=&-T_{D_p}\int d^{p+1}\xi
\Bigl\{e^{-a(p,D)\Phi}\sqrt{-\det\left(G_{mn} + B_{mn} +
2\pi\alpha'F_{mn}\right)}\\ \nn &-&\frac{ \varepsilon^{m_1\ldots
m_{p+1}}}{(p+1)!} \p_{m_1}X^{M_1}\ldots\p_{m_{p+1}}X^{M_{p+1}}
c_{M_1\ldots M_{p+1}}\Bigr\}.\eea $T_{D_P}$=$(2\pi)^{-(p-1)/2}g_s^{-1}T_p$
is the D-brane tension, $g_s$ = $\exp\langle\Phi\rangle$ is the string
coupling expressed by the dilaton vacuum expectation value
$\langle\Phi\rangle$ and $2\pi\alpha'$ is the inverse string tension.
$G_{mn}= \p_m X^M\p_n X^N g_{MN}(X)$, $B_{mn}= \p_m X^M\p_n X^N b_{MN}(X)$
and $\Phi(X)$ are the pullbacks of the background metric, antisymmetric
tensor and dilaton to the D$p$-brane worldvolume, while $F_{mn}(\xi)$ is
the field strength of the worldvolume $U(1)$ gauge field $A_m(\xi)$:
$F_{mn}=2\p_{[m}A_{n]}$. The parameter $a(p,D)$ depends on the brane and
space-time dimensions $p$ and $D$, respectively.

A D$p$-brane action, which generalizes the Polyakov type $p$-brane action,
has been introduced in \cite{AZH97}. Namely, the
action, classically equivalent to (\ref{dbia}), is given by \bea\nn S^{AZH}
&=& -\frac{T_{D_p}}{2}\int d^{p+1} \xi\Bigl\{ e^{-a(p,D)\Phi}
\sqrt{-\mathcal{K}} \left[\mathcal{K}^{mn} \left( G_{mn}+B_{mn}+
2\pi\alpha'F_{mn}\right)-(p-1)\right]\\ \nn &-&2\frac{
\varepsilon^{m_1\ldots m_{p+1}}}{(p+1)!}
\p_{m_1}X^{M_1}\ldots\p_{m_{p+1}}X^{M_{p+1}} c_{M_1\ldots
M_{p+1}}\Bigr\},\eea where $\mathcal{K}$ is the determinant of the matrix
$\mathcal{K}_{mn}$, $\mathcal{K}^{mn}$ is its inverse, and these matrices
have symmetric as well as antisymmetric part \bea\nn
\mathcal{K}^{mn}=\mathcal{K}^{(mn)}+\mathcal{K}^{[mn]},\eea where the symmetric
part $\mathcal{K}^{(mn)}$ is the analogue of the auxiliary metric
$\gamma^{mn}$ in the $p$-brane action (\ref{pa}).

Again, none of these actions satisfy all our requirements. In the same way
as in the $p$-brane case, just considered, one can prove that the action
\bea\label{oda} S_{Dp}&=&\int d^{p+1}\xi\mathcal{L}_{Dp}=\int d^{p+1}\xi
\frac{e^{-a\Phi}}{4\lambda^0}\Bigl[G_{00}-2\lambda^i G_{0i} +
\left(\lambda^i\lambda^j-\kappa^i\kappa^j\right)G_{ij}\\ \nn
&-&\left(2\lambda^0T_{D_p}\right)^2\mathbf{G} +2\kappa^i\left(
\mathcal{F}_{0i}-\lambda^j\mathcal{F}_{ji}\right)
+4\lambda^0T_{D_p}e^{a\Phi} c_{M_0\ldots M_p}\p_0
X^{M_0}\ldots\p_pX^{M_p}\Bigr],\\ \nn &&\mathcal{F}_{mn}=B_{mn} +
2\pi\alpha' F_{mn},\eea which possesses the necessary properties, is
classically equivalent to the action (\ref{dbia}). Here additional
Lagrange multipliers $\kappa^i$ are introduced, in order to linearize the
quadratic term \bea\nn
\left(\mathcal{F}_{0i}-\lambda^k\mathcal{F}_{ki}\right)
\left(G^{-1}\right)^{ij}
\left(\mathcal{F}_{0j}-\lambda^l\mathcal{F}_{lj}\right)\eea arising in the
action. For other actions of this type, see \cite{LU97} - \cite{GL98}.

Varying the action $S_{Dp}$ with respect to Lagrange multipliers $\lambda^m$,
$\kappa^i$, and requiring these variations to vanish,
one obtains the constraints
\bea\label{Dpbic0} &&G_{00}-2\lambda^{j}G_{0j}+
\left(\lambda^{i}\lambda^{j}-\kappa^i\kappa^j\right)G_{ij}
+\left(2\lambda^0T_{Dp}\right)^2\mathbf{G}
+2\kappa^i\left(\mathcal{F}_{0i}-\lambda^j\mathcal{F}_{ji}\right)=0,\\
\label{Dpbicj}
&&G_{0j}-\lambda^{i}G_{ij}=\kappa^i\mathcal{F}_{ij}\\
\label{Dpbick}
&&\mathcal{F}_{0j}-\lambda^i\mathcal{F}_{ij}=\kappa^i G_{ij}.\eea
Instead with the constraint (\ref{Dpbic0}), we will work with the simpler one
\bea\label{Dpbic0'} G_{00}-2\lambda^{j}G_{0j}+
\left(\lambda^{i}\lambda^{j}+\kappa^i\kappa^j\right)G_{ij}
+\left(2\lambda^0T_{Dp}\right)^2\mathbf{G}=0,\eea
which is obtained by inserting (\ref{Dpbick}) into (\ref{Dpbic0}).

We will use the gauge $(\lambda^m, \kappa^i)=constants$ and for simplicity,
we will restrict our considerations to constant dilaton
$\Phi=\Phi_0$ and constant electro-magnetic field $F_{mn}=F^o_{mn}$ on the
D$p$-brane worldvolume. In this case, the equations of motion for $X^M$,
following from (\ref{oda}), are
\bea\nn &&g_{LN}\Bigl[\Di\Dj X^N - \left(2\lambda^0T_{Dp}\right)^2
\p_i\left(\mathbf{G}G^{ij}\p_j X^N\right)
-\kappa^i\kappa^j\p_i\p_j X^N\Bigr]\\
\nn &&+\Gamma_{L,MN}\Bigl[\Di X^M \Dj X^N
\\ \label{Dpbem} &&- \left(2\lambda^0T_{Dp}\right)^2
\mathbf{G}G^{ij}\p_i X^M \p_j X^N-\kappa^i\kappa^j\p_i X^M\p_j X^N\Bigr]\\
\nn &&=2\la^0 T_{Dp}e^{a\Phi_0} H^{\mathbf{c}}_{LM_0\ldots
M_{p}}\p_0 X^{M_0}\ldots \p_p X^{M_p}
+H_{LMN}\kappa^j\Di X^M\p_j X^N,\eea
where $ H^{\mathbf{c}}_{p+2}=dc_{p+1}$ and $H_3=db_2$ are the corresponding
field strengths.

\subsection{Exact solutions in general backgrounds}

The main idea in the mostly used approach for obtaining exact solutions of
the probe branes equations of motion in variable external fields is to
reduce the problem to a particle-like one, and even more - to solving one
dimensional dynamical problem, if possible. To achieve this, one must get
rid of the dependence on the spatial worldvolume coordinates $\xi^i$. To
this end, since the brane actions contain the first derivatives $\p_i
X^M$, the brane coordinates $X^M(\xi^m)$ have to depend on $\xi^i$ at most
linearly: \bea\label{mga} X^M(\xi^0,\xi^i)=\Lambda^M_i \xi^i +
Y^M(\xi^0),\h \Lambda^M_i  \mbox{ are arbitrary constants}.\eea Besides,
the background fields entering the action depend implicitly on $\xi^i$
through their dependence on $X^M$. If we choose $\Lambda^M_i =0$ in
(\ref{mga}), the connection with the $p$-brane setting will be lost. If we
suppose that the background fields do not depend on $X^M$, the result will
be constant background, which is not interesting in the case under
consideration. The {\it compromise} is to accept that the external fields
depend only on part of the coordinates, say $X^a$, and to set namely for
this coordinates $\Lambda^a_i =0$. In other words, we propose the ansatz
($X^M=(X^\mu,X^a)$): \bea\label{ogac} &&X^\mu(\xi^0,\xi^i)=\Lambda^\mu_j
\xi^j + Y^\mu(\xi^0),\h X^a(\xi^0,\xi^i)=Y^a(\xi^0),
\\ \label{ogab} &&\p_\mu g_{MN}=0,\h \p_\mu b_{MN}=0,\h
\p_\mu b_{M_0\ldots M_p}=0,\h \p_\mu c_{M_0\ldots M_p}=0.\eea
The resulting reduced Lagrangian density will
depend only on $\xi^0=\tau$ if the Lagrange multipliers $\lambda^m$,
$\kappa^i$ do not depend on $\xi^i$. Actually, this property follows from
their equations of motion, from where they can be expressed through
quantities depending only on the temporal worldvolume parameter $\tau$.

Thus, we have obtained the general conditions, under which the probe
branes dynamics reduces to the particle-like one. However, we will not
start our considerations relaying on the generic ansatz (\ref{ogac}).
Instead, we will begin in the framework of the commonly used in ten
space-time dimensions {\it static gauge}: $X^m(\xi^n)=\xi^m$. The latter
is a particular case of (\ref{ogac}), obtained under the following
restrictions: \bea\label{sgac} &&\mbox{(1):} \mu = i = 1,\ldots,p;\h
\mbox{(2):} \Lambda^\mu_j=\Lambda^i_j=\delta^i_j;\\ \nn &&\mbox{(3):}
Y^\mu(\tau) = Y^i(\tau) = 0;\h \mbox{(4):} Y^0(\tau)=\tau\in \{Y^a\}.\eea
Therefore, the static gauge is appropriate for backgrounds which may
depend on $X^0=Y^0(\tau)$, but must be independent on $X^i$, $(i=1,\ldots
p)$. Such properties are not satisfactory in the lower dimensions. For
instance, in four dimensional black hole backgrounds, the metric depends
on $X^1$, $X^2$ and the static gauge ansatz does not work. That is why,
our next step is to consider the probe branes dynamics in the framework of
the ansatz \bea\label{imac} X^\mu(\tau,\xi^i)=\Lambda^\mu_m \xi^m=
\Lambda^\mu_0\tau+\Lambda^\mu_i \xi^i,\h X^a(\tau,\xi^i)=Y^a(\tau),\eea
which is obtained from (\ref{ogac}) under the restriction $Y^\mu(\tau) =
\Lambda^\mu_0\tau$. Here, for the sake of symmetry between the worldvolume
coordinates $\xi^0=\tau$ and $\xi^i$, we have included in $X^\mu$ a term
linear in $\tau$. At any time, one can put $\Lambda^\mu_0=0$ and the
corresponding terms in the formulas will disappear. Further, we will refer
to the ansatz (\ref{imac}) as {\it linear gauges}, as far as $X^\mu$ are
linear combinations of $\xi^m$ with {\it arbitrary} constant coefficients.

Finally, we will investigate the classical branes dynamics by using the general
ansatz (\ref{ogac}), rewritten in the form
\bea\label{wgac} X^\mu(\tau,\xi^i)=\Lambda^\mu_0\tau+\Lambda^\mu_i \xi^i +
Y^\mu(\tau),\h X^a(\tau,\xi^i)=Y^a(\tau).\eea
Compared with (\ref{ogac}), here we have separated the linear part of $Y^\mu$
as in the previous ansatz (\ref{imac}). This will allow us to compare the role
of the term $\Lambda^\mu_0\tau$ in these two cases.

\subsubsection{Static gauge dynamics}
Here we begin our analysis of the probe branes dynamics in the framework
of the {\it static gauge} ansatz. In order not to introduce too many type of
indices, we will denote with $Y^a$, $Y^b$, etc., the coordinates, which are
{\it not fixed by the gauge}. However, one have not to forget that {\it by
definition}, $Y^a$ are the coordinates on which the background fields can
depend. In {\it static gauge}, according to (\ref{sgac}), one of this
coordinates, the temporal one $Y^0(\tau)$, is {\it fixed} to coincide with
$\tau$. Therefore, in this gauge, the remaining coordinates $Y^a$ are
{\it spatial} ones in space-times with signature $(-,+,\ldots,+)$.

Let us start with the $p$-branes case.

In static gauge, and under the conditions (\ref{ogab}), the
action (\ref{oa}) reduces to (the over-dot is used for $d/d\tau$)
\bea\label{osga} &&S_p^{SG}=\int d\tau L_p^{SG}(\tau),\h V_p = \int d^p\xi,\\
\nn &&L_p^{SG}(\tau)= \frac{V_p}{4\lambda^0}\Bigl\{g_{ab}(Y^a)\dot{Y^a}\dot{Y^b} +
2\left[g_{0a}(Y^a) -\lambda^i g_{ia}(Y^a)+ 2\lambda^0T_p b_{a1\ldots p}(Y^a)
\right]\dot{Y^a}\\ \nn &&+ g_{00}(Y^a)-2\lambda^i g_{0i}(Y^a)
+ \lambda^i\lambda^j g_{ij}(Y^a)- \left(2\lambda^0 T_p\right)^2
\det(g_{ij}(Y^a))+ 4\lambda^0 T_p b_{01\ldots p}(Y^a)\Bigr\}.\eea
To have finite action, we require the fraction
$V_p/\lambda^0$ to be finite one. For example, in the string case ($p=1$)
and in conformal gauge ($\lambda^1=0, \left(2\lambda^0 T_1\right)^2=1$),
this means that the quantity $V_1/\alpha'= 2\pi V_1 T_1$ must be finite.

The constraints derived from the action (\ref{osga}) are: \bea\label{c0s}
&&g_{ab}\dot{Y^a}\dot{Y^b} + 2\left(g_{0a}-\lambda^i
g_{ia}\right)\dot{Y^a}+ g_{00}-2\lambda^i g_{0i} + \lambda^i\lambda^j
g_{ij}+\left(2\lambda^0T_p\right)^2 \det(g_{ij})=0,\\ \label{cis}
&&g_{ia}\dot{Y^a}+g_{i0}-g_{ij}\lambda^j=0.\eea

The Lagrangian $L_p^{SG}$ does not depend on $\tau$ explicitly, so the
energy $E_p=p^{SG}_a \dot{Y}^a-L_p^{SG}$ is conserved:
\bea\nn &&g_{ab}\dot{Y^a}\dot{Y^b}-
g_{00}+2\lambda^i g_{0i} - \lambda^i\lambda^j g_{ij}+ \left(2\lambda^0
T_p\right)^2 \det(g_{ij})- 4\lambda^0 T_p b_{01\ldots p}= \frac{4\lambda^0
E_p}{V_p}=constant.\eea
With the help of the constraints, we can replace this
equality by the following one \bea\label{p0} g_{0a}\dot{Y^a} +
g_{00}-\lambda^ig_{i0}+2\lambda^0 T_p b_{01\ldots p}=- \frac{2\lambda^0
E_p}{V_p}.\eea

To clarify the physical meaning of the equalities (\ref{cis}) and (\ref{p0}),
we compute the momenta (\ref{nm}) in static gauge
\bea\label{gms} 2\la^0 P^{SG}_M =g_{Ma}\dot{Y}^a+g_{M0}-\la^i g_{Mi}
+2\la^0T_p b_{M1\ldots p}.\eea
The comparison of (\ref{gms}) with (\ref{p0}) and (\ref{cis}) shows that
$P^{SG}_0=-E_p/V_p =const$ and $P^{SG}_i =const=0$. Inserting these conserved
momenta into (\ref{c0s}), we obtain the {\it effective} constraint
\bea\label{ecs} g_{ab}\dot{Y^a}\dot{Y^b}=\mathcal{U}^{S},\eea
where
\bea\nn \mathcal{U}^{S}=-\left(2\lambda^0 T_p\right)^2
\det(g_{ij})+ g_{00}-2\lambda^i g_{0i} + \lambda^i\lambda^j g_{ij}
+ 4\lambda^0\left(T_p b_{01\ldots p}+E_p/V_p\right).\eea

In the gauge $\lambda^m = constants$, the equations
of motion following from $S_p^{SG}$
(or from (\ref{pbem}) after imposing the static gauge) take the form:
\bea\label{ems} g_{ab}\ddot{Y^b}
+\Gamma_{a,bc}\dot{Y^b}\dot{Y^c}=\frac{1}{2}\p_a \mathcal{U}^{S} +
2\p_{[a}\mathcal{A}_{b]}^{S}\dot{Y^b},\eea
where
\bea\nn \mathcal{A}_{a}^{S}=g_{a0}-\lambda^i g_{ai}
+2\lambda^0 T_p b_{a1\ldots p}.\eea
Thus, in general, the time evolution of the reduced dynamical system does not
correspond to a geodesic motion. The deviation from the geodesic trajectory
is due to the appearance of the {\it effective} scalar potential
$\mathcal{U}^{S}$ and of the field strength $2\p_{[a}\mathcal{A}^{S}_{b]}$ of
the {\it effective} $U(1)$-gauge potential $\mathcal{A}_{a}^{S}$. In addition,
our dynamical system is subject to the {\it effective} constraint (\ref{ecs}).


Next, we proceed with the D$p$-branes case.

In static gauge, and for background fields independent of the coordinates
$X^i$ \\(conditions(\ref{ogab})), the reduced Lagrangian, obtained from
(\ref{oda}), is given by \bea\nn L_{Dp}^{SG}(\tau)&=& \frac{V_{Dp}
e^{-a\Phi_0}}{4\lambda^0} \Bigl[g_{ab}\dot{Y^a}\dot{Y^b}+
g_{00}-2\lambda^i g_{0i} +
\left(\lambda^i\lambda^j-\kappa^i\kappa^j\right)g_{ij}\\ \nn &+&
2\left(g_{0a} -\lambda^i g_{ia}+ 2\lambda^0T_{Dp} e^{a\Phi_0}c_{a1\ldots
p} +\kappa^i b_{ai}\right)\dot{Y^a}\\ \nn &-& \left(2\lambda^0
T_{Dp}\right)^2 \det(g_{ij})+ 4\lambda^0 T_{Dp} e^{a\Phi_0}c_{01\ldots
p}\\ \nn &+&2\kappa^i \left(b_{0i}-\lambda^j b_{ji}\right)
+4\pi\alpha'\kappa^i \left(F^o_{0i}-\lambda^j F^o_{ji}\right)\Bigr].\eea As
we already mentioned at the end of Section 2, we restrict our
considerations to the case of constant dilaton $\Phi=\Phi_0$ and constant
electro-magnetic field $F^o_{mn}$ on the D$p$-brane worldvolume.

Now, the constraints (\ref{Dpbic0'}), (\ref{Dpbicj}) and (\ref{Dpbick})
take the form
\bea\label{Dpbic0's} &&g_{ab}\dot{Y^a}\dot{Y^b} + 2\left(g_{0a}-\lambda^i
g_{ia}\right)\dot{Y^a}+ g_{00}-2\lambda^{i}g_{0i}\\ \nn &&+
\left(\lambda^{i}\lambda^{j}+\kappa^i\kappa^j\right)g_{ij}
+\left(2\lambda^0T_{Dp}\right)^2\det(g_{ij})=0,\\
\label{Dpbicjs} &&g_{ja}\dot{Y^a}+g_{0j}-\lambda^{i}g_{ij}
-\kappa^i b_{ij}=2\pi\alpha'\kappa^i F^o_{ij}\\
\nn &&b_{aj}\dot{Y^a}+b_{0j}-\lambda^{i}b_{ij}
-\kappa^i g_{ij}=-2\pi\alpha'\left(F^o_{0j}-\la^i F^o_{ij}\right).\eea

The reduced Lagrangian $L_{Dp}^{SG}$ does not depend on $\tau$ explicitly.
As a consequence, the energy $E_{Dp}$ is conserved: \bea\nn &&g_{ab}\dot{Y^a}
\dot{Y^b}-g_{00}+2\lambda^i g_{0i} -
\left(\lambda^i\lambda^j - \kappa^i\kappa^j\right)g_{ij}+
\left(2\lambda^0 T_{Dp}\right)^2 \det(g_{ij})-
4\lambda^0 T_{Dp}e^{a\Phi_0} c_{01\ldots p}
\\ \nn&&-2\kappa^i\left(b_{0i}-\lambda^j b_{ji}\right) - 4\pi\alpha'\kappa^i
\left(F^o_{0i}-\lambda^j F^o_{ji}\right)
= \frac{4\lambda^0 E_{Dp}}{V_{Dp}}e^{a\Phi_0}=constant.\eea
By using the constraints (\ref{Dpbic0's}) and (\ref{Dpbicjs}),
the above equality can be replaced by the following one
\bea\label{Dp0} g_{0a}\dot{Y^a} +
g_{00}-\lambda^ig_{i0}+2\lambda^0 T_{Dp}e^{a\Phi_0} c_{01\ldots p}
+\kappa^i\left(b_{0i}+2\pi\alpha' F^o_{0i}\right)
=- \frac{2\lambda^0 E_{Dp}}{V_{Dp}}e^{a\Phi_0}.\eea

Now, we compute the momenta, obtained from the initial action (\ref{oda}),
in static gauge
\bea\label{Dgms} 2\la^0 e^{a\Phi_0} P^{SG}_M =g_{Ma}\dot{Y}^a+g_{M0}
-\la^j g_{Mj} +2\la^0T_{Dp}e^{a\Phi_0} c_{M1\ldots p}
+\kappa^j b_{Mj}.\eea
Comparing (\ref{Dgms}) with (\ref{Dp0}) and (\ref{Dpbicjs}), one finds that
\bea\nn &&P_0^{SG}=-\left(\frac{E_{Dp}}{V_{Dp}}+\frac{\pi\alpha'}{\la^0}
e^{-a\Phi_0}\kappa^j F^o_{0j}\right)=constant,
\\ \nn &&P^{SG}_i=-\frac{\pi\alpha'}{\la^0}
e^{-a\Phi_0}\kappa^j F^o_{ij} =constants .\eea
As in the $p$-brane case, not only the energy, but also the spatial components
of the momenta $P^{SG}_i$, along the $X^i$ coordinates, are conserved.
In the D$p$-brane case however, $P^{SG}_i$ are not identically zero due the
existence of a constant worldvolume magnetic field $F^o_{ij}$.

Inserting (\ref{Dpbicjs}) and (\ref{Dp0}) into (\ref{Dpbic0's}),
one obtains the effective constraint
\bea\label{Decs} g_{ab}\dot{Y^a}\dot{Y^b} =\mathcal{U}^{DS},\eea
where
\bea\nn &&\mathcal{U}^{DS}=
-\left(2\lambda^0 T_{Dp}\right)^2 \det(g_{ij})+ g_{00}-2\lambda^i g_{0i} +
\left(\lambda^i\lambda^j-\kappa^i\kappa^j\right)g_{ij}\\ \nn &&+4\lambda^0
e^{a\Phi_0}\left(T_{Dp} c_{01\ldots p}+\frac{E_{Dp}}{V_{Dp}}\right)
+2\kappa^i\left(b_{0i}-\lambda^j b_{ji}\right)+4\pi\alpha'\kappa^i
\left(F^o_{0i}-\lambda^j F^o_{ji}\right).\eea

In the gauge $(\lambda^m,\kappa^i) = constants$, the equations
of motion following from $L_{Dp}^{SG}$
(or from (\ref{Dpbem}) after using the static gauge ansatz) take the form:
\bea\label{Dems} g_{ab}\ddot{Y^b}
+\Gamma_{a,bc}\dot{Y^b}\dot{Y^c}=\frac{1}{2}\p_a \mathcal{U}^{DS} +
2\p_{[a}\mathcal{A}_{b]}^{DS}\dot{Y^b},\eea
where
\bea\nn \mathcal{A}_{a}^{DS}=g_{a0}-\lambda^i g_{ai}
+2\lambda^0 T_{Dp} e^{a\Phi_0}c_{a1\ldots p}+\kappa^i b_{ai}.\eea

It is obvious that the equations of motion (\ref{ems}), (\ref{Dems}) and the
effective constraints (\ref{ecs}), (\ref{Decs}) have the {\it same form} for
$p$-branes and for D$p$-branes. The difference is in the explicit expressions
for the {\it effective} scalar and 1-form gauge potentials.

\subsubsection{Branes dynamics in linear gauges}
Now we will repeat our analysis of the probe branes dynamics in the
framework of the more general {\it linear gauges}, given by the ansatz
(\ref{imac}). The {\it static gauge} is a particular case of the {\it
linear gauges}, corresponding to the following restrictions: \bea\nn
&&\mbox{(1):} \mu = i = 1,\ldots ,p;\h \mbox{(2):}
\Lambda^\mu_0=\Lambda^i_0= 0;\\ \nn &&\mbox{(3):}
\Lambda^\mu_j=\Lambda^i_j=\delta^i_j;\h \mbox{(4):} Y^0(\tau)=\tau\in
\{Y^a\}.\eea

{\bf $P$-branes case}

In linear gauges, and under the conditions (\ref{ogab}), one obtains the
following reduced Lagrangian, arising from the action (\ref{oa})
\bea\label{olga} &&L_p^{LG}(\tau)=
\frac{V_p}{4\lambda^0}\Bigl\{g_{ab}\dot{Y^a}\dot{Y^b} +
2\Bigl[\left(\Lambda_0^\mu-\lambda^i\Lambda_i^\mu\right)g_{\mu a}+
2\lambda^0T_p B_{a1\ldots p}\Bigr]\dot{Y^a}\\
\nn &&+\left(\Lambda_0^\mu-\lambda^i\Lambda_i^\mu\right)
\left(\Lambda_0^\nu-\lambda^j\Lambda_j^\nu\right)g_{\mu\nu}- \left(2\lambda^0
T_p\right)^2 \det(\Lambda_i^\mu\Lambda_j^\nu g_{\mu\nu})\\
\nn &&+ 4\lambda^0T_p\Lambda_0^{\mu}B_{\mu 1\ldots p}\Bigr\},
\h B_{M1\ldots p}\equiv b_{M\mu_1\ldots\mu_p}
\Lambda_1^{\mu_1}\ldots\Lambda_p^{\mu_p}.\eea

The constraints derived from the Lagrangian (\ref{olga}) are:
\bea\label{c0l}
&&g_{ab}\dot{Y^a}\dot{Y^b} +
2\left(\Lambda_0^\mu-\lambda^i\Lambda_i^\mu\right)g_{\mu a}\dot{Y^a}+
\left(\Lambda_0^\mu-\lambda^i\Lambda_i^\mu\right) \\
\nn &&\times
\left(\Lambda_0^\nu-\lambda^j\Lambda_j^\nu\right)g_{\mu\nu}+\left(2\lambda^0
T_p\right)^2 \det(\Lambda_i^\mu\Lambda_j^\nu g_{\mu\nu})=0,\\
\label{cil} &&\Lambda_i^\mu\left[g_{\mu a}\dot{Y^a}+
\left(\Lambda_0^\nu-\lambda^j\Lambda_j^\nu\right)g_{\mu\nu}\right]=0.\eea

The Lagrangian $L_p^{LG}$ does not depend on $\tau$ explicitly, so the
energy $E_p=p^{LG}_a \dot{Y}^a-L_p^{LG}$ is conserved:
\bea\nn &&g_{ab}\dot{Y^a}\dot{Y^b}-\left(\Lambda_0^\mu-
\lambda^i\Lambda_i^\mu\right)\left(\Lambda_0^\nu-\lambda^j\Lambda_j^
\nu\right)g_{\mu\nu}+ \left(2\lambda^0 T_p\right)^2
\det(\Lambda_i^\mu\Lambda_j^\nu g_{\mu\nu})\\ \nn
&&-4\lambda^0T_p\Lambda_0^{\mu}B_{\mu 1\ldots p}=
\frac{4\lambda^0 E_p}{V_p}=constant.\eea
With the help of the constraints (\ref{c0l}) and (\ref{cil}), one can replace
this equality by the following one
\bea\label{p0l} \Lambda_0^\mu\left[g_{\mu a}\dot{Y^a}+
\left(\Lambda_0^\nu-\lambda^j\Lambda_j^\nu\right)g_{\mu\nu}
+2\lambda^0T_p B_{\mu 1\ldots p}\right]=-\frac{2\lambda^0 E_p}{V_p}.\eea

In linear gauges, the momenta (\ref{nm}) take the form
\bea\label{gml} 2\la^0 P^{LG}_M =g_{Ma}\dot{Y}^a+
\left(\Lambda_0^\nu-\lambda^j\Lambda_j^\nu\right)g_{M\nu}
+2\la^0T_p B_{M1\ldots p}.\eea
The comparison of (\ref{gml}) with (\ref{p0l}) and (\ref{cil}) gives
\bea\nn \La^\mu_0 P_\mu^{LG}=-\frac{E_p}{V_p} = constant,\h
\La^\mu_i P_\mu^{LG} = constants=0.\eea
Therefore, in the linear gauges, the projections of the momenta $P^{LG}_\mu$
onto $\La^\mu_n$ are conserved. Moreover, as far as the Lagrangian
(\ref{olga}) does not depend on the coordinates $X^\mu$, the corresponding
conjugated momenta $P^{LG}_\mu$ are also conserved.

Inserting (\ref{p0l}) and (\ref{cil}) into (\ref{c0l}), we obtain the
{\it effective} constraint
\bea\nn g_{ab}\dot{Y^a}\dot{Y^b}=\mathcal{U}^{L},\eea
where the {\it effective} scalar potential is given by
\bea\nn &&\mathcal{U}^{L}=-\left(2\lambda^0 T_p\right)^2
\det(\Lambda_i^\mu\Lambda_j^\nu g_{\mu\nu})+
\left(\Lambda_0^\mu-\lambda^i\Lambda_i^\mu\right)
\left(\Lambda_0^\nu-\lambda^j\Lambda_j^\nu\right)g_{\mu\nu}\\ \nn
&&+ 4\lambda^0\left(T_p \Lambda_0^{\mu}B_{\mu 1\ldots p}
+\frac{E_p}{V_p}\right).\eea

In the gauge $\lambda^m = constants$, the equations
of motion following from $L_p^{LG}$ take the form:
\bea\nn g_{ab}\ddot{Y^b}
+\Gamma_{a,bc}\dot{Y^b}\dot{Y^c}=\frac{1}{2}\p_a \mathcal{U}^{L} +
2\p_{[a}\mathcal{A}_{b]}^{L}\dot{Y^b},\eea
where
\bea\nn \mathcal{A}_{a}^{L}=
\left(\Lambda_0^\mu-\lambda^i\Lambda_i^\mu\right)g_{a\mu}+
2\lambda^0T_p B_{a1\ldots p},\eea
is the {\it effective} 1-form gauge potential, generated by the non-diagonal
components $g_{a\mu}$ of the background metric and by the components
$b_{a\mu_1 \ldots \mu_p}$ of the background $(p+1)$-form gauge field.

{\bf D$p$-branes case}

In linear gauges, and for background fields independent of the coordinates
$X^\mu$\\ (conditions(\ref{ogab})), the reduced Lagrangian, obtained from
(\ref{oda}), is given by \bea\nn L_{Dp}^{LG}(\tau)&=&
\frac{V_{Dp} e^{-a\Phi_0}}{4\lambda^0} \Bigl\{g_{ab}\dot{Y^a}\dot{Y^b}+
\left[\left(\Lambda_0^\mu-\lambda^i\Lambda_i^\mu\right)
\left(\Lambda_0^\nu-\lambda^j\Lambda_j^\nu\right)
-\kappa^i\kappa^j\Lambda_i^\mu\Lambda_j^\nu\right]g_{\mu\nu}\\ \nn &+&
2\left[\left(\Lambda_0^\mu-\lambda^i\Lambda_i^\mu\right)g_{\mu a}
+ 2\lambda^0T_{Dp} e^{a\Phi_0}C_{a1\ldots p}
+\kappa^i\La^\mu_i b_{a\mu}\right]\dot{Y^a}\\
\nn &-& \left(2\lambda^0 T_{Dp}\right)^2
\det(\Lambda_i^\mu\Lambda_j^\nu g_{\mu\nu})+ 4\lambda^0 T_{Dp} e^{a\Phi_0}
\La^\mu_0 C_{\mu 1\ldots p}\\ \nn &-&2\kappa^i\La^\mu_i
\left(\La^\nu_0-\lambda^j \La^\nu_j\right)b_{\mu\nu} +4\pi\alpha'\kappa^i
\left(F^o_{0i}-\lambda^j F^o_{ji}\right)\Bigr\},\eea
where the following shorthand notation has been introduced
\bea\nn C_{M1\ldots p}\equiv
c_{M\mu_1\ldots\mu_p}\La^{\mu_1}_1\ldots\La^{\mu_p}_p.\eea

Now, the constraints (\ref{Dpbic0'}), (\ref{Dpbicj}), and (\ref{Dpbick})
take the form \bea\label{Dpbic0'l} &&g_{ab}\dot{Y^a}\dot{Y^b} +
2\left(\Lambda_0^\mu-\lambda^i\Lambda_i^\mu\right)g_{\mu a}\dot{Y^a}
+\left(2\lambda^0T_{Dp}\right)^2 \det(\Lambda_i^\mu\Lambda_j^\nu
g_{\mu\nu})\\ \nn
&&+\left[\left(\Lambda_0^\mu-\lambda^i\Lambda_i^\mu\right)
\left(\Lambda_0^\nu-\lambda^j\Lambda_j^\nu\right)
+\kappa^i\kappa^j\Lambda_i^\mu\Lambda_j^\nu\right]g_{\mu\nu}=0,
\\ \label{Dpbicjl} &&\Lambda_i^\mu\left[g_{\mu a}\dot{Y^a}+
\left(\Lambda_0^\nu-\lambda^j\Lambda_j^\nu\right)g_{\mu\nu}+
\kappa^j\La^\nu_j b_{\mu\nu}\right]=2\pi\alpha'\kappa^j F^o_{ji}\\
\nn &&\Lambda_i^\mu\left[b_{\mu a}\dot{Y^a}+
\left(\Lambda_0^\nu-\lambda^j\Lambda_j^\nu\right)b_{\mu\nu}+\kappa^j\La^\nu_j
g_{\mu\nu}\right]=2\pi\alpha'\left(F^o_{0i}-\la^j F^o_{ji}\right).\eea

The reduced Lagrangian $L_{Dp}^{LG}$ does not depend on $\tau$ explicitly.
As a consequence, the energy $E_{Dp}$ is conserved: \bea\nn &&g_{ab}\dot{Y^a}
\dot{Y^b}-\left[\left(\Lambda_0^\mu-\lambda^i\Lambda_i^\mu\right)
\left(\Lambda_0^\nu-\lambda^j\Lambda_j^\nu\right)
-\kappa^i\kappa^j\Lambda_i^\mu\Lambda_j^\nu\right]g_{\mu\nu}\\ \nn
&&+\left(2\lambda^0 T_{Dp}\right)^2 \det(\Lambda_i^\mu\Lambda_j^\nu
g_{\mu\nu})- 4\lambda^0 T_{Dp}e^{a\Phi_0}\La^\mu_0 C_{\mu 1\ldots p}
\\ \nn&&-2\kappa^i\La^\mu_i
\left(\La^\nu_0-\lambda^j \La^\nu_j\right)b_{\mu\nu} - 4\pi\alpha'\kappa^i
\left(F^o_{0i}-\lambda^j F^o_{ji}\right)
= \frac{4\lambda^0 E_{Dp}}{V_{Dp}}e^{a\Phi_0}=constant.\eea
By using the constraints (\ref{Dpbic0'l}) and (\ref{Dpbicjl}),
the above equality can be replaced by the following one
\bea\label{Dp0l} &&\Lambda_0^\mu\left[g_{\mu a}\dot{Y^a}+
\left(\Lambda_0^\nu-\lambda^j\Lambda_j^\nu\right)g_{\mu\nu}
+2\lambda^0T_{Dp}e^{a\Phi_0} C_{\mu 1\ldots p}
+\kappa^j\La^\nu_j b_{\mu\nu}\right]\\ \nn
&&+2\pi\alpha'\kappa^i F^o_{0i}
=- \frac{2\lambda^0 E_{Dp}}{V_{Dp}}e^{a\Phi_0}.\eea

In linear gauges, the momenta obtained from the initial action (\ref{oda}),
are
\bea\label{Dgml} 2\la^0 e^{a\Phi_0} P^{LG}_M =
g_{Ma}\dot{Y^a}+
\left(\Lambda_0^\nu-\lambda^j\Lambda_j^\nu\right)g_{M\nu}
+2\lambda^0T_{Dp}e^{a\Phi_0} C_{M1\ldots p}
+\kappa^j\La^\nu_j b_{M\nu}.\eea
Comparing (\ref{Dgml}) with (\ref{Dp0l}) and (\ref{Dpbicjl}), one finds that
the following equalities hold
\bea\nn &&\La^\mu_0 P_\mu^{LG}=-\left(\frac{E_{Dp}}{V_{Dp}}
+\frac{\pi\alpha'}{\la^0}
e^{-a\Phi_0}\kappa^j F^o_{0j}\right)=constant,
\\ \nn &&\La^\mu_i P_\mu^{LG}=-\frac{\pi\alpha'}{\la^0}
e^{-a\Phi_0}\kappa^j F^o_{ij} =constants .\eea
They may be viewed as restrictions on the number of the
arbitrary parameters, presented in the theory.

As in the $p$-brane case, the momenta $P_\mu^{LG}$ are conserved quantities,
due to the independence of the Lagrangian on the coordinates $X^\mu$.

Inserting (\ref{Dpbicjl}) and (\ref{Dp0l}) into (\ref{Dpbic0'l}),
one obtains the effective constraint
\bea\nn g_{ab}\dot{Y^a}\dot{Y^b} =\mathcal{U}^{DL},\eea
where
\bea\nn &&\mathcal{U}^{DL}=
\left[\left(\Lambda_0^\mu-\lambda^i\Lambda_i^\mu\right)
\left(\Lambda_0^\nu-\lambda^j\Lambda_j^\nu\right)
-\kappa^i\kappa^j\Lambda_i^\mu\Lambda_j^\nu\right]g_{\mu\nu}
\\ \nn &&-\left(2\lambda^0 T_{Dp}\right)^2 \det(\Lambda_i^\mu\Lambda_j^\nu
g_{\mu\nu})+4\lambda^0
e^{a\Phi_0}\left(T_{Dp}\La^\mu_0 C_{\mu 1\ldots p}+
\frac{E_{Dp}}{V_{Dp}}\right)\\ \nn
&&+2\kappa^i\La^\mu_i\left(\La^\nu_0-\lambda^j \La^\nu_j\right)b_{\mu\nu}
+4\pi\alpha'\kappa^i\left(F^o_{0i}-\lambda^j F^o_{ji}\right).\eea

In the gauge $(\lambda^m,\kappa^i) = constants$, the equations
of motion following from $L_{Dp}^{LG}$ take the form:
\bea\nn g_{ab}\ddot{Y^b}
+\Gamma_{a,bc}\dot{Y^b}\dot{Y^c}=\frac{1}{2}\p_a \mathcal{U}^{DL} +
2\p_{[a}\mathcal{A}_{b]}^{DL}\dot{Y^b},\eea
where
\bea\nn \mathcal{A}_{a}^{DL}=
\left(\Lambda_0^\mu-\lambda^i\Lambda_i^\mu\right)g_{a\mu}
+ 2\lambda^0T_{Dp} e^{a\Phi_0}C_{a1\ldots p}
+\kappa^i\La^\mu_i b_{a\mu}.\eea

It is clear that the equations of motion and the effective constraints
have the {\it same form} for $p$-branes and for D$p$-branes in linear gauges,
as well as in static gauge. The only difference is in the explicit expressions
for the {\it effective} scalar and 1-form gauge potentials.

\subsubsection{Branes dynamics in the whole space-time}
Working in {\it static gauge} $X^m(\xi^n)=\xi^m$, we actually imply that
the probe branes have no dynamics along the background coordinates $x^m$. The
(proper) time evolution is possible only in the transverse directions,
described by the coordinates $x^a$.

Using the {\it linear gauges}, we have the possibility to place the probe
branes in general position with respect to the coordinates $x^\mu$,
on which the background fields do not depend. However, the real dynamics is
again in the transverse directions only.

Actually, in the framework of our approach, the probe branes can have
'full' dynamical freedom only when the ansatz (\ref{wgac}) is used,
because only then {\it all} of the brane coordinates $X^M$ are allowed to
vary {\it nonlinearly} with the proper time $\tau$. Therefore, with the
help of (\ref{wgac}), we can probe the {\it whole} space-time.

We will use the superscript $A$ to denote that the corresponding quantity is
taken on the ansatz (\ref{wgac}). It is understood that the conditions
(\ref{ogab}) are also fulfilled.


{\bf $P$-branes}

Now, the reduced Lagrangian obtained from the action (\ref{oa}) is given by
\bea\nn &&L_p^{A}(\tau)=
\frac{V_p}{4\lambda^0}\Bigl\{g_{MN}\dot{Y^M}\dot{Y^N} +
2\Bigl[\left(\Lambda_0^\mu-\lambda^i\Lambda_i^\mu\right)g_{\mu N}+
2\lambda^0T_p B_{N1\ldots p}\Bigr]\dot{Y^N}\\
\nn &&+\left(\Lambda_0^\mu-\lambda^i\Lambda_i^\mu\right)
\left(\Lambda_0^\nu-\lambda^j\Lambda_j^\nu\right)g_{\mu\nu}- \left(2\lambda^0
T_p\right)^2 \det(\Lambda_i^\mu\Lambda_j^\nu g_{\mu\nu})\\
\nn &&+ 4\lambda^0T_p\Lambda_0^{\mu}B_{\mu 1\ldots p}\Bigr\}.\eea

The constraints, derived from the above Lagrangian, are:
\bea\label{c0g}
&&g_{MN}\dot{Y^M}\dot{Y^N} +
2\left(\Lambda_0^\mu-\lambda^i\Lambda_i^\mu\right)g_{\mu N}\dot{Y^N}+
\left(\Lambda_0^\mu-\lambda^i\Lambda_i^\mu\right) \\
\nn &&\times
\left(\Lambda_0^\nu-\lambda^j\Lambda_j^\nu\right)g_{\mu\nu}+\left(2\lambda^0
T_p\right)^2 \det(\Lambda_i^\mu\Lambda_j^\nu g_{\mu\nu})=0,\\
\label{cig} &&\Lambda_i^\mu\left[g_{\mu N}\dot{Y^N}+
\left(\Lambda_0^\nu-\lambda^j\Lambda_j^\nu\right)g_{\mu\nu}\right]=0.\eea
The corresponding momenta are ($P_M = P^{A}_M/V_p$)
\bea\nn 2\la^0 P_M =g_{MN}\dot{Y}^N+
\left(\Lambda_0^\nu-\lambda^j\Lambda_j^\nu\right)g_{M\nu}
+2\la^0T_p B_{M1\ldots p},\eea
and part of them, $P_\mu$, are conserved
\bea\label{gmgc} g_{\mu N}\dot{Y}^N+
\left(\Lambda_0^\nu-\lambda^j\Lambda_j^\nu\right)g_{\mu\nu}
+2\la^0T_p B_{\mu 1\ldots p}= 2\la^0 P_\mu = constants ,\eea
because $L_p^{A}$ does not depend on $X^\mu$.
From (\ref{cig}) and (\ref{gmgc}), the compatibility conditions follow
\bea\label{ccs} \La^\mu_i P_\mu =0.\eea
We will regard on (\ref{ccs}) as a solution of the constraints (\ref{cig}),
which restricts the number of the arbitrary parameters
$\La^\mu_i$ and $P_\mu$. That is why from now on, we will deal only with the
constraint (\ref{c0g}).

In the gauge $\lambda^m = constants$, the equations
of motion for $Y^N$, following from $L_p^{A}$, have the form
\bea\label{emg} g_{LN}\ddot{Y^N}
+\Gamma_{L,MN}\dot{Y^M}\dot{Y^N}=\frac{1}{2}\p_L \mathcal{U}^{in} +
2\p_{[L}\mathcal{A}_{N]}^{in}\dot{Y^N},\eea
where
\bea\nn &&\mathcal{U}^{in}=-\left(2\lambda^0 T_p\right)^2
\det(\Lambda_i^\mu\Lambda_j^\nu g_{\mu\nu})+
\left(\Lambda_0^\mu-\lambda^i\Lambda_i^\mu\right)
\left(\Lambda_0^\nu-\lambda^j\Lambda_j^\nu\right)g_{\mu\nu}\\ \nn
&&+ 4\lambda^0T_p \Lambda_0^{\mu}B_{\mu 1\ldots p} ,\\
\nn &&\mathcal{A}_{N}^{in}=
\left(\Lambda_0^\mu-\lambda^i\Lambda_i^\mu\right)g_{N\mu}+
2\lambda^0T_p B_{N1\ldots p}.\eea
Let us first consider this part of the equations of motion (\ref{emg}), which
corresponds to $L=\la$. It follows from (\ref{ogab}) that the connection
coefficients $\Gamma_{\la,MN}$, involved in these equations, are
\bea\nn \Gamma_{\lambda,ab}=\frac{1}{2}\left(\p_ag_{b\lambda}
+\p_bg_{a\lambda}\right), \h \Gamma_{\lambda,\mu
a}=\frac{1}{2}\p_ag_{\mu\lambda},\h \Gamma_{\lambda,\mu\nu}=0.\eea
Inserting these expressions in the part of the differential equations
(\ref{emg}) corresponding to $L=\lambda$ and using that
$\dot{g}_{MN}=\dot{Y}^a\p_ag_{MN}$,
$\dot{B}_{M1\ldots p}=\dot{Y}^a\p_a B_{M1\ldots p}$,
one receives
\bea\nn \frac{d}{d\tau}\left[g_{\mu N}\dot{Y}^N+
\left(\Lambda_0^\nu-\lambda^j\Lambda_j^\nu\right)g_{\mu\nu}
+2\la^0T_p B_{\mu 1\ldots p}\right]=0.\eea
These equalities express the fact that the momenta $P_\mu$ are conserved
(compare with (\ref{gmgc})). Therefore, we have to deal only with the other
part of the equations of motion, corresponding to $L=a$
\bea\label{emga} g_{aN}\ddot{Y^N}
+\Gamma_{a,MN}\dot{Y^M}\dot{Y^N}=\frac{1}{2}\p_a \mathcal{U}^{in} +
2\p_{[a}\mathcal{A}_{N]}^{in}\dot{Y^N}.\eea

Our next task is to {\it separate the variables} $\dot{Y^\mu}$ and
$\dot{Y^a}$ in these equations and in the constraint (\ref{c0g}). To this
end, we will use the conservation laws (\ref{gmgc}) to express
$\dot{Y^\mu}$ through $\dot{Y^a}$. The result is
\bea\label{ymu}
\dot{Y^\mu}=\left(g^{-1}\right)^{\mu\nu} \left[2\la^0(P_\nu - T_p B_{\nu
1\ldots p}) -g_{\nu a}\dot{Y^a}\right] -(\La^\mu_0 - \la^i\La^\mu_i).\eea
We will need also the explicit expressions for the connection coefficients
$\Gamma_{a,\mu b}$ and $\Gamma_{a,\mu\nu}$, which under the conditions
(\ref{ogab}) reduce to
\bea\label{eeg} \Gamma_{a,\mu
b}=-\frac{1}{2}\left(\p_ag_{b\mu}-\p_bg_{a\mu}\right)
=-\p_{[a}g_{b]\mu},\h \Gamma_{a,\mu\nu}=-\frac{1}{2}\p_ag_{\mu\nu}.\eea
By using (\ref{ymu}) and (\ref{eeg}), after some calculations, one rewrites
the equations of motion (\ref{emga}) and the constraint (\ref{c0g}) in the
form
\bea\label{mgemf} &&h_{ab}\ddot{Y}^b +
\Gamma^{\bf{h}}_{a,bc}\dot{Y}^b\dot{Y}^c = \frac{1}{2}\p_a \mathcal{U}^{A}
+ 2\p_{[a}\mathcal{A}^{A}_{b]}\dot{Y}^b,
\\ \label{mgecf}
&&h_{ab}\dot{Y}^a\dot{Y}^b = \mathcal{U}^{A},\eea where a new,
{\it effective metric} appeared
\bea\nn h_{ab} = g_{ab} -
g_{a\mu}(g^{-1})^{\mu\nu}g_{\nu b}.\eea
$\Gamma^{\bf{h}}_{a,bc}$ is the
connection compatible with this metric
\bea\nn
\Gamma^{\bf{h}}_{a,bc}=\frac{1}{2}\left(\p_b h_{ca} +\p_c h_{ba}-\p_a
h_{bc}\right).\eea
The new, {\it effective} scalar and gauge potentials are
given by
\bea\nn &&\mathcal{U}^{A}=-\left(2\lambda^0 T_p\right)^2
\det(\Lambda_i^\mu\Lambda_j^\nu g_{\mu\nu}) - (2\la^0)^2 \left(P_\mu-T_p
B_{\mu 1\ldots p}\right)\left(g^{-1}\right)^{\mu\nu} \left(P_\nu-T_p
B_{\nu 1\ldots p}\right),
\\ \nn &&\mathcal{A}_{a}^{A}=
2\lambda^0\left[g_{a\mu}\left(g^{-1}\right)^{\mu\nu} \left(P_\nu-T_p
B_{\nu 1\ldots p}\right)+ T_p B_{a1\ldots p}\right].\eea
We note that Eqs. (\ref{mgemf}), (\ref{mgecf}), and therefore their solutions, do not depend
on the parameters $\La^\mu_0$ and $\la^i$ in contrast to the previously
considered cases. However, they have the same form as before.

{\bf D$p$-branes}

The reduced Lagrangian, obtained from (\ref{oda}), is given by
\bea\nn  L_{Dp}^{A}(\tau)&=&
\frac{V_{Dp} e^{-a\Phi_0}}{4\lambda^0} \Bigl\{g_{MN}\dot{Y^M}\dot{Y^N}+
\left[\left(\Lambda_0^\mu-\lambda^i\Lambda_i^\mu\right)
\left(\Lambda_0^\nu-\lambda^j\Lambda_j^\nu\right)
-\kappa^i\kappa^j\Lambda_i^\mu\Lambda_j^\nu\right]g_{\mu\nu}
\\ \nn &+&
2\left[\left(\Lambda_0^\mu-\lambda^i\Lambda_i^\mu\right)g_{\mu N}
+ 2\lambda^0T_{Dp} e^{a\Phi_0}C_{N1\ldots p}
+\kappa^i\La^\mu_i b_{N\mu}\right]\dot{Y^N}
\\ \nn &-& \left(2\lambda^0 T_{Dp}\right)^2
\det(\Lambda_i^\mu\Lambda_j^\nu g_{\mu\nu})+ 4\lambda^0 T_{Dp} e^{a\Phi_0}
\La^\mu_0 C_{\mu 1\ldots p}
\\ \nn &-&2\kappa^i\La^\mu_i
\left(\La^\nu_0-\lambda^j \La^\nu_j\right)b_{\mu\nu} +4\pi\alpha'\kappa^i
\left(F^o_{0i}-\lambda^j F^o_{ji}\right)\Bigr\},\eea

The constraints (\ref{Dpbic0'}), (\ref{Dpbicj}), and (\ref{Dpbick}) take
the form
\bea\label{Dpbic0'g} &&g_{MN}\dot{Y^M}\dot{Y^N} +
2\left(\Lambda_0^\mu-\lambda^i\Lambda_i^\mu\right)g_{\mu N}\dot{Y^N}
+\left(2\lambda^0T_{Dp}\right)^2 \det(\Lambda_i^\mu\Lambda_j^\nu
g_{\mu\nu})
\\ \nn
&&+\left[\left(\Lambda_0^\mu-\lambda^i\Lambda_i^\mu\right)
\left(\Lambda_0^\nu-\lambda^j\Lambda_j^\nu\right)
+\kappa^i\kappa^j\Lambda_i^\mu\Lambda_j^\nu\right]g_{\mu\nu}=0,
\\ \label{Dpbicjg} &&\Lambda_i^\mu\left[g_{\mu N}\dot{Y^N}+
\left(\Lambda_0^\nu-\lambda^j\Lambda_j^\nu\right)g_{\mu\nu}+
\kappa^j\La^\nu_j b_{\mu\nu}\right]=2\pi\alpha'\kappa^j F^o_{ji}\\
\label{Dpbickg} &&\Lambda_i^\mu\left[b_{\mu N}\dot{Y^N}+
\left(\Lambda_0^\nu-\lambda^j\Lambda_j^\nu\right)b_{\mu\nu}+\kappa^j\La^\nu_j
g_{\mu\nu}\right]=2\pi\alpha'\left(F^o_{0i}-\la^j F^o_{ji}\right).\eea

Because of the independence of $L_{Dp}^{A}$ on $X^\mu$, the momenta
$P^D_\mu=P_\mu^{DA}/V_{Dp}$
are conserved
\bea\label{gmgcd} 2\la^0 e^{a\Phi_0} P^D_\mu =
g_{\mu N}\dot{Y^N}+
\left(\Lambda_0^\nu-\lambda^j\Lambda_j^\nu\right)g_{\mu\nu}
+2\lambda^0T_{Dp}e^{a\Phi_0} C_{\mu 1\ldots p}
+\kappa^j\La^\nu_j b_{\mu\nu} = constants.\eea
From (\ref{Dpbicjg}) and (\ref{gmgcd}), one obtains the following
compatibility conditions
\bea\nn \La^\mu_j P^D_\mu = \frac{\pi\alpha'}{\la^0}
e^{-a\Phi_0}\kappa^i F^o_{ij},\eea
which we interpret as a solution of the constraints (\ref{Dpbicjg}).

In the gauge $(\lambda^m,\kappa^i) = constants$, the equations
of motion for $Y^N$, following from $L_{Dp}^{A}$, take the form
\bea\label{emgd} g_{LN}\ddot{Y^N}
+\Gamma_{L,MN}\dot{Y^M}\dot{Y^N}=\frac{1}{2}\p_L \mathcal{U}^{Din} +
2\p_{[L}\mathcal{A}_{N]}^{Din}\dot{Y^N},\eea
where
\bea\nn &&\mathcal{U}^{Din}=-\left(2\lambda^0 T_p\right)^2
\det(\Lambda_i^\mu\Lambda_j^\nu g_{\mu\nu})+
\left[\left(\Lambda_0^\mu-\lambda^i\Lambda_i^\mu\right)
\left(\Lambda_0^\nu-\lambda^j\Lambda_j^\nu\right)
-\kappa^i\kappa^j\Lambda_i^\mu\Lambda_j^\nu\right]g_{\mu\nu}
\\ \nn &&+ 4\lambda^0 T_{Dp} e^{a\Phi_0}
\La^\mu_0 C_{\mu 1\ldots p}-2\kappa^i\La^\mu_i
\left(\La^\nu_0-\lambda^j \La^\nu_j\right)b_{\mu\nu},\\
\nn &&\mathcal{A}_{N}^{Din}=
\left(\Lambda_0^\nu-\lambda^j\Lambda_j^\nu\right)g_{N\nu}
+ 2\lambda^0T_{Dp} e^{a\Phi_0}C_{N1\ldots p}
+\kappa^j\La^\nu_j b_{N\nu}.\eea
As in the $p$-brane case, this part of the equations of motion (\ref{emgd}),
which corresponds to $L=\la$, expresses the conservation of the momenta
$P^D_\mu$, in accordance with (\ref{gmgcd}). The remaining equations of
motion, which we have to deal with, are
\bea\label{emgad} g_{aN}\ddot{Y^N}
+\Gamma_{a,MN}\dot{Y^M}\dot{Y^N}=\frac{1}{2}\p_a \mathcal{U}^{Din} +
2\p_{[a}\mathcal{A}_{N]}^{Din}\dot{Y^N}.\eea

To exclude the dependence on $\dot{Y}^\mu$ in the Eqs. (\ref{emgad}) and
in the constraints (\ref{Dpbic0'g}), (\ref{Dpbickg}), we use the
conservation laws (\ref{gmgcd}) to express $\dot{Y^\mu}$ through
$\dot{Y^a}$:
\bea\label{ymud} \dot{Y^\mu}=\left(g^{-1}\right)^{\mu\nu}
\left[2\la^0 e^{a\Phi_0}(P^D_\nu - T_{Dp}C_{\nu 1\ldots p}) -g_{\nu
a}\dot{Y^a} -\kappa^j\La^\rho_j b_{\nu\rho}\right] -(\La^\mu_0 -
\la^i\La^\mu_i).\eea

By using (\ref{ymud}) and (\ref{eeg}), one can rewrite the equations of motion
(\ref{emgad}) and the constraint (\ref{Dpbic0'g}) as
\bea\label{mgemfd} &&h_{ab}\ddot{Y}^b +
\Gamma^{\bf{h}}_{a,bc}\dot{Y}^b\dot{Y}^c
= \frac{1}{2}\p_a \mathcal{U}^{DA}
+ 2\p_{[a}\mathcal{A}^{DA}_{b]}\dot{Y}^b,\\
\label{mgecfd} &&h_{ab}\dot{Y}^a\dot{Y}^b = \mathcal{U}^{DA}.\eea
Now, the {\it effective} scalar and 1-form gauge potentials are given by
\bea\nn \mathcal{U}^{DA}&=&-\left(2\lambda^0 T_p\right)^2
\det(\Lambda_i^\mu\Lambda_j^\nu g_{\mu\nu})
-\kappa^i\kappa^j\Lambda_i^\mu\Lambda_j^\nu g_{\mu\nu}
\\ \nn &-&\left[2\la^0 e^{a\Phi_0}(P^D_\mu
- T_{Dp}C_{\mu 1\ldots p})-\kappa^i\La^\la_i b_{\mu\la}\right]
\left(g^{-1}\right)^{\mu\nu}
\\ \nn&\times&\left[2\la^0 e^{a\Phi_0}(P^D_\nu
- T_{Dp}C_{\nu 1\ldots p})-\kappa^j\La^\rho_j b_{\nu\rho}\right],
\\ \nn
\mathcal{A}_{a}^{DA}&=&
g_{a\mu}\left(g^{-1}\right)^{\mu\nu}\left[2\la^0 e^{a\Phi_0}(P^D_\nu
- T_{Dp}C_{\nu 1\ldots p})-\kappa^j\La^\rho_j b_{\nu\rho}\right]
\\ \nn &+& 2\lambda^0T_{Dp} e^{a\Phi_0}C_{a1\ldots p}
+\kappa^i\La^\mu_i b_{a\mu}.\eea
Eqs. (\ref{mgemfd}), (\ref{mgecfd}), have
the same form as in static and linear gauges, but now they do not depend
on the parameters $\La^\mu_0$ and $\la^i$. Another difference is the
appearance of a new, {\it effective} background metric $h_{ab}$ and the
corresponding connection $\Gamma^{\bf{h}}_{a,bc}$.

In the D-brane case, we have another set of constraints (\ref{Dpbickg}),
generated by the Lagrange multipliers $\kappa^i$. With the help of
(\ref{ymud}), they acquire the form
\bea\nn &&\Bigl\{\left[b_{a\nu}-g_{a\rho}\left(g^{-1}\right)^{\rho\mu}b_{\mu\nu}
\right]\dot{Y}^a +b_{\mu\nu}\left(g^{-1}\right)^{\mu\rho}
\left[2\la^0 e^{a\Phi_0}(P^D_\rho - T_{Dp}C_{\rho 1\ldots p})-
\kappa^j\La^\la_j b_{\rho\la}\right]
\\ \nn
&& -\kappa^i\La^\mu_i g_{\mu\nu}\Bigr\}
\La^\nu_j =-2\pi\alpha'\left(F^o_{0j}-\la^i F^o_{ij}\right).\eea

\subsubsection{Explicit solutions of the equations of motion}
All cases considered so far, have one common feature. The dynamics of the
corresponding reduced particle-like system is described by {\it effective}
equations of motion and one {\it effective} constraint, which have the
{\it same form}, independently of the ansatz used to reduce the $p$-branes
or D$p$-branes dynamics. Our aim here is to find {\it explicit exact
solutions} to them. \footnote{The additional restrictions on the
solutions, depending on the ansatz and on the type of the branes, will be
discussed in the next section.} To be able to describe all cases
simultaneously, let us first introduce some general notations.

We will search for solutions of the following system of {\it nonlinear}
differential equations \bea\label{ee} &&\mathcal{G}_{ab}\ddot{Y}^b +
\Gamma^{\mathcal{G}}_{a,bc}\dot{Y}^b\dot{Y}^c = \frac{1}{2}\p_a
\mathcal{U} + 2\p_{[a}\mathcal{A}_{b]}\dot{Y}^b,\\ \label{ec}
&&\mathcal{G}_{ab}\dot{Y}^a\dot{Y}^b = \mathcal{U},\eea where
$\mathcal{G}_{ab}$, $\Gamma^{\mathcal{G}}_{a,bc}$, $\mathcal{U}$, and
$\mathcal{A}_a$ can be as follows \bea\nn
&&\mathcal{G}_{ab}=\left(g_{ab},h_{ab}\right),\h
\Gamma^{\mathcal{G}}_{a,bc}=\left(\Gamma_{a,bc},\Gamma^{\bf{h}}_{a,bc}\right),
\\ \nn
&&\mathcal{U}=\left(\mathcal{U}^{S},\mathcal{U}^{DS},\mathcal{U}^{L},
\mathcal{U}^{DL},\mathcal{U}^{A},\mathcal{U}^{DA}\right),
\\ \nn
&&\mathcal{A}_a=\left(\mathcal{A}_{a}^{S},\mathcal{A}_{a}^{DS},
\mathcal{A}_{a}^{L},\mathcal{A}_{a}^{DL},\mathcal{A}_{a}^{A},
\mathcal{A}_{a}^{DA}\right),\eea
depending on the ansatz and on the type of the brane ($p$-brane or D$p$-brane).

Let us start with the simplest case, when the background fields depend on
only one coordinate $X^a=Y^a(\tau)$. \footnote{An example of such
background is the generalized Kasner type metric, arising in the
superstring cosmology \cite{LWC99} (see also \cite{CKR02},
\cite{EGJK02}).} In this case the Eqs. (\ref{ee}), (\ref{ec}) simplify to
($d_a\equiv d/dY^a$) \bea\label{ee1}
&&\frac{d}{d\tau}\left(\mathcal{G}_{aa}\dot{Y}^a\right)
-\frac{1}{2}d_a\mathcal{G}_{aa}\left(\dot{Y}^a\right)^2 = \frac{1}{2}d_a
\mathcal{U} ,\\ \label{ec1} &&\mathcal{G}_{aa}\left(\dot{Y}^a\right)^2 =
\mathcal{U},\eea
where we have used that \bea\nn \mathcal{G}_{ab}\ddot{Y}^b
+ \Gamma^{\mathcal{G}}_{a,bc}\dot{Y}^b\dot{Y}^c
=\frac{d}{d\tau}\left(\mathcal{G}_{ab}\dot{Y}^b\right)
-\frac{1}{2}\p_a\mathcal{G}_{bc}\dot{Y}^b\dot{Y}^c.\eea After multiplying
with $2\mathcal{G}_{aa}\dot{Y}^a$ and after using the constraint
(\ref{ec1}),  the Eq. (\ref{ee1}) reduces to \bea\label{fi1}
\frac{d}{d\tau}\left[\left(\mathcal{G}_{aa}\dot{Y}^a\right)^2 -
\mathcal{G}_{aa}\mathcal{U}\right]=0.\eea The solution of (\ref{fi1}),
compatible with (\ref{ec1}), is just the constraint (\ref{ec1}). In other
words, (\ref{ec1}) is first integral of the equation of motion for the
coordinate $Y^a$. By integrating (\ref{ec1}), one obtains the following
exact probe branes solution \bea\label{tsol1}\tau\left(X^a\right)=\tau_0
\pm \int_{X_0^a}^{X^a}
\left(\frac{\mathcal{U}}{\mathcal{G}_{aa}}\right)^{-1/2}dx,\eea where
$\tau_0$ and $X_0^a$ are arbitrary constants.

When one works in the framework of the general ansatz (\ref{wgac}),
one has to also write down the solution for the remaining coordinates $X^\mu$.
It can be obtained as follows. One represents $\dot{Y}^\mu$ as
\bea\nn \dot{Y}^\mu =\frac{dY^\mu}{dY^a}\dot{Y}^a,\eea
and use this and (\ref{ec1}) in (\ref{ymu}) for the $p$-brane,
and in (\ref{ymud}) for the D$p$-brane. The result is a system of ordinary
differential equations of first order with separated variables, which
integration is straightforward. Replacing the obtained solution for
$Y^\mu(X^a)$ in the ansatz (\ref{wgac}), one finally arrives at
\bea\label{xmus1} &&X^\mu(X^a,\xi^i)=X^\mu_0 +
\La^\mu_i\left[\la^i\tau(X^a)+\xi^i\right]
\\ \nn &&-\int_{X^a_0}^{X^a}\left(g^{-1}\right)^{\mu\nu}\left[g_{\nu a}
\mp 2\la^0(P_\nu-T_p B_{\nu 1\ldots p})
\left(\frac{\mathcal{U}^A}{h_{aa}}\right)^{-1/2}\right]dx\eea
for the $p$-brane case, and at
\bea\label{xmusd1} &&X^\mu(X^a,\xi^i)=X^\mu_0 +
\La^\mu_i\left[\la^i\tau(X^a)+\xi^i\right]
\\ \nn &&-\int_{X^a_0}^{X^a}\left(g^{-1}\right)^{\mu\nu}\left\{g_{\nu a}
\mp \left[2\la^0 e^{a\Phi_0}(P^D_\nu
- T_{Dp}C_{\nu 1\ldots p})-\kappa^j\La^\rho_j b_{\nu\rho}\right]
\left(\frac{\mathcal{U}^{DA}}{h_{aa}}\right)^{-1/2}\right\}dx\eea
for the D$p$-brane case correspondingly.
In the above two exact branes solutions,
$X^\mu_0$ are arbitrary constants, and $\tau(X^a)$ is given in (\ref{tsol1}).
We note that the comparison of the solutions $X^\mu(X^a,\xi^i)$ with
the initial ansatz (\ref{wgac}) for $X^\mu$ shows, that the dependence on
$\La^\mu_0$ has disappeared. We will comment on this later on.

Let us turn to the more complicated case, when the background fields depend on
more than one coordinate $X^a=Y^a(\tau)$. We would like to apply the same
procedure for solving the system of differential equations (\ref{ee}),
(\ref{ec}), as in the simplest case just considered.
To be able to do this, we need to suppose that the metric
$\mathcal{G}_{ab}$ is a diagonal one. Then one can rewrite the effective
equations of motion (\ref{ee}) and the effective constraint (\ref{ec})
in the form
\bea\label{eed} &&\frac{d}{d\tau}\left(\mathcal{G}_{aa}\dot{Y}^a\right)^2 -
\dot{Y}^a\p_a\left(\mathcal{G}_{aa}\mathcal{U}\right)
\\ \nn &&+ \dot{Y}^a\sum_{b\ne a}
\left[\p_a\left(\frac{\mathcal{G}_{aa}}{\mathcal{G}_{bb}}\right)
\left(\mathcal{G}_{bb}\dot{Y}^b\right)^2
- 4\p_{[a}\mathcal{A}_{b]}\mathcal{G}_{aa}\dot{Y}^b\right] = 0,
\\ \label{ecd}
&&\mathcal{G}_{aa}\left(\dot{Y}^a\right)^2
+\sum_{b\ne a}\mathcal{G}_{bb}\left(\dot{Y}^b\right)^2 = \mathcal{U}.\eea

To find solutions of the above equations without choosing particular
background, we fix all coordinates $X^a$ except one. Then the exact
probe brane solution of the equations of motion is given
again by the same expression (\ref{tsol1}) for $\tau\left(X^a\right)$.
In the case when one is using the general ansatz (\ref{wgac}),
the solutions (\ref{xmus1}) and (\ref{xmusd1}) still also hold.

To find solutions depending on more than one coordinate, we have to impose
further conditions on the background fields. Let us show, how a number of
{\it sufficient} conditions, which allow us to reduce the order of the
equations of motion by one, can be obtained.

First of all, we split the index $a$ in such a way that $Y^r$ is one of
the coordinates $Y^a$, and $Y^{\alpha}$ are the others. Then we assume
that the effective 1-form gauge field $\mathcal{A}_a$ can be represented
in the form \bea\label{egfr} \mathcal{A}_a
=(\mathcal{A}_r,\mathcal{A}_\alpha)= (\mathcal{A}_r,\p_\alpha f),\eea i.e.,
it is oriented along the coordinate $Y^r$, and the remaining components
$\mathcal{A}_\alpha$ are pure gauges. Now, the Eq.(\ref{eed}) read
\bea\label{eeda} &&\frac{d}{d\tau}\left(\mathcal{G}_{\alpha\alpha}
\dot{Y}^\alpha\right)^2 -
\dot{Y}^\alpha\p_\alpha\left(\mathcal{G}_{\alpha\alpha}\mathcal{U}\right)
\\ \nn &&+\dot{Y}^\alpha\left[\p_\alpha\left(\frac{\mathcal{G}_{\alpha\alpha}}
{\mathcal{G}_{rr}}\right)\left(\mathcal{G}_{rr}\dot{Y}^r\right)^2
-2\mathcal{G}_{\alpha\alpha}\p_\alpha\left(\mathcal{A}_r-\p_r f\right)
\dot{Y}^r\right]
\\ \nn &&+ \dot{Y}^\alpha\sum_{\beta\ne \alpha}
\p_\alpha\left(\frac{\mathcal{G}_{\alpha\alpha}}{\mathcal{G}_{\beta\beta}}
\right)\left(\mathcal{G}_{\beta\beta}\dot{Y}^\beta\right)^2 = 0,
\\ \label{eedr} &&\frac{d}{d\tau}\left(\mathcal{G}_{rr}\dot{Y}^r\right)^2 -
\dot{Y}^r\p_r\left(\mathcal{G}_{rr}\mathcal{U}\right)
\\ \nn &&+ \dot{Y}^r\sum_{\alpha}
\left[\p_r\left(\frac{\mathcal{G}_{rr}}{\mathcal{G}_{\alpha\alpha}}\right)
\left(\mathcal{G}_{\alpha\alpha}\dot{Y}^\alpha\right)^2
+2\mathcal{G}_{rr}\p_{\alpha}\left(\mathcal{A}_r-\p_r f\right)
\dot{Y}^\alpha\right] = 0.\eea
After imposing the conditions \bea\label{ca}
\p_\alpha\left(\frac{\mathcal{G}_{\alpha\alpha}}
{\mathcal{G}_{aa}}\right)=0,
\h\p_\alpha\left(\mathcal{G}_{rr}\dot{Y}^r\right)^2=0,\eea
the Eq.(\ref{eeda}) reduce to \bea\nn
\frac{d}{d\tau}\left(\mathcal{G}_{\alpha\alpha}\dot{Y}^\alpha\right)^2
-\dot{Y}^\alpha\p_\alpha\left\{\mathcal{G}_{\alpha\alpha}\left[\mathcal{U}
+2\left(\mathcal{A}_r-\p_r f\right)\dot{Y}^r\right]\right\}=0,\eea which
are solved by \bea\label{fia}
\left(\mathcal{G}_{\alpha\alpha}\dot{Y}^\alpha\right)^2 =D_{\alpha}
\left(Y^{a\ne\alpha}\right) + \mathcal{G}_{\alpha\alpha}\left[\mathcal{U}
+2\left(\mathcal{A}_r-\p_r f\right)\dot{Y}^r\right]= E_{\alpha}
\left(Y^{\beta}\right)\ge 0,\eea where $D_{\alpha}$, $E_{\alpha}$ are
arbitrary functions of their arguments. ($E_{\alpha}=E_{\alpha}
\left(Y^{\beta}\right)$ follows from (\ref{cr})).

To integrate the Eq. (\ref{eedr}), we impose the condition
\bea\label{cr}
\p_r\left(\mathcal{G}_{\alpha\alpha}\dot{Y}^\alpha\right)^2=0. \eea After
using the second of the conditions (\ref{ca}), the condition (\ref{cr}),
and the already obtained solution (\ref{fia}), the Eq. (\ref{eedr}) can be
recast in the form \bea\nn
&&\frac{d}{d\tau}\left[\left(\mathcal{G}_{rr}\dot{Y}^r\right)^2
+2\mathcal{G}_{rr}\left(\mathcal{A}_r-\p_r f\right)\dot{Y}^r\right]\\ \nn
&&=\dot{Y}^r\p_r\left\{\mathcal{G}_{rr}\left[(1-n_\alpha)\left(\mathcal{U}
+2\left(\mathcal{A}_r-\p_r f\right)\dot{Y}^r\right)
-\sum_{\alpha}\frac{D_{\alpha} \left(Y^{a\ne\alpha}\right)}
{\mathcal{G}_{\alpha\alpha}}\right]\right\},\eea where $n_\alpha$ is the
number of the coordinates $Y^\alpha$. The solution of this equation,
compatible with (\ref{fia}) and with the effective constraint (\ref{ecd}),
is \bea\label{fir} \left(\mathcal{G}_{rr}\dot{Y}^r\right)^2 =
\mathcal{G}_{rr} \left[(1-n_\alpha)\mathcal{U}
-2n_\alpha\left(\mathcal{A}_r-\p_r f\right)\dot{Y}^r
-\sum_{\alpha}\frac{D_{\alpha} \left(Y^{a\ne\alpha}\right)}
{\mathcal{G}_{\alpha\alpha}}\right]= E_r\left(Y^r\right)\ge 0,\eea where
$E_r$ is again an arbitrary function.

Thus, we succeeded to separate the variables $\dot{Y}^a$ and to obtain the
first integrals (\ref{fia}), (\ref{fir}) for the equations of motion
(\ref{eed}), when the conditions (\ref{egfr}), (\ref{ca}), (\ref{cr}) on
the background are fulfilled. \footnote{An example, when the obtained
sufficient conditions are satisfied, is given by the evolution of a
tensionless brane in Kerr space-time. Moreover, in this case, one is able
to find the orbit $r=r(\theta)$ \cite{Bd01}.} Further progress is
possible, when working with particular background configurations, having
additional symmetries (see for instance, \cite{KK99}).

{\bf To summarize}, we addressed here the problem of obtaining {\it explicit exact}
solutions for probe branes moving in general string theory backgrounds. We
concentrated our attention to the {\it common} properties of the
$p$-branes and D$p$-branes dynamics and tried to formulate an approach,
which is effective for different embeddings, for arbitrary worldvolume and
space-time dimensions, for different variable background fields, for
tensile and tensionless branes. To achieve this, we first performed an
analysis with the aim to choose brane actions, which are
most appropriate for our purposes.

Next, we formulated the frameworks in which to search for exact
probe branes solutions. The guiding idea is the reduction of the brane
dynamics to a particle-like one. In view of the existing practice, we
first consider the case of {\it static gauge} embedding, which is the
mostly used one in higher dimensions. Then we turn to the more general
case of {\it linear embeddings}, which are appropriate for lower
dimensions too. After that, we consider the branes dynamics by using a more general ansatz,
allowing for its reduction to particle-like one. The
obtained results reveal one common property in all the cases considered.
The {\it effective} equations of motion and one of the constraints, the
{\it effective} constraint, have the {\it same form} independently of the
ansatz used to reduce the $p$-branes or D$p$-branes dynamics. In general,
the effective equations of motion do not coincide with the geodesic ones.
The deviation from the geodesic motion is due to the appearance of {\it
effective} scalar and 1-form gauge potentials. The same scalar potential
arises in the effective constraint.

Also, we considered the problem of obtaining {\it
explicit exact} solutions of the effective equations of motion and the
effective constraint, without using the explicit structure of the
effective potentials.

In the case when the background fields depend on only one coordinate
$x^a=X^a(\tau)$, we showed that these equations can always be integrated and
give the probe brane solution in the form $\tau=\tau(X^a)$, where $\tau$
is the worldvolume temporal parameter. We also give the explicit solutions
for the brane coordinates $X^\mu$ in the form $X^\mu=X^\mu(X^a,\xi^i)$.
They are nontrivial when one uses the more general ansatz (\ref{wgac}).
Let us remind that $x^\mu$ are the coordinates, on which the
background fields do not depend.

In the case when the background fields depend on more than one coordinate,
and we fix all brane coordinates $X^a$ except one, the exact
solutions are given by the same expressions as in the case considered before, if
the metric $\mathcal{G}_{ab}$ is a diagonal one. In this way, we have realized
the possibility to obtain probe brane solutions as functions of {\it every}
single one coordinate, on which the background depends. In the case when none
of the brane coordinates is kept fixed, we were able to find {\it sufficient}
conditions, which ensure the separation of the variables
$\dot{X}^a=\dot{Y}^a(\tau)$. As a result, we have found the manifest
expressions for $n_a$ first integrals of the equations of motion, where
$n_a$ is the number of the brane coordinates $Y^a$.

In obtaining the solutions described above, it was not taken into account that
some restrictions on them can arise, depending on the ansatz used and on the
type of the branes considered. As far as we are interested here in the
{\it common} properties of the probe branes dynamics, we will not
make an exhaustive investigation of all possible peculiarities, which can arise
in different particular cases. Nevertheless, we will point out some
specific properties, characterizing the dynamics of the different type
of branes for different embeddings.

We note that in static gauge, the brane coordinates $X^a$ figuring in our
solutions, are spatial ones. This is so, because in this gauge the background
temporal coordinate, on which the background fields can depend, is
identified with the worldvolume time $\tau$.

The solutions $X^\mu(X^a,\xi^i)$, given by (\ref{xmus1}) for the $p$-brane and
by (\ref{xmusd1}) for the D$p$-brane, depend on the worldvolume parameters
$(\tau,\xi^i)$ through the specific combination
$\La^\mu_i\left(\la^i\tau+\xi^i\right)$. It is interesting to understand if
its origin has some physical meaning. To this end, let us consider the
$p$-branes equations of motion (\ref{pbem}) and constraints (\ref{pbic0}),
(\ref{pbicj}) in the tensionless limit $T_p\to 0$, when they take the form
\cite{1,2}
\bea\nn &&g_{LN}\Di\Dj X^N+\Gamma_{L,MN}\Di X^M \Dj X^N = 0 ,\\ \label{tec}
&&g_{MN}
\left(\p_0-\lambda^i\p_i\right) X^M\left(\p_0-\lambda^j\p_j\right)X^N=0,\\ \nn
&&g_{MN}\left(\p_0-\lambda^i\p_i\right) X^M\p_j X^N=0.\eea
It is easy to check that in $D$-dimensional space-time, any $D$ arbitrary
functions of the type $F^M=F^M\left(\la^i\tau+\xi^i\right)$ solve this system
of partial differential equations. Hence, the linear part of the {\it tensile}
$p$-brane and D$p$-brane solutions (\ref{xmus1}) and (\ref{xmusd1}), is a
background independent solution of the {\it tensionless} $p$-brane equations
of motion and constraints.

Let us also point out here that by construction, the actions used in our
considerations allow for taking the tensionless limit $T_p\to 0$
($T_{Dp}\to 0$). Moreover, from the explicit form of the obtained exact probe
branes solutions it is clear that the opposite limit $T_p\to\infty$
($T_{Dp}\to\infty$) can be also taken.

We have obtained solutions of the probe branes equations of motion and one
of the constrains, which have the same form for all of the considered cases.
Now, let us see how we can satisfy the other constraints present in the theory.
These are $p$ constraints, obtained by varying the corresponding actions with
respect to the Lagrange multipliers $\la^i$. For the D$p$-brane, we have $p$
additional constraints, obtained by varying the action with respect to the
Lagrange multipliers $\kappa^i$. Actually, the constraints generated by the
$\la^i$-multipliers are satisfied. Due to the conservation of the corresponding
momenta, they just restrict the number of the {\it independent} parameters
present in the solutions. The only exception is the $p$-brane in static gauge
case, where the momenta $P_i^{SG}$ must be zero. Let us give an example how
the problem can be resolved in a particular situation, which is nevertheless
general enough. Let the background metric along the probe $p$-brane be a
diagonal one. Then from (\ref{cis}) and (\ref{gms}) it follows that the
momenta $P_i^{SG}$ will be identically zero, if we work in the gauge
$\la^i=0$. In the general case, and this is also valid for the $\kappa^i$-
generated constraints, we have to insert the obtained solution of the equations
of motion into the unresolved constraints. The result will be a number of
algebraic relations between the background fields. If they are not satisfied
(on the solution) at least for some particular values of the free parameters
in the solution, it would be fair to say that our approach does not work
properly in this case, and some modification is needed.

Finally, let us say a few words about some possible generalizations of the
obtained results.

As is known, the branes charges are restricted up to a sign to be equal to
the branes tensions from the condition for space-time supersymmetry of the
corresponding actions. In our computations, however, the coefficients in
front of the background antisymmetric fields do not play any special role.
That is why, to account for nonsupersymmetric probe branes, it is enough
to make the replacements \bea\nn T_p b_{p+1} \to Q_p b_{p+1},\h T_{Dp}
c_{p+1} \to Q_{Dp} c_{p+1}.\eea

In our D$p$-brane action (\ref{oda}), we have included only the leading
Wess-Zumino term of the possible D$p$-brane couplings. It is easy to see that
our results can be generalized to include other interaction terms just by the
replacement
\bea\nn  c_{p+1} \to c_{p+1}+c_{p-1}\wedge b_2 + \ldots .\eea
This is a consequence of the fact that we do not used the explicit form of the
background field $c_{M_0\ldots M_p}$. We have used only its antisymmetry and
its independence on part of the background coordinates.


\subsection{Particular cases}

\subsubsection{Tensionless string in Demianski-Newman background}
The metric for Demianski-Newman space-time is of the following type: \bea\nn
ds^2&=&g_{00}(dx^0)^2+2g_{01}dx^0 dx^1+2g_{03}dx^0 dx^3+2g_{13}dx^1
dx^3+g_{22}(dx^2)^2+ g_{33}(dx^3)^2 .\eea It can be put in a form in
which the manifest expressions for $g_{\mu\nu}$ are given by the
equalities (all other components are zero):  \bea\nn
g_{00}&=&- \exp(+2U) \h,\h g_{11}=
\Biggl(1-\frac{a^2\sin^2\theta}{{\cal R}^2}\Biggr)\exp(-2U),
\\ \label{dnm}
g_{22}&=& \left({\cal R}^2 - a^2\sin^2\theta\right)\exp(-2U),\\ \nn
g_{33}&=& {\cal R}^2\sin^2\theta\exp(-2U)
-\Biggl[\frac{2(Mr+l)a\sin^2\theta + 2{\cal R}^{2}l\cos\theta}{{\cal
R}^2 - a^2\sin^2\theta}\Biggr]^2\exp(+2U) ,
\\ \nn
g_{03}&=&- \frac{2(Mr+l)a\sin^2\theta + 2{\cal
R}^{2}l\cos\theta}{{\cal R}^2 - a^2\sin^2\theta}\exp(+2U), \eea where
\bea\nn \exp(\pm 2U) = \Biggl[1-2\frac{Mr + l\left(a\cos\theta +
l\right)}{r^2 + \left(a\cos\theta + l\right)^2}\Biggr]^{\pm 1},\h
{\cal R}^2 = r^2 - 2Mr + a^2 - l^2 , \eea $M$ is a mass parameter, $a$
is an angular momentum per unit mass and $l$ is the NUT parameter.
The metric (\ref{dnm}) is a stationary axisymmetric metric. It
belongs to the vacuum solutions of the Einstein field equations,
which are of type D under Petrov's classification. The Kerr and NUT
space-times are particular cases of the considered metric and can be
obtained by putting $l=0$ or $a=0$ in (\ref{dnm}). The case $l=0,a=0$
obviously corresponds to the Schwarzschild solution.

Solving the equations of motion and constraints (\ref{tec}) for the string case ($p=1$)
in Demianski-Newman background, one can find the following solution \cite{1}
\bea\nn t-t_0&=&\pm\int\limits_{r_0}^{r}dr\Biggl[\frac{\left(C^0 + C^3
{\cal A}_0\right){\cal A}_0}{{\cal R}^2\sin^2\theta_0}\exp(+2U_0) -
C^3\exp(-2U_0)\Biggr]W^{-1/2} ,
\\ \label{GSC}
\varphi - \varphi_{0}&=&\mp\int\limits_{r_0}^{r}dr \frac{C^0 + C^3
{\cal A}_0}{{\cal R}^2\sin^2\theta_0}\exp(+2U_0)W^{-1/2} ,
\\ \nn
C_1 (\tau -\tau_{0})&=& \pm\int\limits_{r_0}^{r}drW^{-1/2} ,\\ \nn
W&=&\left({\cal R}^2-a^2\sin^2\theta_0\right)^{-1}
\Biggl[\left(C^3\right)^2{\cal R}^2 -  \frac{\left(C^0 + C^3 {\cal
A}_0\right)^2}{\sin^2\theta_0}\exp(+4U_0)\Biggr], \\ \nn {\cal
A}_0&=&\frac{2(Mr+l)a\sin^2\theta_0 + 2{\cal
R}^{2}l\cos\theta_0}{{\cal R}^2 - a^2\sin^2\theta_0},\h
U_0=U_{|\theta=\theta_0},
\\ \nn
t_0, r_0, \varphi_{0}, \tau_{0} &-& constants. \eea

\subsubsection{Tensionless $p$-branes in a solitonic background}
The solitonic $(\tilde{d}-1)$-brane background is given by
\bea\nn ds^2 =
g_{MN}dx^M dx^N = \exp\left(2A\right)\eta_{\mu\nu}dx^{\mu}dx^{\nu} +
\exp\left(2B\right)\left(dr^2 + r^2d\Omega^2_{D-\tilde{d}-1}\right),
\\ \nn \exp\left(2A\right) =
\left(1+\frac{k_{\tilde{d}}}{r^d}\right)^{-\frac{d}{d+\tilde{d}}}, \h
\exp\left(2B\right) =
\left(1+\frac{k_{\tilde{d}}}{r^d}\right)^{+\frac{\tilde{d}}{d+\tilde{d}}},
\h k_{\tilde{d}}=const, \\ \nn d + \tilde{d} = D-2,\h
\eta_{\mu\nu}=\mbox{diag}(-,+,...,+), \h \mu, \nu =
0,1,...,\tilde{d}-1. \eea The $(D-\tilde{d}-1)$-dimensional sphere
$\mathbf{S}^{D-\tilde{d}-1}$ is supposed to be parameterized so that
\bea\nn g_{kk} &=&
\exp\left(2B\right)r^2\prod_{n=1}^{D-k-1}\sin^2\theta_{n},\h D-k-1 =
1,2,...,D-\tilde{d}-2, \\ \nn g_{D-1,D-1} &=&
\exp\left(2B\right)r^2.\eea

In this background the following tensionless $p$-brane solutions
of the equations of motion and constraints (\ref{tec}) do exist
\cite{2}
\bea\nn y^{\mu} = y^{\mu}_0 \pm E^{\mu}
\int_{r_0}^{r}du\left(1+\frac{k_{\tilde{d}}}{u^d}\right)\left(
\mathcal{E}-\frac{\left(C^{D-1}\right)^2}{u^2}+\frac{\mathcal{E}k_{\tilde{d}}}{u^d}
\right)^{-1/2},\h r\equiv y^{\tilde{d}},\\ \nn E^{\mu}=
\left(E^0,E^1,\ldots,E^{\tilde{d}-1}\right)=
\left(CC^{\tilde{d}-1},C^1,\ldots,CC^0\right); \\ \label{exsdv}
\varphi = \varphi_0 \pm C^{D-1} \int_{r_0}^{r}\frac{du}{u^2}\left(
\mathcal{E}- \frac{\left(C^{D-1}\right)^2}{u^2} +
\frac{\mathcal{E}k_{\tilde{d}}}{u^d} \right)^{-1/2},\h
\varphi\equiv y^{D-1};\\ \nn \tau = \tau_0 \pm
\int_{r_0}^{r}du\left(1+\frac{k_{\tilde{d}}}{u^d}\right)^
{\frac{\tilde{d}}{d+\tilde{d}}} \left(
\mathcal{E}-\frac{\left(C^{D-1}\right)^2}{u^2} +
\frac{\mathcal{E}k_{\tilde{d}}}{u^d}
\right)^{-1/2};\h\h\hspace{.5cm}
\\ \nn \mathcal{E}\equiv - E^{\mu}E^{\nu}\eta_{\mu\nu}
= \left(CC^{\tilde{d}-1}\right)^2 -
\left(CC^0\right)^2 - \sum_{\alpha =
1}^{\tilde{d}-2}\left(C^{\alpha}\right)^2\geq 0.\eea

Let us restrict ourselves to the particular case of ten dimensional
solitonic 5-brane background. The corresponding values of the
parameters $D$, $\tilde{d}$ and $d$ are $D=10$, $\tilde{d}=6$, $d=2$.
Taking this into account and performing the integration in
(\ref{exsdv}), one obtains the following explicit exact solution of
the equations of motion and constraints for a tensionless $p$-brane
living in such curved space-time ($\mathcal{C} = k_6 - \left(C^{9}\right)^2/\mathcal{E} > 0$)
\bea\nn y^{\mu} =
y^{\mu}_0 \mp \frac{k_6 E^{\mu}}{\left(\mathcal{CE}\right)^{1/2}}
\ln\left(\frac{\frac{\mathcal{C}^{1/2}}{r} + \left(1 +
\frac{\mathcal{C}}{r^2}\right)^{1/2}}{\frac{\mathcal{C}^{1/2}}{r_0} +
\left(1 + \frac{\mathcal{C}}{r_{0}^{2}}\right)^{1/2}}\right) \\ \nn
\pm \frac{E^{\mu}}{\mathcal{E}^{1/2}}\left[\left(\mathcal{C} +
r^2\right)^{1/2} - \left(\mathcal{C} + r_{0}^{2}\right)^{1/2}\right],
\\ \nn \varphi = \varphi_0 \mp
\frac{C^9}{\left(\mathcal{CE}\right)^{1/2}}
\ln\left(\frac{\frac{\mathcal{C}^{1/2}}{r} + \left(1 +
\frac{\mathcal{C}}{r^2}\right)^{1/2}}{\frac{\mathcal{C}^{1/2}}{r_0} +
\left(1 + \frac{\mathcal{C}}{r_{0}^{2}}\right)^{1/2}}\right), \\
\label{expls} \tau = \tau_0 \mp \left(\frac{k_6^3}{\mathcal{C}^4
\mathcal{E}^2}\right)^{1/4}r^{3/2}\left(1 +
\frac{\mathcal{C}}{r^2}\right)^{1/2}F_2\left(3/4,1,-3/4;3/2,3/4;1 +
\frac{r^2}{\mathcal{C}},-\frac{r^2}{k_6}\right)\\ \nn \mp
2\left(\frac{k_6^3 r_0^2}{\mathcal{C}^2 \mathcal{E}^2}\right)^{1/4}
F_1\left(1/4,1/2,-3/4;5/4;-\frac{r_0^2}{\mathcal{C}},
-\frac{r_0^2}{k_6}\right)\\ \nn \pm
\frac{\Gamma(1/4)k_6}{4\Gamma(3/4)}\sqrt{\frac{\pi}{\mathcal{CE}}}
\ {}_2 F_{1}\left(1/4,1/2;-1/4;1 -
\frac{k_6}{\mathcal{C}}\right)\\ \nn \mp
\frac{\Gamma(1/4)k_6}{2\Gamma(3/4)}\sqrt{\frac{\pi}{\mathcal{CE}}}
\left(1 - \frac{k_6}{\mathcal{C}}
\right)^{-1/4}{}_2F_{1}\left(1/4,3/2;3/4;\left(1 -
\frac{k_6}{\mathcal{C}}\right)^{-1}\right). \eea

\subsubsection{String and D-string solutions in other backgrounds}
Let us first give an explicit example of exact solution for a
string moving in four dimensional cosmological Kasner type background.
Namely, the line element is ($x^0 \equiv t$) \bea\label{km} ds^2 &=&
g_{MN}dx^M dx^N = -(d t)^2 + \sum_{\mu=1}^{3}t^{2q_{\mu}}(dx^{\mu})^2,\\
\label{Kc} &&\sum_{\mu=1}^{3}q_{\mu}=1,\h \sum_{\mu=1}^{3}q^2_{\mu}=1.\eea
For definiteness, we choose $q_{\mu}=\left(2/3,2/3,-1/3\right)$. Then, one can obtain
the following exact solution of the equations of motion and constraints
in the considered particular metric \cite{3}
\bea\label{Ks}
X^{\mu}\left(t,\s\right)= X^{\mu}_0 +
C^{\mu}_{\pm}\left(\lambda^1\tau+\sigma\right) \pm
A^{\pm}_{\mu}I^{\mu}(t),\h \tau(t)=\tau_0 \pm I^0(t),\\ \nn I^M (t)\equiv
\int_{t_0}^{t}d u u^{-2q_M}\left(V^{\pm}\right)^{-1/2},\h q_M =
(0,2/3,2/3,-1/3),\eea
and $V^{\pm}$ is the corresponding effective scalar potential.

Although we have chosen relatively simple background
metric, the expressions for $I^M$ are too complicated. Because of that, we
shall write down here only the formulas for the two limiting cases $T=0$
and $T\to\infty$ for $t>t_0\ge 0$. The former corresponds to considering
tensionless strings (high energy string limit).

When $T=0$, $I^M$ reads \bea\nn I^M =
\frac{1}{2\left[\left(A_1^{\pm}\right)^2 +
\left(A_2^{\pm}\right)^2\right]^{1/2}}\Biggl[
\frac{t^{2/3-2q_M}}{\left(q_M-1/3\right)A}\ {}_2F_1\left(1/2,q_M-1/3;q_M+2/3;
-\frac{1}{A^2t^2}\right)\\ \nn + \frac{t_0^{5/3-2q_M}}{q_M-5/6}\
{}_2F_1\left(1/2,5/6-q_M;11/6-q_M;-A^2t_0^2\right) +
\frac{\Gamma\left(q_M-1/3\right)\Gamma\left(5/6-q_M\right)}
{\sqrt{\pi}A^{5/3-2q_M}}\Biggr],\eea where \bea\nn A^2\equiv
\frac{\left(A_3^{\pm}\right)^2}{\left(A_1^{\pm}\right)^2 +
\left(A_2^{\pm}\right)^2}.\eea

When $T\to\infty$, $I^M$ are given by the equalities \bea\nn I^{0} &=&
\pm\frac{1}{4\lambda^0TC^3_{\pm}}\Biggl[\frac{6}{C}t^{1/3}
{}_2F_1\left(1/2,-1/6;5/6;-\frac{1}{C^2t^2}\right)\\ \nn
&-&\frac{3}{2}t_0^{4/3}{}_2F_1\left(1/2,2/3;5/3; -C^2t_0^2\right) +
\frac{\Gamma\left(-1/6\right)\Gamma\left(2/3\right)}
{\sqrt{\pi}C^{4/3}}\Biggr],\\ \nn I^{1,2} &=&
\pm\frac{1}{4\lambda^0TC^3_{\pm}}\left[\ln\left|
\frac{\left(1+C^2t^2\right)^{1/2}-1}{\left(1+C^2t^2\right)^{1/2}+1}\right|
- \ln\left|\frac{\left(1+C^2t_0^2\right)^{1/2}-1}
{\left(1+C^2t_0^2\right)^{1/2}+1}\right|\right],\\ \nn I^3 &=&
\pm\frac{1}{2\lambda^0TC^3_{\pm}C^2}\left[\left(1+C^2t^2\right)^{1/2} -
\left(1+C^2t_0^2\right)^{1/2}\right],\eea where \bea\nn C^2\equiv
\frac{\left(C^1_{\pm}\right)^2 + \left(C^2_{\pm}\right)^2}
{\left(C^3_{\pm}\right)^2}.\eea

Our choice of the scale factors $\left(t^{2/3},t^{2/3},t^{-1/3}\right)$
was dictated only by the simplicity of the solution. However, this is a
very special case of a Kasner type metric. Actually, this is one of the
two solutions of the constraints (\ref{Kc}) (up to renaming of the
coordinates $x^\mu$) for which two of the exponents $q_\mu$ are equal. The
other such solution is $q_{\mu}=\left(0,0,1\right)$ and it corresponds to
flat space-time. Now, we will write down the exact tensionless string solution
$(T=0)$ for a gravity background with arbitrary, but different $q_{\mu}$.
It is given by (\ref{Ks}), where \bea\nn &&I^M(t)= constant
-\frac{\sqrt{\pi}}{A_2^{\pm}}\sum_{k=0}^{\infty}
\frac{\left(A_3^{\pm}/A_2^{\pm}\right)^{2k}}{k!\Gamma\left(1/2-k\right)}
\frac{t^{\mathcal{P}}}{\mathcal{P}}\times \\ \nn &&{}_2F_1\left(1/2+k,\frac{
\mathcal{P}}{2(q_2-q_1)}; \frac{2(q_2-q_3)k+3q_2-2q_1+1-2q_M}{2(q_2-q_1)};
-\left(\frac{A_1^{\pm}}{A_2^{\pm}}\right)^2 t^{2(q_2-q_1)}\right),\\ \nn
&& \mathcal{P}\equiv 2(q_2-q_3)k+q_2+1-2q_M ,\h q_M = (0,q_1,q_2,q_3),\h
\mbox{for $q_1 > q_2$,} \eea  and \bea\nn &&I^M(t)= constant
+\frac{\sqrt{\pi}}{A_1^{\pm}}\sum_{k=0}^{\infty}
\frac{\left(A_3^{\pm}/A_1^{\pm}\right)^{2k}}{k!\Gamma\left(1/2-k\right)}
\frac{t^{\mathcal{Q}}}{\mathcal{Q}}\times \\ \nn &&{}_2F_1\left(1/2+k,\frac{
\mathcal{Q}}{2(q_1-q_2)}; \frac{2(q_1-q_3)k+3q_1-2q_2+1-2q_M}{2(q_1-q_2)};
-\left(\frac{A_2^{\pm}}{A_1^{\pm}}\right)^2 t^{2(q_1-q_2)}\right),\\ \nn
&& \mathcal{Q}\equiv 2(q_1-q_3)k+q_1+1-2q_M,\h\mbox{for $q_1 < q_2$}.\eea
Because there are no restrictions on $q_\mu$, except $q_1\neq q_2\neq
q_3$, the above probe string solution is also valid in generalized Kasner
type backgrounds arising in superstring cosmology. In string
frame, the effective Kasner constraints for the four dimensional
dilaton-moduli-vacuum solution are \bea\nn
&&\sum_{\mu=1}^{3}q_{\mu}=1+\mathcal{K},\h
\sum_{\mu=1}^{3}q^2_{\mu}=1-\mathcal{B}^2,\\ \nn
&&-1-\sqrt{3\left(1-\mathcal{B}^2\right)}\le \mathcal{K}\le
-1+\sqrt{3\left(1-\mathcal{B}^2\right)},\h
\mathcal{B}^2\in\left[0,1\right].\eea In Einstein frame, the metric has the
same form, but in new, rescaled coordinates and with new powers
$\tilde{q}_{\mu}$ of the scale factors. The generalized Kasner constraints
are also modified as follows \bea\nn &&\sum_{\mu=1}^{3}\tilde{q}_{\mu}=1,\h
\sum_{\mu=1}^{3}\tilde{q}^2_{\mu}=1-\tilde{\mathcal{B}}^2 -
\frac{1}{2}\tilde{\mathcal{K}}^2,\h \tilde{\mathcal{B}}^2 +
\frac{1}{2}\tilde{\mathcal{K}}^2\in \left[0,1\right] .\eea Actually, the
obtained tensionless string solution is also relevant to considerations
within a {\it pre-big bang} context, because there exist a class of models
for pre-big bang cosmology, which is a particular case of the given
generalized Kasner backgrounds.

Our next example is for a string moving in the following ten dimensional
supergravity background given in Einstein frame \bea\nn &&ds^2
= g^E_{MN}dx^Mdx^N = \exp(2A)\eta_{mn}dx^mdx^n + \exp(2B)\left(dr^2 +
r^2d\Omega^2_7\right),\\ \nn &&\exp{[-2(\phi-\phi_0)]}= 1+\frac{k}{r^6},\h
\phi_0, k=constants,\\ \nn &&A=\frac{3}{4}(\phi-\phi_0),\h
B=-\frac{1}{4}(\phi-\phi_0),\h B_{01} =
-\exp\left[-2\left(\phi-\frac{3}{4}\phi_0\right)\right]. \eea All other
components of $B_{MN}$ as well as all components of the gravitino $\psi_M$
and dilatino $\lambda$ are zero. If we parameterize the sphere $S^7$ so
that \bea\nn g^E_{10-j,10-j}= \exp(2B)r^2\prod_{l=1}^{j-1}\sin^2x^{10-l},\h
j=2,3,...,7,\h g^E_{99}=\exp(2B)r^2,\eea the metric $g^E_{MN}$ does not
depend on $x^0$, $x^1$ and $x^3$, i.e. $\mu=0,1,3$. Then we set \cite{3} $y^\alpha
= y^\alpha_0 = constants$ for $\alpha=4,...,9$ and obtain a solution of
the equations of motion and constraints as a function of the radial
coordinate $r$: \bea\nn &&X^{\mu}\left(r,\s\right)= X^{\mu}_0 +
C^{\mu}_{\pm}\left(\lambda^1\tau+\sigma\right) \pm I^{\mu}(r),\h
\tau(r)=\tau_0 \pm I(r),\\ \nn &&I^{m}(r)= \eta^{mn} \int_{r_0}^{r}d u
\left[B^{\pm}_{n}+2\lambda^0
T\varepsilon_{nk}C^k_{\pm}\exp\left(-\phi_0/2\right)
\left(1+\frac{k}{u^6}\right)\right]\left(1+\frac{k}{u^6}\right)
W_{\pm}^{-1/2},\\ \nn &&I^{3}(r)= \frac{B_{3}^{\pm}}{s^2}\int_{r_0}^{r}
\frac{d u}{u^2} W_{\pm}^{-1/2},\h s\equiv \prod_{l=1}^{6}\sin y_0^{10-l},
\h I(r)=\exp\left(\phi_0/2\right)\int_{r_0}^{r}d u W_{\pm}^{-1/2},\eea
where \bea\nn &&W_{\pm}=\Biggl\{\left[B^{\pm}_{0}+2\lambda^0 TC^1_{\pm}
\exp\left(-\phi_0/2\right)\left(1+\frac{k}{u^6}\right)\right]^2 \\ \nn &&
- \left[B^{\pm}_{1}-2\lambda^0 TC^0_{\pm}\exp\left(-\phi_0/2\right)
\left(1+\frac{k}{u^6}\right)\right]^2 \Biggr\}
\left(1+\frac{k}{u^6}\right)-\left(\frac{B_3^\pm}{s}\right)^2\frac{1}{u^2}
\\ \nn &&+ \left(2\lambda^0T\right)^2\exp\left(\phi_0\right)
\Biggl\{\left[\left(C^0_\pm\right)^2 - \left(C^1_\pm\right)^2\right]
\left(1+\frac{k}{u^6}\right)^{-1}-\left(C_\pm^3 s\right)^2 u^2
\Biggr\}.\eea This is the solution also in the string frame, because we
have one and the same metric in the action expressed in two different
ways.

The above solution extremely simplifies in the tensionless limit $T\to 0$.
Let us give the manifest expressions for this case. For $r_0<r$, they are:
\bea\nn &&\lim_{T\to 0}I^m (r) = \eta^{mn}B_m^{\pm} \left(J^0 +
kJ^6\right),\h \lim_{T\to 0}I^3 (r) = \frac{B_3^\pm}{s^2}J^2,\h \lim_{T\to
0}I(r) = J^0,\eea where \bea\nn &&J^{\beta}(r) =
-\sqrt{\frac{\pi}{\left(B_0^\pm\right)^2 - \left(B_1^\pm\right)^2}}\\ \nn
&&\times \Biggl\{
\frac{1}{r^{\beta-1}}\sum_{n=0}^{\infty}\frac{\Gamma\left(-
\frac{6n+\beta-5}{4}\right)\left(k/r^6\right)^n}{\left(6n+\beta-1\right)
\Gamma\left( \frac{1-2n}{2}\right)\Gamma\left(
-\frac{2n+\beta-5}{4}\right)} P_{n}^{\left(-\frac{6n+\beta-1}{4},
-n-1\right)} \left(1-2\frac{\delta}{k}r^4\right)
 \\ \nn &&-\frac{1}{r_0^{\beta-1}}\sum_{n=0}^{\infty}\frac{\Gamma\left(
\frac{2n+\beta+3}{4}\right)\left(-\delta/r_0^2
\right)^n}{\left(2n+\beta-1\right) \Gamma\left(
\frac{1-2n}{2}\right)\Gamma\left( \frac{6n+\beta+3}{4}\right)}
P_n^{\left(\frac{2n+\beta-1}{4},
-n-1\right)}\left(1-2\frac{k}{\delta}r_0^{-4}\right)\Biggr\},\\ \nn
&&\delta\equiv \frac{\left(B_3^{\pm}/s\right)^2}{\left(B_0^\pm\right)^2 -
\left(B_1^\pm\right)^2},\eea and $P_n^{(\alpha,\beta)}(z)$ are the Jacobi
polynomials. To obtain the solution for $r_0>r$, one has to exchange $r$
and $r_0$ in the expression for $J^{\beta}$.

Now let us turn to the case of a $D$-string living in five dimensional
{\it anti de Sitter} space-time. The corresponding metric may be written
as \bea\nn &&g_{00}=-\left(1+\frac{r^2}{R^2}\right),\h g_{11}=
\left(1+\frac{r^2}{R^2}\right)^{-1}, \\ \nn &&g_{22}=
r^2\sin^2x^3\sin^2x^4,\h g_{33}=  r^2\sin^2x^4,\h g_{44}=r^2,\eea where $K=
-1/R^2$ is the constant curvature.

The exact string solution as a function of $r$ found in \cite{3}
is \bea\nn &&X^{0}\left(r,\s\right)= X^{0}_0 +
C^{0}_{\pm}\left(\lambda^1\tau+\sigma\right) \mp
B^\pm_0\int_{r_0}^{r}d u
\left(1+\frac{u^2}{R^2}\right)^{-1}\left(g_{00}V_D^{\pm0}\right)^{-1/2},
\\ \nn &&X^{2}\left(r,\s\right)= X^{2}_0 +
C^{2}_{\pm}\left(\lambda^1\tau+\sigma\right) \pm
\frac{B^\pm_2}{c^2}\int_{r_0}^{r}\frac{d u}{u^2}
\left(g_{00}V_D^{\pm0}\right)^{-1/2}, \\ \nn &&\tau(r)= \tau_0
\pm\int_{r_0}^{r}d u\left(g_{00}V_D^{\pm0}\right)^{-1/2},\h c\equiv\sin
x_0^3\sin x_0^4,\\ \nn &&F_{01}(r)=-g_sT_D\lambda^2 \Biggl\{\left[\left(
\frac{C_\pm^0}{R}\right)^2 - \left(cC_\pm^2\right)^2 \right]r^2 +
\left(C_\pm^0\right)^2 \Biggr\},\eea where \bea\nn &&g_{00}V_D^{\pm0} =
\left[\left(B_0^\pm\right)^2 - \left(\frac{B^\pm_2}{c R}\right)^2 +
\left(A_\pm C_\pm^0\right)^2\right] - \left( \frac{B^\pm_2}{c}\right)^2
\frac{1}{u^2}\\ \nn &&+A_\pm^2\left[2\left( \frac{C_\pm^0}{R}\right)^2 -
\left(cC_\pm^2\right)^2 \right]u^2 + \left( \frac{A_\pm}{R}\right)^2
\left[\left( \frac{C_\pm^0}{R}\right)^2 - \left(cC_\pm^2\right)^2
\right]u^4.\eea This solution describes a $D$-string evolving in the
subspace ($x^0,x^1,x^2$).

Alternatively, we could fix the coordinates $r=r_0$,
$x^4=x^4_0\equiv\psi_0$ and obtain a solution as a function of the
coordinate $x^3\equiv\theta$. In this case, the result is the following \cite{3}
\bea\nn &&X^{0}\left(\theta,\s\right)= X^{0}_0 +
C^{0}_{\pm}\left(\lambda^1\tau+\sigma\right) \mp \frac{B^\pm_0
\varrho}{g^0_{00}} \int_{\theta_0}^{\theta}d u
\left(-V_D^{\pm0}\right)^{-1/2},
\\ \nn &&X^{2}\left(\theta,\s\right)= X^{2}_0 +
C^{2}_{\pm}\left(\lambda^1\tau+\sigma\right) \pm
\frac{B^\pm_2}{\varrho}\int_{\theta_0}^{\theta}\frac{d u}{\sin^2u}
\left(-V_D^{\pm0}\right)^{-1/2}, \\ \nn &&\tau(\theta)= \tau_0 \pm \varrho
\int_{\theta_0}^{\theta}d u\left(-V_D^{\pm0}\right)^{-1/2},\h
\varrho\equiv r_0\sin\psi_0,\\ \nn &&F_{01}(\theta)=g_sT_D\lambda^2
\left[\left( C_\pm^0\right)^2 g^0_{00} + \left(\varrho C_\pm^2\right)^2
\sin^2\theta\right] ,\eea where \bea\nn -V_D^{\pm0} =
\left[\left(B_0^\pm\right)^2 g_{11}^0 - \left(A_\pm C_\pm^0\right)^2
g^0_{00} \right] -  \left(A_\pm C_\pm^2 \varrho\right)^2\sin^2u -
\frac{\left(B_2^\pm/\varrho\right)^2}{\sin^2u}.\eea This is a solution for
$D$-string placed in the subspace described by the coordinates
($x^0,x^2,x^3$).

Finally, we will give an example of exact solution for a $D$-string moving
in a non-diagonal metric. To this end, let us consider the ten dimensional
black hole solution of \cite{HMS96}. In string frame metric, it can be
written as \cite{M98} \bea\nn &&ds^2 = \left(1+\frac{r_0^2
\sinh^2\alpha}{r^2}\right)^{-1/2} \left(1+\frac{r_0^2
\sinh^2\gamma}{r^2}\right)^{-1/2}\\ \nn &&\times \Biggl\{-dt^2 + (dx^9)^2
+ \frac{r_0^2}{r^2}\left(\cosh\chi d t+\sinh\chi dx^9\right)^2 \\ \nn &&+
\left(1+\frac{r_0^2 \sinh^2\alpha}{r^2}\right)\left[(dx^5)^2 + (dx^6)^2 +
(dx^7)^2 +(dx^8)^2 \right]\Biggr\}\\ \nn &&+ \left(1+\frac{r_0^2
\sinh^2\alpha}{r^2}\right)^{1/2} \left(1+\frac{r_0^2 \sinh^2\gamma}{r^2}
\right)^{1/2}\left[\left(1-\frac{r_0^2}{r^2}\right)^{-1} dr^2 +
r^2d\Omega_3^2 \right],\\ \label{bhs}
&&\exp\left[-2(\phi-\phi_\infty)\right] = \left(1+\frac{r_0^2
\sinh^2\alpha}{r^2}\right)^{-1} \left(1+\frac{r_0^2 \sinh^2\gamma}{r^2}
\right).\eea The equalities (\ref{bhs}) define a solution of type IIB
string theory, which low energy action in Einstein frame contains the
terms \bea\label{leea} \int d^{10}x\sqrt{-g}\left[R -
\frac{1}{2}\left(\nabla\phi\right)^2 - \frac{1}{12}\exp(\phi)H'^{2}
\right],\eea where $H'$ is the Ramond-Ramond three-form field strength. The
Neveu-Schwarz 3-form field strength, the selfdual 5-form field strength
and the second scalar are set to zero.
After some simplifications \cite{3}, the exact $D$-string solution as a function of the
radial coordinate $r$ reads \bea\nn &&X^{\mu}\left(r,\s\right)= X^{\mu}_0 +
C^{\mu}_{\pm}\left(\lambda^1\tau+\sigma\right) \pm I^{\mu}(r),\h \mu =
0,2,5,6,7,8,9, \\ \nn &&X^{3,4}= X_0^{3,4}=constants,\h \tau(r)=\tau_0 \pm
I(r),\eea where \bea\nn &&I^0 = -\int_{r_0}^{r}d u \left[B_0^\pm g_{99} -
B_9^\pm g_{09} \right]g_{11}\mathcal{W}^{-1/2},\\ \nn &&I^9 =
\int_{r_0}^{r}d u \left[B_0^\pm g_{09} - B_9^\pm g_{00}
\right]g_{11}\mathcal{W}^{-1/2},\\ \nn &&I^l= B_l^\pm
\int_{r_0}^{r}\frac{d u}{g_{ll}^0}\mathcal{W}^{-1/2},\h l=2,5,6,7,8,\h
I=\int_{r_0}^{r}d u \mathcal{W}^{-1/2},\\ \nn &&F_{01} = g_s T_D \lambda^2
C^\mu_\pm C^\nu_\pm g_{\mu\nu}^0 (r),\\ \nn &&\mathcal{W}=
\left[\left(B_9^\pm\right)^2 - \frac{\left(A_\pm C_\pm^0\right)^2}{g_{11}}
\right]  g_{00} + \left[\left(B_0^\pm\right)^2 - \frac{\left(A_\pm
C_\pm^9\right)^2}{g_{11}}\right]g_{99} \\ \nn &&- 2\left(B_0^\pm B_9^\pm +
\frac{A_\pm^2 C_\pm^0C_\pm^9}{g_{11}} \right)g_{09} \\ \nn
&&-\frac{1}{g_{11}}\Biggl\{ \sum_{l=5}^{8} \left[\left(A_\pm
C_\pm^l\right)^2 + \left(B_\pm^l\right)^2\right] + \left(
\frac{B_\pm^2}{c}\right)^2 \frac{1}{u^2} + \left(A_\pm C_\pm^2 c\right)^2
u^2 \Biggr\}. \eea

All previous considerations are based on the ansatz (\ref{ogac}).
However, there exist other embeddings which allow for
simplification of the dynamics and as a consequence for obtaining
exact solutions of the equations of motion and constraints. An
example is the following one \bea\label{sLA}
X^\mu(\xi^m)=\Lambda^\mu_m \xi^m +
Z^\mu(\alpha\sigma+\beta\tau),\h
X^a(\xi^m)=Z^a(\alpha\sigma+\beta\tau),\h \alpha, \beta
=constants,\eea where $\tau=\xi^0$, $\sigma$ is one of the
worldvolume spatial coordinates $\xi^{i}$,  and $Z^\mu$, $Z^a$ are
arbitrary functions. This ansatz will be used further on.

\setcounter{equation}{0}
\section{AdS/CFT}
The AdS/CFT duality \cite{AdS/CFT} between string/M-theory on
curved space-times with Anti-de Sitter subspaces and conformal
field theories in different dimensions has been actively
investigated in the last years. A lot of impressive progresses
have been made in this field of research based mainly on the
integrability structures discovered on both sides of the
correspondence. The most studied example is the duality between
type IIB string theory on ${\rm AdS}_5\times S^5$ target space and
the ${\cal N}=4$ super Yang-Mills theory (SYM) in four space-time
dimensions. However, many other cases are also of interest, and
have been investigated intensively (for recent review on the
AdS/CFT duality, see \cite{RO}).

Different classical string/M-theory solutions play important role
in checking and understanding the AdS/CFT correspondence
\cite{AAT1012}. To establish relations with the dual gauge theory,
one has to take the semiclassical limit of {\it large} conserved
charges \cite{GKP02}.

An interesting issue to solve is to find the {\it finite-size
effects}, related to the wrapping interactions in the dual field
theory \cite{Janikii}.

\subsection{Classical string solutions and string/field theory duality}

In \cite{5} and \cite{6} the string dynamics in general string
theory target space-times is considered by using the Polyakov
action (\ref{pa}) (for strings $p=1$). Exact solutions of the
equations of motion and Virasoro constraints are found. This is
done in the covariant worldsheet gauge $\gamma^{mn}=constants$.
The considerations in \cite{5} are based on the ansatz
(\ref{ogac}), while in \cite{6} a particular case of the ansatz
(\ref{sLA}) is used, corresponding to $\alpha=1,\ \beta=0$. Then,
the general results are applied for several string backgrounds
having dual field theory description. Namely, $AdS_5 \times S^5$
with field theory dual ${\cal N}=4$ SYM, $AdS_5$ black hole with
field theory dual {\it finite temperature} ${\cal N}=4$ SYM, and
$AdS_3\times S^3\times\mathcal{M}$, with NS-NS 2-form gauge field.
In accordance with the $AdS/CFT$ duality, the string theory on
$AdS_3\times S^3\times\mathcal{M}$ is dual to a superconformal
field theory on a cylinder, which is the boundary of $AdS_3$ in
global coordinates.

Analogous considerations have been made in \cite{27}. The
difference is that the most general form of the embedding
(\ref{sLA}) is applied for strings. Let us describe the general
results obtained there.

In what follow we will  use {\it conformal gauge}
$\gamma^{mn}=\eta^{mn}=diag(-1,1)$ in which the string Lagrangian,
the Virasoro constraints and the equations of motion take the
following form: \bea\label{CG} &&\mathcal{L}
=\frac{T}{2}\left(G_{00}-G_{11}+2 B_{01}\right),\\ \nn
&&G_{00}+G_{11}=0,\h G_{01}=0, \\ \nn &&
g_{LK}\left[\left(\p_0^2-\p_1^2\right)X^K+ \Gamma_{MN}^K\left(\p_0
X^M \p_0 X^N-\p_1 X^M \p_1
X^N\right)\right]=H_{LMN}\p_0X^M\p_1X^N.\eea

Now, let us {\it suppose} that there exist some number of
commuting Killing vector fields along part of $X^M$ coordinates
and split $X^M$ into two parts \bea\nn X^M=(X^\mu,X^a),\eea where
$X^\mu$ are the isometric coordinates, while $X^a$ are the
non-isometric ones. The existence of isometric coordinates leads
to the following conditions on the background fields:
\bea\label{cbf} \p_\mu g_{MN}=0,\h \p_\mu b_{MN}=0.\eea Then from
the string action, we can compute the conserved charges
\bea\label{CC} Q_\mu=\int d\sigma \frac{\p \mathcal{L}}{\p(\p_0
X^\mu)}\eea under the above conditions.

Next, we introduce the following ansatz for the string embedding
\bea\label{A} X^{\mu}(\tau,\sigma)=\Lambda^{\mu}\tau
+\tilde{X}^{\mu}(\alpha\sigma+\beta\tau),\h
X^{a}(\tau,\sigma)=\tilde{X}^{a}(\alpha\sigma+\beta\tau),\eea
where $\Lambda^{\mu}$, $\alpha$, $\beta$ are arbitrary parameters.
Further on, we will use the notation $\xi=\alpha\sigma+\beta\tau
$. Applying this ansatz, one can find that the equalities
(\ref{CG}), (\ref{CC}) become \bea\label{LA} \mathcal{L}=
\frac{T}{2}\Big[-(\alpha^2-\beta^2)g_{MN}\frac{d\tilde{X}^M}{d\xi}\frac{d\tilde{X}^N}{d\xi}
+2\Lambda^\mu\left(\beta g_{\mu N}+\alpha b_{\mu N}\right)
\frac{d\tilde{X}^N}{d\xi} +\Lambda^\mu\Lambda^\nu g_{\mu
\nu}\Big],\eea

\bea\label{V1}
G_{00}+G_{11}=(\alpha^2+\beta^2)g_{MN}\frac{d\tilde{X}^M}{d\xi}\frac{d\tilde{X}^N}{d\xi}
+2\beta\Lambda^\mu g_{\mu N} \frac{d\tilde{X}^N}{d\xi}
+\Lambda^\mu\Lambda^\nu g_{\mu \nu} =0,\eea

\bea\label{V2}  G_{01}&=&\alpha\beta
g_{MN}\frac{d\tilde{X}^M}{d\xi}\frac{d\tilde{X}^N}{d\xi}+\alpha\Lambda^\mu
g_{\mu N}\frac{d\tilde{X}^N}{d\xi} =0,\eea

\bea\nn
&&-(\alpha^2-\beta^2)\left[g_{LK}\frac{d^2\tilde{X}^K}{d\xi^2}+
\Gamma_{L,MN}\frac{d\tilde{X}^M}{d\xi}\frac{d\tilde{X}^N}{d\xi}\right]
+2\beta\Lambda^\mu\Gamma_{L,\mu N}\frac{d\tilde{X}^N}{d\xi}
+\Lambda^\mu\Lambda^\nu \Gamma_{L,\mu\nu}
\\ \label{EM}
&&= \alpha\Lambda^\mu H_{L\mu N}\frac{d\tilde{X}^N}{d\xi},\eea

\bea\label{Q} &&Q_{\mu}= \frac{T}{\alpha}\int
d\xi\left[\left(\beta g_{\mu N}+\alpha b_{\mu
N}\right)\frac{d\tilde{X}^N}{d\xi}+\Lambda^\nu g_{\mu
\nu}\right].\eea

Our next task is to try to solve the equations of motion
(\ref{EM}) for the isometric coordinates, i.e. for $L=\lambda$.
Due to the conditions (\ref{cbf}) imposed on the background
fields, we obtain that \bea\nn &&\Gamma_{\lambda,a
b}=\frac{1}{2}\left(\p_a g_{b \lambda}+\p_b
 g_{a \lambda}\right),\h  \Gamma_{\lambda,\mu a}=\frac{1}{2}\p_a g_{\mu\lambda}\h
 \Gamma_{\lambda,\mu \nu}=0,
 \\ \nn && H_{\lambda a b}=\p_a b_{b \lambda}+\p_b b_{\lambda a}, \h
 H_{\lambda \mu a}=\p_a b_{\lambda\mu}, \h  H_{\lambda \mu \nu}=0.
 \eea
By using this, one can find the following first integrals for
$\tilde{X}^\mu$: \bea\label{FIM} \frac{d\tilde{X}^{\mu}}{d\xi}=
\frac{1}{\alpha^2-\beta^2}\left[g^{\mu\nu}
\left(C_\nu-\alpha\Lambda^\rho
b_{\nu\rho}\right)+\beta\Lambda^\mu\right] -g^{\mu\nu}g_{\nu
a}\frac{d\tilde{X}^{a}}{d\xi},\eea where $C_\nu$ are arbitrary
integration constants. Therefore, according to our ansatz
(\ref{A}), the solutions for the string coordinates $X^\mu$ can be
written as \bea\label{MS} X^{\mu}(\tau,\sigma)=\Lambda^{\mu}\tau
+\frac{1}{\alpha^2-\beta^2}\int d\xi \left[g^{\mu\nu}
\left(C_\nu-\alpha\Lambda^\rho
b_{\nu\rho}\right)+\beta\Lambda^\mu\right] -\int g^{\mu\nu}g_{\nu
a}d\tilde{X}^{a}(\xi).\eea

Now, let us turn to the remaining equations of motion
corresponding to $L=a$, where \bea\nn &&\Gamma_{a,\mu
b}=-\frac{1}{2}(\p_a g_{b \mu}-\p_b g_{a \mu}),\h \Gamma_{a,\mu
\nu}=-\frac{1}{2}\p_a g_{\mu\nu},
\\ \nn &&H_{a \mu\nu}=\p_a b_{\mu\nu},\h H_{a \mu b}=-\p_a b_{b \mu}+\p_b b_{a \mu}.\eea
Taking this into account and replacing the first integrals for
$\tilde{X}^{\mu}$ already found, one can write these equations in
the form (prime is used for $d/d\xi$) \bea\label{Ea}
(\alpha^2-\beta^2)\left[h_{a
b}\tilde{X}^{b''}+\Gamma^{h}_{a,bc}\tilde{X}^{b'}\tilde{X}^{c'}\right]
= 2\p_{[a} A_{b]}\tilde{X}^{b'}-\p_a U, \eea where \bea\label{Ma}
&&h_{a b}= g_{a b}-g_{a \mu}g^{\mu\nu}g_{\nu b},\h
\Gamma^{h}_{a,bc}=\frac{1}{2}\left(\p_b h_{ca}+\p_c h_{ba}-\p_a
h_{bc}\right)
\\ \label{VPa} &&A_a= g_{a\mu}g^{\mu\nu}
\left(C_\nu-\alpha\Lambda^\rho
b_{\nu\rho}\right)+\alpha\Lambda^\mu b_{a\mu},
\\ \label{SP} &&U=
\frac{1/2}{\alpha^2-\beta^2}\left[\left(C_\mu-\alpha\Lambda^\rho
b_{\mu\rho}\right)g^{\mu\nu} \left(C_\nu-\alpha\Lambda^\lambda
b_{\nu\lambda}\right)+\alpha^2\Lambda^\mu\Lambda^\nu
g_{\mu\nu}\right].\eea

One can show that the above equations for $\tilde{X}^{a}$ can be
derived from the effective Lagrangian \bea\nn
\mathcal{L}^{eff}(\xi) =\frac{1}{2}(\alpha^2-\beta^2) h_{a
b}\tilde{X}^{a'}\tilde{X}^{b'}+A_a \tilde{X}^{a'}-U .\eea The
corresponding effective Hamiltonian is \bea\nn
\mathcal{H}^{eff}(\xi) =\frac{1}{2}(\alpha^2-\beta^2) h_{a
b}\tilde{X}^{a'}\tilde{X}^{b'}+U ,\eea or in terms of the momenta
$p_a$  conjugated to $\tilde{X}^a$ \bea\nn \mathcal{H}^{eff}(\xi)
=\frac{1}{2}(\alpha^2-\beta^2) h^{a
b}\left(p_a-A_a\right)\left(p_b-A_b\right)+U .\eea

The Virasoro constraints (\ref{V1}), (\ref{V2}) become:
\bea\label{V12} \frac{1}{2}(\alpha^2-\beta^2) h_{a
b}\tilde{X}^{a'}\tilde{X}^{b'}+U=0,\h \alpha\Lambda^\mu C_\mu
=0.\eea

Finally, let us write down the expressions for the conserved
charges (\ref{Q}) \bea\nn Q_\mu &=&\frac{T}{\alpha^2-\beta^2}\int
d\xi \left[\frac{\beta}{\alpha}C_\mu+\alpha\Lambda^\nu g_{\mu\nu}
+ b_{\mu\nu}g^{\nu\rho} \left(C_\rho-\alpha\Lambda^\lambda
b_{\rho\lambda}\right)\right. \\ \label{Qmu}
&+&\left.(\alpha^2-\beta^2) \left(b_{\mu
a}-b_{\mu\nu}g^{\nu\rho}g_{\rho
a}\right)\tilde{X}^{a'}\right].\eea

\subsubsection{Rotating strings in type IIA reduction of
M-theory\\ on $G_2$ manifold and their semiclassical limits}
The type IIA background, in which we will search for rotating
string solutions, has the form \cite{HN0210} \bea\nn ds_{10}^2 &=&
r_0^{1/2}C \left\{-(dx^0)^2 + \delta_{IJ}dx^I dx^J + A^2 \left[
(g^1)^2 + (g^2)^2 \right]\right.
\\ \nn &+& \left. B^2 \left[ (g^3)^2
+ (g^4)^2 \right] + D^2 (g^5)^2 \right\} +
r_0^{1/2}\frac{dr^2}{C},\h (I,J=1,2,3),\h r_0=const,
\\ \label{10db} &&e^\Phi = r_0^{3/4}C^{3/2},\h
F_2 = \sin \theta_1 d\phi_1 \wedge d\theta_1 - \sin \theta_2
d\phi_2 \wedge d\theta_2.\eea Here, $g^1$,...,$g^5$ are given by
\bea\nn
&&g^1=-\sin\theta_1d\phi_1-\cos\psi_1\sin\theta_2d\phi_2+\sin\psi_1
d\theta_2,\\ \nn &&g^2=d\theta_1-\sin\psi_1\sin\theta_2
d\phi_2-\cos\psi_1 d\theta_2,\\ \nn
&&g^3=-\sin\theta_1d\phi_1+\cos\psi_1\sin\theta_2d\phi_2-\sin\psi_1
d\theta_2,\\ \nn &&g^4=d\theta_1+\sin\psi_1\sin\theta_2
d\phi_2+\cos\psi_1 d\theta_2,\\ \nn &&g^5=d\psi_1+\cos\theta_1
d\phi_1+\cos\theta_2 d\phi_2,\eea and the functions $A$, $B$, $C$
and $D$ depend on the radial coordinate $r$ only: \bea\nn
A&=&\frac{1}{\sqrt{12}} \sqrt{(r - 3 r_0/2)(r + 9 r_0/2)},\h
B=\frac{1}{\sqrt{12}} \sqrt{(r + 3 r_0/2)(r - 9 r_0/2)},
\\ \label{G2-3} C&=&\sqrt{\frac{(r - 9 r_0/2)(r + 9 r_0/2)}{(r - 3 r_0/2)(r + 3 r_0/2)}},\h D=r/3.\eea
In (\ref{10db}), $\Phi$ and $F_2$ are the Type IIA dilaton and the
field strength of the Ramond-Ramond one-form gauge field
respectively.

The above ten dimensional background arises as dimensional
reduction of M-theory on a $G_2$ manifold with field theory dual
four dimensional $\mathcal{N}=1$ SYM.

The type IIA solution (\ref{10db}) describes a D6-brane wrapping
the ${\bf S}^3$ in the deformed conifold geometry. For
$r\to\infty$, the metric becomes that of a singular conifold, the
dilaton is constant, and the flux is through the ${\bf S}^2$
surrounding the wrapped D6-brane. For $r - 9r_0/2 = \epsilon \to
0$, the string coupling $e^\Phi$ goes to zero like
$\epsilon^{3/4}$, whereas the curvature blows up as $\epsilon^{-
{3/2}}$ just like in the near horizon region of a flat D6-brane.
This means that classical supergravity is valid for sufficiently
large radius. However, the singularity in the interior is the same
as the one of flat D6 branes, as expected. On the other hand, the
dilaton continuously decreases from a finite value at infinity to
zero, so that for small $r_0$ classical string theory is valid
everywhere. As explained in \cite{NPB0106034}, the global geometry
is that of a warped product of flat Minkowski space and a
non-compact space, $Y_6$, which for large radius is simply the
conifold since the backreaction of the wrapped D6 brane becomes
less and less important. However, in the interior, the
backreaction induces changes on $Y_6$ away from the conifold
geometry. For $r \to 9 r_0/2$, the  ${\bf S}^2$ shrinks to zero
size, whereas an ${\bf S}^3$ of finite size remains. This behavior
is similar to that of the deformed conifold but the two metrics
are different.

Three types of rotating string solutions have been found in
\cite{10}.

The first one is given by ($\Delta r=r-3l$, $\Delta r_1=r_1-3l$)
\bea\label{ssG1} &&\xi^1(r)=
\frac{8}{\left(\Lambda_+^2+\Lambda_-^2\right)^{1/2}}\left[\frac{l\Delta
r}{(3l-r_2)\Delta r_1}\right]^{1/2}\times
\\ \nn &&F_D^{(5)}\left(1/2;-1/2,-1/2,1/2,1/2,1/2;3/2;
-\frac{\Delta r}{2l},-\frac{\Delta r}{4l},-\frac{\Delta
r}{6l},-\frac{\Delta r}{3l-r_2}, \frac{\Delta r}{\Delta
r_1}\right).\eea

Now, we can compute the conserved momenta on the obtained
solution. They are: \bea\label{EP1}
\frac{E}{\Lambda_0^0}=\frac{P_I}{\Lambda_0^I}&=&
T\left[\frac{2^7l\Delta r_1}{\left(\Lambda_+^2+
\Lambda_-^2\right)(3l-r_2)}\right]^{1/2}\left(1+\frac{\Delta
r_1}{3l-r_2}\right)^{-1/2}
\\ \nn &&\times F_D^{(1)}\left(1/2;1/2;3/2;\frac{1}{1+\frac{3l-r_2}{\Delta
r_1}}\right),\eea
\bea\label{pt12}
&&P_{\theta_1}=\left(\Lambda_0^{\theta_1}-\Lambda_0^{\theta_2}\cos\psi_1^0\right)I_A
+\left(\Lambda_0^{\theta_1}+\Lambda_0^{\theta_2}\cos\psi_1^0\right)I_B,\\
\nn
&&P_{\theta_2}=\left(\Lambda_0^{\theta_2}-\Lambda_0^{\theta_1}\cos\psi_1^0\right)I_A
+\left(\Lambda_0^{\theta_2}+\Lambda_0^{\theta_1}\cos\psi_1^0\right)I_B,\eea
where \bea\label{IA} I_A&=&T\left[\frac{2^7l^5\Delta
r_1}{\left(\Lambda_+^2+
\Lambda_-^2\right)(3l-r_2)}\right]^{1/2}\left(1+\frac{\Delta
r_1}{2l}\right)\left(1+\frac{\Delta r_1}{6l}\right)
\left(1+\frac{\Delta r_1}{3l-r_2}\right)^{-1/2}
\\ \nn &&\times F_D^{(3)}\left(1/2;-1,-1,1/2;3/2;\frac{1}{1+\frac{2l}{\Delta
r_1}},\frac{1}{1+\frac{6l}{\Delta
r_1}},\frac{1}{1+\frac{3l-r_2}{\Delta r_1}}\right),\eea
\bea\label{IB} I_B&=&\frac{T}{9}\left[\frac{2^9\left(l\Delta
r_1\right)^3}{\left(\Lambda_+^2+
\Lambda_-^2\right)(3l-r_2)}\right]^{1/2}\left(1+\frac{\Delta
r_1}{4l}\right) \left(1+\frac{\Delta r_1}{3l-r_2}\right)^{-1/2}
\\ \nn &&\times F_D^{(2)}\left(1/2;-1,1/2;5/2;\frac{1}{1+\frac{4l}{\Delta
r_1}},\frac{1}{1+\frac{3l-r_2}{\Delta r_1}}\right).\eea

Our next task is to find the relation between the energy $E$ and
the other conserved quantities $P_I$, $P_{\theta_1}$,
$P_{\theta_2}$, in the semiclassical limit (large conserved
charges), which corresponds to $r_1\to\infty$. In this limit,
\bea\nn \frac{E}{\Lambda_0^0}=\frac{P_I}{\Lambda_0^I}= \frac{\pi
T\left(2^3l\right)^{1/2}}{\left(\Lambda_+^2+\Lambda_-^2\right)^{1/2}},
\h I_A=I_B=\frac{\pi
T\left(2l\right)^{1/2}v_0^2}{\left(\Lambda_+^2+\Lambda_-^2\right)^{3/2}},
\eea which leads to the following energy-charge
relation
\bea\label{EG1} E^2=\mathbf{P}^2+2\pi T
\left(6r_0\right)^{1/2}\left(P_{\theta_1}^2+P_{\theta_2}^2\right)^{1/2},\h
\mathbf{P}^2=\delta_{IJ}P_IP_J.\eea

The second type of rotating string solution can be written as
\bea\label{ssG3'} &&\xi^1(r)=
\frac{8}{\left(\bar{\Lambda}_+^2+\bar{\Lambda}_-^2
+4\Lambda_D^2/3\right)^{1/2}}\left[\frac{l\Delta r}{(3l-r_2)\Delta
r_1}\right]^{1/2}\times
\\ \nn &&F_D^{(5)}\left(1/2;-1/2,-1/2,1/2,1/2,1/2;3/2;
-\frac{\Delta r}{2l},-\frac{\Delta r}{4l},-\frac{\Delta
r}{6l},-\frac{\Delta r}{3l-r_2}, \frac{\Delta r}{\Delta
r_1}\right).\eea For $E$ and $P_I$ we have
\bea\label{EP2}
\frac{E}{\Lambda_0^0}=\frac{P_I}{\Lambda_0^I}&=&
8T\left[\frac{2l\Delta
r_1}{\left(\bar{\Lambda}_+^2+\bar{\Lambda}_-^2
+4\Lambda_D^2/3\right)(3l-r_2)}\right]^{1/2}\left(1+\frac{\Delta
r_1}{3l-r_2}\right)^{-1/2}
\\ \nn &&\times F_D^{(1)}\left(1/2;1/2;3/2;\frac{1}{1+\frac{3l-r_2}{\Delta
r_1}}\right).\eea For the conserved angular momenta $P_\theta$ and
$P_\phi$ one finds \bea\label{am12}
P_{\theta}&=&\left(\Lambda_0^{\theta}-\Lambda_0^{\phi}\sin\psi_1^0\sin\theta_2^0\right)J_A
+\left(\Lambda_0^{\theta}+\Lambda_0^{\phi}\sin\psi_1^0\sin\theta_2^0\right)J_B,
\\ \nn
P_{\phi}&=&\left(\Lambda_0^{\phi}\sin\theta_2^0-\Lambda_0^{\theta}\sin\psi_1^0\right)\sin\theta_2^0J_A
+\left(\Lambda_0^{\phi}\sin\theta_2^0+\Lambda_0^{\theta}\sin\psi_1^0\right)\sin\theta_2^0J_B
\\ \nn &&+\Lambda_0^{\phi}\cos^2\theta_2^0 J_D, \eea where
\bea\label{JA} J_A &=& 8T\left[\frac{2l^5\Delta
r_1}{\left(\bar{\Lambda}_+^2+\bar{\Lambda}_-^2
+4\Lambda_D^2/3\right)(3l-r_2)}\right]^{1/2}\left(1+\frac{\Delta
r_1}{2l}\right)\left(1+\frac{\Delta r_1}{6l}\right)
\\ \nn &&\times\left(1+\frac{\Delta r_1}{3l-r_2}\right)^{-1/2}
F_D^{(3)}\left(1/2;-1,-1,1/2;3/2;\frac{1}{1+\frac{2l}{\Delta
r_1}},\frac{1}{1+\frac{6l}{\Delta
r_1}},\frac{1}{1+\frac{3l-r_2}{\Delta r_1}}\right),\eea
\bea\nn
J_B&=& \frac{16}{9}T\left[\frac{2(l\Delta
r_1)^3}{\left(\bar{\Lambda}_+^2+\bar{\Lambda}_-^2
+4\Lambda_D^2/3\right)(3l-r_2)}\right]^{1/2}\left(1+\frac{\Delta
r_1}{4l}\right)\left(1+\frac{\Delta r_1}{3l-r_2}\right)^{-1/2}
\\ \label{JB} &&\times F_D^{(2)}\left(1/2;-1,1/2;5/2;\frac{1}{1+\frac{4l}{\Delta
r_1}},\frac{1}{1+\frac{3l-r_2}{\Delta r_1}}\right),\eea
\bea\nn
J_D &=& 8T\left[\frac{2l^5\Delta
r_1}{\left(\bar{\Lambda}_+^2+\bar{\Lambda}_-^2
+4\Lambda_D^2/3\right)(3l-r_2)}\right]^{1/2} \left(1+\frac{\Delta
r_1}{3l}\right)\left(1+\frac{\Delta r_1}{3l-r_2}\right)^{-1/2}
\\ \label{JD} &&\times F_D^{(2)}\left(1/2;-2,1/2;3/2;\frac{1}{1+\frac{3l}{\Delta
r_1}},\frac{1}{1+\frac{3l-r_2}{\Delta r_1}}\right).\eea

In the semiclassical limit $r_1\to\infty$, one gets the following
dependence of the energy on the charges $P_I$, $P_{\theta}$ and
$P_{\phi}$ \bea\label{EG3'} E^2=\mathbf{P}^2+2\pi T
\left(6r_0\right)^{1/2}\left(P_{\theta}^2+
\frac{3P_{\phi}^2}{3-\cos^2\theta_2^0}\right)^{1/2}.\eea

The third solution and the conserved charges can be computed in
the same way. These computations lead to the following
energy-charge relation after taking the semiclassical limit
\bea\label{EG2'} E^2&=&\mathbf{P}^2+2\pi T
\left(\frac{6r_0}{\Delta}\right)^{1/2}\times\\ \nn
&&\left[\left(3-\cos^2\theta_2^0\right)P_{\phi_1}^2+
\left(3-\cos^2\theta_1^0\right)P_{\phi_2}^2
-4P_{\phi_1}P_{\phi_2}\cos\theta_1^0\cos\theta_2^0\right]^{1/2},\eea
where
\bea\nn \Delta= 3-\cos^2\theta_1^0-
\cos^2\theta_2^0-\cos^2\theta_1^0\cos^2\theta_2^0.\eea

\subsubsection{Strings in $AdS_5\times S^5$, integrable systems
and \\finite-size effects for single spikes}

We use the reduction of the string dynamics on $R_t\times S^3$
subspace of $AdS_5\times S^5$ to the Neumann-Rosochatius (NR)
integrable system to map {\it all} string solutions described by
this dynamical system onto solutions of the complex sine-Gordon
(CSG) integrable model. This mapping relates the parameters in the
solutions on both sides of the correspondence. Then, we find
finite-size string solutions, their images in the (complex)
sine-Gordon (SG) system, and the leading finite-size effects of
the single spike ``$E-\Delta\varphi$'' relation for both
$R_t\times S^2$ and $R_t\times S^3$ cases \cite{14}.

{\bf Strings on $R_t\times S^3$ and the NR integrable system}

We choose to work in {\it conformal gauge} in which the string
Lagrangian and the Virasoro constraints take the form (\ref{CG})
(in $AdS_5\times S^5$ the 2-form $B$-field is zero) \bea\label{l}
&&\mathcal{L}_s=\frac{T}{2}\left(G_{00}-G_{11}\right) \\
\label{00} && G_{00}+G_{11}=0,\qquad G_{01}=0.\eea

We embed the string in $R_t\times S^3$ subspace of $AdS_5\times
S^5$ as follows \bea\nn Z_0=Re^{it(\tau,\sigma)},\h
W_j=Rr_j(\tau,\sigma)e^{i\varphi_j(\tau,\sigma)},\h
\sum_{j=1}^{2}W_j\bar{W}_j=R^2,\eea where $R$ is the common radius
of $AdS_5$ and $S^5$, and $t$ is the $AdS$ time. For this
embedding, the metric induced on the string worldsheet is given by
\bea\nn G_{ab}=-\p_{(a}Z_0\p_{b)}\bar{Z}_0
+\sum_{j=1}^{2}\p_{(a}W_j\p_{b)}\bar{W}_j=R^2\left[-\p_at\p_bt +
\sum_{j=1}^{2}\left(\p_ar_j\p_br_j +
r_j^2\p_a\varphi_j\p_b\varphi_j\right)\right].\eea The
corresponding string Lagrangian becomes \bea\nn
\mathcal{L}=\mathcal{L}_s +
\Lambda_s\left(\sum_{j=1}^{2}r_j^2-1\right),\eea where $\Lambda_s$
is a Lagrange multiplier. In the case at hand, the background
metric does not depend on $t$ and $\varphi_j$. Therefore, the
conserved quantities are the string energy $E_s$ and two angular
momenta $J_j$, given by \bea\label{gcqs} E_s=-\int
d\sigma\frac{\p\mathcal{L}_s}{\p(\p_0 t)},\h J_j=\int
d\sigma\frac{\p\mathcal{L}_s}{\p(\p_0\varphi_j)}.\eea

It is known that restricting ourselves to the case \bea\label{NRA}
&&t(\tau,\sigma)=\kappa\tau,\h r_j(\tau,\sigma)=r_j(\xi),\h
\varphi_j(\tau,\sigma)=\omega_j\tau+f_j(\xi),\\ \nn
&&\xi=\alpha\sigma+\beta\tau,\h \kappa, \omega_j, \alpha,
\beta=constants,\eea reduces the problem to solving the NR
integrable system \cite{KRT06}. For the case under consideration,
the NR Lagrangian reads (prime is used for $d/d\xi$)
\bea\label{LNR} L_{NR}=(\alpha^2-\beta^2)
\sum_{j=1}^{2}\left[r_j'^2-\frac{1}{(\alpha^2-\beta^2)^2}
\left(\frac{C_j^2}{r_j^2} + \alpha^2\omega_j^2r_j^2\right)\right]
+\Lambda_s\left(\sum_{j=1}^{2}r_j^2-1\right),\eea where the
parameters $C_j$ are integration constants after single time
integration of the equations of motion for $f_j(\xi)$: \bea
f_j'=\frac{1}{\alpha^2-\beta^2}\left(\frac{C_j}{r_j^2} +
\beta\omega_j\right).\label{fjprime} \eea The constraints
(\ref{00}) give the conserved Hamiltonian $H_{NR}$ and a relation
between the embedding parameters and the arbitrary constants
$C_j$: \bea\label{HNR} &&H_{NR}=(\alpha^2-\beta^2)
\sum_{j=1}^{2}\left[r_j'^2+\frac{1}{(\alpha^2-\beta^2)^2}
\left(\frac{C_j^2}{r_j^2} + \alpha^2\omega_j^2r_j^2\right)\right]
=\frac{\alpha^2+\beta^2}{\alpha^2-\beta^2}\kappa^2,
\\ \label{01R} &&\sum_{j=1}^{2}C_j\omega_j + \beta\kappa^2=0.\eea
For closed strings, $r_j$ and $f_j$ satisfy the following
periodicity conditions \bea r_j(\xi+2\pi\alpha)=r_j(\xi),\h
f_j(\xi+2\pi\alpha)=f_j(\xi)+2\pi n_\alpha,\label{pbc} \eea where
$n_\alpha$ are integer winding numbers. On the ansatz (\ref{NRA}),
$E_s$ and $J_j$ introduced in (\ref{gcqs}) take the form
\bea\label{cqs} E_s=
\frac{\sqrt{\lambda}}{2\pi}\frac{\kappa}{\alpha}\int d\xi,\h J_j=
\frac{\sqrt{\lambda}}{2\pi}\frac{1}{\alpha^2-\beta^2}\int d\xi
\left(\frac{\beta}{\alpha}C_j+\alpha\omega_j r_j^2\right),\eea
where we have used that the string tension and the 't Hooft
coupling constant $\lambda$ are related by
$TR^2=\frac{\sqrt{\lambda}}{2\pi}$.

In order to identically satisfy the embedding condition \bea\nn
\sum_{j=1}^{2}r_j^2-1=0,\eea we introduce a new variable
$\theta(\xi)$ by \bea r_1(\xi)=\sin{\theta(\xi)},\h
r_2(\xi)=\cos{\theta(\xi)}.\label{sincos}\eea Then, Eq.(\ref{HNR})
leads to \bea\label{tsol}
\theta'(\xi)&=&\pm\frac{1}{\alpha^2-\beta^2}
\left[(\alpha^2+\beta^2)\kappa^2 - \frac{C_1^2}{\sin^2{\theta}} -
\frac{C_2^2}{\cos^2{\theta}} -
\alpha^2\left(\omega_1^2\sin^2{\theta}
+\omega_2^2\cos^2{\theta}\right)\right]^{1/2}\\
&\equiv& \pm\frac{1}{\alpha^2-\beta^2}\ \Theta(\theta),\nonumber
\eea which can be integrated to give \bea \xi(\theta)=
\pm(\alpha^2-\beta^2)\int\frac{d\theta}{\Theta(\theta)}. \eea From
Eqs.(\ref{fjprime}) and (\ref{sincos}), we can obtain
\bea\label{f1s} &&f_1=\frac{\beta\omega_1\xi}{\alpha^2-\beta^2}\pm
C_1\int\frac{d\theta}{\sin^2\theta\ \Theta(\theta)},\\ \label{f2s}
&&f_2=\frac{\beta\omega_2\xi}{\alpha^2-\beta^2}\pm
C_2\int\frac{d\theta}{\cos^2\theta\ \Theta(\theta)}. \eea Let us
also point out that the solutions for $\xi(\theta)$ and $f_j$ must
satisfy the conditions (\ref{01R}) and (\ref{pbc}). All these
solve formally the NR system for the present case.

{\bf Relationship between the NR and CSG integrable systems}

Due to Pohlmeyer \cite{P76}, we know that the string dynamics on
$R_t\times S^3$ can be described by the CSG equation. Here, we
derive the relation between the solutions of the two integrable
systems - NR and CSG.

The CSG system is defined by the Lagrangian \bea\nn
\mathcal{L}(\psi) =
\frac{\eta^{ab}\p_a\bar{\psi}\p_b\psi}{1-\bar{\psi}\psi} +
M^2\bar{\psi}\psi \eea which give the equation of motion \bea\nn
\p_a\p^a\psi +\bar{\psi}\frac{\p_a\psi\p^a\psi}{1-\bar{\psi}\psi}
- M^2(1-\bar{\psi}\psi)\psi=0.\eea If we represent $\psi$ in the
form \bea\nn \psi=\sin(\phi/2)\exp(i\chi/2),\eea the Lagrangian
can be expressed as \bea\nn
\mathcal{L}(\phi,\chi)=\frac{1}{4}\left[\p_a\phi\p^a\phi +
\tan^2(\phi/2)\p_a\chi\p^a\chi + (2M)^2\sin^2(\phi/2)\right],\eea
along with the equations of motion \bea\label{fem} &&\p_a\p^a\phi
- \frac{1}{2}\frac{\sin(\phi/2)}{\cos^3(\phi/2)}\p_a\chi\p^a\chi -
M^2\sin\phi=0,\\ \label{kem} &&\p_a\p^a\chi +
\frac{2}{\sin\phi}\p_a\phi\p^a\chi=0.\eea The SG system
corresponds to a particular case of $\chi=0$.

To relate the NR system with the CSG integrable system, we
consider the case \bea\nn \phi=\phi(\xi),\h \chi=A\sigma+B\tau +
\tilde{\chi}(\xi), \eea where $\phi$ and $\tilde{\chi}$ depend on
only one variable $\xi=\alpha\sigma+\beta\tau$ in the same way as
in our NR ansatz (\ref{NRA}). Then the equations of motion
(\ref{fem}), (\ref{kem}) reduce to \bea\label{fer} &&\phi'' -
\frac{1}{2}\frac{\sin(\phi/2)}{\cos^3(\phi/2)}
\left[\tilde{\chi}'^2 +
2\frac{A\alpha-B\beta}{\alpha^2-\beta^2}\tilde{\chi}' +
\frac{A^2-B^2}{\alpha^2-\beta^2}\right]
- \frac{M^2\sin\phi}{\alpha^2-\beta^2}=0,\\
\label{ker} &&\tilde{\chi}'' +
\frac{2\phi'}{\sin\phi}\left(\tilde{\chi}' +
\frac{A\alpha-B\beta}{\alpha^2-\beta^2}\right)=0.\eea

We further restrict ourselves to the case of $A\alpha=B\beta$. A
trivial solution of Eq.(\ref{ker}) is $\tilde{\chi}=constant$,
which corresponds to the solutions of the CSG equations considered
in \cite{CDO06,OS06} for a GM string on $R_t\times S^3$. More
nontrivial solution of (\ref{ker}) is \bea\label{kfi}
\tilde{\chi}= C_\chi\int \frac{d\xi}{\tan^2(\phi/2)}.\eea The
replacement of the above into (\ref{fer}) gives
\bea\label{fef}\phi''=\frac{M^2\sin\phi}{\alpha^2-\beta^2} +
\frac{1}{2}\left[C_\chi^2\frac{\cos(\phi/2)}{\sin^3(\phi/2)}
-\frac{A^2}{\beta^2}\frac{\sin(\phi/2)}{\cos^3(\phi/2)}\right].\eea
Integrating once, we obtain \bea\label{ffi}
\phi'&=&\pm\left[\left(C_\phi -
\frac{2M^2}{\alpha^2-\beta^2}\right) +
\frac{4M^2}{\alpha^2-\beta^2}\sin^2(\phi/2)
-\frac{A^2/\beta^2}{1-\sin^2(\phi/2)} -
\frac{C_\chi^2}{\sin^2(\phi/2)}\right]^{1/2}\\
&\equiv&\pm\Phi(\phi),\nonumber \eea from which we get \bea\nn
\xi(\phi)=\pm\int \frac{d\phi}{\Phi(\phi)},\qquad\chi(\phi)=
\frac{A}{\beta}\left(\beta\sigma+\alpha\tau\right)\pm C_\chi\int
\frac{d\phi}{\tan^2(\phi/2)\Phi(\phi)}.\eea All these solve the
CSG system for the considered particular case. It is clear from
(\ref{ffi}) that the expression inside the square root must be
positive.

Now we are ready to establish a correspondence between the NR and
CSG integrable systems described above. To this end, we make the
following identification \bea\label{equiv} \sin^2(\phi/2)\equiv
\frac{\sqrt{-G}}{K^2}\eea where $G$ is the determinant of the
induced metric $G_{ab}$ computed on the constraints (\ref{00}) and
$K^2$ is a parameter which will be fixed later\footnote{For
$K^2=\kappa^2$, this definition of the angle $\phi$ coincides with
the one used in \cite{CDO06}, which is based on the Pohlmeyer's
reduction procedure \cite{P76}.}. For our NR system, $\sqrt{-G}$
is given by \bea\label{equiv1} \sqrt{-G}
=\frac{R^2\alpha^2}{\alpha^2-\beta^2}\left[(\kappa^2-\omega_1^2) +
(\omega_1^2-\omega_2^2)\cos^2\theta\right].\eea

We want the field $\phi$, defined in (\ref{equiv}) through NR
quantities, to {\it identically} satisfy (\ref{ffi}) derived from
the CSG equations. This imposes relations between the parameters
involved, which are given in appendix A. In this way, we mapped
{\it all} string solutions on $R_t\times S^3$ (in particular on
$R_t\times S^2$) described by the NR integrable system onto
solutions of the CSG (in particular SG) equations. From
(\ref{equiv2}) one can see that the parameters $A$ and $C_\chi$
are nonzero in general on $R_t\times S^2$ where $\omega_2=C_2=0$.
This means that there exist string solutions on $R_t\times S^2$
which correspond to solutions of the CSG system. Only when
$M^2=\kappa^2$, all string solutions on $R_t\times S^2$ are
represented by solutions of the SG equation.

For the GM and SS solutions, which we are interested in, the
relations between the NR and CSG parameters simplify a lot. Let us
write them explicitly. The GM solutions correspond to $C_2=0$,
$\kappa^2=\omega_1^2$. This leads to \bea\label{GMp} &&C_\phi=
\frac{2}{\alpha^2-\beta^2}\left[3M^2- 2\left(\omega_1^2
-\frac{\omega_2^2}{1-\beta^2/\alpha^2}\right)\right],\h
K^2=R^2M^2,
\\ \nn &&A^2=\frac{4}{\alpha^2/\beta^2-1}\left(M^2-\omega_1^2
+ \frac{\omega_2^2}{1-\beta^2/\alpha^2}\right), \h C_{\chi}=0.\eea
Therefore, for all GM strings the field $\chi$ is linear function
at most. Since $A=C_{\chi}=0$ implies $\chi=0$, it follows from
here that there exist GM string solutions on $R_t\times S^3$,
which are mapped not on CSG solutions but on SG solutions instead.
This happens exactly when \bea\nn M^2=\omega_1^2 -
\frac{\omega_2^2}{1-\beta^2/\alpha^2}.\eea In that case the
nonzero parameters are \bea\nn K^2=R^2\left(\omega_1^2 -
\frac{\omega_2^2}{1-\beta^2/\alpha^2}\right),\h C_\phi=
\frac{2}{\alpha^2-\beta^2}\left(\omega_1^2 -
\frac{\omega_2^2}{1-\beta^2/\alpha^2}\right),\eea and the
corresponding solution of the SG equation  can be found from
(\ref{ffi}) to be \bea\label{S3SG} \sin(\phi/2)=
\frac{1}{\cosh\left[\sqrt{\frac{\omega_1^2-\omega_2^2/
\left(1-\beta^2/\alpha^2\right)}{1-\beta^2/\alpha^2}}
\left(\sigma+ \frac{\beta}{\alpha}\tau\right)-\eta_0\right]},\h
\eta_0=const.\eea Replacing (\ref{S3SG}) in (\ref{equiv}),
(\ref{equiv1}), one obtains the GM solution (\ref{gmsol}) as it
should be.

For the SS solutions $C_2=0$,
$\kappa^2=\omega_1^2\alpha^2/\beta^2$. This results in \bea\nn
&&C_\phi= \frac{2}{\beta^2-\alpha^2}\left[
2\left(2\omega_1^2\alpha^2/\beta^2
+\frac{\omega_2^2}{\beta^2/\alpha^2-1}\right)-3M^2\right], \\
\label{SSp}
&&A^2=\frac{4}{M^4(1-\alpha^2/\beta^2)}\left(\omega_1^2\alpha^2/\beta^2
-M^2\right)^2
\left(\frac{\omega_2^2}{\beta^2/\alpha^2-1}-M^2\right), \\ \nn
&&C_{\chi}=\frac{2\omega_1^2\omega_2\alpha^3}
{M^2(\beta^2-\alpha^2)\beta^2},\h K^2=R^2M^2.\eea We want to point
out that $C_{\chi}$ is always nonzero on $S^3$ contrary to the GM
case, which makes $\chi$ also non-vanishing. To our knowledge, the
CSG solutions corresponding to the SS on $R_t\times S^3$ are not
given in the literature. To study this problem, we will consider
the case when $A=0$. $A$ can be zero when \bea\label{asszo}
M^2=\kappa^2=\omega_1^2\alpha^2/\beta^2\h \mbox{or}\h
M^2=\frac{\omega_2^2}{\beta^2/\alpha^2-1},\eea As is seen from
(\ref{asszo}), we have two options, and we restrict ourselves to
the first one\footnote{It turns out that the second option does
not allow real solutions.}. Replacing
$M^2=\omega_1^2\alpha^2/\beta^2$ in (\ref{SSp}) and using the
resulting expressions for $C_\phi$ and $C_{\chi}$ in (\ref{ffi}),
one obtains the simplified equation \bea\nn \phi'^2=\frac{4}
{\beta^2-\alpha^2}
\left[\omega_1^2\frac{\alpha^2}{\beta^2}\cos^2(\phi/2) -
\frac{\omega_2^2}{\beta^2/\alpha^2-1}\cot^2(\phi/2)\right]\eea
with solution \bea\label{phis} \sin^2(\phi/2)=
\tanh^2\left(C\xi\right) +
\frac{\omega_2^2}{\omega_1^2\left(1-\alpha^2/\beta^2\right)\cosh^2
\left(C\xi\right)},\eea where \bea\nn
C=\frac{\alpha\omega_1\sqrt{1-\alpha^2/\beta^2-\omega_2^2/\omega_1^2}}
{\beta^2\left(1-\alpha^2/\beta^2\right)}.\eea This agrees with
Eqs. (\ref{equiv}) and (\ref{equiv1}). By inserting (\ref{phis})
into (\ref{kfi}) one can find \bea\nn
\chi=\tilde{\chi}=2\arctan\left[\frac{\omega_1}{\omega_2}
\sqrt{1-\alpha^2/\beta^2-\omega_2^2/\omega_1^2}
\tanh\left(C\xi\right)\right].\eea Hence, the CSG field $\psi$ for
the case at hand is given by \bea\label{CSGsol}
\psi&=&\sqrt{\tanh^2\left(C\xi\right) +
\frac{\omega_2^2}{\omega_1^2\left(1-\alpha^2/\beta^2\right)\cosh^2
\left(C\xi\right)}}\\
\nn &&\times \exp\left\{i\arctan\left[\frac{\omega_1}{\omega_2}
\sqrt{1-\alpha^2/\beta^2-\omega_2^2/\omega_1^2}
\tanh\left(C\xi\right)\right]\right\}.\eea Here we have set the
integration constants $\phi_0$, $\chi_0$ equal to zero. Several
examples, which illustrate the established NR - CSG
correspondence, are considered in an Appendix.

{\bf Finite-size effects for single spike string}

Here, we will give finite-size single spike string solutions, the
corresponding conserved quantities, and the leading corrections to
the SS ``$E-\Delta\varphi$'' relation: first for the $R_t\times
S^2$ case, then for the SS string with two angular momenta.

The solution for the SS on $R_t\times S^2$ can be written as
($\alpha^2<\beta^2$) \bea\nn
&&W_1=R\sqrt{1-\left(1-\kappa^2/\omega_1^2\right)dn^2\left(C\xi|m\right)}
\label{fssss}\\
&&\times\exp\left\{-i\omega_1\frac{\alpha/\beta}{1-\alpha^2/\beta^2}
\left(\sigma+\frac{\alpha}{\beta}\tau\right)\pm
i\frac{\beta/\alpha}{\sqrt{1-\kappa^2/\omega_1^2}}
\Pi\left(am(C\xi),\beta^2/\alpha^2-1|m \right)\right\} \nn,\\ \nn
&&W_2=R\sqrt{1-\kappa^2/\omega_1^2}dn\left(C\xi|m\right),\qquad
Z_0=R\exp(i\kappa\tau) \\ \nn
&&C=\pm\frac{\alpha\omega_1\sqrt{1-\kappa^2/\omega_1^2}}{\beta^2(1-\alpha^2/\beta^2)},
\h m\equiv\frac{\beta^2/\alpha^2-1}{\omega_1^2/\kappa^2-1}.\eea

The conserved quantities for the present string solution are given
by \bea\nn \mathcal{E}_s&=&2\frac{\kappa}{\alpha}
\int_{\theta_{min}}^{\theta_{max}}\frac{d \theta}{\theta'}
=2\frac{\kappa(\beta^2/\alpha^2-1)}
{\omega_1\sqrt{1-\kappa^2/\omega_1^2}}\mathbf{K}(m),\nn\\
\mathcal{J}&=& \frac{2}{\alpha}
\int_{\theta_{min}}^{\theta_{max}}\frac{d
\theta}{\theta'}\sin^2\theta\left(\beta f'_1+\omega_1\right)
=2\sqrt{1-\kappa^2/\omega_1^2} \left[\mathbf{E}(m)
-\frac{1-\beta^2\kappa^2/\alpha^2\omega_1^2}{1-\kappa^2/\omega_1^2}\mathbf{K}
(m)\right].\nn \eea In addition, we compute $\Delta\varphi_1$
\bea\nn
\Delta\varphi\equiv\Delta\varphi_1=2\int_{\theta_{min}}^{\theta_{max}}\frac{d
\theta}{\theta'}f'_1=-2\frac{\beta/\alpha}{\sqrt{1-\kappa^2/\omega_1^2}}
\left[\Pi\left(1-\frac{\beta^2}{\alpha^2}|
m\right)-\mathbf{K}(m)\right].\eea Defining parameters \bea\nn
\epsilon\equiv 1-m,\h v\equiv \beta/\alpha,\eea we can rewrite
these as \bea\nn
&&\mathcal{E}_s=2\sqrt{(v^2-1)(1-\epsilon)}\mathbf{K}(1-\epsilon),\h
\mathcal{J}=2\sqrt{\frac{v^2-1}{v^2-\epsilon}}
\left[\mathbf{E}(1-\epsilon)-\epsilon\mathbf{K}(1-\epsilon)\right],\\
\nn &&\Delta\varphi=-2v\sqrt{\frac{v^2-\epsilon}{v^2-1}}
\left[\Pi\left(1-v^2|
1-\epsilon\right)-\mathbf{K}(1-\epsilon)\right]\\ \nn
&&\mathcal{E}_s-\Delta\varphi=2v\sqrt{\frac{v^2-\epsilon}{v^2-1}}
\left[\Pi\left(1-v^2|
1-\epsilon\right)-\left(1-\frac{(v^2-1)\sqrt{1-\epsilon}}{v\sqrt{v^2-\epsilon}}\right)
\mathbf{K}(1-\epsilon)\right].\eea

Now we make small $\epsilon$ expansion of the above expressions by
using the following representation for $v$ \bea\nn
v(\epsilon)=v_0(p)+v_1(p)\epsilon +
v_2(p)\epsilon\log(\epsilon)\eea
and obtain \bea\nn
\mathcal{J}=2\sqrt{1-\frac{1}{v_0^2}},\h
v_1=\frac{(v_0^2-1)\left[v_0^2(1+\log(16))-2\right]}{4v_0},\h
v_2=-\frac{v_0(v_0^2-1)}{4}.\eea From the expansion for
$\Delta\varphi$, we obtain $\epsilon$ as a function of
$\Delta\varphi$ and $\mathcal{J}$ \bea\nn \epsilon=16
\exp\left(-\frac{\sqrt{4-\mathcal{J}^2}}{\mathcal{J}}\left[\Delta\varphi+
\arcsin\left(\frac{\mathcal{J}}{2}\sqrt{4-\mathcal{J}^2}\right)\right]
\right).\eea Using these results in the expansion for
$\mathcal{E}_s-\Delta\varphi$, one can see that the divergent
terms cancel each other for $\mathcal{J}^2<2$ and the finite
result is \bea\nn E_s-\frac{\sqrt{\lambda}}{2\pi}\Delta\varphi &=&
\frac{\sqrt{\lambda}}{\pi}
\left[\frac{1}{2}\arcsin\left(\frac{\mathcal{J}}{2}\sqrt{4-\mathcal{J}^2}\right)
+\frac{\mathcal{J}^3}{16\sqrt{4-\mathcal{J}^2}}
\ \epsilon\right]\\
\label{ssS2c} &=&\frac{\sqrt{\lambda}}{\pi}
\left[\frac{p}{2}+4\sin^2\frac{p}{2}\tan\frac{p}{2}
\exp\left(-\frac{\Delta\varphi+ p}
{\tan\frac{p}{2}}\right)\right],\eea where we used the
identification \bea\nn
\arcsin\left(\mathcal{J}/2\right)=\frac{p}{2}=\bar{\theta}
=\pi/2-\arcsin\frac{\kappa}{\omega_1}.\eea
 This includes the
leading finite-size correction to the SS ``$E-\Delta\varphi$''
relation (the term proportional to $\epsilon$). Let us also note
that to the leading order, the length $L$ of this SS string can be
computed to be \bea\nn L=\frac{\alpha}{\kappa}\left(\Delta\varphi+
p\right).\eea

The full string solution on $R_t\times S^3$
is given by \bea\nn &&Z_0=R\exp(i\kappa\tau),\\
\nn &&W_1=R\sqrt{1-z^2_+
dn^2\left(C\xi|m\right)}\exp\left\{i\omega_1\tau +
\frac{2i\beta/\alpha}{z_+\sqrt{1-\omega_2^2/\omega_1^2}}\right.\\
\nn &&\times\left.\left[F\left(am(C\xi)|m\right) -
\frac{\kappa^2/\omega_1^2}{1-z^2_+}\Pi\left(am(C\xi),-\frac{z^2_+
-z^2_-}{1-z^2_+}|m\right)\right]\right\} ,\\
\label{fssS3} &&W_2=Rz_+
dn\left(C\xi|m\right)\exp\left\{i\omega_2\tau
+\frac{2i\beta\omega_2/\alpha\omega_1}
{z_+\sqrt{1-\omega_2^2/\omega_1^2}}
F\left(am(C\xi)|m\right)\right\} .\eea

The CSG solution related to (\ref{fssS3}) can be written as
\bea\label{fisS3}
\sin^2(\phi/2)=\frac{\omega_1^2/M^2}{\beta^2/\alpha^2-1}
\left[\left(1-\kappa^2/\omega_1^2\right) -
\left(1-\omega_2^2/\omega_1^2\right)\left(z^2_+ cn^2(C\xi|m) +
z^2_- sn^2(C\xi|m)\right)\right].\eea After that, we use
(\ref{fisS3}) in (\ref{kfi}) and integrate. The result is
\bea\label{chiS3} \chi=\frac{A}{\beta}(\beta\sigma+\alpha\tau) -
C_\chi(\alpha\sigma+\beta\tau) + \frac{C_\chi}{C
D}\Pi\left(am(C\xi),n|m\right),\eea where $A/\beta$ and $C_\chi$
are given in (\ref{equiv2}), $C_2=0$, and \bea\nn
D=\frac{\omega_1^2/M^2}{\beta^2/\alpha^2-1}
\left[\left(1-\kappa^2/\omega_1^2\right) -
\left(1-\omega_2^2/\omega_1^2\right)z^2_+\right],\h
n=\frac{\left(1-\omega_2^2/\omega_1^2\right)
(z^2_+-z^2_-)}{\left(1-\kappa^2/\omega_1^2\right) -
\left(1-\omega_2^2/\omega_1^2\right)z^2_+}.\eea Hence for the
present case, the CSG field $\psi=\sin(\phi/2)\exp(i\chi/2)$ is
defined by (\ref{fisS3}) and (\ref{chiS3}).

The computation of the conserved quantities (\ref{cqs}) and
$\Delta\varphi_1$ now gives \bea\nn &&\mathcal{E}_s
=\frac{2\kappa(\beta^2/\alpha^2-1)}
{\omega_1\sqrt{1-\omega_2^2/\omega_1^2}z_+}\mathbf{K}
\left(1-z^2_-/z^2_+\right), \\ \nn &&\mathcal{J}_1= \frac{2
z_+}{\sqrt{1-\omega_2^2/\omega_1^2}} \left[\mathbf{E}
\left(1-z^2_-/z^2_+\right)
-\frac{1-\beta^2\kappa^2/\alpha^2\omega_1^2}{z^2_+}\mathbf{K}
\left(1-z^2_-/z^2_+\right)\right], \\ \nn &&\mathcal{J}_2=
-\frac{2 z_+ \omega_2/\omega_1
}{\sqrt{1-\omega_2^2/\omega_1^2}}\mathbf{E}
\left(1-z^2_-/z^2_+\right), \\ \nn &&\Delta\varphi=
-\frac{2\beta/\alpha}{\sqrt{1-\omega_2^2/\omega_1^2}z_+}
\left[\frac{\kappa^2/\omega_1^2}{1-z^2_+}\Pi\left(-\frac{z^2_+ -
z^2_-}{1-z^2_+}|1-z^2_-/z^2_+\right) -\mathbf{K}
\left(1-z^2_-/z^2_+\right)\right].\eea

Our next step is to introduce the new parameters \bea\nn
\epsilon\equiv z^2_-/z^2_+,\h v\equiv\beta/\alpha,\h
u\equiv\omega^2_2/\omega^2_1 ,\eea and to expand the above
conserved quantities about $\epsilon=0$. We also need to consider
the $\epsilon$-expansion for $u$ and $v$ as follows: \bea\nn
v(\epsilon)=v_0+v_1\epsilon + v_2\epsilon\log(\epsilon),\h
u(\epsilon)=u_0+u_1\epsilon + u_2\epsilon\log(\epsilon).\eea The
coefficients can be determined by the condition that
$\mathcal{J}_1$ and $\mathcal{J}_2$ should be finite,
\bea\label{j1j2}
&&v_0=\frac{2\mathcal{J}_1}{\sqrt{\left(\mathcal{J}_1^2-\mathcal{J}_2^2\right)
\left[4-\left(\mathcal{J}_1^2-\mathcal{J}_2^2\right)\right]}},\h
u_0=\frac{\mathcal{J}_2^2}{\mathcal{J}_1^2},
\\ \nn
&&v_1=\frac{(1-u_0)v_0^2-1}{4(u_0-1)(v_0^2-1)v_0}
\left\{(u_0-1)v_0^4(1+\log(16))-2\right.\\ \label{uvs} &&+ \left.
v_0^2\left[3+\log(16)+u_0(\log(4096)-5)\right]\right\},
\\ \nn
&&v_2=-\frac{v_0\left[1-(1-u_0)v_0^2\right]\left[1+3u_0-(1-u_0)v_0^2\right]}
{4(1-u_0)(v_0^2-1)},\\ \nn
&&u_1=\frac{u_0\left[1-(1-u_0)v_0^2\right]\log(16)}{v_0^2-1},\h
u_2=-\frac{u_0\left[1-(1-u_0)v_0^2\right]}{v_0^2-1}. \eea The
parameter $\epsilon$ can be obtained from $\Delta\varphi$ and to
the leading order one finds: \bea\label{es} \epsilon=16
\exp\left(-\frac{\sqrt{(1-u_0)v_0^2-1}}{v_0^2-1}\left[\Delta\varphi
+ \arcsin\left(\frac{2\sqrt{(1-u_0)v_0^2-1}}
{(1-u_0)v_0^2}\right)\right]\right).\eea From Eqs.(\ref{j1j2}),
(\ref{uvs}) and (\ref{es}), $\mathcal{E}_s-\Delta\varphi$ can be
derived to be \bea\label{DiffJ12}\mathcal{E}_s-\Delta\varphi&=&
\arcsin N(\mathcal{J}_1,\mathcal{J}_2) +
2\left(\mathcal{J}_1^2-\mathcal{J}_2^2\right)
\sqrt{\frac{4}{\left[4-\left(\mathcal{J}_1^2-\mathcal{J}_2^2\right)\right]}-1}\\
&\times&\exp\left[-\frac{2\left(\mathcal{J}_1^2-\mathcal{J}_2^2\right)
N(\mathcal{J}_1,\mathcal{J}_2)}
{\left(\mathcal{J}_1^2-\mathcal{J}_2^2\right)^2 +
4\mathcal{J}_2^2}\left[\Delta\varphi
+\arcsin N(\mathcal{J}_1,\mathcal{J}_2)\right]\right],\\
N(\mathcal{J}_1,\mathcal{J}_2)&\equiv&
\frac{1}{2}\left[4-\left(\mathcal{J}_1^2-\mathcal{J}_2^2\right)\right]
\sqrt{\frac{4}{\left[4-\left(\mathcal{J}_1^2-\mathcal{J}_2^2\right)\right]}-1}.\eea
Here $\mathcal{J}_1^2-\mathcal{J}_2^2<2$ is assumed. Finally, by
using the SS relation between the angular momenta \bea\nn
\mathcal{J}_1=\sqrt{\mathcal{J}_2^2+4\sin^2(p/2)},\eea we obtain
($-\pi/2\le p\le\pi/2$) \bea\label{ssS3c}
E_s-\frac{\sqrt{\lambda}}{2\pi}\Delta\varphi=
\frac{\sqrt{\lambda}}{\pi}
\left[\frac{p}{2}+4\sin^2\frac{p}{2}\tan\frac{p}{2}
\exp\left(-\frac{\tan\frac{p}{2}(\Delta\varphi+ p)}
{\tan^2\frac{p}{2} + \mathcal{J}_2^2 \csc^2p}\right)\right].\eea
This is our final result including the leading finite-size
correction to the ``$E-\Delta\varphi$'' relation for the SS string
with two angular momenta. It is obvious that for $\mathcal{J}_2=0$
(\ref{ssS3c}) reduces to (\ref{ssS2c}) as it should be.

\subsubsection{Finite-size effect of the dyonic giant magnons\\ in
$\mathcal{N}=6$ super Chern-Simons-matter theory}
The AdS/CFT correspondence between type IIB string theory on
$AdS_5\times S^5$ and ${\cal N}=4$ SYM theory led to many exciting
developments and to understanding non-perturbative sturctures of
the string and gauge theories. Another exciting possibility is
that the same type of duality does exist. The promising candidate
for the three-dimensional conformal field theory is ${\cal N}=6$
super Chern-Simons (CS) theory with $SU(N)\times SU(N)$ gauge
symmetry and level $k$ \cite{ABJM}. In the planar limit of $N,
k\to\infty$ with a fixed value of 't Hooft coupling $\lambda=N/k$,
the ${\cal N}=6$ CS is believed to be dual to type IIA superstring
theory on $AdS_4\times \mathbb{CP}^3$.

In \cite{17} we consider finite-size effects for the dyonic giant
magnon of the type IIA string theory on $AdS_4\times
\mathbb{CP}^3$ by applying L\"uscher $\mu$-term formula which is
derived from a proposed $S$-matrix for the ${\cal N}=6$ super
Chern-Simons theory. We compute explicitly the effect for the case
of a symmetric configuration where the two external bound states,
each of $A$ and $B$ particles, have the same momentum $p$ and spin
$J_2$. We compare this with the classical string theory result
which we computed by reducing it to the Neumann-Rosochatius
integrable system. The two results match perfectly.

{\bf Classical string analysis}

Let us consider a classical string moving in $R_t\times\cp$. Using
the complex coordinates \bea\nn z=y^0+iy^4,\h w_1=x^1+ix^2,\h
w_2=x^3+ix^4,\h w_3=x^5+ix^6,\h w_4=x^7+ix^8,\eea we embed the
string as follows \cite{ABR} \bea\nn
&&z=Z(\tau,\sigma)=\frac{R}{2}e^{it(\tau,\sigma)},\h
w_a=W_a(\tau,\sigma)=R
r_a(\tau,\sigma)e^{i\varphi_a(\tau,\sigma)}. \eea Here $t$ is the
$AdS$ time. These complex coordinates should satisfy \bea \nn
&&\sum_{a=1}^{4}W_a\bar{W}_a=R^2,\h
\sum_{a=1}^{4}\left(W_a\p_m\bar{W}_a-\bar{W}_a\p_m W_a\right)=0,
\eea or \bea \sum_{a=1}^{4}r_a^2=1,\h
\sum_{a=1}^{4}r_a^2\p_m\varphi_a=0,\quad m=0,1. \label{cp3cond}
\eea

{\it NR reduction}

In order to reduce the string dynamics on $R_t\times\cp$ to the NR
integrable system, we use the ansatz \bea\nn
&&t(\tau,\sigma)=\kappa\tau,\h r_a(\tau,\sigma)=r_a(\xi),\h
\varphi_a(\tau,\sigma)=\omega_a\tau+f_a(\xi),\\ \label{Anz}
&&\xi=\alpha\sigma+\beta\tau,\h \kappa, \omega_a, \alpha,
\beta={\rm constants}.\eea It can be shown \cite{ABR} that after
integration of the equations of motion for $f_a$, which gives
\bea\label{fafi} f'_a=\frac{1}{\alpha^2-\beta^2}
\left(\frac{C_a}{r_a^2}+\beta\omega_a\right),\h C_a=constants,
\eea one ends up with the following effective Lagrangian for the
coordinates $r_a$ \bea \label{NRL-} L_{NR}&=&(\alpha^2-\beta^2)
\sum_{a=1}^{4}\left[r^{'2}_a-\frac{1}{(\alpha^2-\beta^2)^2}
\left(\frac{C_a^2}{r_a^2} + \alpha^2\omega_a^2r_a^2\right)\right]
\\ \nn &-&\Lambda\left(\sum_{a=1}^{4}r_a^2-1\right) .\eea
This is the Lagrangian for the NR integrable system \cite{KRT06}.
In addition, the $\cp$ embedding conditions in (\ref{cp3cond})
lead to \bea \sum_{a=1}^{4}\omega_a r_a^2=0,\h
\sum_{a=1}^{4}C_a=0.\label{aec} \eea

The Virasoro constraints give the conserved Hamiltonian $H_{NR}$
and a relation between the embedding parameters and the arbitrary
constants $C_a$: \bea\label{HNR+} &&H_{NR}=(\alpha^2-\beta^2)
\sum_{a=1}^{4}\left[r_a'^2+\frac{1}{(\alpha^2-\beta^2)^2}
\left(\frac{C_a^2}{r_a^2} + \alpha^2\omega_a^2r_a^2\right)\right]
=\frac{\alpha^2+\beta^2}{\alpha^2-\beta^2}\frac{\kappa^2}{4},
\\ \label{01R-} &&\sum_{a=1}^{4}C_a\omega_a + \beta(\kappa/2)^2=0.\eea

The conserved charges can be defined by \bea\nn E_s=-\int
d\sigma\frac{\p\mathcal{L}}{\p(\p_0 t)},\h J_a=\int
d\sigma\frac{\p\mathcal{L}}{\p(\p_0\varphi_a)},\h a=1,2,3,4,\eea
where $\mathcal{L}$ is the Polyakov string Lagrangian taken in
conformal gauge. Using the ansatz (\ref{Anz}) and (\ref{fafi}), we
can find \bea\label{cqs-} E_s=
\frac{\kappa\sqrt{2\lambda}}{2\alpha}\int d\xi,\h J_a=
\frac{2\sqrt{2\lambda}}{\alpha^2-\beta^2}\int d\xi
\left(\frac{\beta}{\alpha}C_a+\alpha\omega_a r_a^2\right).\eea In
view of (\ref{aec}), one obtains \bea\label{Jc}
\sum_{a=1}^{4}J_a=0.\eea

{\it Dyonic giant magnon solution}

We are interested in finding string configurations corresponding
to the following particular solution of (\ref{aec}) \bea\nn
r_1=r_3=\frac{1}{\sqrt{2}}\sin\theta,\h
r_2=r_4=\frac{1}{\sqrt{2}}\cos\theta,\h \omega_1=-\omega_3,\h
\omega_2=-\omega_4.\eea The two frequencies $\omega_1, \omega_2$
are independent and lead to strings moving in $\cp$ with two
angular momenta. From the NR Hamiltonian (\ref{HNR}) one finds
\bea\nn \theta'^2(\xi)&=&\frac{1}{(\alpha^2-\beta^2)^2}
\Bigg[\frac{\kappa^2}{4}(\alpha^2+\beta^2) -
2\left(\frac{C_1^2+C_3^2}{\sin^2{\theta}} +
\frac{C_2^2+C_4^2}{\cos^2{\theta}}\right)
\\ \nn &-&
\alpha^2\left(\omega_1^2\sin^2{\theta}
+\omega_2^2\cos^2{\theta}\right)\Bigg]. \eea We further restrict
ourselves to $C_2=C_4=0$ to search for GM string configurations.
Eqs. (\ref{aec}) and (\ref{01R-}) give \bea\nn
C_1=-C_3=-\frac{\beta\kappa^2}{8\omega_1}. \eea In this case, the
above equation for $\theta'$ can be rewritten in the form
\bea\label{S3eq}
(\cos\theta)'=\mp\frac{\alpha\sqrt{\omega_1^2-\omega_2^2}}{\alpha^2-\beta^2}
\sqrt{(z_+^2-\cos^2\theta)(\cos^2\theta-z_-^2)},\eea where \bea\nn
&&z^2_\pm=\frac{1}{2(1-\frac{\omega_2^2}{\omega_1^2})}
\left\{y_1+y_2-\frac{\omega_2^2}{\omega_1^2}
\pm\sqrt{(y_1-y_2)^2-\left[2\left(y_1+y_2-2y_1
y_2\right)-\frac{\omega_2^2}{\omega_1^2}\right]
\frac{\omega_2^2}{\omega_1^2}}\right\}, \\ \nn
&&y_1=1-\frac{\kappa^2}{4\omega_1^2},\h
y_2=1-\frac{\beta^2}{\alpha^2}\frac{\kappa^2}{4\omega_1^2}.\eea
The solution of (\ref{S3eq}) is given by \bea\label{S3sol}
\cos\theta=z_+ \mathbf{dn}\left(C\xi|m\right),\h
C=\mp\frac{\alpha\sqrt{\omega_1^2-\omega_2^2}}{\alpha^2-\beta^2}
z_+,\h m\equiv 1-z^2_-/z^2_+ .\eea

To find the full string solution, we also need to obtain the
explicit expressions for the functions $f_a$ from (\ref{fafi})
\bea\nn f_a=\frac{1}{\alpha^2-\beta^2}\int
d\xi\left(\frac{C_a}{r_a^2}+\beta\omega_a\right).\eea Using the
solution (\ref{S3sol}) for $\theta(\xi)$, we can find \bea\nn
&&f_1=-f_3=\frac{\beta/\alpha}{z_+\sqrt{1-\omega_2^2/\omega_1^2}}
\left[C\xi -
\frac{2(\kappa/2)^2/\omega_1^2}{1-z^2_+}\Pi\left(am(C\xi),-\frac{z^2_+
-z^2_-}{1-z^2_+}|m\right)\right],\\ \nn
&&f_2=-f_4=\frac{\beta\omega_2}{\alpha^2-\beta^2}\xi.\eea As a
consequence, the string solution can be written as \bea \nn
&&W_1=\frac{R}{\sqrt{2}}\sqrt{1-z_+^2\mathbf{dn}^2\left(C\xi|m\right)}\
e^{i(\omega_1\tau+f_1)},\\ \label{fss} &&W_2=\frac{R}{\sqrt{2}}z_+
\mathbf{dn}\left(C\xi|m\right)\ e^{i(\omega_2\tau+f_2)},\\ \nn
&&W_3=\frac{R}{\sqrt{2}}\sqrt{1-z_+^2\mathbf{dn}^2\left(C\xi|m\right)}\
e^{-i(\omega_1\tau+f_1)},\\ \nn &&W_4=\frac{R}{\sqrt{2}}z_+
\mathbf{dn}\left(C\xi|m\right)\ e^{-i(\omega_2\tau+f_2)} .\eea

The GM in infinite volume can be obtained by taking $z_-\to 0$. In
this limit, the solution for $\theta$ reduces to \bea\nn
\cos\theta=\frac{\sin\frac{p}{2}}{\cosh(C\xi)}, \eea where the
constant $z_{+}\equiv\sin p/2$ is given by \bea\nn
z^2_{+}=\frac{y_2-\omega_2^2/\omega_1^2}{1-\omega_2^2/\omega_1^2}.
\eea One spin solution corresponds $\omega_2=0$.
Inserting this into (\ref{cqs-}), one can find the energy-charge
dispersion relation. For the {\it single} DGM, the energy and
angular momentum $J_1$ become infinite but their difference
remains finite:
\bea E_s-J_1=\sqrt{\frac{J_2^2}{4}+2\lambda\sin^2\frac{p}{2}}.
\label{infdis} \eea

{\it Finite-size effects}

Using the most general solutions (\ref{fss}), we can calculate the
finite-size corrections to the energy-charge relation
(\ref{infdis}) in the limit when the string energy $E_s\to\infty$.
Here we consider the case of $\alpha^2>\beta^2$ only since it
corresponds to the GM case. We obtain from (\ref{cqs-}) the
following expressions for the conserved string energy $E_s$ and
the angular momenta $J_a$ \bea\nn &&\mathcal{E}
=\frac{2\kappa(1-\beta^2/\alpha^2)} {\omega_1
z_+\sqrt{1-\omega_2^2/\omega_1^2}}\mathbf{K}
\left(1-z^2_-/z^2_+\right), \\ \label{cqsGM} &&\mathcal{J}_1=
\frac{2 z_+}{\sqrt{1-\omega_2^2/\omega_1^2}} \left[
\frac{1-\beta^2(\kappa/2)^2/\alpha^2\omega_1^2}{z^2_+}\mathbf{K}
\left(1-z^2_-/z^2_+\right)-\mathbf{E}
\left(1-z^2_-/z^2_+\right)\right], \\ \nn &&\mathcal{J}_2= \frac{2
z_+ \omega_2/\omega_1 }{\sqrt{1-\omega_2^2/\omega_1^2}}\mathbf{E}
\left(1-z^2_-/z^2_+\right),\h \mathcal{J}_3=-\mathcal{J}_1,\h
\mathcal{J}_4=-\mathcal{J}_2.\eea As a result, the condition
(\ref{Jc}) is identically satisfied. Here, we introduced the
notations \bea\label{not} \mathcal{E}=\frac{E_s}{\sqrt{2\lambda}}
,\h \mathcal{J}_a=\frac{J_a}{\sqrt{2\lambda}}.\eea The computation
of $\Delta\varphi_1$ gives \bea\label{pws} p\equiv\Delta\varphi_1
&=& 2\int_{\theta_{min}}^{\theta_{max}}\frac{d
\theta}{\theta'}f'_1=
\\ \nn &-&\frac{2\beta/\alpha}{z_+\sqrt{1-\omega_2^2/\omega_1^2}}
\left[\frac{(\kappa/2)^2/\omega_1^2}{1-z^2_+}\mathbf{\Pi}\left(-\frac{z^2_+
- z^2_-}{1-z^2_+}\bigg\vert 1-z^2_-/z^2_+\right) -\mathbf{K}
\left(1-z^2_-/z^2_+\right)\right].\eea

Expanding the elliptic integrals, we obtain \bea
E-J_1&=& 2\sqrt{\frac{J_2^2}{4}+2\lambda\sin^2\frac{p}{2}}\label{stringDGM}\\
&-&\frac{32\lambda\sin^4\frac{p}{2}}
{\sqrt{J_2^2+8\lambda\sin^2\frac{p}{2}}}\exp\left[-\frac{2\sin^2\frac{p}{2}\left(J_1
+ \sqrt{J_2^2+8\lambda\sin^2\frac{p}{2}}\right)
\sqrt{J_2^2+8\lambda\sin^2\frac{p}{2}}}{J_2^2+8\lambda\sin^4\frac{p}{2}}
\right].\nn \eea This also gives the finite-size effect for
ordinary GM by taking $J_2\to 0$ \bea E-J_1=
2\sqrt{2\lambda}\sin\frac{p}{2}-16\sqrt{\frac{\lambda}{2}}\sin^3\frac{p}{2}
\exp\left[-\frac{J_1}{\sqrt{2\lambda}\sin\frac{p}{2}}-2\right].
\eea

{\bf Finite-size effects from the $S$-matrix}

The ${\cal N}=6$ CS theory has two sets of excitations, namely
$A$-particles and $B$-particles, each of which form a
four-dimensional representation of $SU(2|2)$ \cite{GGY,AN}. We
propose an $S$-matrix with the following structure: \bea\nn
S^{AA}(p_1,p_2)&=&S^{BB}(p_1,p_2)=S_0(p_1,p_2){\widehat
S}(p_1,p_2)
\\ \nn
S^{AB}(p_1,p_2)&=&S^{BA}(p_1,p_2)={\tilde S}_0(p_1,p_2){\widehat
S}(p_1,p_2), \eea where $\widehat S$ is the matrix part determined
by the $SU(2|2)$ symmetry, and is essentially the same as that
found for ${\cal N}=4$ SYM in \cite{Be,AFZ}. An important
difference arises in the dressing phases $S_0, {\tilde S}_0$ due
to the fact that the $A$- and $B$-particles are related by complex
conjugation.

{\it L\"uscher $\mu$-term formula}

Here we want to generalize multi-particle L\"uscher formula
\cite{BajJan,HatSuzii} to the case of the bound states. Consider
$M_A$ number of A-type DGMs, $\vert Q_1,\ldots Q_{M_A}\rangle$,
and $M_B$ number of B-type DGMs, $\vert {\tilde Q}_1,\ldots
{\tilde Q}_{M_B}\rangle$. We use $\alpha_k$ for the $SU(2|2)$
quantum numbers carried by the DGMs and $C_k$ for $A$ or $B$, the
two types of particles. Then we propose the multi-particle
L\"uscher formula for generic DGM states as follows: \bea &&\delta
E_{\mu}=-i\sum_{b=1}^4\left\{\sum_{l=1}^{M_A}(-1)^{F_b}
\left(1-\frac{\epsilon'_{Q_{l}}(p_l)}{\epsilon'_{1}({\tilde
q}^*)}\right) e^{-i{\tilde q}^* L}\left[\mathop {\rm Res}
\limits_{q^*={\tilde q}^*}
{S^{AA}}^{b\alpha_{l}}_{b\alpha_{l}}(q^*,p_l)\right]\right.
\\ \nn &&
\left.\times\prod_{k\neq
l}^{M_A+M_B}{S^{AC_k}}^{b\alpha_{k}}_{b\alpha_{k}}(q^*,p_k)\right.\\
\nn &&+\left.\sum_{l=1}^{M_B}(-1)^{F_b}
\left(1-\frac{\epsilon'_{{\tilde
Q}_{l}}(p_l)}{\epsilon'_{1}({\tilde q}^*)}\right) e^{-i{\tilde
q}^* L}\left[\mathop {\rm Res} \limits_{q^*={\tilde q}^*}
{S^{BB}}^{b\alpha_{l}}_{b\alpha_{l}}(q^*,p_l)\right] \prod_{k\neq
l}^{M_A+M_B}{S^{BC_k}}^{b\alpha_{k}}_{b\alpha_{k}}(q^*,p_k)\right\}.
\label{luscher} \eea Here, the energy dispersion relation for the
DGM is given by \bea
\epsilon_Q(p)=\sqrt{\frac{Q^2}{4}+4g^2\sin^2\frac{p}{2}}.
\label{disper} \eea The coupling constant $g=h(\lambda)$ is still
unknown function of $\lambda$ which behaves as $h(\lambda) \sim
\lambda$ for small $\lambda$, and $h(\lambda) \sim
\sqrt{\lambda/2}$ for large $\lambda$.

{\it $S$-matrix elements for the dyonic GM}

The $S$-matrix elements for the DGM are in general complicated.
However, we can consider a simplest case of the DGMs composed of
only A-type $\phi_1$'s which are the first bosonic particle in the
fundamental representation of $SU(2|2)$. It is obvious that these
bound states do exist since the elementary $S$-matrix element
$S{^{AA}}^{11}_{11}$ does have a pole. The same holds for the
B-type DGMs. However, the hybrid type DGMs are not possible
because the $S^{AB}$ $S$-matrix does not have any bound-state
pole.

The L\"uscher correction needs only those $S$-matrix elements
which have the same incoming and outgoing $SU(2|2)$ quantum
numbers after scattering with a virtual particle. In particular,
we can easily compute the matrix elements between an elementary
magnon and a the bound-state made of only $\phi_1$'s ($Q$ of them)
denoted by ${\mathbf 1}_Q$ \cite{HatSuzi} \bea
{S^{AA}}^{b\mathbf{1}_Q }_{b {\mathbf
1}_Q}(y,{X^{(Q)}})=\prod_{k=1}^Q {S^{AA}}^{b 1}_{b 1}(y,x_k)
=\prod_{k=1}^Q\left[
\frac{1-\frac{1}{y^+x^-_k}}{1-\frac{1}{y^-x^+_k}}\sigma_{\rm
BES}(y,x_k) {\tilde a}_b(y,x_k)\right], \eea where ${\tilde a}_b$
are given by \cite{Be,AFZ} \bea {\tilde a}_1(y,x)&=&a_1(y,x),\quad
{\tilde a}_2(y,x)=a_1(y,x)+a_2(y,x),
\\ \nn &&\quad {\tilde
a}_3(y,x)={\tilde a}_4(y,x)=a_6(y,x)\\ \nn
a_1(y,x)&=&\frac{x^--y^+}{x^+-y^-}\frac{\eta(x)\eta(y)}{{\tilde\eta}(x){\tilde\eta}(y)}
\\ \nn
a_2(y,x)&=&\frac{(y^--y^+)(x^--x^+)(x^--y^+)}{(y^--x^+)(x^-y^--x^+y^+)}
\frac{\eta(x)\eta(y)}{{\tilde\eta}(x){\tilde\eta}(y)}
\\ \nn
a_6(y,x)&=&\frac{y^+-x^+}{y^--x^+}\frac{\eta(y)}{{\tilde\eta}(y)}.
\eea

As noticed in \cite{HatSuzi}, $a_2/a_1$ and $a_6/a_1$ are
negligible ${\cal O}(1/g)$ corrections in the classical limit
$g>>1$. Therefore, the $S$-matrix with $b=1$ is a most important
factor for our computation which can be written as \bea\label{saa}
{S^{AA}}^{1 \mathbf{1}_{Q}}_{1
\mathbf{1}_{Q}}(y,{X^{(Q)}})&=&\sigma_{\rm BES}(y,X^{(Q)})
\prod_{k=1}^Q\left[\frac{1-\frac{1}{y^+x^-_k}}{1-\frac{1}{y^-x^+_k}}\cdot
\frac{x^-_k-y^+}{x^+_k-y^-}\frac{\eta(x_k)\eta(y)}{{\tilde\eta}(x_k){\tilde\eta}(y)}\right]
\\ \nn
&=&\sigma_{\rm BES}(y,X^{(Q)})S_{\rm BDS}(y,X^{(Q)})
\frac{\eta(X^{(Q)})}{{\tilde\eta}(X^{(Q)})}\left(\frac{\eta(y)}{{\tilde\eta}(y)}\right)^Q,
\\ \label{sab}
{S^{AB}}^{1 \mathbf{1}_{Q}}_{1
\mathbf{1}_{Q}}(y,{X^{(Q)}})&=&\sigma_{\rm BES}(y,X^{(Q)})
\frac{\eta(X^{(Q)})}{{\tilde\eta}(X^{(Q)})}\left(\frac{\eta(y)}{{\tilde\eta}(y)}\right)^Q,
\eea where the BDS  $S$-matrix is defined by \bea S_{\rm
BDS}(y,x)\equiv\frac{1-\frac{1}{y^+x^-}}{1-\frac{1}{y^-x^+}}\cdot
\frac{x^--y^+}{x^+-y^-}. \eea The spectral parameter $X^{(Q)}$ for
the DGM is defined by \bea {X^{(Q)}}^{\pm}=\frac{e^{\pm i
p/2}}{4g\sin\frac{p}{2}}\left(Q
+\sqrt{Q^2+16g^2\sin^2\frac{p}{2}}\right)\equiv e^{(\theta\pm i
p)/2}, \eea where we introduce $\theta$ defined by \bea
\sinh\frac{\theta}{2}&\equiv&\frac{Q}{4g\sin\frac{p}{2}}. \eea The
frame factors $\eta$ and ${\tilde\eta}$ are given by \cite{AFZ}
\bea
\frac{\eta(x_1)}{{\tilde\eta}(x_1)}=\frac{\eta(x_2)}{{\tilde\eta}(x_2)}=1
\eea for the spin-chain frame and \bea
\frac{\eta(x_1)}{{\tilde\eta}(x_1)}=\sqrt{\frac{x^+_2}{x^-_2}},\quad
\frac{\eta(x_2)}{{\tilde\eta}(x_2)}=\sqrt{\frac{x^-_1}{x^+_1}}
\eea for the string frame.

{\it Symmetric DGM state}

The classical two spins solution is a symmetric DGM configuration
for both of $S^2$ subspaces. Corresponding L\"uscher formula is
given by Eq.(\ref{luscher}) with $M_A=M_B=1$, which can be much
simplified as \bea\nn \delta E_{\mu}&=&-i\sum_{b=1}^4
(-1)^{F_b}e^{-i{\tilde q}^* L}\left\{
\left(1-\frac{\epsilon'_{Q}(p_1)}{\epsilon'_{1}({\tilde
q}^*)}\right) \left[\mathop {\rm Res} \limits_{q^*={\tilde q}^*}
{S^{AA}}^{b\mathbf{1}_Q}_{b\mathbf{1}_Q}(q^*,p_1)\right]
{S^{AB}}^{b\mathbf{1}_{\tilde Q}}_{b\mathbf{1}_{\tilde
Q}}(q^*,p_2)\right.
\\ \label{luscher1}
&+&\left. \left(1-\frac{\epsilon'_{{\tilde
Q}}(p_2)}{\epsilon'_{1}({\tilde q}^*)}\right) \left[\mathop {\rm
Res} \limits_{q^*={\tilde q}^*} {S^{AA}}^{b\mathbf{1}_{\tilde
Q}}_{b\mathbf{1}_{\tilde Q}}(q^*,p_2)\right]
{S^{AB}}^{b\mathbf{1}_Q}_{b\mathbf{1}_Q}(q^*,p_1)\right\}.
 \eea

As mentioned earlier, only the two cases of $b=1,2$ contributes
equally in the sum of Eq.(\ref{luscher1}) since these elements
contain $a_1$. Instead of the summation, we can multiply a factor
2 for the case of $b=1$. In that case, we can compute easily each
term using the $S$-matrix elements (\ref{saa}) and (\ref{sab}).
Furthermore, we restrict ourselves for the case where the two DGMs
are symmetric in both spheres, namely, $p_1=p_2$ and $Q={\tilde
Q}$. This leads to \bea \delta E_{\mu}=-4i e^{-i{\tilde q}^* L}
\left(1-\frac{\epsilon'_{Q}(p)}{\epsilon'_{1}({\tilde
q}^*)}\right) \left[\mathop {\rm Res} \limits_{q^*={\tilde q}^*}
{S^{AA}}^{1\mathbf{1}_Q}_{1\mathbf{1}_Q}(q^*,p)\right]
{S^{AB}}^{1\mathbf{1}_Q}_{1\mathbf{1}_Q}(q^*,p). \label{luscher2}
\eea

Explicit computations of each factor in (\ref{luscher2}) are
exactly the same as those in \cite{HatSuzi}. There are two types
of poles of $S_{\rm BDS}(y,X^{(Q)})$. The $s$-channel pole which
describe $(Q+1)$-DGM arises at $y^-={X^{(Q)}}^+$ while the
$t$-channel pole for $(Q-1)$-DGM (for $Q\ge 2$) at
$y^+={X^{(Q)}}^+$. We consider the $s$-channel pole first. Using
the location of the pole, we can find \bea {\tilde
q}^*=-\frac{i}{2g\sin\left(\frac{p-i\theta}{2}\right)}\quad
\to\quad e^{-i{\tilde q}^* L}\approx\exp\left[
-\frac{L}{2g\sin\left(\frac{p-i\theta}{2}\right)}\right]. \eea
From Eq.(\ref{disper}), one can also obtain \bea
1-\frac{\epsilon'_{Q}(p)}{\epsilon'_{1}({\tilde q}^*)}\approx
\frac{\sin\frac{p}{2}\sin\frac{p-i\theta}{2}}{\cosh\frac{\theta}{2}}.
\eea Furthermore, one can notice from Eqs.(\ref{saa}) and
(\ref{sab}) \bea \left[\mathop {\rm Res} \limits_{q^*={\tilde
q}^*} {S^{AA}}^{1\mathbf{1}_Q}_{1\mathbf{1}_Q}(q^*,p)\right]
{S^{AB}}^{1\mathbf{1}_Q}_{1\mathbf{1}_Q}(q^*,p)= \mathop {\rm Res}
\limits_{q^*={\tilde q}^*} {S_{\rm
SYM}}^{1\mathbf{1}_Q}_{1\mathbf{1}_Q}(q^*,p) \eea where $S_{\rm
SYM}$ is the $S$-matrix of the ${\cal N}=4$ SYM theory. Explicit
evaluation of the residue term becomes in the leading order \bea
-\frac{8ig e^{-ip}\sin^2\frac{p}{2}}{\sin\frac{p-i\theta}{2}}
\exp\left[-\frac{2e^{-\theta/2}\sin\frac{p}{2}}{\sin\frac{p-i\theta}{2}}\right]
\left(\frac{\eta(X^{(Q)})}{{\tilde\eta}(X^{(Q)})}\right)^2
\left(\frac{\eta(y)}{{\tilde\eta}(y)}\right)^{2Q}. \eea Combining
all these together, we get \bea \delta E_{\mu}&=&-\frac{8g
e^{-ip}\sin^3\frac{p}{2}}{\cosh\frac{\theta}{2}}
\exp\left[-\frac{2e^{-\theta/2}\sin\frac{p}{2}}{\sin\frac{p-i\theta}{2}}
-\frac{L}{2g\sin\left(\frac{p-i\theta}{2}\right)}\right]
\left(\frac{\eta(X^{(Q)})}{{\tilde\eta}(X^{(Q)})}\right)^2
\left(\frac{\eta(y)}{{\tilde\eta}(y)}\right)^{2Q}
\\ \nn
&=&-\frac{32g\sin^3\frac{p}{2}\
e^{i\alpha}}{\cosh\frac{\theta}{2}}\exp\left[
-\frac{2\sin^2\frac{p}{2}\cosh^2\frac{\theta}{2}}{\sin^2\frac{p}{2}+\sinh^2\frac{\theta}{2}}
\left(\frac{L-Q}{2g\sin\frac{p}{2}\cosh\frac{\theta}{2}}+1\right)\right]
\\ \nn
&=&-\frac{32g^2\sin^4\frac{p}{2}\
e^{i\alpha}}{\sqrt{Q^2+16g^2\sin^2\frac{p}{2}}}
\exp\left[-\frac{2\sin^2\frac{p}{2}\left(L+\sqrt{Q^2+16g^2\sin^2\frac{p}{2}}\right)
\sqrt{Q^2+16g^2\sin^2\frac{p}{2}}}{Q^2+16g^2\sin^4\frac{p}{2}}\right].
\eea The phase factor $e^{i\alpha}$ includes various phases
arising in the computation as well as the frame dependence of
$\eta$. As argued in \cite{HatSuzi}, we will drop this phase
assuming that this cancels out with appropriate prescription for
the L\"uscher formula.

The $t$-channel pole at $y^+={X^{(Q)}}^+$ gives exactly the same
contribution up to a phase factor. Therefore, combining together,
we finally obtain the finite-size effect of the two symmetric DGM
configuration as follows: \bea\nn \delta
E_{\mu}=-\frac{64g^2\sin^4\frac{p}{2}}{\sqrt{Q^2+16g^2\sin^2\frac{p}{2}}}
\exp\left[-\frac{2\sin^2\frac{p}{2}\left(L+\sqrt{Q^2+16g^2\sin^2\frac{p}{2}}\right)
\sqrt{Q^2+16g^2\sin^2\frac{p}{2}}}{Q^2+16g^2\sin^4\frac{p}{2}}\right].
\eea This is exactly what we have derived in Eq.(\ref{stringDGM})
if we identify $J_1=L,\ J_2=Q$ and $g=\sqrt{\lambda/2}$.

\subsubsection{Finite-size dyonic giant magnons in TsT-transformed
$AdS_5 \times S^5$}

Investigations on AdS/CFT duality for the cases with reduced or
without supersymmetry is of obvious interest and importance. An
interesting example of such correspondence between gauge and
string theory models with reduced supersymmetry is provided by an
exactly marginal deformation of $\mathcal{N} = 4$ SYM theory
\cite{LS95} and string theory on a $\beta$-deformed $AdS_5\times
S^5$ background suggested in \cite{LM05}. When $\beta\equiv\gamma$
is real, the deformed background can be obtained from $AdS_5\times
S^5$ by the so-called TsT transformation. It includes T-duality on
one angle variable, a shift of another isometry variable, then a
second T-duality on the first angle \cite{LM05,F05}. Taking into
account that the five-sphere has three isometric coordinates, one
can consider generalization of the above procedure, consisting of
chain of three TsT transformations. The result is a regular
three-parameter deformation of $AdS_5\times S^5$ string
background, dual to a non-supersymmetric deformation of
$\mathcal{N} = 4$ SYM \cite{F05}, which is conformal in the planar
limit to any order of perturbation theory \cite{AKS06}. The action
for this $\gamma_i$-deformed $(i=1,2,3)$ gauge theory can be
obtained from the initial one after replacement of the usual
product with associative $*$-product \cite{LM05,F05,BR05}.

An essential property of the TsT transformation is that it
preserves the classical integrability of string theory on
$AdS_5\times S^5$ \cite{F05}. The $\gamma$-dependence enters only
through the {\it twisted} boundary conditions and the {\it
level-matching} condition. The last one is modified since a closed
string in the deformed background corresponds to an open string on
$AdS_5\times S^5$ in general.

The finite-size correction to the GM energy-charge relation, in
the $\gamma$-deformed background, has been found in \cite{BF08},
by using conformal gauge and the string sigma model reduced to
$R_t\times S^3$. For the deformed case, this is the smallest
consistent reduction due to the {\it twisted} boundary conditions.
It turns out that even for the three-parameter deformation, the
reduced model depends only on one of them - $\gamma_3$. As far as
there are two isometry angles $\phi_1$, $\phi_2$ on $S^3$, the
solution can carry two non-vanishing angular momenta $J_1$, $J_2$.
Then, the GM is an open string solution with only one charge
$J_1\ne 0$. The momentum $p$ of the magnon excitation in the
corresponding spin chain is identified with the angular difference
$\Delta\phi_1$ between the end-points of the string. The other
angle satisfies the following {\it twisted} boundary conditions
\cite{BF08} \bea\nn \Delta\phi_2=2\pi(n_2-\gamma_3 J_1),\eea where
$n_2$ is an integer winding number of the string in the second
isometry direction of the deformed sphere  $S_\gamma^3$.

An interesting extension of this study is the dyonic giant magnon.
This state corresponds to bound states of the fundamental magnons
and stable even in the deformed theory. Understanding its string
theory analog in the strong coupling limit can be helpful to
extend the AdS/CFT duality to the deformed theories.

In \cite{18} we investigated dyonic giant magnons propagating on
$\gamma$-deformed $AdS_5\times S^5$ by Neumann-Rosochatius
reduction method with twisted boundary conditions. We compute
finite-size effect of the dispersion relations of dyonic giant
magnons, which generalizes the previously known case of the giant
magnons with one angular momentum found by Bykov and Frolov.

The bosonic part of the Green-Schwarz action for strings on the
$\gamma$-deformed $AdS_5\times S_\gamma^5$ reduced to $R_t\times
S_\gamma^5$ can be written as (the common radius $R$ of $AdS_5$
and $S_\gamma^5$ is set to 1) \cite{AAF05} \bea\label{BGS}
S&=&-\frac{T}{2}\int d\tau
d\sigma\left\{\sqrt{-\gamma}\gamma^{ab}\left[-\p_a t\p_b t+\p_a
r_i\p_br_i+Gr_i^2\p_a\varphi_i\p_b\varphi_i \right.\right.\\
\nn &&+\left.\left.Gr_1^2r_2^2r_3^2
\left(\hat{\gamma}_i\p_a\varphi_i\right)
\left(\hat{\gamma}_j\p_b\varphi_j\right) \right]\right.\\ \nn
&&-2G\left.\epsilon^{ab}\left(\hat{\gamma}_3r_1^2r_2^2\p_a\varphi_1\p_b\varphi_2
+\hat{\gamma}_1r_2^2r_3^2\p_a\varphi_2\p_b\varphi_3
+\hat{\gamma}_2r_3^2r_1^2\p_a\varphi_3\p_b\varphi_1\right)\right\}
,\eea where $\varphi_i$  are the three isometry angles of the
deformed $S_\gamma^5$, and \bea\label{roG}
\sum_{i=1}^{3}r_i^2=1,\h
G^{-1}=1+\hat{\gamma}_3r_1^2r_2^2+\hat{\gamma}_1r_2^2r_3^2
+\hat{\gamma}_2r_1^2r_3^2.\eea The deformation parameters
$\hat{\gamma}_i$ are related to $\gamma_i$ which appear in the
dual gauge theory as follows \bea\nn \hat{\gamma}_i = 2\pi T
\gamma_i = \sqrt{\lambda} \gamma_i .\eea When
$\hat{\gamma}_i=\hat{\gamma}$ this becomes the supersymmetric
background of \cite{LM05}, and the deformation parameter $\gamma$
enters the $\mathcal{N}=1$ SYM superpotential in the following way
\bea\nn W\propto
tr\left(e^{i\pi\gamma}\Phi_1\Phi_2\Phi_3-e^{-i\pi\gamma}\Phi_1\Phi_3\Phi_2\right).\eea

By using the TsT transformations which map the string theory on
$AdS_5\times S^5$ to the $\gamma_i$-deformed theory, one can
relate the angle variables $\phi_i$ on $S^5$ to the angles
$\varphi_i$ of the $\gamma_i$-deformed geometry \cite{F05}:
\bea\label{r} p_i=\pi_i,\h r_i^2\phi'_i=
r_i^2\left(\varphi'_i-2\pi\epsilon_{ijk}\gamma_j p_k\right),\h
i=1,2,3,\eea where $p_i$, $\pi_i$ are the momenta conjugated to
$\phi_i$, $\varphi_i$ respectively, and the summation is over
$j,k$. The equality $ p_i=\pi_i$ implies that the charges \bea\nn
J_i=\int d\sigma p_i \eea are invariant under the TsT
transformation.

If none of the variables $r_i$ is vanishing on a given string
solution, from (\ref{r}) one gets \bea\nn \phi'_i=
\varphi'_i-2\pi\epsilon_{ijk}\gamma_j p_k.\eea Integrating the
above equations and taking into account that for a closed string
in the $\gamma$-deformed background \bea\nn
\Delta\varphi_i=\varphi_i(r)-\varphi_i(-r)=2\pi n_i,\h n_i\in
\mathbb{Z} ,\eea one finds the {\it twisted} boundary conditions
for the angles $\phi_i$ on the original $S^5$ space \bea\nn
\Delta\phi_i=\phi_i(r)-\phi_i(-r)=2\pi\left(n_i-\nu_i\right),\h
\nu_i=\varepsilon_{ijk}\gamma_jJ_k.\eea It is obvious that if the
{\it twists} $\nu_i$ are not integer, then a closed string on the
deformed  background is mapped to an open string on $AdS_5\times
S^5$.

As we already explained, instead of considering strings on the
$\gamma$-deformed background $AdS_5\times S_\gamma^5$, we can
consider strings on the original $AdS_5\times S^5$ space, but with
{\it twisted} boundary conditions. Actually, here we are
interested in string configurations living in the $R_t\times S^3$
subspace, which can be described by the NR integrable system. Afer
the NR reduction one obtains the following expressions for the
conserved charges \bea\nn
&&\mathcal{E}=\frac{\kappa}{\alpha}\int_{-r}^{r}d\xi
=\frac{(1-v^2)w}{\sqrt{1-u^2}} \int_{\chi_{min}}^{\chi_{max}}
\frac{d\chi}{\sqrt{(\chi_{max}-\chi)(\chi-\chi_{min})(\chi-\chi_{n})}},
\\ \label{CQS} &&\mathcal{J}_1=\frac{1}{\sqrt{1-u^2}} \int_{\chi_{min}}^{\chi_{max}}
\frac{\left[1-v^2\left(w^2-u^2j\right)-\chi\right]d\chi}{\sqrt{(\chi_{max}-\chi)(\chi-\chi_{min})(\chi-\chi_{n})}},
\\ \nn &&\mathcal{J}_2=\frac{u}{\sqrt{1-u^2}} \int_{\chi_{min}}^{\chi_{max}}
\frac{\left(\chi-v^2j\right)d\chi}{\sqrt{(\chi_{max}-\chi)(\chi-\chi_{min})(\chi-\chi_{n})}},\eea
and for the angular differences \bea\nn
p=\Delta\phi_1=\phi_1(r)-\phi_1(-r),\h
\delta=\Delta\phi_2=\phi_2(r)-\phi_2(-r) =2\pi\left(n_2-\gamma_3
J_1\right).\eea \bea\label{dp}&&p=\int_{-r}^{r}d\xi
f'_1=\frac{\beta\omega_1}{\alpha^2(1-v^2)}\int_{-r}^{r}\left(1-\frac{w^2-u^2j}{r_1^2}\right)d\xi
\\ \nn &&=\frac{v}{\sqrt{1-u^2}}\int_{\chi_{min}}^{\chi_{max}}
\left(\frac{w^2-u^2j}{1-\chi}-1\right)\frac{d\chi}{\sqrt{(\chi_{max}-\chi)(\chi-\chi_{min})(\chi-\chi_{n})}},
\eea \bea\label{dd} &&\delta=\int_{-r}^{r}d\xi
f'_2=\frac{\beta\omega_2}{\alpha^2(1-v^2)}\int_{-r}^{r}\left(1-\frac{j}{r_2^2}\right)d\xi
\\ \nn &&=\frac{uv}{\sqrt{1-u^2}}\int_{\chi_{min}}^{\chi_{max}}
\left(\frac{j}{\chi}-1\right)\frac{d\chi}{\sqrt{(\chi_{max}-\chi)(\chi-\chi_{min})(\chi-\chi_{n})}}
.\eea

By using that \bea\nn &&\int_{\chi_{min}}^{\chi_{max}}
\frac{d\chi}{\sqrt{(\chi_{max}-\chi)(\chi-\chi_{min})(\chi-\chi_{n})}}
=\frac{2}{\sqrt{\chi_{max}-\chi_{n}}}\mathbf{K}(1-\epsilon),\\ \nn
&&\int_{\chi_{min}}^{\chi_{max}} \frac{\chi
d\chi}{\sqrt{(\chi_{max}-\chi)(\chi-\chi_{min})(\chi-\chi_{n})}}
\\ \nn &&
=\frac{2\chi_{n}}{\sqrt{\chi_{max}-\chi_{n}}}\mathbf{K}(1-\epsilon)+2\sqrt{\chi_{max}-\chi_{n}}
\mathbf{E}(1-\epsilon),
\\ \nn &&\int_{\chi_{min}}^{\chi_{max}}
\frac{d\chi}{\chi\sqrt{(\chi_{max}-\chi)(\chi-\chi_{min})(\chi-\chi_{n})}}
=\frac{2}{\chi_{max}\sqrt{\chi_{max}-\chi_{n}}}
\mathbf{\Pi}\left(1-\frac{\chi_{min}}{\chi_{max}}\vert
1-\epsilon\right),
\\ \nn&&\int_{\chi_{min}}^{\chi_{max}}
\frac{d\chi}{\left(1-\chi\right)\sqrt{(\chi_{max}-\chi)(\chi-\chi_{min})(\chi-\chi_{n})}}
\\ \nn &&=\frac{2}{\left(1-\chi_{max}\right)\sqrt{\chi_{max}-\chi_{n}}}
\mathbf{\Pi}\left(-\frac{\chi_{max}-\chi_{min}}{1-\chi_{max}}\vert
1-\epsilon\right),\eea where \bea\nn
\epsilon=\frac{\chi_{min}-\chi_{n}}{\chi_{max}-\chi_{n}},\eea from
(\ref{CQS}), (\ref{dp}) and (\ref{dd}) one finds \bea\nn
&&\mathcal{E}=\frac{4\tilde{\kappa}}{\sqrt{(1-\chi_n)(1-\tilde{v}^2)}}\mathbf{K}(1-\epsilon),
\\ \nn
&&\mathcal{J}_1=\frac{4\tilde{\kappa}}{(1-v^2)\sqrt{(1-\chi_n)(1-\tilde{v}^2)}}
\left[\left(\omega (1-\chi_n)-\frac{v^2}{\omega}(1+\nu
A_2)\right)\mathbf{K}(1-\epsilon)\right. \\ \nn
&&-\left.\omega(1-\chi_n)
(1-\tilde{v}^2)\mathbf{E}(1-\epsilon)\right],
\\ \label{fexpr}
&&\mathcal{J}_2=\frac{4\tilde{\kappa}}{(1-v^2)\sqrt{(1-\chi_n)(1-\tilde{v}^2)}}
\left[\left(v^2 A_2+\nu\chi_n)\right)\mathbf{K}(1-\epsilon)\right. \\
\nn &&+\left.\nu(1-\chi_n)
(1-\tilde{v}^2)\mathbf{E}(1-\epsilon)\right],
\\ \nn &&p=\frac{4\tilde{\kappa}v}{(1-v^2)\sqrt{(1-\chi_n)(1-\tilde{v}^2)}}
\left[\frac{1+\nu A_2}{\omega(1-\chi_n)\tilde{v}^2}
\mathbf{\Pi}\left(\frac{\tilde{v}^2-1}{\tilde{v}^2}(1-\epsilon)\vert
1-\epsilon\right)-\omega\mathbf{K}(1-\epsilon)\right],
\\ \nn &&\delta=-\frac{2\tilde{\kappa}v}{(1-v^2)\sqrt{(1-\chi_n)(1-\tilde{v}^2)}}
\left[\frac{A_2}{(1-\tilde{v}^2)\left(1+\chi_n\frac{\tilde{v}^2}{1-\tilde{v}^2}\right)}
\mathbf{\Pi}\left(\frac{1-\chi_n}{1+\chi_n\frac{\tilde{v}^2}{1-\tilde{v}^2}}(1-\epsilon)\vert
1-\epsilon\right)\right. \\ \nn
&&+\left.\nu\mathbf{K}(1-\epsilon)\right], \\ \nn
&&\tilde{\kappa}=\frac{1-v^2}{2\sqrt{\omega^2-\nu^2}}.\eea In the
above equalities we introduced the new parameters \bea\nn
\tilde{v}^2=\frac{1-\chi_{max}}{1-\chi_{n}},\h
\epsilon=\frac{\chi_{min}-\chi_{n}}{\chi_{max}-\chi_{n}}\eea
instead of $\chi_{max}$ and $\chi_{min}$.

In order to obtain the finite-size correction to the energy-charge
relation, we have to consider the limit $\epsilon\to 0$ in
(\ref{fexpr}). For the parameters in (\ref{fexpr}), we make the
following ansatz \bea\nn
&&v=v_0+v_1\epsilon+v_2\epsilon\log(\epsilon),\h
\tilde{v}=\tilde{v}_0+\tilde{v}_1\epsilon+\tilde{v}_2\epsilon\log(\epsilon),\h
\omega=1+\omega_1\epsilon,\\ \label{pex}
&&\nu=\nu_0+\nu_1\epsilon+\nu_2\epsilon\log(\epsilon),\h
A_2=A_{21}\epsilon,\h \chi_n=\chi_{n1}\epsilon.\eea We insert all
these expansions into (\ref{fexpr}) and impose the conditions:
\begin{enumerate}
\item{p - finite} \item{$\mathcal{J}_2$ - finite}
\item{$\mathcal{E}-\mathcal{J}_1=\frac{2\sqrt{1-v_0^2-\nu_0^2}}{1-\nu_0^2}
- \frac{(1-v_0^2-\nu_0^2)^{3/2}}{2(1-\nu_0^2)}\cos(\Phi)\epsilon$}
\end{enumerate}
From the first two conditions, we obtain the relations
\bea\label{ctpj2}
p=\arcsin\left(\frac{2v_0\sqrt{1-v_0^2-\nu_0^2}}{1-\nu_0^2}\right),
\h \tilde{v}_0=\frac{v_0}{\sqrt{1-\nu_0^2}},\h
\mathcal{J}_2=\frac{2\nu_0\sqrt{1-v_0^2-\nu_0^2}}{1-\nu_0^2},\eea
as well as six more equations. The third condition gives another
two equations for the coefficients in (\ref{pex}). Thus, we have a
system of eight equations, from which we can find all remaining
coefficients in (\ref{pex}), except $A_{21}$. $A_{21}$ can be
found from the equation for $\delta$ to be \bea\nn
A_{21}=-\Lambda\frac{(1-v_0^2-\nu_0^2)^{3/2}}{v_0(1-\nu_0^2)}\sin(\Phi),\eea
where $\Lambda$ is constant with respect to $\Phi$ (actually,
$\Lambda$ can be fixed to 1). The equations (\ref{ctpj2}) are
solved by \bea\label{zms}
v_0=\frac{\sin(p)}{\sqrt{\mathcal{J}_2^2+4\sin^2(p/2)}},\h
\tilde{v}_0=\cos(p/2),\h
\nu_0=\frac{\mathcal{J}_2}{\sqrt{\mathcal{J}_2^2+4\sin^2(p/2)}}.\eea
Replacing (\ref{zms}) into the solutions for the other
coefficients, one can obtain the expressions for the remaining
parameters in terms of physical quantities.

To the leading order, the equation for $\mathcal{J}_1$ gives
\bea\nn \epsilon=16\exp\left[-\frac{2\left(\mathcal{J}_1 +
\sqrt{\mathcal{J}_2^2+4\sin^2(p/2)}\right)
\sqrt{\mathcal{J}_2^2+4\sin^2(p/2)}\sin^2(p/2)}{\mathcal{J}_2^2+4\sin^4(p/2)}
\right].\eea Accordingly, to the leading order again, the equation
for $\delta$ reads \bea\label{fd}
2\pi\left(n_2-\gamma_3\frac{\sqrt{\lambda}}{2\pi}\mathcal{J}_1\right)+
\mathcal{J}_2\frac{\mathcal{J}_1 +
\sqrt{\mathcal{J}_2^2+4\sin^2(p/2)}}{\mathcal{J}_2^2
+4\sin^4(p/2)}\sin(p)=\Lambda\Phi .\eea Finally, the dispersion
relation, including the leading finite-size correction, takes the
form \bea\label{IEJ1} &&\mathcal{E}-\mathcal{J}_1 =
\sqrt{\mathcal{J}_2^2+4\sin^2(p/2)} - \frac{16 \sin^4(p/2)}
{\sqrt{\mathcal{J}_2^2+4\sin^2(p/2)}}\cos(\Phi)\\ \nn
&&\exp\left[-\frac{2\left(\mathcal{J}_1 +
\sqrt{\mathcal{J}_2^2+4\sin^2(p/2)}\right)
\sqrt{\mathcal{J}_2^2+4\sin^2(p/2)}\sin^2(p/2)}{\mathcal{J}_2^2+4\sin^4(p/2)}
\right].\eea For $\mathcal{J}_2=0$, (\ref{IEJ1}) reduces to the
result found in \cite{BF08}\footnote{We want to point out that our
result is different from \cite{BF08} which has extra $\cos^3(p/4)$
in the denominator of the phase $\Phi$.}.

\subsubsection{Finite-size giant magnons on $AdS_4 \times CP^3_{\gamma}$}

In \cite{21} we investigated finite-size giant magnons propagating
on $\gamma$-deformed  $AdS_4 \times CP^3_{\gamma}$ type IIA string
theory background, dual to one parameter deformation of the
$\mathcal{N}=6$ super Chern-Simoms-matter theory (ABJM theory)
\cite{ABJM}. The resulting theory has $\mathcal{N}=2$
supersymmetry and the modified superpotential is
\cite{EI0808}\bea\label{dsp} W_\gamma\propto
Tr\left(e^{-i\pi\gamma/2}A_1 B_1A_2B_2-e^{i\pi\gamma/2}A_1
B_2A_2B_1\right).\eea Here the chiral superfields $A_i$, $B_i$,
$(i=1,2)$ represent the matter part of the theory. As in the
$\mathcal{N} = 4$ SYM case, the marginality of the deformation
translates into the fact that $AdS_4$ part of the background is
untouched. Taking into account that $CP^3$ has three isometric
coordinates, one can consider a chain of three TsT
transformations. The result is a regular three-parameter
deformation of $AdS_4 \times CP^3$ string background, dual to a
non-supersymmetric deformation of ABJM theory, which reduces to
the supersymmetric one by putting $\gamma_1=\gamma_2=0$ and
$\gamma_3=\gamma$ \cite{EI0808}.

The dispersion relation for the GM in the $\gamma$-deformed $AdS_4
\times CP^3_{\gamma}$ background, carrying two nonzero angular
momenta, has been found in \cite{SR09}. Here we are interested in
obtaining the finite-size correction to it. Analyzing the
finite-size effect on the dispersion relation, we found that it is
modified compared to the undeformed case, acquiring $\gamma$
dependence.

Let us first write down the deformed background. It is given by
\cite{EI0808}\footnote{There are also nontrivial dilaton and
fluxes $F_2$, $F_4$, but since the fundamental string does not
interact with them at the classical level, we do not need to know
the corresponding expressions.} \bea\nn &&
ds^2_{IIA}=R^2\left(\frac{1}{4}ds^2_{AdS_4}+ds^2_{CP^3_{\gamma}}\right),
\\ \nn && ds^2_{CP^3_{\gamma}}= d\psi^2+G\sin^2 \psi \cos^2 \psi
\left(\frac{1}{2}\cos\theta_1 d\phi_1-\frac{1}{2}\cos\theta_2
d\phi_2+d\phi_3\right)^2 \\ \nn &&+\frac{1}{4}\cos^2 \psi
\left(d\theta_1^2+G\sin^2\theta_1 d\phi_1^2\right)
+\frac{1}{4}\sin^2 \psi \left(d\theta_2^2+G\sin^2\theta_2
d\phi_2^2\right) \\ \nn && +\tilde{\gamma}G\sin^4 \psi \cos^4 \psi
\sin^2\theta_1\sin^2\theta_2 d\phi_3^2,\eea \bea \nn && B_2=-R^2
\tilde{\gamma}G\sin^2 \psi \cos^2 \psi
\\\nn &&\times\left[\frac{1}{2}\cos^2 \psi\sin^2\theta_1\cos\theta_2 d\phi_3\wedge
d\phi_1+\frac{1}{2}\sin^2 \psi\sin^2\theta_2\cos\theta_1
d\phi_3\wedge d\phi_2 \right.
\\ \nn &&
+\left.\frac{1}{4}\left(\sin^2\theta_1\sin^2\theta_2+ \cos^2
\psi\sin^2\theta_1\cos^2\theta_2 +\sin^2
\psi\sin^2\theta_2\cos^2\theta_1\right)d\phi_1\wedge
d\phi_2\right] ,\eea where \bea\nn G^{-1}=1+\tilde{\gamma}^2\sin^2
\psi \cos^2 \psi \left(\sin^2\theta_1\sin^2\theta_2+ \cos^2
\psi\sin^2\theta_1\cos^2\theta_2 +\sin^2
\psi\sin^2\theta_2\cos^2\theta_1\right).\eea The deformation
parameter $\tilde{\gamma}$ above is given by
$\tilde{\gamma}=\frac{R^2}{4}\gamma$, where $\gamma$ appears in
the dual field theory superpotential (\ref{dsp}).

Further on, we restrict our attention to the $R_t\times
RP^3_{\gamma}$ subspace of $AdS_4 \times CP^3_{\gamma}$, where
$\theta_1=\theta_2=\pi/2$, $\phi_3=0$, and \bea\nn &&
ds^2=R^2\left(-\frac{1}{4}dt^2+d\psi^2 + \frac{G}{4}\cos^2\psi
d\phi_1^2 +\frac{G}{4}\sin^2\psi d\phi_2^2\right),
\\ \nn && B_2 =b_{\phi_1\phi_2}d\phi_1\wedge d\phi_2
=-\frac{R^2}{4} \tilde{\gamma}G\sin^2 \psi \cos^2 \psi
d\phi_1\wedge d\phi_2,
\\ \nn && G^{-1}=1+\tilde{\gamma}^2\sin^2 \psi \cos^2 \psi.\eea

To find the string solutions we are interested in, we use the
ansatz ($j=1,2$) \bea\label{gNRA} &&t(\tau,\sigma)=\kappa\tau,\h
\psi(\tau,\sigma)=\psi(\xi),\h
\phi_j(\tau,\sigma)=\omega_j\tau+f_j(\xi),\\ \nn
&&\xi=\alpha\sigma+\beta\tau,\h \kappa, \omega_j, \alpha,
\beta={\rm constants}.\eea It leads to reduction of the string
dynamics to the one of the $\gamma$-deformed NR system. The string
Lagrangian becomes (prime is used for $d/d\xi$) \bea\nn
&&\mathcal{L}_s=-\frac{TR^2}{2}(\alpha^2-\beta^2) \left[\psi'^2+
\frac{G}{4}\cos^2\psi\left(f'_1-\frac{\beta\omega_1}{\alpha^2-\beta^2}\right)^2
+\frac{G}{4}\sin^2\psi\left(f'_2-\frac{\beta\omega_2}{\alpha^2-\beta^2}\right)^2\right.
\\ \label{rl} &&-\left.\frac{G\alpha^2}{4(\alpha^2-\beta^2)^2}
\left(\omega_1^2\cos^2\psi+\omega_2^2\sin^2\psi\right)
+\frac{\alpha\tilde{\gamma}G}{2}\sin^2 \psi \cos^2 \psi
\frac{\omega_1 f'_2-\omega_2 f'_1}{\alpha^2-\beta^2}\right], \eea
while the Virasoro constraints acquire the form \bea\nn &&
\psi'^2+
\frac{G}{4}\cos^2\psi\left(f'^2_1+\frac{2\beta\omega_1}{\alpha^2+\beta^2}f'_1+\frac{\omega_1^2}{\alpha^2+\beta^2}\right)
\\ \label{rc} &&+\frac{G}{4}\sin^2\psi\left(f'^2_2+\frac{2\beta\omega_2}{\alpha^2+\beta^2}f'_2+\frac{\omega_2^2}{\alpha^2+\beta^2}\right)
=\frac{\kappa^2/4}{\alpha^2+\beta^2},
\\ \nn && \psi'^2+
\frac{G}{4}\cos^2\psi\left(f'^2_1+\frac{\omega_1}{\beta}f'_1\right)
+\frac{G}{4}\sin^2\psi\left(f'^2_2+\frac{\omega_2}{\beta}f'_2\right)
=0 .\eea The equations of motion for $f_j(\xi)$ following from
(\ref{rl}) can be integrated once to give \bea\label{fjs}  &&
f'_1=\frac{1}{\alpha^2-\beta^2} \left[\frac{C_1}{\cos^2 \psi}
+\beta\omega_1+\tilde{\gamma}\left(\alpha\omega_2+\tilde{\gamma}C_1\right)\sin^2\psi\right],
\\ \nn && f'_2=\frac{1}{\alpha^2-\beta^2} \left[\frac{C_2}{\sin^2 \psi}
+\beta\omega_2-\tilde{\gamma}\left(\alpha\omega_1-\tilde{\gamma}C_2\right)\cos^2\psi\right]
,\eea where $C_j$ are constants. Replacing (\ref{fjs}) into
(\ref{rc}), one can rewrite the Virasoro constraints as
\bea\label{00r} &&\psi'^2=\frac{1}{4(\alpha^2-\beta^2)^2}
\Bigg[(\alpha^2+\beta^2)\kappa^2 -\frac{C_1^2}{\cos^2\psi}
-\frac{C_2^2}{\sin^2\psi}
\\ \nn &&-\left(\alpha\omega_1-\tilde{\gamma}C_2\right)^2\cos^2\psi
-\left(\alpha\omega_2+\tilde{\gamma}C_1\right)^2\sin^2\psi\Bigg],
\\ \label{01r} && \omega_1C_1+\omega_2C_2+\beta\kappa^2=0.\eea
Let us point out that (\ref{00r}) is the first integral of the
equation of motion for $\psi$. Integrating (\ref{fjs}) and
(\ref{00r}), one can find string solutions with very different
properties. Particular examples are (dyonic) giant magnons and
single-spike strings.

In the case at hand, the background metric does not depend on $t$
and $\phi_j$. The corresponding conserved quantities are the
string energy $E_s$ and two angular momenta $J_j$, given by
\bea\label{cqsg} &&E_s= \frac{TR^2}{4}\frac{\kappa}{\alpha}\int
d\xi,\\ \nn &&J_1= \frac{TR^2}{4}\frac{1}{\alpha^2-\beta^2}\int
d\xi
\left[\frac{\beta}{\alpha}C_1+\left(\alpha\omega_1-\tilde{\gamma}C_2
\right)\cos^2\psi \right], \\ \nn &&J_2=
\frac{TR^2}{4}\frac{1}{\alpha^2-\beta^2}\int d\xi
\left[\frac{\beta}{\alpha}C_2+\left(\alpha\omega_2+\tilde{\gamma}C_1
\right)\sin^2\psi \right].\eea Let us remind that the relation
between the string tension $T$ and the t'Hooft coupling constant
$\lambda$ for the $\mathcal{N}=6$ super Chern-Simoms-matter theory
is given by \bea\nn TR^2=2\sqrt{2\lambda}.\eea

If we introduce the variable \bea\nn \chi=\cos^2\psi,\eea and use
(\ref{01r}), the first integral (\ref{00r}) can be rewritten as
\bea\nn \chi'^{2}
=\frac{\Omega_2^2\left(1-u^{2}\right)}{\alpha^2(1-v^2)^2}
(\chi_{p}-\chi)(\chi-\chi_{m})(\chi-\chi_{n}) ,\eea where \bea\nn
&&\chi_p+\chi_m+\chi_n=\frac{2-(1+v^2)W-u^2}{1
-u^2},\\
\label{3eqs} &&\chi_p \chi_m+\chi_p \chi_n+\chi_m
\chi_n=\frac{1-(1+v^2)W+(v W-u K)^2-K^2}{1 -u^2},\\ \nn && \chi_p
\chi_m \chi_n=- \frac{K^2}{1 -u^2},\eea and \bea\nn &&
v=-\frac{\beta}{\alpha},\h u=\frac{\Omega_1}{\Omega_2},\h
W=\left(\frac{\kappa}{\Omega_2}\right)^2,\h
K=\frac{C_1}{\alpha\Omega_2},
\\ \nn &&\Omega_1=\omega_1\left(1-\tilde{\gamma}\frac{C_2}{\alpha\omega_1}\right), \h
\Omega_2=\omega_2\left(1+\tilde{\gamma}\frac{C_1}{\alpha\omega_2}\right).\eea
We are interested in the case \bea\nn 0<\chi_{m}<\chi<
\chi_{p}<1,\h \chi_{n}<0,\eea which corresponds to the finite-size
giant magnons.

In terms of the newly introduced variables, the conserved
quantities (\ref{cqsg}) and the angular differences
\bea\label{dad} p_1\equiv\Delta\phi_1=\phi_1(r)-\phi_1(-r),\h
p_2\equiv\Delta\phi_2=\phi_2(r)-\phi_2(-r) ,\eea transform to
\bea\label{E} &&\mathcal{E}\equiv \frac{E_s}{TR^2}
=\frac{(1-v^2)\sqrt{W}}{\sqrt{1-u^2}}\frac{
\mathbf{K}(1-\epsilon)}{\sqrt{\chi_{p}-\chi_{n}}},
\\ \label{J1} &&\mathcal{J}_1\equiv\frac{J_1}{TR^2}=\frac{1}{\sqrt{1-u^2}}
\left[\frac{u\chi_n-v K}{\sqrt{\chi_{p}-\chi_{n}}}
\mathbf{K}(1-\epsilon) +u\sqrt{\chi_{p}-\chi_{n}}
\mathbf{E}(1-\epsilon)\right],
\\ \label{J2} &&\mathcal{J}_2\equiv\frac{J_2}{TR^2}=\frac{1}{\sqrt{1-u^2}}
\Big[\frac{1-\chi_n-v\left(v W-u
K\right)}{\sqrt{\chi_{p}-\chi_{n}}}\mathbf{K}(1-\epsilon)
\\ \nn
&&-\sqrt{\chi_{p}-\chi_{n}} \mathbf{E}(1-\epsilon)\Big],\eea
\bea
\label{gp1}&&p_1=\frac{4}{\sqrt{1-u^2}}
\\ \nn &&\times\Bigg\{\frac{K}{\chi_p\sqrt{\chi_{p}-\chi_{n}}}
\mathbf{\Pi}\left(1-\frac{\chi_{m}}{\chi_{p}}\vert
1-\epsilon\right) - \left[uv+\tilde{\gamma}v\left(v W-u
K\right)-\tilde{\gamma}\left(1-\chi_n\right)\right]
\frac{\mathbf{K}(1-\epsilon)}{\sqrt{\chi_{p}-\chi_{n}}}
\\ \nn &&-\tilde{\gamma}
\sqrt{\chi_{p}-\chi_{n}} \mathbf{E}(1-\epsilon)\Bigg\} ,\eea
\bea
\label{gp2} &&p_2=\frac{4}{\sqrt{1-u^2}}
\\ \nn &&\times\Bigg\{\frac{v W-u
K}{(1-\chi_p)\sqrt{\chi_{p}-\chi_{n}}}
\mathbf{\Pi}\left(-\frac{\chi_{p}-\chi_{m}}{1-\chi_{p}}\vert
1-\epsilon\right)-\left[v\left(1-\tilde{\gamma}K\right)+\tilde{\gamma}u\chi_{n}\right]
\frac{\mathbf{K}(1-\epsilon)}{\sqrt{\chi_{p}-\chi_{n}}}  \\
\nn &&-\tilde{\gamma}u\sqrt{\chi_{p}-\chi_{n}}
\mathbf{E}(1-\epsilon)\Bigg\},\eea where $\epsilon$ is given by
\bea\label{de}
\epsilon=\frac{\chi_{m}-\chi_{n}}{\chi_{p}-\chi_{n}}.\eea

From (\ref{E})-(\ref{J2}) one can see that the conserved charges
are not affected by the $\gamma$-deformation as it should be. Only
the angular differences are shifted.

Further on, we will consider the case when $\mathcal{E}$,
$\mathcal{J}_2$ and $p_1$ are large, while
$\mathcal{E}-\mathcal{J}_2$, $\mathcal{J}_1$ and $p_2$ are finite.
To this end, we will introduce appropriate expansions.

{\bf Expansions}

In order to find the leading finite-size correction to the
energy-charge relation, we have to consider the limit $\epsilon\to
0$ in (\ref{3eqs}), (\ref{E})-(\ref{de})). We will use the
following ansatz for the parameters
$(\chi_p,\chi_m,\chi_n,v,u,W,K)$ \bea\nn
&&\chi_p=\chi_{p0}+\left(\chi_{p1}+\chi_{p2}\log(\epsilon)\right)\epsilon,
\\ \nn &&\chi_m=\chi_{m0}+\left(\chi_{m1}+\chi_{m2}\log(\epsilon)\right)\epsilon,
\\ \nn &&\chi_n=\chi_{n0}+\left(\chi_{n1}+\chi_{n2}\log(\epsilon)\right)\epsilon,
\\
\label{Dpars} &&v=v_0+\left(v_1+v_2\log(\epsilon)\right)\epsilon, \\
\nn &&u=u_0+\left(u_1+u_2\log(\epsilon)\right)\epsilon,
\\ \nn &&W=W_0+\left(W_1+W_2\log(\epsilon)\right)\epsilon,
\\ \nn &&K=K_0+\left(K_1+K_2\log(\epsilon)\right)\epsilon .\eea
A few comments are in order. To be able to reproduce the
dispersion relation for the infinite-size giant magnons, we set
\bea\label{is} \chi_{m0}=\chi_{n0}=K_0=0,\h W_0=1.\eea Also to
reproduce the undeformed case \cite{ABR} in the ${\tilde\gamma}\to
0$ limit, we need to fix \bea\label{k2} \chi_{m2}=
\chi_{n2}=W_2=K_2=0.\eea

Replacing (\ref{Dpars}) into (\ref{3eqs}) and (\ref{de}), one
finds six equations for the coefficients in the expansions of
$\chi_p$, $\chi_m$, $\chi_n$ and $W$.
They are solved by \bea\label{chi} &&\chi_{p0}=1-\frac{v_0^2}{1-u_0^2}, \\
\nn &&\chi_{p1}=
\frac{v_0}{\left(1-v_0^2\right)\left(1-u_0^2\right)(1-v_0^2-u_0^2)}
\Big\{-2v_0u_0(1-v_0^2)(1-v_0^2-u_0^2)u_1
\\ \nn &&+2\left(1-u_0^2\right)(1-v_0^2-u_0^2)
\left[K_1u_0(1+v_0^2)-(1-v_0^2)v_1\right]
\\ \nn
&&+v_0(1-v_0^2-2u_0^2)\sqrt{(1-u_0^2-v_0^2)^4-4K_1^2(1-u_0^2)^2(1-u_0^2-v_0^2)}\Big\},\eea
\bea \nn &&\chi_{p2}=
-2v_0\frac{v_2+(v_0u_2-u_0v_2)u_0}{\left(1-u_0^2\right)^2} \\
\nn &&\chi_{m1}= \frac{u_0^4
-2u_0^2(1-v_0^2)+(1-v_0^2)^2+\sqrt{(1-u_0^2-v_0^2)^4-4K_1^2(1-u_0^2)^2(1-u_0^2-v_0^2)}}
{2(1-u_0^2)(1-v_0^2-u_0^2)}, \\
\nn &&\chi_{n1}= -\frac{u_0^4
-2u_0^2(1-v_0^2)+(1-v_0^2)^2-\sqrt{(1-u_0^2-v_0^2)^4-4K_1^2(1-u_0^2)^2(1-u_0^2-v_0^2)}}
{2(1-u_0^2)(1-v_0^2-u_0^2)},
\\ \nn &&W_1=-\frac{2K_1u_0v_0(1-u_0^2)+\sqrt{(1-u_0^2-v_0^2)^4-4K_1^2(1-u_0^2)^2(1-u_0^2-v_0^2)}}
{(1-u_0^2)(1-v_0^2)}.\eea

As a next step, we impose the conditions for $\mathcal{J}_1$,
$p_2$ to be independent of $\epsilon$. By expanding r.h.s. of
(\ref{J1}), (\ref{p2}) in $\epsilon$, one gets \bea \label{j2ex}
\mathcal{J}_1=\frac{ u_0\sqrt{1-v_0^2-u_0^2}}{1-u_0^2},\eea
\bea\label{p2ex} p_2=2\arcsin\left(\frac{2
v_0\sqrt{1-v_0^2-u_0^2}}{1-u_0^2}\right) -4\tilde{\gamma}u_0
\frac{\sqrt{1-v_0^2-u_0^2}}{1-u_0^2},\eea along with four more
equations from the coefficients of $\epsilon$ and $\epsilon\log
\epsilon$. The equalities (\ref{j2ex}), (\ref{p2ex}) lead to
\bea\label{zmsg}
v_0=\frac{\sin\Psi}{2\sqrt{\mathcal{J}_1^2+\sin^2(\Psi/2)}},\h
u_0=\frac{\mathcal{J}_1}{\sqrt{\mathcal{J}_1^2+\sin^2(\Psi/2)}},
\h p_2=2\left(\Psi-2\tilde{\gamma}\mathcal{J}_1\right),\eea where
the angle $\Psi$ is defined as \bea\nn \Psi=\arcsin\left(\frac{2
v_0\sqrt{1-v_0^2-u_0^2}}{1-u_0^2}\right).\eea After the
replacement of (\ref{chi}) into the remaining four equations, they
can be solved with respect to $v_1$, $v_2$, $u_1$, $u_2$, leading
to the following form of the dispersion relation in the considered
approximation \bea\label{echr1}
\mathcal{E}-\mathcal{J}_2=\frac{\sqrt{1-v_0^2-u_0^2}}{1-u_0^2}
-\frac{1}{4}\frac{\sqrt{(1-v_0^2-u_0^2)^3-4K_1^2(1-u_0^2)^2}}{1-u_0^2}
\epsilon .\eea To the leading order, the expansion for
$\mathcal{J}_2$ gives \bea\label{eps}
\epsilon=16\exp\left[-\frac{2}{1-v_0^2}\left(1-\frac{
v_0^2}{1-u_0^2}+\mathcal{J}_2\sqrt{1-v_0^2-u_0^2}\right)\right].\eea
By using (\ref{zmsg}) and (\ref{eps}), (\ref{echr1}) can be
rewritten as \bea\label{IEJ1i} &&\mathcal{E}-\mathcal{J}_2 =
\sqrt{\mathcal{J}_1^2+\sin^2(\Psi/2)} - 4\sqrt{\frac{\sin^8(\Psi/2)}{\mathcal{J}_1^2+\sin^2(\Psi/2)}-4K_1^2}\\
\nn &&\exp\left[-\frac{2\left(\mathcal{J}_2 +
\sqrt{\mathcal{J}_1^2+\sin^2(\Psi/2)}\right)
\sqrt{\mathcal{J}_1^2+\sin^2(\Psi/2)}\sin^2(\Psi/2)}{\mathcal{J}_1^2+\sin^4(\Psi/2)}
\right].\eea

The parameter $K_1$ in (\ref{IEJ1i}) can be related to the angular
difference $p_1$. To see that, let us consider the leading order
in the $\epsilon$-expansion for it: \bea\nn &&p_1=\frac{4K_1
\arctan
\sqrt{\frac{\chi_{p0}}{\chi_{m1}}-1}}{\sqrt{(1-u_0^2)\chi_{p0}\chi_{m1}(\chi_{p0}-\chi_{m1})}}
\\ \label{pt1} &&-\frac{2}{\sqrt{(1-u_0^2)\chi_{p0}}} \left[u_0v_0\log(16)
+\tilde{\gamma}\left(2\chi_{p0}-(1-v_0^2)\log(16)\right)\right]
\\ \nn &&+\frac{2}{\sqrt{(1-u_0^2)\chi_{p0}}} \left[u_0v_0
-\tilde{\gamma}(1-v_0^2)\right]\log(\epsilon).\eea So, it is
natural to introduce the angle $\Phi$ as \bea\label{ltd}
\frac{\Phi}{2}= \arctan \sqrt{\frac{\chi_{p0}}{\chi_{m1}}-1}.\eea
On the solution for the other parameters this gives
\bea\label{kn1f}
K_1=\frac{(1-v_0^2-u_0^2)^{3/2}}{2(1-u_0^2)}\sin(\Phi)
=\frac{\sin^4(\Psi/2)}{2\sqrt{\mathcal{J}_1^2+\sin^2(\Psi/2)}}\sin(\Phi)
.\eea As a result, the relation (\ref{pt1}) between the angles
$p_1$ and $\Phi$ becomes \bea\label{Phi} &&\Phi=
\frac{p_1}{2}-\left(2\tilde{\gamma}-\mathcal{J}_1\frac{\sin\Psi}{\mathcal{J}_1^2+\sin^4(\Psi/2)}\right)\mathcal{J}_2+
\mathcal{J}_1\frac{\sin\Psi\sqrt{\mathcal{J}_1^2+\sin^2(\Psi/2)}}{\mathcal{J}_1^2+\sin^4(\Psi/2)}
,\eea where due to the periodicity condition we should set \bea\nn
p_1=2\pi n_1,\h n_1 \in \mathbb{Z}.\eea

Finally, in view of (\ref{kn1f}), the dispersion relation
(\ref{IEJ1i}) for the dyonic giant magnons acquires the form
\bea\label{IEJ1f} &&\mathcal{E}-\mathcal{J}_2 =
\sqrt{\mathcal{J}_1^2+\sin^2(\Psi/2)} - \frac{4
\sin^4(\Psi/2)}{\sqrt{\mathcal{J}_1^2+\sin^2(\Psi/2)}}\cos\Phi\\
\nn &&\exp\left[-\frac{2\left(\mathcal{J}_2 +
\sqrt{\mathcal{J}_1^2+\sin^2(\Psi/2)}\right)
\sqrt{\mathcal{J}_1^2+\sin^2(\Psi/2)}\sin^2(\Psi/2)}{\mathcal{J}_1^2+\sin^4(\Psi/2)}
\right].\eea Based on the L\"uscher $\mu$-term formula for the
undeformed case \cite{17}, we propose to identify the angle
$\Psi\left(=\frac{p_2}{2}+2\tilde{\gamma}\mathcal{J}_1\right)$
with the momentum $p$ of the magnon exitations in the dual spin
chain.

Let us point out that (\ref{IEJ1f}) has the same form as the
dispersion relation for dyonic giant magnons on $R_t\times
S^3_{\gamma}$ subspace of the  $\gamma$-deformed $AdS_5\times S^5$
\cite{17}. Actually, the two energy-charge relations coincide
after appropriate normalization of the charges and after exchange
of the indices 1 and 2. The only remaining difference is in the
first terms in the expressions for the angle $\Phi$: \bea\nn
&&R_t\times RP^3_{\gamma}: \Phi=\frac{p_1}{2}+\ldots
\\ \nn &&R_t\times S^3_{\gamma}: \Phi= p_2+\ldots .\eea

All of the above results simplify a lot when one consider giant
magnons with one angular momentum, i.e. $\mathcal{J}_1=0$. In
particular, the energy-charge relation (\ref{IEJ1f}) reduces to
\bea\label{IEJ10} &&\mathcal{E}-\mathcal{J}_2 =\sin\frac{p}{2}
\left[1-4\sin^2\frac{p}{2} \cos\left(\pi
n_1-2\tilde{\gamma}\mathcal{J}_2\right)
e^{-2-2\mathcal{J}_2\csc\frac{p}{2}}\right].\eea

\subsubsection{String solutions in $AdS_3 \times S^3 \times T^4$
with NS-NS B-field}

In \cite{27} we developed an approach for solving the string
equations of motion and Virasoro constraints in any background
which has some (unfixed) number of commuting Killing vector fields
(see the beginning of Section {\bf 3.1}). It is based on a
specific ansatz for the string embedding, which is the one given
in (\ref{A}).

Here, we apply the above mentioned approach for strings moving in
$AdS_3\times S^3\times T^4$ with 2-form NS-NS B-field. We
succeeded to find solutions for a large class of string
configurations on this background. In particular, we derive dyonic
giant magnon solutions in the $R_t \times S^3$ subspace, and
obtain the leading finite-size correction to the dispersion
relation.

{\bf Strings in $AdS_3\times S^3\times T^4$ with NS-NS B-field}

The background geometry of this target space can be written in the
following form \footnote{The common radius $R$ of the three
subspaces is set to 1, and $q$ is the parameter used in
\cite{HST1311}.}: \bea\nn &&ds^2_{AdS_3}=
-(1+r^2)dt^2+(1+r^2)^{-1}dr^2+r^2d\phi^2,\h b_{t\phi}=q r^2,
\\ \nn &&ds^2_{S^3}=
d\theta^2+\sin^2\theta d\phi_1^2+\cos^2\theta d\phi_2^2,\h
b_{\phi_1\phi_2}=-q\cos^2\theta, \\ \nn
&&ds^2_{T^4}=(d\varphi^i)^2,\h i=1,2,3,4.\eea

According to our notations \bea\nn
&&X^\mu=\left(t,\phi,\phi_1,\phi_2,\varphi^i\right),\h
X^a=\left(r,\theta\right),
\\ \nn &&g_{\mu\nu}=\left(g_{t t},g_{\phi
\phi},g_{\phi_1\phi_1},g_{\phi_2\phi_2},g_{ij}\right),\h
g_{ab}=\left(g_{r r},g_{\theta\theta}\right),\h g_{a \mu}=0,\h
h_{ab}=g_{ab},
\\ \nn &&b_{\mu\nu}=(b_{t \phi},b_{\phi_1\phi_2}),\h b_{a \nu}=0,
\\ \label{idc} &&A_a=0,\eea
where \bea\nn &&g_{t t}=(g^{t t})^{-1}=-(1+r^2),\h g_{\phi
\phi}=(g^{\phi \phi})^{-1}=r^2, \h g_{\phi_1 \phi_1}=(g^{\phi_1
\phi_1})^{-1}=\sin^2\theta,
\\ \nn &&g_{\phi_2 \phi_2}=(g^{\phi_2 \phi_2})^{-1}=\cos^2\theta,
\h g_{ij}=\left(g^{ij}\right)^{-1}= \delta _{ij},
\\ \nn
&&g_{r r}=(g^{r r})^{-1}=(1+r^2)^{-1}, \h g_{\theta\theta}=1,
\\ \label{bfs} &&b_{t\phi}=q r^2,\h b_{\phi_1\phi_2}=-q\cos^2\theta .\eea

Since $g_{a \mu}=0$, the solutions (\ref{MS}) for the coordinates
$X^\mu$ are simplified to \bea\label{Xmu}
X^{\mu}(\tau,\sigma)=\Lambda^{\mu}\tau +\tilde{X}^\mu(\xi)
=\Lambda^{\mu}\tau +\frac{1}{\alpha^2-\beta^2}\int d\xi
\left[g^{\mu\nu} \left(C_\nu-\alpha\Lambda^\rho
b_{\nu\rho}\right)+\beta\Lambda^\mu\right] ,\eea where
$g^{\mu\nu}$ and $b_{\nu\rho}$ must be replaced from above.

Now, we want to find the solutions for the non-isometric string
coordinates $X^a$. To this end we have to solve the equations
(\ref{Ea}), which in the case at hand reduce to \bea\label{Eqa}
(\alpha^2-\beta^2)\left[g_{a
b}\tilde{X}^{b''}+\Gamma_{a,bc}\tilde{X}^{b'}\tilde{X}^{c'}\right]
+\p_a \sum_{b=r,\theta} U_b=0,\eea where the scalar potential $U$
in (\ref{SP}) is represented as a sum of two parts: $ U_r= U_r(r)$
for the $AdS_3$ subspace and $ U_\theta= U_\theta(\theta)$ for the
$S^3$ subspace of the background.

Taking into account that the metric $g_{a b}$ is diagonal, one can
find the following two first integrals of (\ref{Eqa})
\bea\label{FIa} \tilde{X}^{a'}=
\sqrt{\frac{C_a-2U_a}{(\alpha^2-\beta^2)g_{aa}}}.\eea It follows
from here that \bea\label{xi} d\xi=
\frac{d\tilde{X}^a}{\sqrt{\frac{C_a-2U_a}{(\alpha^2-\beta^2)g_{aa}}}}.\eea
So, we have two different expressions for $ d\xi$, which obviously
must coincide. This is a condition for self-consistency. It leads
to \bea\label{O} \int\frac{dr}{\sqrt{\frac{C_r-2U_r}{g_{rr}}}}=
\int\frac{d\theta}{\sqrt{\frac{C_\theta-2U_\theta}{g_{\theta\theta}}}},\eea
which actually gives implicitly the ``orbit'' $r(\theta)$, i.e.
how the radial coordinate $r$ on $AdS_3$ depends on the angle
$\theta$ in $S^3$.

Now, we have to check if the first integrals for
$\tilde{X}^a(\xi)$ are compatible with the Virasoro constraints
(\ref{V12}). Replacing $\tilde{X}^{a'}$ in the first of them, one
finds \bea\nn C_r+C_\theta=0.\eea

Thus, we found all first integrals of the string equations of
motion, compatible with the Virasoro constraints, which reduce to
algebraic relations between the embedding parameters and the
integration constants.

Now, let us give the expressions for the conserved charges
(\ref{Qmu}), corresponding to the isometric coordinates.
\bea\label{EsS} &&-Q_t\equiv
E_s=\frac{T}{\alpha^2-\beta^2}\left[\left(\alpha
\Lambda^t-\frac{\beta}{\alpha}C_t-q \ C_\phi\right)\int
d\xi+\alpha (1-q^2)\Lambda^t\int d\xi r^2\right],
\\ \nn &&Q_\phi\equiv
S=\frac{T}{\alpha^2-\beta^2}\Bigg[\left(\frac{\beta}{\alpha}C_\phi+q
C_t+ q^2 \alpha \Lambda^\phi\right)\int d\xi
+(1-q^{2})\alpha\Lambda^{\phi}\int d\xi r^2
\\ \label{Sf} &&-\left(q C_t+q^2\alpha
\Lambda^\phi\right)\int \frac{d\xi}{1+r^2}\Bigg], \eea
\bea\label{Qt} &&Q_{\phi_1}\equiv
J_{1}=\frac{T}{\alpha^2-\beta^2}\Bigg[\left(\frac{\beta}{\alpha}C_{\phi_1}+
\alpha\Lambda^{\phi_1}-q C_{\phi_2}\right)\int d\xi
\\ \nn &&-(1-q^2)\alpha \Lambda^{\phi_1}\int \cos^2\theta d\xi\Bigg],
\\ \nn &&Q_{\phi_2}\equiv
J_{2}=\frac{T}{\alpha^2-\beta^2}\Bigg[\left(\frac{\beta}{\alpha}C_{\phi_2}
-q\left(C_{\phi_1}+q\alpha\Lambda^{\phi_2}\right)\right)\int d\xi \\
\nn &&+(1-q^2)\alpha\Lambda^{\phi_2}\int \cos^2\theta d\xi
+q\left(C_{\phi_1}+q\alpha\Lambda^{\phi_2}\right)\int
\frac{d\xi}{1-\cos^2\theta}\Bigg],\eea \bea \label{Ji}
&&Q_{i}\equiv J_i^T= \frac{T}{\alpha^2-\beta^2}
\left(\frac{\beta}{\alpha}C_i +\alpha \Lambda^j
\delta_{ij}\right)\int d\xi.\eea Here we used the following
notations: $E_s$ is the string energy, $S$ is the spin in $AdS_3$,
$J_{1}$ and $J_{2}$ are the two angular momenta in $S^3$, while
$J_i^T$ are the four angular momenta on $T^4$.

The explicit expressions for the string coordinates, the ``orbit''
$r(\theta)$, and the conserved charges in this background are
given in Appendix C.

{\bf Giant magnon solutions}

The giant magnon string solution was found in \cite{HM06}. It is a
specific string configuration, living in the $R_t\times S^2$
subspace of $AdS_5\times S^5$ with an angular momentum $J_1$ which
goes to $\infty$. A similar configuration, dyonic giant magnon,
has been obtained in \cite{CDO06} which moves in $R_t\times S^3$
subspace with two angular momenta $J_1,\ J_2$ with $J_1\to\infty$.
These classical configurations have played an important role in
understanding exact, quantum aspects of the AdS/CFT
correspondence. In particular, corrections due to a large but
finite $J_1$ obtained in \cite{AFZ06} and \cite{HS08} can provide
a nontrivial check for the exact worldsheet $S$-matrix.

Here we provide similar string solutions in $AdS_3\times S^3\times
T^4$ with NS-NS B-field for a large but finite $J_1$. Dyonic giant
magnon solution with infinite angular momentum $J_1$ has been
constructed in \cite{HST1311} with the following dispersion
relation \footnote{The terms proportional to $q$ are due to the
nonzero B-field on $S^3$.} \bea\label{Tdr} E_s-J_1= \sqrt{(J_2-q T
\Delta\phi_1)^2 +4 T^2(1-q^2) \sin^2\frac{\Delta\phi_1}{2}}.\eea
This relation is already quite different from those for the
ordinary (dyonic) giant magnons. We will show that there exist
even bigger differences for the finite-size corrections.

{\it Exact results}

In order to consider dyonic giant magnon solutions, we restrict
our general ansatz (\ref{A}) in the following way: \bea\nn
&&X^t\equiv t=\kappa\tau,\h \mbox{i.e.}\h \Lambda^t=\kappa,\h
\tilde{X}^t(\xi)=0,
\\ \nn &&X^\phi\equiv \phi=0,\h \mbox{i.e.}\h
\Lambda^\phi=0,\h \tilde{X}^\phi(\xi)=0
\\ \nn &&X^r \equiv r=\tilde{X}^r(\xi)=0,
\\ \nn &&X^{\phi_1}\equiv\phi_1=\omega_1\tau
+\tilde{X}^{\phi_1}(\xi),\h \mbox{i.e.}\h
\Lambda^{\phi_1}=\omega_1,
\\ \nn &&X^{\phi_2}\equiv\phi_2=\omega_2\tau
+\tilde{X}^{\phi_2}(\xi),\h \mbox{i.e.}\h
\Lambda^{\phi_2}=\omega_2,
\\ \nn &&X^\theta \equiv \theta =\tilde{X}^\theta(\xi),\h X^{\varphi^i}\equiv\varphi^i=0.\eea

As a result, we can claim that \bea\nn C_t=\beta\kappa ,\eea which
comes from $\frac{d\tilde{X}^{t}}{d\xi}=0$.

Now, we can rewrite the first integrals for $\tilde{X}^\mu$ on
$S^3$ as \bea\label{FmuR}
 &&\frac{d\tilde{X}^{{\phi_1}}}{d\xi}=\frac{1}{\alpha^2-\beta^2}
\left[\left(C_{\phi_1}+q\alpha\omega_2\right)\frac{1}{1-\chi}
+\beta\omega_1-q\alpha\omega_2\right],
\\ \nn &&\frac{d\tilde{X}^{{\phi_2}}}{d\xi}=\frac{1}{\alpha^2-\beta^2}\left(\frac{C_{\phi_2}}{\chi}
+\beta\omega_2-q\alpha\omega_1\right),\eea where
$\chi=\cos^2\theta$.

The first Virasoro constraint, which in the case under
consideration is the first integral of the equation of motion for
$\theta$, reduces to \bea\label{cp2R}
\left(\frac{d\chi}{d\xi}\right)^2 &=&
\frac{4}{(\alpha^2-\beta^2)^2}\chi(1-\chi)
\Bigg[(\alpha^2+\beta^2)\kappa^2-\frac{\left(C_{\phi_1}+q\alpha\omega_2\chi\right)^2}{1-\chi}
\\ \nn && -\frac{\left(C_{\phi_2}
-q\alpha\omega_1\chi\right)^2}{\chi}
-\alpha^2(\omega_2^2-\omega_1^2)\chi-\alpha^2\omega_1^2\Bigg].\eea
Also, the second Virasoro constraint becomes \bea\label{V2R}
\omega_1 C_{\phi_1}+\omega_2 C_{\phi_2}+\beta\kappa^2=0.\eea

Taking (\ref{V2R}) into account, we can rewrite (\ref{cp2R}) as
\bea\label{V1f}  \left(\frac{d\chi}{d\xi}\right)^2=4(1-q^{2})
\frac{\omega_1^2}{\alpha^2}
\frac{1-u^2}{(1-v^2)^2}(\chi_p-\chi)(\chi-\chi_m)(\chi-\chi_n),\eea
where \bea\label{hes} &&\chi_p+\chi_m+\chi_n =
\frac{-\left(v^2W+\left(W+u^2-2+q^2)\right)
+2q\left(uvW+K(1-u^2)\right)\right)}{(1-q^{2})(1-u^2)},
 \\ \nn
&&\chi_p\chi_m+\chi_p\chi_n+\chi_m\chi_n=
-\frac{\left(1+v^2\right)W+K^2-\left(v W-u K\right)^2 -1+2q
K}{(1-q^{2})(1-u^2)},
\\ \nn &&\chi_p\chi_m\chi_n= -\frac{K^2}{(1-q^2) (1-u^2)},\eea and
we introduced the notations \bea\nn v=-\frac{\beta}{\alpha},\h
u=\frac{\omega_2}{\omega_1},\h
W=\left(\frac{\kappa}{\omega_1}\right)^2,\h
K=\frac{C_{\phi_2}}{\alpha\omega_1}.\eea This leads to
\bea\label{GMsdxi}
d\xi=\frac{\alpha}{2\omega_1}\frac{1-v^2}{\sqrt{(1-q^{2})(1-u^2)}}
\frac{d\chi}{\sqrt{(\chi_p-\chi)(\chi-\chi_m)(\chi-\chi_n)}}.\eea

Integrating (\ref{GMsdxi}) and inverting $\xi(\chi)$ to
$\chi(\xi)\equiv \cos^2[\theta(\xi)]$, one finds the following
explicit solution \bea\label{cossol} \chi = (\chi_p-\chi_n)\
\mathbf{dn}^2\left[\frac{\sqrt{(1-q^{2})(1-u^2)(\chi_p-\chi_n)}}{1-v^2}
\ \omega_1(\sigma-v
\tau),\frac{\chi_p-\chi_m}{\chi_p-\chi_n}\right]+\chi_n .\eea

Next, we integrate (\ref{FmuR}), and according to our ansatz,
obtain that the solutions for the isometric angles on $S^3$ are
given by \bea\label{1one} &&\phi_1= \omega_1 \tau+
\frac{2}{\sqrt{(1-q^{2})(1-u^2)(\chi_p-\chi_n)}}
\\ \nn &&\left[\frac{v W-K u+q u}{1-\chi_p}\ \Pi\left(\arcsin\sqrt{\frac{\chi_p-\chi}
{\chi_p-\chi_m}},-\frac{\chi_p-\chi_m}{1-\chi_p},\frac{\chi_p-\chi_m}{\chi_p-\chi_n}\right)\right.
\\ \nn &&\left.
-(v+q u)F\left(\arcsin\sqrt{\frac{\chi_p-\chi}
{\chi_p-\chi_m}},\frac{\chi_p-\chi_m}{\chi_p-\chi_n}\right)\right]\eea
\bea\label{Isols} &&\phi_2=\omega_2
\tau+\frac{2}{\sqrt{(1-q^{2})(1-u^2)(\chi_p-\chi_n)}}
\\ \nn &&\left[\frac{K}{\chi_p}\ \Pi\left(\arcsin\sqrt{\frac{\chi_p-\chi}
{\chi_p-\chi_m}},1-\frac{\chi_m}{\chi_p},\frac{\chi_p-\chi_m}{\chi_p-\chi_n}\right)\right.
\\ \nn &&\left.
-(u v+q)F\left(\arcsin\sqrt{\frac{\chi_p-\chi}
{\chi_p-\chi_m}},\frac{\chi_p-\chi_m}{\chi_p-\chi_n}\right)\right].\eea

By using (\ref{GMsdxi}), one can find also the conserved
quantities, namely, the string energy $E_s$ and the two angular
momenta $J_1$, $J_2$ : \bea\label{GMsEs} E_s=
2T\frac{(1-v^2)\sqrt{W}}{\sqrt{(1-q^{2})(1-u^2)(\chi_p-\chi_n)}}\
\mathbf{K}(1-\epsilon),\eea

\bea\nn &&J_1= \frac{2T}{\sqrt{(1-q^{2})(1-u^2)(\chi_p-\chi_n)}}
\Big\{\left[1-v^2 W+K(uv-q)\right]\mathbf{K}(1-\epsilon)
\\ \label{Jtf}
&&-(1-q^2)\left[\chi_n \ \mathbf{K}(1-\epsilon)+(\chi_p-\chi_n)\
\mathbf{E}(1-\epsilon)\right]\Big\},\eea

\bea\nn &&J_2= \frac{2T}{\sqrt{(1-q^{2})(1-u^2)(\chi_p-\chi_n)}}
\Big\{(1-q^{2}) u\left[\chi_n \
\mathbf{K}(1-\epsilon)+(\chi_p-\chi_n)\
\mathbf{E}(1-\epsilon)\right]
\\ \label{Jf} &&-\left[K v+q\left(v W-K u\right)+q^{2}u\right]\ \mathbf{K}(1-\epsilon)
\\ \nn &&+q\frac{v W-K u+q u}{1-\chi_p}
\mathbf{\Pi}\left(-\frac{\chi_p-\chi_n}{1-\chi_p}(1-\epsilon),1-\epsilon\right)
\Big\}.\eea where $\epsilon$ is defined as
\bea\label{epsi}
\epsilon=\frac{\chi_m-\chi_n}{\chi_p-\chi_n}.\eea

We will need also the expression for the angular difference
$\Delta\phi_1$. It can be found to be \bea\label{dtt}
&&\Delta\phi_1 = \frac{2}{\sqrt{(1-q^{2})(1-u^2)(\chi_p-\chi_n)}}
\\ \nn &&\left[\frac{v W-K u+q u}{1-\chi_p}\
\mathbf{\Pi}\left(-\frac{\chi_p-\chi_n}{1-\chi_p}(1-\epsilon),1-\epsilon\right)
-(v+q u)\ \mathbf{K}\left(1-\epsilon\right)\right].\eea

The expressions (\ref{GMsEs}), (\ref{Jtf}), (\ref{Jf}),
(\ref{dtt}) are for the finite-size dyonic strings living in the
$R_t\times S^3$ subspace of $AdS_3\times S^3\times T^4$.

{\it Leading finite-size effect on the dispersion relation}

In order to find the leading finite-size effect on the dispersion
relation, we have to consider the limit $\epsilon\to 0$, since
$\epsilon= 0$ corresponds to the infinite-size case. In this
subsection we restrict ourselves to the particular case when
$\chi_n = K=0$\footnote{As we will see later on, this choice allow
us to reproduce the dispersion relation in the infinite volume
limit \cite{HST1311}.}. Then the third equation in (\ref{hes}) is
satisfied identically, while the other two simplify to
\bea\label{sce} &&\chi_p+\chi_m = \frac{2-(1+v^2)W-u^2-2q(u v
W+\frac{q}{2})}{(1-q^{2})(1-u^2)},
\\ \nn
&&\chi_p\chi_m= \frac{(1-W)(1-v^2W)}{(1-q^{2})(1-u^2)},\eea and
$\epsilon$ becomes \bea\label{epss}
\epsilon=\frac{\chi_m}{\chi_p}.\eea The relevant expansions of the
parameters are \bea\nn
&&\chi_p=\chi_{p0}+\left(\chi_{p1}+\chi_{p2}\log(\epsilon)\right)\epsilon,
\h W=1+W_{1}\epsilon ,
\\
\label{pars} &&v=v_0+\left(v_1+v_2\log(\epsilon)\right)\epsilon,
\h u=u_0+\left(u_1+u_2\log(\epsilon)\right)\epsilon.\eea Replacing
(\ref{epss}), (\ref{pars}) into (\ref{sce}), one finds the
following solutions in the small $\epsilon$ limit \bea\label{4sol}
&&\chi_{p0}= \frac{1-v^2_0-u^2_0-2q(u_0
v_0+\frac{q}{2})}{(1-q^{2})(1-u_0^2)},
\\ \nn &&\chi_{p1}= -\frac{v_0+q u_0}{(1-q^{2})^2(1-v_0^2)(1-u_0^2)^2}\times
\\ \nn &&\Big[\Big(1-v_0^2-u_0^2-2q(u_0v_0+\frac{q}{2})\Big)\Big(v_0^3
+q u_0(1+3v_0^2)-v_0(1-2u_0^2-2q^2)\Big)
\\ \nn &&+2(1-q^{2})(1-v_0^2)\left((1-u_0^2)v_1+(u_0v_0+q)u_1\right)\Big]
\\ \nn &&\chi_{p2}= -\frac{2(v_0+q u_0)\left((1-u_0^2)v_2+(u_0 v_0+q)u_2\right)}
{(1-q^{2})(1-u_0^2)^2}
\\ \nn &&W_1= -\frac{\left(1-v^2_0-u^2_0-2q(u_0 v_0+\frac{q}{2})\right)^2}
{(1-q^{2})(1-u_0^2)(1-v_0^2)}.\eea

The coefficients in the expansions of $v$ and $u$, will be
obtained by imposing the conditions that $J_2$ and $\Delta\phi_1$
do not depend on $\epsilon$, as in the cases without $B$-field
($AdS_5\times S^5$ and $AdS_4\times CP^3$) and their
$TsT$-deformations, where the $B$-field is nonzero, but its
contribution is different.

Expanding (\ref{Jf}) and (\ref{dtt}) to the leading order in
$\epsilon$ (now $\chi_n = K =0$), one finds that on the solutions
(\ref{4sol}) \bea\label{J} &&J_2=2T\left(\frac{u_0
\sqrt{1-u_0^2-v_0^2-2q(u_0v_0+\frac{q}{2})}}{1-u_0^2}\right.
\\ \nn &&\left.+q\arcsin\sqrt{\frac{1-u_0^2-v_0^2-2q(u_0v_0+\frac{q}{2})}{(1-q^2)(1-u_0^2)}}\right),\eea

\bea\label{Dtt}
\Delta\phi_1=2\arcsin\sqrt{\frac{1-u_0^2-v_0^2-2q(u_0v_0+\frac{q}{2})}{(1-q^2)(1-u_0^2)}}
,\eea \bea\label{u1} &&u_1=
\frac{1-u_0^2-v_0^2-2q(u_0v_0+\frac{q}{2})}{4(1-q^2)(1-u_0^2)}
\\ \nn &&\times
\left[u_0\left(1-\log 16 -v_0^2(1+\log 16)\right)-2q v_0\log
16)\right],\eea \bea\label{v1} &&v_1 =
\frac{1-u_0^2-v_0^2-2q(u_0v_0+\frac{q}{2})}{4(1-q^2)(1-u_0^2)(1-v_0^2)}
\\ \nn &&\times
\left[v_0\left((1-4q^2)(1-\log 16)-u_0^2(5-\log
4096)\right)\right.
\\ \nn && \left.-v_0^3\left(1-\log 16-u_0^2(1+\log 16)\right)
-4q u_0\left(1-\log 4+v_0^2(1-\log 64)\right)\right],\eea
\bea\label{u2} &&u_2 = \frac{\left(u_0(1+v_0^2)+2q v_0\right)
\left(1-u_0^2-v_0^2-2q(u_0v_0+\frac{q}{2})\right)}{4
(1-q^2)(1-v_0^2)},\eea \bea\label{v2} &&v_2 =
\frac{1-u_0^2-v_0^2-2q(u_0v_0+\frac{q}{2})}{4(1-q^2)(1-u_0^2)(1-v_0^2)}
\\ \nn &&\times
\left[v_0\left(1-v_0^2-u_0^2(3+v_0^2)\right)
-2q\left(u_0(1+3v_0^2)+2q v_0\right)\right].\eea

Now, let us turn to the energy-charge relation. Expanding
(\ref{GMsEs}) and (\ref{Jtf}) in $\epsilon$ and taking into
account the solutions (\ref{4sol}), (\ref{u1}) -(\ref{v2}), we
obtain \bea\label{EsJtp} E_s-J_1=
2T\frac{\sqrt{1-u_0^2-v_0^2-2q(u_0v_0+\frac{q}{2})}}{1-u_0^2}
\left(1-\frac{1-u_0^2-v_0^2-2q(u_0v_0+\frac{q}{2})}{4(1-q^2)}\
\epsilon\right).\eea The expression for $\epsilon$ can be found
from the expansion of $J_1$. To the leading order, it is given by
\bea\label{seps} \epsilon= 16 \exp\left[-\frac{J_1}{T}
\frac{\sqrt{1-u_0^2-v_0^2-2q(u_0v_0+\frac{q}{2})}}{1-v_0^2}
-2\frac{1-u_0^2-v_0^2-2q(u_0v_0+\frac{q}{2})}{(1-v_0^2)(1-u_0^2)}\right].\eea

Next, we would like to express the right hand side of
(\ref{EsJtp}) in terms of $J_2$ and $\Delta\phi_1$. To this end,
we solve (\ref{J}), (\ref{Dtt}) with respect to $u_0$, $v_0$. The
result is \bea\label{u0} && u_0= \frac{J_2-q T \Delta\phi_1}
{\sqrt{(J_2-q T \Delta\phi_1)^2+4(1-q^2) T^2
\sin^2\frac{\Delta\phi_1}{2}}},
\\ \label{v0} && v_0= \frac{T(1-q^2)\sin \Delta\phi_1-q(J-q T \Delta\phi_1)}
{\sqrt{(J_2-q T \Delta\phi_1)^2+4(1-q^2) T^2
\sin^2\frac{\Delta\phi_1}{2}}}.\eea

Replacing (\ref{u0}), (\ref{v0}) into (\ref{EsJtp}), (\ref{seps}),
one finds \bea\label{EsJtpf} &&E_s-J_1=\sqrt{(J_2-q T
\Delta\phi_1)^2+4(1-q^2) T^2 \sin^2\frac{\Delta\phi_1}{2}}
\\ \nn
&&\left(1-\frac{(1-q^2) T^2 \sin^4\frac{\Delta\phi_1}{2}}{(J_2-q T
\Delta\phi_1)^2+4(1-q^2) T^2 \sin^2\frac{\Delta\phi_1}{2}}\
\epsilon\right),\eea where \bea\label{sepsf} \epsilon=16\
e^{-\frac{2\left(J_1+\sqrt{(J_2-q T \Delta\phi_1)^2+4(1-q^2) T^2
\sin^2\frac{\Delta\phi_1}{2}}\right) \sqrt{(J_2-q T
\Delta\phi_1)^2+4(1-q^2) T^2
\sin^2\frac{\Delta\phi_1}{2}}\sin^2\frac{\Delta\phi_1}{2}} {(J_2-q
T \Delta\phi_1)^2+4T^2\sin^4\frac{\Delta\phi_1}{2} +2q
T\sin\Delta\phi_1\left((J_2-q T \Delta\phi_1)+\frac{q}{2}
T\sin\Delta\phi_1\right)}}.\eea

Our result \footnote{Eqs.(\ref{EsJtpf}) and (\ref{sepsf}) have
been confirmed by an independent analysis based on algebraic curve
method \cite{BDO} after our result has been appeared in the
arXiv.} matches with that of  \cite{HST1311} in (\ref{Tdr}) when
we take $\epsilon\to 0$ limit by sending $J_1\to\infty$. This
dispersion relation is different from the ordinary giant magnon's
one.

The dispersion relation for the ordinary giant magnon with one
nonzero angular momentum cam be obtained by setting $J_2=1$ and
taking the limit $T\to\infty$. To take into account the {\it
leading} finite-size effect only, we restrict ourselves to the
case when $\frac{J_1}{T}>>1$. The result is the following:
\bea\label{1spin} E_s-J_1=T \sqrt{p^2 q^2+4(1-q^2)\sin^2
\frac{p}{2}}\left(
1-\frac{\left(1-q^2\right)\sin^4\frac{p}{2}}{p^2
q^2+4(1-q^2)\sin^2 \frac{p}{2}}\ \epsilon\right),\eea where
\bea\nn &&\epsilon =16 \exp\left[\frac{-2}{q^2(p-\sin p)^2+4
\sin^4\frac{p}{2}} \left(\frac{J_1}{T}+\sqrt{p^2
q^2+4(1-q^2)\sin^2 \frac{p}{2}}\right)\right.
\\ \nn && \left.\sqrt{p^2 q^2+4(1-q^2)\sin^2 \frac{p}{2}}\  \sin^2 \frac{p}{2}\right].\eea

Our results on the leading finite-size correction to the
dispersion relation can provide an important check for the exact
integrability conjecture and $S$-matrix elements based on it.

\subsubsection{Finite-size giant magnons on $\eta$-deformed $AdS_5 \times S^5$}

A new integrable deformation of the type IIB $AdS_5\times S^5$
superstring action, depending on one real parameter $\eta$, has
been found recently in \cite{DMV0913}. The bosonic part of the
superstring sigma model Lagrangian on this $\eta$-deformed
background was determined in \cite{ABF1312}. Then the authors of
\cite{ABF1312} used it to compute the perturbative $S$-matrix of
bosonic particles in the model.

Interesting new developments were made in \cite{ALT0314}. There
the spectrum of a string moving on $\eta$-deformed $AdS_5\times
S^5$ is considered. This is done by treating the corresponding
worldsheet theory as integrable field theory. In particular, it
was found that the dispersion relation for the infinite-size giant
magnons \cite{HM06} on this background, in the large string
tension limit $g\to\infty$ is given by \bea\label{14}
E=\frac{2g\sqrt{1+\tilde{\eta}^2}}{\tilde{\eta}}
\mbox{arcsinh}\left(\tilde{\eta} \sin\frac{p}{2}\right),\eea where
$\tilde{\eta}$ is related to the deformation parameter $\eta$
according to \bea\label{ek} \tilde{\eta}=\frac{2
\eta}{1-\eta^2}.\eea Here, we are going to extend the result
(\ref{14}) to the case of finite-size giant magnons \cite{28}.

{\bf String Lagrangian and background fields}

The bosonic part of the string Lagrangian $\mathcal{L}$ on the
$\eta$-deformed $AdS_5\times S^5$ found in \cite{ABF1312} is given
by a sum of the Lagrangians $\mathcal{L}_a$ and $\mathcal{L}_s$,
for the $AdS$ and sphere subspaces. Since there is nonzero
$B$-field on both subspaces, which leads to the appearance of
Wess-Zumino terms, these Lagrangians can be further decomposed as
\bea\label{Leta} \mathcal{L}_a=L_a^g+L_a^{WZ},\h
\mathcal{L}_s=L_s^g+L_s^{WZ},\eea where the superscript ``g'' is
related to the dependence on the background metric. The explicit
expressions for the Lagrangians in (\ref{Leta}) are as follows
\cite{ABF1312}

\bea\label{Lag} L_a^g &=&-\frac{T}{2}\gamma^{\alpha\beta}
\left[-\frac{(1+\rho^2)\p_\alpha t \p_\beta t} {1-\tilde{\eta}^2
\rho^2}+ \frac{\p_\alpha \rho \p_\beta \rho}
{(1+\rho^2)(1-\tilde{\eta}^2 \rho^2)}+ \frac{\rho^2 \p_\alpha
\zeta \p_\beta \zeta}{1+\tilde{\eta}^2 \rho^4 \sin^2\zeta} \right.
\\ \nn && \left. +\frac{\rho^2 \cos^2 \zeta \ \p_\alpha \psi_1 \p_\beta \psi_1}
{1+\tilde{\eta}^2 \rho^4 \sin^2\zeta}+\rho^2 \sin^2\zeta \
\p_\alpha \psi_2 \p_\beta \psi_2\right],\eea

\bea\label{LaWZ} L_a^{WZ}=\frac{T}{2}\tilde{\eta} \
\epsilon^{\alpha\beta} \frac{\rho^4 \sin 2\zeta}{1+\tilde{\eta}^2
\rho^4 \sin^2\zeta}\ \p_\alpha\psi_1 \p_\beta\zeta,\eea

\bea\label{sag} L_s^g &=&-\frac{T}{2}\gamma^{\alpha\beta}
\left[\frac{(1-r^2)\p_\alpha \phi \p_\beta \phi} {1+\tilde{\eta}^2
r^2}+ \frac{\p_\alpha r \p_\beta r} {(1-r^2)(1+\tilde{\eta}^2
r^2)}+ \frac{r^2 \p_\alpha \xi \p_\beta \xi}{1+\tilde{\eta}^2 r^4
\sin^2\xi} \right.
\\ \nn && \left. +\frac{r^2 \cos^2 \xi \ \p_\alpha \phi_1 \p_\beta \phi_1}
{1+\tilde{\eta}^2 r^4 \sin^2\xi}+r^2 \sin^2\xi \ \p_\alpha \phi_2
\p_\beta \phi_2\right],\eea

\bea\label{LsWZ} L_s^{WZ}=-\frac{T}{2}\tilde{\eta} \
\epsilon^{\alpha\beta} \frac{r^4 \sin 2\xi}{1+\tilde{\eta}^2 r^4
\sin^2\xi}\ \p_\alpha\phi_1 \p_\beta\xi,\eea where we introduced
the notation \bea\label{T} T=g\sqrt{1+\tilde{\eta}^2}.\eea

Comparing (\ref{Lag})-(\ref{LsWZ}) with the Polyakov string
Lagrangian, one can extract the components of the background
fields. They are given by \bea\label{bfsa} &&
g_{tt}=-\frac{1+\rho^2}{1-\tilde{\eta}^2 \rho^2},\h
g_{\rho\rho}=\frac{1}{(1+\rho^2)(1-\tilde{\eta}^2 \rho^2)},\h
g_{\zeta\zeta}= \frac{\rho^2}{1+\tilde{\eta}^2 \rho^4 \sin^2
\zeta}
\\ \nn && g_{\psi_1 \psi_1}= \frac{\rho^2\cos^2\zeta}{1+\tilde{\eta}^2 \rho^4\sin^2 \zeta},\h
g_{\psi_2 \psi_2}=\rho^2 \sin^2\zeta,\h
b_{\psi_1\zeta}=\tilde{\eta} \frac{\rho^4 \sin 2\zeta}
{1+\tilde{\eta}^2 \rho^4\sin^2 \zeta}.\eea

\bea\label{bfss} && g_{\phi\phi}=\frac{1-r^2}{1+\tilde{\eta}^2
r^2},\h g_{rr}=\frac{1}{(1-r^2)(1+\tilde{\eta}^2 r^2)} ,\h
g_{\xi\xi}=\frac{r^2}{1+\tilde{\eta}^2 r^4 \sin^2\xi}
\\ \nn && g_{\phi_1\phi_1}= \frac{r^2 \cos^2 \xi}{1+\tilde{\eta}^2 r^4 \sin^2\xi}.\h
g_{\phi_2\phi_2}= r^2 \sin^2\xi,\h b_{\phi_1\xi}=-\tilde{\eta}
\frac{r^4 \sin 2\xi}{1+\tilde{\eta}^2 r^4 \sin^2\xi}.\eea

{\bf GM solutions}

Since we are going to consider giant magnon solutions, we restrict
ourselves to the $R_t\times S^3_\eta$ subspace, which corresponds
to the following choice in $AdS_\eta$ \bea\nn \rho=0,\h \zeta=0,\h
\psi_1=\psi_2=0 \ \Rightarrow\ b_{\psi_1\zeta}=0.\eea On
$S^5_\eta$ we first introduce the angle $\tilde{\theta}$ in the
following way \bea\nn r=\sin\tilde{\theta},\eea which leads to
\bea\nn ds^2_{S^5_\eta}&=& \frac{\cos^2
\tilde{\theta}}{1+\tilde{\eta}^2 \sin^2 \tilde{\theta}}\ d\phi^2
+\frac{d \tilde{\theta}^2}{1+\tilde{\eta}^2 \sin^2 \tilde{\theta}}
+\frac{\sin^2 \tilde{\theta}}{1+\tilde{\eta}^2 \sin^4
\tilde{\theta}\sin^2 \xi}\ d \xi^2
\\ \nn &&+\frac{\sin^2 \tilde{\theta} \cos^2 \xi}{1+\tilde{\eta}^2 \sin^4 \tilde{\theta}\sin^2 \xi}\ d \phi_1^2
+\sin^2 \tilde{\theta}\ \sin^2 \xi\ d \phi_2^2,\eea \bea\nn
b_{\phi_1\xi}=-\tilde{\eta} \frac{\sin^4\tilde{\theta} \sin
2\xi}{1+\tilde{\eta}^2 \sin^4\tilde{\theta} \sin^2\xi}.\eea Now,
to go to $S^3_\eta$, we can safely set $\phi=0$,
$\tilde{\theta}=\frac{\pi}{2}$ (we also exchange $\phi_1$ and
$\phi_2$ and replace $\xi$ with $\theta$). Thus, the background
seen by the string moving in the $R_t\times S^3_\eta$ subspace can
be written as \bea\nn &&g_{tt}=-1,\h
g_{\phi_1\phi_1}=\sin^2\theta,\h
g_{\phi_2\phi_2}=\frac{\cos^2\theta}{1+\tilde{\eta}^2
\sin^2\theta},
\\ \label{fb} &&g_{\theta\theta}=\frac{1}{1+\tilde{\eta}^2 \sin^2\theta},\h b_{\phi_2\theta}=
-\tilde{\eta} \frac{\sin 2\theta}{1+\tilde{\eta}^2
\sin^2\theta}.\eea

Working in conformal wordsheet gauge, we impose the following
ansatz for the string embedding \bea\label{Az}
t(\tau,\sigma)=\kappa \tau,\h \phi_i(\tau,\sigma)=\omega_i
\tau+F_i(\xi), \h \theta(\tau,\sigma)=\theta(\xi),\h
\xi=\alpha\sigma+\beta\tau ,\ i=1,2,\eea where $\tau$ and $\sigma$
are the string world-sheet coordinates, $F_i(\xi)$, $\theta(\xi)$
are arbitrary functions of $\xi$, and $\kappa, \omega_i, \alpha,
\beta$ are parameters.

Then one can find the following solutions of the equations of
motion for $\phi_i(\tau,\sigma)$ (we introduced the notation
$\chi\equiv \cos^2\theta$) \bea\label{f1}
&&\phi_1(\tau,\sigma)=\omega_1 \tau+\frac{1}{\alpha^2-\beta^2}\int
d\xi\left(\frac{C_1}{1-\chi} +\beta \omega_1\right),
\\\label{f2} &&\phi_2(\tau,\sigma)=\omega_2 \tau+\frac{1}{\alpha^2-\beta^2}\int d\xi \left[\frac{(1+\tilde{\eta}^2)C_2}{\chi}
+\beta \omega_2-\tilde{\eta}^2 C_2\right],\eea where $C_1$, $C_2$
are integration constants.

By using (\ref{f1}), (\ref{f2}), one can show that the Virasoro
constraints take the form \bea\label{V1e} \left(\frac{d \chi}{d
\xi}\right)^2 &=& \frac{4
\chi(1-\chi)\left[1+\tilde{\eta}^2(1-\chi)\right]}{(\alpha^2-\beta^2)^2}
\left[(\alpha^2+\beta^2)\kappa^2-\frac{C_1^2}{1-\chi}-C_2^2\frac{1+\tilde{\eta}^2(1-\chi)}
{\chi}\right.
\\ \nn &&-\left.\alpha^2 \omega_1^2(1-\chi)-
\alpha^2 \omega_2^2
\frac{\chi}{1+\tilde{\eta}^2(1-\chi)}\right],\eea \bea\label{V2e}
\omega_1 C_1+\omega_2 C_2+\beta \kappa^2=0.\eea Next, we solve
(\ref{V2e}) with respect to $C_1$ and replace the solution into
(\ref{V1e}). The result is \bea\label{cpf} \left(\frac{d \chi}{d
\xi}\right)^2=
\frac{4}{(\alpha^2-\beta^2)^2}\alpha^2\tilde{\eta}^2\omega_1^2
(\chi_\eta-\chi)(\chi_p-\chi)(\chi-\chi_m)(\chi-\chi_n),\eea where
\bea\label{roots1} \chi_\eta+\chi_p+\chi_m+\chi_n=
-\frac{\alpha^2\left[\omega_2^2-\omega_1^2+\tilde{\eta}^2(\kappa^2-3
\omega_1^2) \right]+\tilde{\eta}^2\beta^2 \kappa^2+\tilde{\eta}^4
C_2^2}{\alpha^2\tilde{\eta}^2\omega_1^2},\eea \bea\label{roots2}
&&\chi_p \chi_\eta+(\chi_p + \chi_\eta)\chi_n+ \chi_m (\chi_p +
\chi_\eta +\chi_n) =
\\ \nn && \frac{1}{\tilde{\eta}^2\alpha^2 \omega_1^4}
\left\{\beta^2\kappa^2\left[\tilde{\eta}^2(\kappa^2-2\omega_1^2)-\omega_1^2\right]
+2C_2\beta\tilde{\eta}^2\kappa^2\omega_2\right.
\\ \nn &&+\left. \alpha^2\omega_1^2\left[\left(2+3\tilde{\eta}^2\right)\omega_1^2
-\omega_2^2-\left(1+2\tilde{\eta}^2\right)\kappa^2\right]\right.
\\ \nn &&+\left. C_2^2\tilde{\eta}^2\left(\omega_2^2-\left(2+3\tilde{\eta}^2\right)\omega_1^2\right)\right\},\eea
\bea\label{roots3}
&&\chi_m\chi_n\chi_p+\chi_m\chi_n\chi_\eta+\chi_m\chi_p\chi_\eta+\chi_n\chi_p\chi_\eta=
\\ \nn &&-\frac{1+\tilde{\eta}^2}{\tilde{\eta}^2\alpha^2 \omega_1^4}
\left[C_2^2(1+3\tilde{\eta}^2)\omega_1^2-2C_2\beta\kappa^2\omega_2-C_2^2\omega_2^2
-(\kappa^2-\omega_1^2)(\beta^2\kappa^2-\alpha^2\omega_1^2)\right],\eea
\bea\label{roots4} \chi_m\chi_n\chi_p\chi_\eta=
-\frac{C_2^2(1+\tilde{\eta}^2)^2}{\tilde{\eta}^2\alpha^2
\omega_1^2}.\eea

The solution $\xi(\chi)$ of (\ref{cpf}) is \bea\label{sxi}
&&\xi(\chi) =
\frac{\alpha^2-\beta^2}{\tilde{\eta}\alpha\omega_1\sqrt{(\chi_\eta-\chi_m)(\chi_p-\chi_n)}}\times
\\ \nn
&&
F\left(\arcsin\sqrt{\frac{(\chi_\eta-\chi_m)(\chi_p-\chi)}{(\chi_p-\chi_m)(\chi_\eta-\chi)}},
\frac{(\chi_p-\chi_m)(\chi_\eta-\chi_n)}{(\chi_\eta-\chi_m)(\chi_p-\chi_n)}\right),\eea
where \bea\nn \chi_\eta >\chi_p>\chi>\chi_m>\chi_n.\eea Inverting
$\xi(\chi)$ to $\chi(\xi)$, one finds \bea\label{chise} \chi(\xi)=
\frac{\chi_\eta(\chi_p-\chi_n)\ \mathbf{dn}^2(x\vert m)
+(\chi_\eta-\chi_p)\chi_n}{(\chi_p-\chi_n)\ \mathbf{dn}^2(x\vert
m)+\chi_\eta-\chi_p},\eea where \bea\nn &&x=\frac{\tilde{\eta}
\alpha \omega_1\sqrt{(\chi_\eta-\chi_m)(\chi_p-\chi_n)}}
{\alpha^2-\beta^2}\ \xi ,
\\ \nn &&m= \frac{(\chi_p-\chi_m)(\chi_\eta-\chi_n)}{(\chi_\eta-\chi_m)(\chi_p-\chi_n)}.\eea

By using (\ref{cpf}) we can find the explicit solutions for the
isometric angles $\phi_1$, $\phi_2$. They are given by
\bea\label{f1se} &&\phi_1(\tau,\sigma)= \omega_1
\tau+\frac{1}{\tilde{\eta}\alpha\omega_1^2(\chi_\eta-1)
\sqrt{(\chi_\eta-\chi_m)(\chi_p-\chi_n)}} \times
\\ \nn &&\Bigg\{\Big[\beta\left(\kappa^2+\omega_1^2(\chi_\eta-1)+C_2 \omega_2\right)
\Big]\
F\left(\arcsin\sqrt{\frac{(\chi_\eta-\chi_m)(\chi_p-\chi)}{(\chi_p-\chi_m)(\chi_\eta-\chi)}},
m\right)
\\ \nn &&-\frac{(\chi_\eta-\chi_p)(\beta\kappa^2+C_2 \omega_2)}{1-\chi_p}\
\Pi\left(\arcsin\sqrt{\frac{(\chi_\eta-\chi_m)(\chi_p-\chi)}{(\chi_p-\chi_m)(\chi_\eta-\chi)}}
,-\frac{(\chi_\eta-1)(\chi_p-\chi_m)}{(1-\chi_p)(\chi_\eta-\chi_m)},m\right)\Bigg\}.\eea

\bea\label{f2se} &&\phi_2(\tau,\sigma)=\omega_2\tau
+\frac{1}{\tilde{\eta}\alpha\omega_1\chi_\eta
\sqrt{(\chi_\eta-\chi_m)(\chi_p-\chi_n)}}
 \times
\\ \nn &&\Bigg\{\Big[C_2\left(1-\tilde{\eta}^2(\chi_\eta-1)\right)+\beta \omega_2 \chi_\eta
\Big]\
F\left(\arcsin\sqrt{\frac{(\chi_\eta-\chi_m)(\chi_p-\chi)}{(\chi_p-\chi_m)(\chi_\eta-\chi)}},
m\right)
\\ \nn &&+\frac{C_2(1+\tilde{\eta}^2)(\chi_\eta-\chi_p)}{\chi_p}\times
\\ \nn
&&\Pi\left(\arcsin\sqrt{\frac{(\chi_\eta-\chi_m)(\chi_p-\chi)}{(\chi_p-\chi_m)(\chi_\eta-\chi)}}
,\frac{\chi_\eta(\chi_p-\chi_m)}{(\chi_\eta-\chi_m)\chi_p},m\right)\Bigg\}.\eea

Now, let us go to the computations of the conserved charges
$Q_\mu$, i.e. the string energy $E_s$ and the two angular momenta
$J_1$, $J_2$. Starting with \bea\nn Q_\mu=\int d\sigma\frac{\p
\mathcal{L}}{\p \left(\p_\tau X^{\mu}\right)},\h
X^{\mu}=(t,\phi_1,\phi_2),\eea and applying the ansatz (\ref{A}),
one finds \bea\label{Esi}
E_s=\frac{T}{\tilde{\eta}}\left(1-\frac{\beta^2}{\alpha^2}\right)
\frac{\kappa}{\omega_1} \int_{\chi_m}^{\chi_p}
\frac{d\chi}{\sqrt{(\chi_\eta-\chi)(\chi_p-\chi)(\chi-\chi_m)(\chi-\chi_n)}},\eea
\bea\label{J1i} J_1&=&\frac{T}{\tilde{\eta}}
\left[\left(1-\frac{\beta(\beta\kappa^2+C_2\omega_2)}{\alpha^2\omega_1^2}\right)
\int_{\chi_m}^{\chi_p}
\frac{d\chi}{\sqrt{(\chi_\eta-\chi)(\chi_p-\chi)(\chi-\chi_m)(\chi-\chi_n)}}\right.
\\ \nn &&-\left.\int_{\chi_m}^{\chi_p}
\frac{\chi
d\chi}{\sqrt{(\chi_\eta-\chi)(\chi_p-\chi)(\chi-\chi_m)(\chi-\chi_n)}}\right],\eea
\bea\nn J_2&=&\frac{T}{\tilde{\eta}^3}
\Bigg[\left(1+\frac{1}{\tilde{\eta}^2}\right)\frac{\omega_2}{\omega_1}
\int_{\chi_m}^{\chi_p}
\frac{d\chi}{\left(1+\frac{1}{\tilde{\eta}^2}-\chi\right)\sqrt{(\chi_\eta-\chi)(\chi_p-\chi)(\chi-\chi_m)(\chi-\chi_n)}}
\\ \label{J2i} &&-\left(\frac{\omega_2}{\omega_1}-\tilde{\eta}^2 \frac{\beta C_2}{\alpha^2\omega_1}\right)
\int_{\chi_m}^{\chi_p}
\frac{d\chi}{\sqrt{(\chi_\eta-\chi)(\chi_p-\chi)(\chi-\chi_m)(\chi-\chi_n)}}\Bigg].\eea

We will need also the expression for the angular difference
$\Delta\phi_1$. The computations give the following result
\bea\label{adi} \Delta\phi_1 &=&\frac{1}{\tilde{\eta}}
\Bigg[\frac{\beta}{\alpha}\int_{\chi_m}^{\chi_p}
\frac{d\chi}{\sqrt{(\chi_\eta-\chi)(\chi_p-\chi)(\chi-\chi_m)(\chi-\chi_n)}}
\\ \nn &&-\left(\frac{\beta\kappa^2}{\alpha\omega_1^2}+
\frac{\omega_2C_2}{\alpha\omega_1^2}\right)\int_{\chi_m}^{\chi_p}
\frac{d\chi}{(1-\chi)\sqrt{(\chi_\eta-\chi)(\chi_p-\chi)(\chi-\chi_m)(\chi-\chi_n)}}\Bigg].\eea
Solving the integrals in (\ref{Esi})-(\ref{adi}) and introducing
the notations \bea\label{note} v=-\frac{\beta}{\alpha},\
u=\frac{\omega_2}{\omega_1}, \ W=\frac{\kappa^2}{\omega_1^2},\
K_2=\frac{C_2}{\alpha\omega_1},\
\epsilon=\frac{(\chi_\eta-\chi_p)(\chi_m-\chi_n)}{(\chi_\eta-\chi_m)(\chi_p-\chi_n)},\eea
we finally obtain \bea\label{Esf} E_s=\frac{2T}{\tilde{\eta}}
\frac{(1-v^2)\sqrt{W}}{\sqrt{(\chi_\eta-\chi_m)(\chi_p-\chi_n)}} \
\mathbf{K}(1-\epsilon),\eea \bea\label{J1f}
J_1&=&\frac{2T}{\tilde{\eta}\sqrt{(\chi_\eta-\chi_m)(\chi_p-\chi_n)}}
\Bigg[\left(1-v^2W+K_2 u v-\chi_\eta\right)\
\mathbf{K}(1-\epsilon)
\\ \nn &&+(\chi_\eta-\chi_p)\ \mathbf{\Pi}\left(\frac{\chi_p-\chi_m}{\chi_\eta-\chi_m},
1-\epsilon\right)\Bigg],\eea \bea\label{J2f} J_2&=&
\frac{2T}{\tilde{\eta}^3\sqrt{(\chi_\eta-\chi_m)(\chi_p-\chi_n)}}
\Bigg\{\frac{\left(1+\frac{1}{\tilde{\eta}^2}\right)u}
{\left(1+\frac{1}{\tilde{\eta}^2}-\chi_\eta\right)}\times
\\ \nn &&\Bigg[\mathbf{K}(1-\epsilon)-\frac{\chi_\eta-\chi_p}{1+\frac{1}{\tilde{\eta}^2}-\chi_p}
\
\mathbf{\Pi}\left(\frac{(\chi_p-\chi_m)\left(1+\frac{1}{\tilde{\eta}^2}-\chi_\eta\right)}
{(\chi_\eta-\chi_m)\left(1+\frac{1}{\tilde{\eta}^2}-\chi_p\right)},1-\epsilon\right)\Bigg]
\\ \nn &&-\left(u+\tilde{\eta}^2 K_2 v\right)\ \mathbf{K}(1-\epsilon)\Bigg\},\eea
\bea\label{adf} \Delta\phi_1 &=&
\frac{2}{\tilde{\eta}\sqrt{(\chi_\eta-\chi_m)(\chi_p-\chi_n)}}\times
\\ \nn &&\Bigg\{\frac{v W-K_2 u}{(\chi_\eta-1)(1-\chi_p)}\Bigg[(\chi_\eta-\chi_p)\
\mathbf{\Pi}\left(-\frac{(\chi_\eta-1)(\chi_p-\chi_m)}{(\chi_\eta-\chi_m)(1-\chi_p)},
1-\epsilon\right)
\\ \nn &&-(1-\chi_p)\ \mathbf{K}(1-\epsilon)\Bigg]-v \ \mathbf{K}(1-\epsilon)\Bigg\}.\eea

{\bf Small $\epsilon$-expansions and dispersion relation}

In this section we restrict ourselves to the simpler case of giant
magnons with one nonzero angular momentum. To this end, we set the
second isometric angle $\phi_2=0$. From the solution (\ref{f2se})
it is clear that $\phi_2$ is zero when \bea\nn \omega_2=C_2=0,\eea
or equivalently (see (\ref{note})) \bea\nn u=K_2=0.\eea Then it
follows from (\ref{roots4}) that $\chi_n=0$ because
$\chi_\eta>\chi_p>\chi_m>0$ for the finite-size case. In addition,
we express $\chi_m$ through the other parameters \bea\nn
\chi_m=\frac{\chi_\eta \chi_p}{\chi_\eta-(1-\epsilon)\chi_p}\
\epsilon.\eea As a consequence (\ref{roots1})-(\ref{roots3}) take
the form \bea\label{roots1r}
\frac{(1-\epsilon)\chi_p^2-2\epsilon\chi_p\chi_\eta-\chi_\eta^2}{\chi_\eta-(1-\epsilon)\chi_p}
+3-(1+v^2)W+\frac{1}{\tilde{\eta}^2}=0,\eea \bea\label{roots2r}
\chi_p\chi_\eta+\frac{\epsilon\chi_p\chi_\eta(\chi_p+\chi_\eta)}{\chi_\eta-(1-\epsilon)\chi_p}
-\frac{2-(1+v^2)W+\left(3-\left(2+v^2(2-W)\right)W\right)\tilde{\eta}^2}{\tilde{\eta}^2}=0,\eea
\bea\label{roots3r}
\frac{\epsilon\chi_p^2\chi_\eta^2}{\chi_\eta-(1-\epsilon)\chi_p}
-\frac{(1+\tilde{\eta}^2)(1-W)(1-v^2 W)}{\tilde{\eta}^2}=0.\eea

In order to obtain the leading finite-size effect on the
dispersion relation, we consider the limit $\epsilon\to 0$ in
(\ref{roots1r})-(\ref{roots3r}) first. We will use the following
small $\epsilon$-expansions for the remaining parameters
\bea\label{chiseps} &&\chi_{\eta}= \chi_{\eta 0}+(\chi_{\eta
1}+\chi_{\eta 2}\log\epsilon)\epsilon
\\ \nn &&\chi_{p}= \chi_{p 0}+(\chi_{p 1}+\chi_{p 2}\log\epsilon)\epsilon,
\\ \nn &&v= v_0+(v_1+v_2\log\epsilon)\epsilon,
\\ \nn &&W= 1+W_1 \epsilon .\eea
Replacing (\ref{chiseps}) into (\ref{roots1r})-(\ref{roots3r}) and
expanding in $\epsilon$ one finds the following solution of the
resulting equations \bea\label{chises} &&\chi_{p 0}=1-v_0^2,\h
\chi_{p 1}=1-v_0^2-2v_0 v_1- \frac{(1-v_0^2)^2}{1+\tilde{\eta}^2
v_0^2},\h \chi_{p 2}=-2v_0 v_2,
\\ \nn &&\chi_{\eta 0}=1+\frac{1}{\tilde{\eta}^2},\h \chi_{\eta 1}=\chi_{\eta 2}=0,
\\ \nn &&W_1=-\frac{(1+\tilde{\eta}^2)(1-v_0^2)}{1+\tilde{\eta}^2 v_0^2}.\eea

Next, we expand $\Delta\phi_1$ in $\epsilon$ and impose the
condition that the resulting expression does not depend on
$\epsilon$. After using (\ref{chises}) this gives \bea\label{df1}
\Delta\phi_1=2\
\mbox{arccot}\left(v_0\sqrt{\frac{1+\tilde{\eta}^2}{1-v_0^2}}\right)
\eea and two equations with solution \bea\label{v1v2}
v_1=\frac{v_0(1-v_0^2)\left[1-\log 16 +\tilde{\eta}^2
\left(2-v_0^2(1+\log 16 )\right)\right]}{4(1+\tilde{\eta}^2
v_0^2)}, \h v_2=\frac{1}{4}v_0(1-v_0^2).\eea Solving (\ref{df1})
with respect to $v_0$ one finds \bea\label{v0sol}
v_0=\frac{\cot\frac{\Delta\phi_1}{2}}{\sqrt{\tilde{\eta}^2+\csc^2\frac{\Delta\phi_1}{2}}}.\eea

Now let us go to the $\epsilon$-expansion of the difference
$E_s-J_1$. Taking into account the solutions for the parameters,
it can be written as \bea\label{fr} E_s-J_1= 2 g
\sqrt{1+\tilde{\eta}^2} \left[\frac{1}{\tilde{\eta}}
\mbox{arcsinh}\left(\tilde{\eta}
\sin\frac{p}{2}\right)-\frac{(1+\tilde{\eta}^2)
\sin^3\frac{p}{2}}{4 \sqrt{1+\tilde{\eta}^2 \sin^2\frac{p}{2}}}\
\epsilon\right].\eea where the expression for $\epsilon$ can be
found from the expansion of $J_1$. To the leading order, the
result is \bea\label{epse} \epsilon =16\
\exp\left[-\left(\frac{J_1}{g}
+\frac{2\sqrt{1+\tilde{\eta}^2}}{\tilde{\eta}}\mbox{arcsinh}\left(\tilde{\eta}
\sin\frac{p}{2}\right)
\right)\sqrt{\frac{1+\tilde{\eta}^2\sin^2\frac{p}{2}}{\left(1+\tilde{\eta}^2\right)\sin^2\frac{p}{2}}}
\right].\eea In writing (\ref{fr}), (\ref{epse}), we used
(\ref{T}) and identified the angular difference $\Delta\phi_1$
with the magnon momentum $p$ in the dual spin chain.

For $\epsilon=0$, (\ref{fr}) reduces to the dispersion relation
for the infinite-size giant magnon obtained in \cite{ALT0314} for
the large $g$ case. In the limit $\tilde{\eta}\to 0$, (\ref{fr})
gives the correct result for the undeformed case found in
\cite{AFZ06}.

\subsection{Membrane results}

\subsubsection{M2-brane solutions in $AdS_7 \times S^4$}
In \cite{7} different M2-brane configurations in the M-theory
$AdS_7\times S^4$ background with field theory dual $A_{N-1}(2,0)$
$SCFT$ have been considered. New membrane solutions are found and
compared with the known ones. Here we will give an example of such
solution chosen among the ones obtained there.

We use the following coordinates for the $AdS_7 \times S^4$ metric
\bea\nn l_p^{-2}ds^2_{AdS_{7}\times S^4} &=&
4R^2\left\{-\cosh^2\rho dt^2 + d\rho^2 +
\sinh^2\rho\left(d\psi_1^2 + \cos^2\psi_1 d\psi_2^2+\sin^2\psi_1
d\Omega^2_3\right)\right.\\ \label{AdS7S4m}
&+&\left.\frac{1}{4}\left[d\alpha^2 + \cos^2\alpha
d\theta^2+\sin^2\alpha\left(d\beta^2 + \cos^2\beta d\gamma^2
\right)\right]\right\},\\ \nn d\Omega^2_3 &=& d\psi_3^2 +
\cos^2\psi_3 d\psi_4^2 + \cos^2\psi_3\cos^2\psi_4 d\psi_5^2,\eea
and embed the membrane according to the ansatz
\bea\nn
&&X^0(\tau,\delta,\s)\equiv t(\tau,\delta,\s)=
\La_1^0\delta+\La_2^0\s+Y^0(\tau),
\\ \nn &&X^1(\tau,\delta,\s)=Y^1(\tau)=\rho(\tau),
\\ \label{ga2} &&X^2(\tau,\delta,\s)\equiv \psi_2(\tau,\delta,\s)=
\La_1^2\delta+\La_2^2\s+Y^2(\tau),
\\ \nn &&X^3(\tau,\delta,\s)\equiv \psi_5(\tau,\delta,\s)=
\La_1^3\delta+\La_2^3\s+Y^3(\tau),
\\ \nn &&X^4(\tau,\delta,\s)=Y^4(\tau)=\alpha(\tau),
\\ \nn &&X^5(\tau,\delta,\s)\equiv \theta(\tau,\delta,\s)=
\La_1^5\delta+\La_2^5\s+Y^5(\tau) ,
\\ \nn &&X^\mu=X^{0,2,3,5}, X^a=X^{1,4}.\eea
Then the background seen by the membrane is ($\psi_1=\pi/4$)
\bea\label{orb2} ds^2 = (2l_p R)^2\left[-\cosh^2\rho dt^2 +
d\rho^2 + \frac{1}{2}\sinh^2\rho\left(d\psi_2^2+d\psi^2_5\right)
+\frac{1}{4}\left(d\alpha^2 + \cos^2\alpha
d\theta^2\right)\right],\eea and in our notations
$X^\mu=X^{0,2,3,5}, X^a=X^{1,4}$.

Performing the necessary computations, one finds the following two
first integrals of the equations of motion for $\rho(\tau)$ and
$\alpha(\tau)$
\bea\label{fir3} &&\left(g_{11}\dot{\rho}\right)^2=
\frac{(2\la^0)^2}{(2\pi)^4}\left[\frac{E^2}{\cosh^2\rho}
+\frac{2(p_2^2+p_3^2)}{\sinh^2\rho}\right] +(2l_p
R)^6\left(\la^0T_2\right)^2
\\ \nn &&\times \left[2\left(\Delta^2_{02}
+\Delta^2_{03}\right)\cosh^2\rho -
\Delta^2_{23}\sinh^2\rho\right]\sinh^2\rho-4d \equiv F_1(\rho)\ge
0,
\\ \label{fia3}
&&\left(g_{44}\dot{\alpha}\right)^2= d+(l_pR)^6\left(4\la^0
T_2\Delta_{05}\right)^2\cos^2\alpha
-\frac{(2\la^0p_5)^2}{(2\pi)^4\cos^2\alpha} \equiv F_4(\alpha)\ge
0.\eea

The general solutions of the above two equations are \bea\nn
\tau(\rho)=(2l_pR)^2\int\frac{d\rho}{\sqrt{F_1(\rho)}},\h
\tau(\alpha)=(l_pR)^2\int\frac{d\alpha}{\sqrt{F_4(\alpha)}}.\eea
One can also find the orbit $\rho=\rho(\alpha)$: \bea\label{orb}
4\int\frac{d\rho}{\sqrt{F_1(\rho)}}=
\int\frac{d\alpha}{\sqrt{F_4(\alpha)}}.\eea

The solutions for the M2-brane coordinates $X^\mu$ are given
by
\bea\nn &&X^0(\rho,\alpha;\delta,\s)\equiv
t(\rho,\alpha;\delta,\s)=
\La_1^0\left[\la^1\tau(\rho)+\delta\right]
+\La_2^0\left[\la^2\tau(\rho)+\s\right]
\\ \nn &&\hspace{4.5cm}+\frac{2\la^0E}{(2\pi)^2}\int
\frac{d\rho}{\cosh^2\rho\sqrt{F_1(\rho)}} + f^0[C(\rho,\alpha)],
\\ \nn &&X^2(\rho,\alpha;\delta,\s)\equiv \psi_2(\rho,\alpha;\delta,\s)=
\La_1^2\left[\la^1\tau(\rho)+\delta\right]
+\La_2^2\left[\la^2\tau(\rho)+\s\right]
\\ \nn &&\hspace{4.5cm}+\frac{4\la^0p_2}{(2\pi)^2}\int
\frac{d\rho}{\sinh^2\rho\sqrt{F_1(\rho)}} + f^2[C(\rho,\alpha)],
\\ \nn &&X^3(\rho,\alpha;\delta,\s)\equiv \psi_5(\rho,\alpha;\delta,\s)=
\La_1^3\left[\la^1\tau(\rho)+\delta\right]
+\La_2^3\left[\la^2\tau(\rho)+\s\right]
\\ \nn &&\hspace{4.5cm}+\frac{4\la^0p_3}{(2\pi)^2}\int
\frac{d\rho}{\sinh^2\rho\sqrt{F_1(\rho)}} + f^3[C(\rho,\alpha)],
\\ \nn &&X^5(\rho,\alpha;\delta,\s)\equiv \theta(\rho,\alpha;\delta,\s)=
\frac{1}{p_5}\left\{\left(\La_1^0 E-\La_1^2 p_2-\La_1^3 p_3\right)
\left[\la^1\tau(\alpha)+\delta\right]\right.
\\ \nn &&\hspace{4.5cm}+\left.\left(\La_2^0 E - \La_2^2 p_2-\La_2^3 p_3\right)
\left[\la^2\tau(\alpha)+\s\right]\right\}
\\ \nn &&\hspace{4.5cm}+\frac{2\la^0p_5}{(2\pi)^2}\int
\frac{d\alpha}{\cos^2\alpha\sqrt{F_4(\alpha)}} +
f^5[C(\rho,\alpha)],\eea where $f^\mu[C(\rho,\alpha)]$ are
arbitrary functions of $C(\rho,\alpha)$. In turn, $C(\rho,\alpha)$
is the first integral of the equation (\ref{orb}).

\subsubsection{M2-brane solutions in $AdS_4 \times S^7$}
The metric and the three-form gauge field are given by
\bea\nn
&&ds^2_{AdS_{4}\times S^7} = l_{11}^2\left[-\cosh^2\rho dt^2 +
d\rho^2 + \sinh^2\rho \left(d\alpha^2 + \sin^2\alpha
d\beta^2\right)+B^2ds^2_7\right],
\\ \nn &&b_3=-\frac{k}{3}\sinh^3\rho \sin\alpha dt\wedge d\alpha\wedge d\beta, \h k=const ,\eea
where $B$ is the relative radius of $AdS_4$ with respect to the
seven-sphere. We choose to parameterize $S^7$ as \bea\nn
ds^2_7=4d\xi^2&+&\cos^2\xi\left(d\theta^2+d\phi^2+d\psi^2+2\cos\theta
d\phi d\psi\right)\\ \nn &+&
\sin^2\xi\left(d\theta^2_1+d\phi^2_1+d\psi^2_1+2\cos\theta_1
d\phi_1 d\psi_1\right).\eea Now, consider the following membrane
embedding \cite{8}
\bea\nn &&X^0(\tau,\delta,\s)\equiv
t(\tau,\delta,\s)=\Lambda_0^0\tau,
\\ \nn &&X^1(\tau,\delta,\s)=Z^1(\s)=\rho(\s),
\\ \nn &&X^2(\tau,\delta,\s)=Z^2(\s)=\alpha(\s),
\\ \nn &&X^3(\tau,\delta,\s)\equiv \beta(\tau,\delta,\s)=\Lambda_0^3\tau + \La_1^3\delta+\La_2^3\s,
\\ \nn &&X^4(\tau,\delta,\s)\equiv \phi(\tau,\delta,\s)=\Lambda_0^4\tau+\La_1^4\delta+\La_2^4\s,
\\ \nn &&X^5(\tau,\delta,\s)\equiv \psi(\tau,\delta,\s)=\Lambda_0^5\tau+\La_1^5\delta+\La_2^5\s ,\eea
where in our notations $\mu=0, 3, 4, 5$, $a=1, 2$. The relevant
background seen by the membrane is
\bea\nn ds^2 &=&
l_{11}^2\left[-\cosh^2\rho dt^2 + d\rho^2 + \sinh^2\rho
\left(d\alpha^2 + \sin^2\alpha d\beta^2\right)\right.
\\ \nn &+&\left.B^2\left(d\phi^2+d\psi^2+2d\phi d\psi\right)\right]
\\ \nn &&b_{023}=-\frac{k}{3}\sinh^3\rho\sin\alpha.\eea

For simplicity, we choose to work in diagonal worldvolume gauge
$\lambda^i=0$ in which we must have $G_{0i}=0$. There exist four
types of solutions for these constraints, for the membrane
embedding used:
\bea\label{Ia} \Lambda_{i}^{3}=0,\h
\Lambda_{i}^{5}= -\Lambda_{i}^{4},
\\ \label{Ib}  \Lambda_{i}^{3}=0,\h \Lambda_{0}^{5}= -\Lambda_{0}^{4},
\\ \nn \Lambda_{0}^{3}=0,\h \Lambda_{i}^{5}= -\Lambda_{i}^{4},
\\ \nn \Lambda_{0}^{3}=0,\h \Lambda_{0}^{5}= -\Lambda_{0}^{4}.\eea
Let us note that in the first two cases, which will be considered
here, the induced $B$-field is zero, while for the last two, it is
not.

Working in the framework of ansatz (\ref{Ia}), one obtains that
$\det{G_{mn}}=\mathbf{G}=0$, i.e. this case corresponds to
tensionless membrane. Then one finds that the membrane trajectory
$\rho=\rho(\alpha)$ is given by \bea\nn \sinh\rho(\alpha)=
\left[\frac{\left(B/\Lambda_0^0\right)^2\left(\Lambda_0^4+\Lambda_0^5\right)^2-1}
{1-\left(\Lambda_0^3/\Lambda_0^0\right)^2\sin^2\alpha}\right]^{1/2}.\eea
In the case under consideration the conserved charges are
connected with each other by the equality \bea\nn \Lambda_0^0 E =
\Lambda_0^3 S + \frac{2\lambda^0}{l_{11}^2B^2} J^2,\eea where $E$,
$S$ and $J$ are the membrane energy, spin and angular momentum
respectively.

In the framework of ansatz (\ref{Ib}), one obtains that the
conserved quantities read \bea\nn
&&E=\frac{l_{11}^2}{2\lambda^0}\Lambda_0^0 \int
d^{2}\xi\cosh^2\rho,
\\ \nn &&S=\frac{l_{11}^2}{2\lambda^0}\Lambda_0^3 \int d^{2}\xi\sinh^2\rho\sin^2\alpha,
\\ \nn &&J= 0.\eea

Taking $\alpha=0$, we find the solution \bea\nn \rho(\sigma)=
\ln\tan\left(\sigma/2A_{\rho}\right),\h
A_{\rho}=2\lambda^0T_2l_{11}B\frac{\Lambda_{1}^{4}+\Lambda_{1}^{5}}{\Lambda_{0}^{0}}.\eea
For $\alpha=\alpha_0\ne 0,\pi$, there exist two different
solutions: \bea\nn &&\rho(\sigma)=
\frac{1}{2}\ln\frac{1+\mathbf{sn}(\s/A_\rho)}{1-\mathbf{sn}(\s/A_\rho)},
\h k^2 = 1-\left(\Lambda_0^3/\Lambda_0^0\right)^2\sin^2\alpha_0
\in(0,1);
\\ \nn && \tanh\rho(\sigma)=\frac{1}{\sqrt{1+k^2}}\mathbf{sn}\left(\frac{\sqrt{1+k^2}}{A_\rho}\sigma\right),
\h k^2= \left(\Lambda_0^3/\Lambda_0^0\right)^2\sin^2\alpha_0 - 1
\in(0,1),\eea

Fixing $\rho=\rho_0\ne 0$, one obtains \bea\nn &&\alpha(\s)=
\arcsin\left[\mathbf{sn}\left(\s/A_{\alpha}\right)\right],\h
\alpha\in (-\pi/2,\pi/2),
\\ \nn &&A_{\alpha}=A_{\rho}\tanh\rho_0,\h k^2=\left(\Lambda_0^3/\Lambda_0^0\right)^2\tanh^2\rho_0\in (0,1).\eea

Now, let us turn to the general case, when none of the coordinates
$\rho$ and $\alpha$ are kept fixed. In order to be able to give
{\it explicit} solution, we set $\Lambda_0^3=0$ and find \bea\nn
\cosh\rho(\s)=\frac{A}{\mathbf{cn}(C\s)}.\eea \bea\nn
A=\sqrt{\frac{1}{2}\left[1+\sqrt{1+\left(\frac{2d}{l_{11}\Lambda_0^0
K}\right)^2}\right]}, \h
C=\frac{l_{11}\Lambda_0^0}{K}\left[1+\left(\frac{2d}{l_{11}\Lambda_0^0
K}\right)^2\right]^{1/4},\eea and $d$ is an arbitrary constant.
The solution for the membrane trajectory is the following \bea\nn
\alpha(\rho)=\frac{d}{\left(A^2-1\right)CK^2}
\left[A^2\Pi\left(\varphi, -\frac{1}{A^2-1}, k\right) -
\left(A^2-1\right)F\left(\varphi, k\right)\right],\eea where
\bea\nn \varphi=\arccos\left(\frac{A}{\cosh\rho}\right),\h
k=\left[1+\left(\frac{2d}{l_{11}\Lambda_0^0
K}\right)^2\right]^{-1/4}
\sqrt{\frac{1}{2}\left[\sqrt{1+\left(\frac{2d}{l_{11}\Lambda_0^0
K}\right)^2}-1\right]}.\eea

The condition $\Lambda_0^3=0$ leads to $S=0$, and the membrane
energy $E$ remains the only nontrivial conserved quantity. On the
obtained solution, it is given by \bea\nn E= A^2E_0 + \frac{\pi
CK^2}{2\lambda^0\Lambda_0^0} \left[\frac{\mathbf{sn}(2\pi
C)\mathbf{dn}(2\pi C)}{\mathbf{cn}(2\pi C)} -
\mathbf{E}(k)\right],\h E_0=E_{\rho=0}.\eea

\subsubsection{Exact rotating membrane solutions on a $G_2$
manifold\\
and their semiclassical limits}

We obtain exact rotating membrane solutions and explicit
expressions for the conserved charges on a manifold with exactly
known metric of $G_2$ holonomy in M-theory, with four dimensional
$\mathcal{N}=1$ gauge theory dual. After that, we investigate
their semiclassical limits and derive different relations between
the energy and the other conserved quantities, which is a step
towards M-theory lift of the semiclassical string/gauge theory
correspondence for $\mathcal{N}=1$ field theories \cite{9}.

To our knowledge, the only paper devoted to rotating membranes on
$G_2$ manifolds at that time is \cite{HN0210}, where various
membrane configurations on different $G_2$ holonomy backgrounds
have been studied systematically, but not exactly. In the
semiclassical limit (large conserved charges), the following
relations between the energy and the corresponding conserved
charge $K$ have been obtained: $E\sim K^{1/2}$, $E\sim K^{2/3}$,
$E-K\sim K^{1/3}$, $E-K\sim \ln K$.

Here, our approach will be different. Taking into account that
only a small number of $G_2$ holonomy metrics are known {\it
exactly}, we choose to search for rotating membrane solutions on
one of these metrics. Namely, the one discovered in
\cite{NPB0106034}. First, we describe the $G_2$ holonomy
background of \cite{NPB0106034}. Second, we obtain  a number of
exact rotating membrane solutions and the explicit expressions for
the corresponding conserved charges. Then, we take the
semiclassical limit and derive different energy-charge relations.
They reproduce and generalize part of the results obtained in
\cite{HN0210}, for the case of more than two conserved quantities.

The background we are interested in is a one-parameter family of
$G_2$ holonomy metrics (parameterized by $r_0$), which play an
important role as supergravity dual of the large $N$ limit of four
dimensional $\mathcal{N}=1$ SYM. These metrics describe the
M~theory lift of the supergravity solution corresponding to a
collection of D6-branes wrapping the supersymmetric three-cycle of
the deformed conifold geometry for any value of the string
coupling constant. The explicit expression for the metric with
$SU(2)\times SU(2)\times U(1)\times Z_2$ symmetry is given by
\cite{NPB0106034}
\bea\label{G21} ds^2_7 = \sum_{a=1}^7 e^a\otimes
e^a,\eea with the following vielbeins \bea\nn e^1 & =
&A(r) (\sigma_1-\Sigma_1) ~,~~e^2 = A(r) (\sigma_2-\Sigma_2) ~, \\
\nn e^3 & = &D(r) (\sigma_3-\Sigma_3) ~,~~e^4 = B(r)
(\sigma_1+\Sigma_1) ~, \\ \nn e^5 & = &B(r)
(\sigma_2+\Sigma_2) ~,~~e^6 = r_0 C(r)(\sigma_3+\Sigma_3) ~, \\
\label{G22} e^7 & = &dr/C(r),\eea where
\bea\nn
A&=&\frac{1}{\sqrt{12}} \sqrt{(r - 3 r_0/2)(r + 9 r_0/2)},\h
B=\frac{1}{\sqrt{12}} \sqrt{(r + 3 r_0/2)(r - 9 r_0/2)},
\\ \label{G23} C&=&\sqrt{\frac{(r - 9 r_0/2)(r + 9 r_0/2)}{(r - 3 r_0/2)(r + 3 r_0/2)}},\h D=r/3,\eea
and \bea\nn &&\sigma_1 = \sin \psi \sin \theta d \phi+\cos \psi d
\theta, \h\Sigma_1 = \sin \tilde \psi \sin \tilde \theta d \tilde
\phi+\cos \tilde \psi d \tilde \theta , \\ \nn &&\sigma_2 =\cos
\psi \sin \theta d \phi - \sin \psi d \theta, \h \Sigma_2 = \cos
\tilde \psi \sin \tilde \theta d \tilde \phi- \sin \tilde \psi d
\tilde \theta , \\ \label{G24} &&\sigma_3 = \cos \theta d \phi+d
\psi,\hspace{2.2cm} \Sigma_3 = \cos \tilde \theta d \tilde \phi+d
\tilde \psi.\eea This metric is Ricci flat and complete for $r
\geq 9r_0/2$. It has a $G_2$-structure given by the following
covariantly constant three-form \bea\nn \Phi&=&{9r_0^3\over
16}\epsilon_{abc}\;\left(\sigma_a\wedge\sigma_b\wedge\sigma_c-
\Sigma_a\wedge\Sigma_b\wedge\Sigma_c\right)
\\ \nn &+& d\left[{r\over 18}\left(r^2-{27r_0^2\over 4}\right)\left(\sigma_1\wedge \Sigma_1+ \sigma_2\wedge
\Sigma_2\right)+ {r_0\over 3}\left(r^2-{81r_0^2\over
8}\right)\sigma_3\wedge \Sigma_3\right],\eea which guarantees the
existence of a unique covariantly constant spinor
\cite{NPB0106034}.

The M-theory background for our case can be written as
\bea\label{11db} l_{11}^{-2}ds_{11}^{2}=-dt^2 + \delta_{IJ}dx^I
dx^J + ds_{7}^{2},\eea where $l_{11}$ is the eleven dimensional
Planck length, \small{({\it I,J=1,2,3})} and $ds_{7}^{2}$ is given
in (\ref{G21})-(\ref{G24}). In other words, the background is
direct product of flat, four dimensional space-time, and a seven
dimensional $G_2$ manifold.

We will search for solutions, for which the background felt by the
membrane depends on only one coordinate. This will be the radial
coordinate $r$, i.e. the rotating membrane embedding along this
coordinate has the form $r=r(\sigma)$. Then, the remaining
membrane coordinates, which are not fixed, will depend linearly on
the worldvolume coordinates $\tau$, $\delta$ and $\sigma$. The
membrane configurations considered below are all for which, we
were able to obtain {\it exact} solutions under the described
conditions.

{\bf First type of membrane embedding \cite{9}}

Let us consider the following membrane configuration: \bea\nn
&&X^0\equiv
t=\Lambda_0^0\tau+\frac{1}{\Lambda_0^0}\left[\left(\mathbf{\Lambda}_0.\mathbf{\Lambda}_1\right)\delta
+ \left(\mathbf{\Lambda}_0.\mathbf{\Lambda}_2\right)\sigma\right],
\h X^I=\Lambda_0^I\tau + \Lambda_1^I\delta + \Lambda_2^I\sigma,
\\ \label{A1} &&X^4\equiv r(\s),\h X^6\equiv \theta=\Lambda_0^6\tau,\h X^9\equiv\tilde{\theta}=\Lambda_0^9\tau;
\h
\left(\mathbf{\Lambda}_0.\mathbf{\Lambda}_i\right)=\delta_{IJ}\Lambda_0^I\Lambda_i^J.\eea
It corresponds to membrane extended in the radial direction $r$,
and rotating in the planes given by the angles  $\theta$ and
$\tilde{\theta}$. In addition, it is nontrivially spanned along
$X^0$ and $X^I$. The relations between the parameters in $X^0$ and
$X^I$ guarantee that the constraints are identically satisfied. At
the same time, the membrane moves along $t$-coordinate with
constant energy $E$, and along $X^I$ with constant momenta
$P_{I}$. In this case, the target space metric seen by the
membrane becomes \bea\nn &&g_{00}\equiv g_{tt}=-l_{11}^{2},\h
g_{IJ}=l_{11}^{2}\delta_{IJ}, \h g_{44}\equiv
g_{rr}=\frac{l_{11}^{2}}{C^2(r)},
\\ \nn &&g_{66}\equiv g_{\theta\theta}=l_{11}^{2}\left[A^2(r)+B^2(r)\right],\h
g_{99}\equiv
g_{\tilde{\theta}\tilde{\theta}}=l_{11}^{2}\left[A^2(r)+B^2(r)\right],\\
\label{b1} &&g_{69}\equiv
g_{\theta\tilde{\theta}}=-l_{11}^{2}\left[A^2(r)-B^2(r)\right].\eea
Therefore, in our notations, we have
$\mu=(0,I,6,9)\equiv(t,I,\theta,\tilde{\theta})$, $a=4\equiv r$.
The metric induced on the membrane worldvolume is \bea\nn
&&G_{00}=-l_{11}^{2}\left[(\Lambda_0^0)^2-\mathbf{\Lambda}_0^2 -
(\Lambda_0^-)^2 A^2-(\Lambda_0^+)^2 B^2\right],\\ \nn
&&G_{11}=l_{11}^{2}M_{11},\h G_{12}=l_{11}^{2}M_{12},\h
G_{22}=l_{11}^{2}\left[M_{22} + \frac{r'^2}{C^2}\right],\eea where
\bea \label{DM} M_{ij}=
\left(\mathbf{\Lambda}_i.\mathbf{\Lambda}_j\right)
-\frac{\left(\mathbf{\Lambda}_0.\mathbf{\Lambda}_i\right)
\left(\mathbf{\Lambda}_0.\mathbf{\Lambda}_j\right)}{\left(\Lambda_0^0\right)^2},
\h \Lambda_0^{\pm}=\Lambda_0^6\pm\Lambda_0^9.\eea The constants of
the motion $\mathcal{P}^2_\mu$, introduced in \cite{8}, are given
by \bea\label{cm1} &&\mathcal{P}^2_0=-\frac{2\lambda^0 T_2^2
l_{11}^4}{\Lambda_0^0}
\left[\left(\mathbf{\Lambda}_0.\mathbf{\Lambda}_1\right)M_{12} -
\left(\mathbf{\Lambda}_0.\mathbf{\Lambda}_2\right)M_{11}\right],\\
\nn &&\mathcal{P}^2_I=2\lambda^0 T_2^2 l_{11}^4\left(\Lambda_1^I
M_{12}-\Lambda_2^I M_{11}\right),\h
\mathcal{P}^2_6=\mathcal{P}^2_9=0.\eea The membrane Lagrangian
takes the form \bea\nn &&\mathcal{L}^{A}(\sigma)
=\frac{1}{4\lambda^0}\left(K_{rr}r'^2 - V\right),\h
K_{rr}=-(2\lambda^0 T_2 l_{11}^2)^2\frac{M_{11}}{C^2},\\ \nn
&&V=(2\lambda^0 T_2 l_{11}^2)^2 \det M_{ij} +
l_{11}^{2}\left[(\Lambda_0^0)^2-\mathbf{\Lambda}_0^2 -
(\Lambda_0^-)^2 A^2-(\Lambda_0^+)^2 B^2\right].\eea

Let us first consider the particular case when $\Lambda_0^-=0$,
i.e. $\theta=\tilde{\theta}$. From the first integral of the
equation of motion for $r(\sigma)$ \bea\nn K_{rr}r'^2 + U=0,\h U=V
+ 4\lambda^0\Lambda^\mu_2\mathcal{P}^2_\mu ,\eea one obtains the
turning points of the effective one-dimensional periodic motion by
solving the equation $r'=0$. In the case under consideration, the
result is \bea\nn &&r_{min}=3l,\h
r_{max}=r_1=l\left(2\sqrt{1+\frac{3u_0^2}{l^2(\Lambda_0^+)^2}}+1\right)>3l,
\\ \nn &&r_2=-l\left(2\sqrt{1+\frac{3u_0^2}{l^2(\Lambda_0^+)^2}}-1\right)<0, \h l=3r_0/2,\eea
where we have introduced the notation \bea\label{u02}
u_0^2&=&(2\lambda^0 T_2 l_{11})^2\det M_{ij} +
(\Lambda_0^0)^2-\mathbf{\Lambda}_0^2 +
4\lambda^0\Lambda^\mu_2\mathcal{P}^2_\mu/l_{11}^2
\\ \nn &=&(\Lambda_0^0)^2-\mathbf{\Lambda}_0^2 - (2\lambda^0 T_2 l_{11})^2\det M_{ij}. \eea

Now, we can write down the following expression for the membrane
solution ($\Delta r=r-3l$) \bea\nn
&&\sigma(r)=\int_{3l}^{r}\left[-\frac{K_{rr}(t)}{U(t)}\right]^{1/2}dt=
\frac{16\lambda^0 T_2
l_{11}}{\Lambda_0^+}\left[\frac{M_{11}l\Delta
r}{\left(r_1-3l\right)\left(3l-r_2\right)}\right]^{1/2}\times
\\ \label{s1} &&F_D^{(5)}\left(1/2;-1/2,-1/2,1/2,1/2,1/2;3/2;-\frac{\Delta r}{2l},-\frac{\Delta r}{4l},
-\frac{\Delta r}{6l},-\frac{\Delta r}{3l-r_2},\frac{\Delta
r}{r_1-3l}\right),\eea where the following normalization condition
must be satisfied ($\Delta r_1=r_1-3l$) \cite{9} \bea\nn
&&2\pi=2\int_{3l}^{r_1}\left[-\frac{K_{rr}(t)}{U(t)}\right]^{1/2}dt=
\frac{32\lambda^0 T_2
l_{11}\left(M_{11}l\right)^{1/2}}{\Lambda_0^+\left(3l-r_2\right)^{1/2}}\times
\\ \nn &&F_D^{(5)}\left(1/2;-1/2,-1/2,1/2,1/2,1/2;3/2;-\frac{\Delta r_1}{2l},-\frac{\Delta r_1}{4l},
-\frac{\Delta r_1}{6l},-\frac{\Delta r_1}{3l-r_2},1\right)=
\\ \nn &&\frac{16\pi\lambda^0 T_2 l_{11}\left(M_{11}l\right)^{1/2}}{\Lambda_0^+\left(3l-r_2\right)^{1/2}}
F_D^{(4)}\left(1/2;-1/2,-1/2,1/2,1/2,;1;-\frac{\Delta
r_1}{2l},-\frac{\Delta r_1}{4l},-\frac{\Delta
r_1}{6l},-\frac{\Delta r_1}{3l-r_2}\right)
\\ \nn &&=\frac{16\pi\lambda^0 T_2 l_{11}\left(M_{11}l\right)^{1/2}}{\Lambda_0^+\left(3l-r_2\right)^{1/2}}
\left(1+\frac{\Delta r_1}{2l}\right)^{1/2}\left(1+\frac{\Delta
r_1}{4l}\right)^{1/2} \left(1+\frac{\Delta
r_1}{6l}\right)^{-1/2}\left(1+\frac{\Delta
r_1}{3l-r_2}\right)^{-1/2}
\\ \label{nc1} &&\times F_D^{(4)}\left(1/2;-1/2,-1/2,1/2,1/2,;1;\frac{1}{1+\frac{2l}{\Delta r_1}},
\frac{1}{1+\frac{4l}{\Delta r_1}}, \frac{1}{1+\frac{6l}{\Delta
r_1}},\frac{1}{1+\frac{3l-r_2}{\Delta r_1}}\right).\eea

Now, we can compute the conserved charges on the obtained
solution. They are: \bea\label{EP}
&&E=-P_0=\frac{\pi^2l_{11}^2}{\lambda^0}\Lambda_0^0,\h
\mathbf{P}=\frac{\pi^2l_{11}^2}{\lambda^0}\mathbf{\Lambda}_0,
\\ \nn &&P_{\theta}=P_{\tilde{\theta}}=\frac{\pi l_{11}^2}{\lambda^0}\Lambda_0^+ \int_{3l}^{r_1}
\left[-\frac{K_{rr}(t)}{U(t)}\right]^{1/2}B^2(t)dt=
\frac{4\pi^2T_2
l_{11}^3\left(M_{11}l^3\right)^{1/2}}{3\left(3l-r_2\right)^{1/2}}\times
\\ \nn &&\Delta r_1 F_D^{(4)}\left(3/2;-1/2,-3/2,1/2,1/2,;2;-\frac{\Delta r_1}{2l},-\frac{\Delta r_1}{4l},
-\frac{\Delta r_1}{6l},-\frac{\Delta r_1}{3l-r_2}\right)
\\ \nn &&=\frac{4\pi^2T_2 l_{11}^3\left(M_{11}l^3\right)^{1/2}}{3\left(3l-r_2\right)^{1/2}}\Delta r_1
\left(1+\frac{\Delta r_1}{2l}\right)^{1/2}\left(1+\frac{\Delta
r_1}{4l}\right)^{3/2} \left(1+\frac{\Delta
r_1}{6l}\right)^{-1/2}\left(1+\frac{\Delta
r_1}{3l-r_2}\right)^{-1/2}
\\ \label{cmom1} &&\times F_D^{(4)}\left(1/2;-1/2,-3/2,1/2,1/2,;2;\frac{1}{1+\frac{2l}{\Delta r_1}},
\frac{1}{1+\frac{4l}{\Delta r_1}}, \frac{1}{1+\frac{6l}{\Delta
r_1}}, \frac{1}{1+\frac{3l-r_2}{\Delta r_1}}\right).\eea

Our next task is to find the relation between the energy $E$ and
the other conserved quantities $\mathbf{P}$,
$P_{\theta}=P_{\tilde{\theta}}$ in the semiclassical limit (large
conserved charges). This corresponds to $r_1\to\infty$, which in
the present case leads to $3u_0^2/[l^2(\Lambda_0^+)^2]\to\infty$.
In this limit, the condition (\ref{nc1}) reduces to \bea\nn
\Lambda_0^+ = 2\sqrt{3}\lambda^0 T_2 l_{11}M_{11}^{1/2},\eea while
the expression (\ref{cmom1}) for the momentum $P_{\theta}$, takes
the form \bea\nn P_{\theta}=P_{\tilde{\theta}}= \sqrt{3}\pi^2 T_2
l_{11}^3 M_{11}^{1/2}\frac{u_0^2}{(\Lambda_0^+)^2}.\eea Combining
these results with (\ref{EP1}), one obtains \bea\label{Eg1}
&&\left\{E^2\left(E^2-\mathbf{P}^2\right) - (2\pi^2 T_2
l_{11}^3)^2\left\{\left(\mathbf{\Lambda}_1\times\mathbf{\Lambda}_2\right)^2
E^2 -
\left[\left(\mathbf{\Lambda}_1\times\mathbf{\Lambda}_2\right)\times\mathbf{P}\right]^2\right\}\right\}^2
\\ \nn &&-(4\sqrt{3}\pi^2 T_2 l_{11}^3)^2E^2\left[\mathbf{\Lambda}_1^2 E^2
 - \left(\mathbf{\Lambda}_1.\mathbf{P}\right)^2\right]P_{\theta}^2 = 0,
\h \left(\mathbf{\Lambda}_1\times\mathbf{\Lambda}_2\right)_I =
\varepsilon_{IJK}\Lambda_1^J\Lambda_2^K.\eea This is {\it fourth}
order algebraic equation for $E^2$. Its positive solutions give
the explicit dependence of the energy on $\mathbf{P}$ and
$P_{\theta}$: $E^2=E^2(\mathbf{P}, P_{\theta})$.

Let us consider a few particular cases. In the simplest case, when
$\Lambda_0^I=0$, i.e. $\mathbf{P}=0$, and
$\Lambda_2^I=c\Lambda_1^I$, which corresponds to the membrane
embedding (see (\ref{A1})) \bea\nn X^0\equiv t=\Lambda_0^0\tau, \h
X^I=\Lambda_1^I(\delta + c\sigma),\h X^4\equiv r(\s), \h X^6\equiv
\theta=\Lambda_0^6\tau =
X^9\equiv\tilde{\theta}=\Lambda_0^9\tau,\eea (\ref{Eg1})
simplifies to \bea\label{p1} E^2=4\sqrt{3}\pi^2 T_2
l_{11}^3\mid\mathbf{\Lambda}_1\mid P_{\theta}.\eea This is the
relation $E\sim K^{1/2}$ obtained for $G_2$-manifolds in
\cite{27}. If we impose only the conditions $\Lambda_0^I=0$, and
$\Lambda_i^I$ remain independent, (\ref{Eg1}) gives \bea\label{p2}
E^2 =(2\pi^2 T_2
l_{11}^3)^2\left(\mathbf{\Lambda}_1\times\mathbf{\Lambda}_2\right)^2
+ 4\sqrt{3}\pi^2 T_2 l_{11}^3\mid\mathbf{\Lambda}_1\mid
P_{\theta}.\eea Now, let us take $\Lambda_0^I\ne 0$,
$\Lambda_2^I=c\Lambda_1^I$. Then, (\ref{Eg1}) reduces to \bea\nn
E^2\left[\left(E^2-\mathbf{P}^2\right)^2 - (4\sqrt{3}\pi^2 T_2
l_{11}^3)^2\mathbf{\Lambda}_1^2P_{\theta}^2\right]
+(4\sqrt{3}\pi^2 T_2 l_{11}^3)^2
\left(\mathbf{\Lambda}_1.\mathbf{P}\right)^2P_{\theta}^2=0 ,\eea
which is {\it third} order algebraic equation for $E^2$. If the
three-dimensional vectors $\mathbf{\Lambda}_1$ and $\mathbf{P}$
are orthogonal to each other, i.e.
$\left(\mathbf{\Lambda}_1.\mathbf{P}\right)=0$, the above relation
simplifies to \bea\label{p3} E^2=\mathbf{P}^2 + 4\sqrt{3}\pi^2 T_2
l_{11}^3\mid\mathbf{\Lambda}_1\mid P_{\theta}.\eea The obvious
conclusion is that in the framework of a given embedding, one can
obtain different relations between the energy and the other
conserved charges, depending on the choice of the embedding
parameters.

Now, we will consider the general case, when $\Lambda_0^-\ne 0$,
i.e. $\theta\ne\tilde{\theta}$. The turning points are given by
\bea\nn &&r_{min}=3l,\h
r_{max}=r_1=l\left[2\sqrt{\frac{k^2+3}{4}+\frac{3u_0^2}{l^2\left((\Lambda_0^+)^2+(\Lambda_0^-)^2\right)}}+k\right],
\\ \nn &&r_2=-l\left[2\sqrt{\frac{k^2+3}{4}+\frac{3u_0^2}{l^2\left((\Lambda_0^+)^2+(\Lambda_0^-)^2\right)}}-k\right],
\h
k=\frac{(\Lambda_0^+)^2-(\Lambda_0^-)^2}{(\Lambda_0^+)^2+(\Lambda_0^-)^2}\in
[0,1].\eea

Now the solution for $\sigma(r)$ is \bea\nn
&&\sigma(r)=\int_{3l}^{r}\left[-\frac{K_{rr}(t)}{U(t)}\right]^{1/2}dt=
\frac{16\lambda^0 T_2
l_{11}}{\left[(\Lambda_0^+)^2+(\Lambda_0^-)^2\right]^{1/2}}\left[\frac{M_{11}l\Delta
r}{\left(r_1-3l\right)\left(3l-r_2\right)}\right]^{1/2}\times
\\ \label{sg1} &&F_D^{(5)}\left(1/2;-1/2,-1/2,1/2,1/2,1/2;3/2;-\frac{\Delta r}{2l},-\frac{\Delta r}{4l},
-\frac{\Delta r}{6l},-\frac{\Delta r}{3l-r_2},\frac{\Delta
r}{r_1-3l}\right).\eea

The normalization condition reads \bea\nn &&\frac{8\lambda^0 T_2
l_{11}\left(M_{11}l\right)^{1/2}}{\left[(\Lambda_0^+)^2+(\Lambda_0^-)^2\right]^{1/2}\left(3l-r_2\right)^{1/2}}\times
\\ \nn &&F_D^{(4)}\left(1/2;-1/2,-1/2,1/2,1/2,;1;-\frac{\Delta r_1}{2l},-\frac{\Delta r_1}{4l},-\frac{\Delta r_1}{6l},
-\frac{\Delta r_1}{3l-r_2}\right)=
\\ \nn &&\frac{8\lambda^0 T_2 l_{11}\left(M_{11}l\right)^{1/2}}{\left[(\Lambda_0^+)^2+(\Lambda_0^-)^2\right]^{1/2}\left(3l-r_2\right)^{1/2}}
\times
\\ \nn &&\left(1+\frac{\Delta r_1}{2l}\right)^{1/2}\left(1+\frac{\Delta r_1}{4l}\right)^{1/2} \left(1+\frac{\Delta r_1}{6l}\right)^{-1/2}
\left(1+\frac{\Delta r_1}{3l-r_2}\right)^{-1/2}\times
\\ \label{ncg1} &&F_D^{(4)}\left(1/2;-1/2,-1/2,1/2,1/2,;1;\frac{1}{1+\frac{2l}{\Delta r_1}},
\frac{1}{1+\frac{4l}{\Delta r_1}}, \frac{1}{1+\frac{6l}{\Delta
r_1}},\frac{1}{1+\frac{3l-r_2}{\Delta r_1}}\right)=1.\eea

Computing the conserved momenta, one obtains the same expressions
for $E$ and $\mathbf{P}$ as in (\ref{EP1})\footnote{Actually,
these expressions for $E$ and $\mathbf{P}$ are always valid  for
the background we use.}, and \bea\nn
&&\frac{1}{2}\left(P_{\theta}+P_{\tilde{\theta}}\right)=
\frac{4\pi^2T_2 l_{11}^3\Lambda_0^+\left(M_{11}l^3\right)^{1/2}}
{3\left[(\Lambda_0^+)^2+(\Lambda_0^-)^2\right]^{1/2}\left(3l-r_2\right)^{1/2}}\times
\\ \nn &&\Delta r_1 F_D^{(4)}\left(3/2;-1/2,-3/2,1/2,1/2,;2;-\frac{\Delta r_1}{2l},-\frac{\Delta r_1}{4l},
-\frac{\Delta r_1}{6l},-\frac{\Delta r_1}{3l-r_2}\right)
\\ \nn &&=\frac{4\pi^2T_2 l_{11}^3\Lambda_0^+\left(M_{11}l^3\right)^{1/2}}
{3\left[(\Lambda_0^+)^2+(\Lambda_0^-)^2\right]^{1/2}\left(3l-r_2\right)^{1/2}}\times
\\ \nn &&\Delta r_1 \left(1+\frac{\Delta r_1}{2l}\right)^{1/2}\left(1+\frac{\Delta r_1}{4l}\right)^{3/2}
\left(1+\frac{\Delta r_1}{6l}\right)^{-1/2}\left(1+\frac{\Delta r_1}{3l-r_2}\right)^{-1/2}\times
\\ \label{cmomg1t1} &&F_D^{(4)}\left(1/2;-1/2,-3/2,1/2,1/2,;2;\frac{1}{1+\frac{2l}{\Delta r_1}}, \frac{1}{1+\frac{4l}{\Delta r_1}},
\frac{1}{1+\frac{6l}{\Delta r_1}}, \frac{1}{1+\frac{3l-r_2}{\Delta
r_1}}\right),\eea

\bea\nn &&\frac{1}{2}\left(P_{\theta}-P_{\tilde{\theta}}\right)=
\frac{8\pi^2T_2 l_{11}^3\Lambda_0^-\left(M_{11}l^5\right)^{1/2}}
{\left[(\Lambda_0^+)^2+(\Lambda_0^-)^2\right]^{1/2}\left(3l-r_2\right)^{1/2}}\times
\\ \nn &&F_D^{(4)}\left(1/2;-3/2,-1/2,-1/2,1/2,;1;-\frac{\Delta r_1}{2l},-\frac{\Delta r_1}{4l},
-\frac{\Delta r_1}{6l},-\frac{\Delta r_1}{3l-r_2}\right)
\\ \nn &&=\frac{8\pi^2T_2 l_{11}^3\Lambda_0^-\left(M_{11}l^5\right)^{1/2}}
{\left[(\Lambda_0^+)^2+(\Lambda_0^-)^2\right]^{1/2}\left(3l-r_2\right)^{1/2}}\times
\\ \nn &&\left(1+\frac{\Delta r_1}{2l}\right)^{3/2}\left(1+\frac{\Delta r_1}{4l}\right)^{1/2} \left(1+\frac{\Delta r_1}{6l}\right)^{1/2}
\left(1+\frac{\Delta r_1}{3l-r_2}\right)^{-1/2}\times
\\ \label{cmomg1t2} &&F_D^{(4)}\left(1/2;-3/2,-1/2,-1/2,1/2,;1;\frac{1}{1+\frac{2l}{\Delta r_1}}, \frac{1}{1+\frac{4l}{\Delta r_1}},
\frac{1}{1+\frac{6l}{\Delta r_1}}, \frac{1}{1+\frac{3l-r_2}{\Delta
r_1}}\right).\eea

Now, we go to the semiclassical limit $r_1\to\infty$. The
normalization condition gives \bea\nn
\left[(\Lambda_0^+)^2+(\Lambda_0^-)^2\right]^{1/2} =
2\sqrt{3}\lambda^0 T_2 l_{11}M_{11}^{1/2},\eea whereas
(\ref{cmomg1t1}) and (\ref{cmomg1t2}) take the form \bea\nn
\frac{1}{2}\left(P_{\theta}\pm P_{\tilde{\theta}}\right)=
\frac{\sqrt{3}\pi^2T_2 l_{11}^3\Lambda_0^\pm M_{11}^{1/2}u_0^2}
{\left[(\Lambda_0^+)^2+(\Lambda_0^-)^2\right]^{3/2}}.\eea The
above expressions, together with (\ref{EP1}), lead to the
following connection between the energy and the conserved momenta
\bea\label{Egg1} &&\left\{E^2\left(E^2-\mathbf{P}^2\right) -
(2\pi^2 T_2
l_{11}^3)^2\left\{\left(\mathbf{\Lambda}_1\times\mathbf{\Lambda}_2\right)^2
E^2 -
\left[\left(\mathbf{\Lambda}_1\times\mathbf{\Lambda}_2\right)\times\mathbf{P}\right]^2\right\}\right\}^2
\\ \nn &&-6(2\pi^2 T_2 l_{11}^3)^2E^2\left[\mathbf{\Lambda}_1^2 E^2 -
\left(\mathbf{\Lambda}_1.\mathbf{P}\right)^2\right]\left(P^2_{\theta}+P^2_{\tilde{\theta}}\right)
= 0.\eea Obviously, (\ref{Egg1}) is the generalization of
(\ref{Eg1}) for the case $P_{\theta}\ne P_{\tilde{\theta}}$ and
for $P_{\theta}=P_{\tilde{\theta}}$ coincides with it, as it
should be. The particular cases (\ref{p1}), (\ref{p2}) and
(\ref{p3}) now generalize to \bea\nn &&E^2=2\sqrt{6}\pi^2 T_2
l_{11}^3\mid\mathbf{\Lambda}_1\mid
\left(P^2_{\theta}+P^2_{\tilde{\theta}}\right)^{1/2},\\ \nn &&E^2
=(2\pi^2 T_2
l_{11}^3)^2\left(\mathbf{\Lambda}_1\times\mathbf{\Lambda}_2\right)^2
+ 2\sqrt{6}\pi^2 T_2 l_{11}^3\mid\mathbf{\Lambda}_1\mid
\left(P^2_{\theta}+P^2_{\tilde{\theta}}\right)^{1/2},\\
\label{sscb} &&E^2=\mathbf{P}^2 + 2\sqrt{6}\pi^2 T_2
l_{11}^3\mid\mathbf{\Lambda}_1\mid
\left(P^2_{\theta}+P^2_{\tilde{\theta}}\right)^{1/2}.\eea

Finally, let us give the semiclassical limit of the membrane
solution (\ref{sg1}), which is \bea\label{sg1SCl}
\sigma_{scl}(r)&=& \left\{\frac{32(4\pi^2T_2
l_{11}^3)^2\left[\mathbf{\Lambda}_1^2 E^2 -
\left(\mathbf{\Lambda}_1.\mathbf{P}\right)^2\right]}{27
E^2\left(P^2_{\theta}+P^2_{\tilde{\theta}}\right)}\right\}^{1/4}(l\Delta
r)^{1/2}
\\ \nn &\times & F_D^{(3)}\left(1/2;-1/2,-1/2,1/2;3/2;-\frac{\Delta r}{2l},-\frac{\Delta r}{4l},-\frac{\Delta r}{6l}\right)
\\ \nn &=& \left\{\frac{32(4\pi^2T_2 l_{11}^3)^2\left[\mathbf{\Lambda}_1^2 E^2 - \left(\mathbf{\Lambda}_1.\mathbf{P}\right)^2\right]}
{27
E^2\left(P^2_{\theta}+P^2_{\tilde{\theta}}\right)}\right\}^{1/4}(l\Delta
r)^{1/2}\\ \nn &\times & \left(1+\frac{\Delta
r}{2l}\right)^{1/2}\left(1+\frac{\Delta r}{4l}\right)^{1/2}
\left(1+\frac{\Delta r}{6l}\right)^{-1/2}
\\ \nn &&F_D^{(3)}\left(1;-1/2,-1/2,1/2,;3/2;\frac{1}{1+\frac{2l}{\Delta r}}, \frac{1}{1+\frac{4l}{\Delta r}},
\frac{1}{1+\frac{6l}{\Delta r}}\right).\eea

{\bf Second type of membrane embedding \cite{9}}

Let us consider membrane, which is extended along the radial
direction $r$ and rotates in the planes defined by the angles
$\theta$ and $\tilde{\theta}$, with angular momenta $P_{\theta}$
and $P_{\tilde{\theta}}$. Now we want to have nontrivial wrapping
along $X^6$ and $X^9$. The embedding parameters in $X^6$ and $X^9$
have to be chosen in such a way that the constraints are satisfied
identically. It turns out that the angular momenta $P_{\theta}$
and $P_{\tilde{\theta}}$ must be equal, and the constants of the
motion $\mathcal{P}^2_\mu$ are identically zero for this case. In
addition, we want the membrane to move along $X^0$ and $X^I$ with
constant energy $E$ and constant momenta $P_{I}$ respectively. All
this leads to the following ansatz: \bea\nn &&X^0\equiv
t=\Lambda_0^0\tau,\h X^I=\Lambda_0^I\tau ,\h X^4\equiv r(\s),
\\ \label{A2} &&X^6\equiv \theta=\Lambda_0^6\tau+\Lambda_1^6\delta + \Lambda_2^6\sigma,
\h X^9\equiv\tilde{\theta}=\Lambda_0^6\tau-(\Lambda_1^6\delta +
\Lambda_2^6\sigma).\eea

The background felt by the membrane is the same as in (\ref{b1}),
but the metric induced on the membrane worldvolume is different
and is given by \bea\nn
&&G_{00}=-l_{11}^{2}\left[(\Lambda_0^0)^2-\mathbf{\Lambda}_0^2-(\Lambda_0^+)^2
B^2\right], \h G_{11}=4l_{11}^{2}(\Lambda_1^6)^2 A^2,
\\ \nn &&G_{12}=4l_{11}^{2}\Lambda_1^6\Lambda_2^6 A^2,\h G_{22}=l_{11}^{2}\left[\frac{r'^2}{C^2} + 4(\Lambda_2^6)^2 A^2\right].\eea

For the present case, the membrane Lagrangian reduces to \bea\nn
&&\mathcal{L}^{A}(\sigma) =\frac{1}{4\lambda^0}\left(K_{rr}r'^2 -
V\right),\h K_{rr}=-(4\lambda^0 T_2
l_{11}^2)^2(\Lambda_1^6)^2\frac{A^2}{C^2},\\ \nn
&&V=U=l_{11}^{2}\left[(\Lambda_0^0)^2-\mathbf{\Lambda}_0^2
-(\Lambda_0^+)^2 B^2\right].\eea The turning points of the
effective one-dimensional periodic motion \bea\nn K_{rr}r'^2 +
V=0,\eea are given by \bea\nn &&r_{min}=3l,\h
r_{max}=r_1=l\left(2\sqrt{1+\frac{3v_0^2}{l^2(\Lambda_0^+)^2}}+1\right)>3l,
\\ \label{r'II} &&r_2=-l\left(2\sqrt{1+\frac{3v_0^2}{l^2(\Lambda_0^+)^2}}-1\right)<0, \h v_0^2=(\Lambda_0^0)^2-\mathbf{\Lambda}_0^2.\eea

Now, the membrane solution reads: \bea\nn
&&\sigma(r)=\int_{3l}^{r}\left[-\frac{K_{rr}(t)}{V(t)}\right]^{1/2}dt=
\frac{32\lambda^0 T_2 l_{11}\Lambda_1^6}{\Lambda_0^+}
\left[\frac{l^3\Delta
r}{\left(r_1-3l\right)\left(3l-r_2\right)}\right]^{1/2}\times
\\ \label{s2} &&F_D^{(4)}\left(1/2;-1,-1/2,1/2,1/2;3/2;-\frac{\Delta r}{2l},-\frac{\Delta r}{4l},
-\frac{\Delta r}{3l-r_2},\frac{\Delta r}{r_1-3l}\right).\eea The
normalization condition leads to the following relation between
the parameters \bea\nn &&\frac{16\lambda^0 T_2 l_{11}\Lambda_1^6
l^{3/2}}{\Lambda_0^+\left(3l-r_2\right)^{1/2}}
F_D^{(3)}\left(1/2;-1,-1/2,1/2;1;-\frac{\Delta
r_1}{2l},-\frac{\Delta r_1}{4l},-\frac{\Delta r_1}{3l-r_2}\right)
\\ \nn &&=\frac{16\lambda^0 T_2 l_{11}\Lambda_1^6 l^{3/2}}{\Lambda_0^+\left(3l-r_2\right)^{1/2}}
\left(1+\frac{\Delta r_1}{2l}\right)\left(1+\frac{\Delta
r_1}{4l}\right)^{1/2} \left(1+\frac{\Delta
r_1}{3l-r_2}\right)^{-1/2}\\ \label{nc2} &&\times
F_D^{(3)}\left(1/2;-1,-1/2,1/2,;1;\frac{1}{1+\frac{2l}{\Delta
r_1}}, \frac{1}{1+\frac{4l}{\Delta r_1}},
\frac{1}{1+\frac{3l-r_2}{\Delta r_1}}\right)=1.\eea In the case
under consideration, the conserved quantities are $E$,
$\mathbf{P}$ and $P_{\theta}=P_{\tilde{\theta}}$. We derive the
following result for $P_{\theta}=P_{\tilde{\theta}}$ \bea\nn
&&P_{\theta}=P_{\tilde{\theta}}= \frac{8\pi^2T_2
l_{11}^3\Lambda_1^6 l^{5/2}}{3\left(3l-r_2\right)^{1/2}}\Delta r_1
F_D^{(3)}\left(3/2;-1,-3/2,1/2;2;-\frac{\Delta
r_1}{2l},-\frac{\Delta r_1}{4l},-\frac{\Delta r_1}{3l-r_2}\right)
\\ \nn &&=\frac{8\pi^2T_2 l_{11}^3\Lambda_1^6 l^{5/2}}{3\left(3l-r_2\right)^{1/2}}\Delta r_1
\left(1+\frac{\Delta r_1}{2l}\right)\left(1+\frac{\Delta
r_1}{4l}\right)^{3/2} \left(1+\frac{\Delta
r_1}{3l-r_2}\right)^{-1/2}\\ \label{cmom2} &&\times
F_D^{(3)}\left(1/2;-1,-3/2,1/2,;2;\frac{1}{1+\frac{2l}{\Delta
r_1}}, \frac{1}{1+\frac{4l}{\Delta r_1}},
\frac{1}{1+\frac{3l-r_2}{\Delta r_1}}\right).\eea

In the semiclassical limit, (\ref{nc2}) and (\ref{cmom2}) reduce
to \bea\nn (\Lambda_0^+)^2=\frac{8\sqrt{3}}{\pi}\lambda^0 T_2
l_{11}\Lambda_1^6\left[(\Lambda_0^0)^2-\mathbf{\Lambda}_0^2\right]^{1/2},
\h P_{\theta}=P_{\tilde{\theta}}=\frac{16\pi T_2
l_{11}^3\Lambda_1^6 }{\sqrt{3}(\Lambda_0^+)^3}
\left[(\Lambda_0^0)^2-\mathbf{\Lambda}_0^2\right]^{3/2}.\eea From
here and (\ref{EP}), one obtains the relation \bea\label{EPc2}
E^2=\mathbf{P}^2+3^{5/3}(2\pi T_2
l_{11}^3\Lambda_1^6)^{2/3}P_{\theta}^{4/3}.\eea In the particular
case when $\mathbf{P}=0$, (\ref{EPc2}) coincides with the
energy-charge relation $E\sim K^{2/3}$, first obtained for
$G_2$-manifolds in \cite{27}. For the given embedding (\ref{A2}),
the semiclassical limit of the membrane solution (\ref{s2}) is as
follows \bea\nn &&\sigma_{scl}(r)=8\pi^{1/3}\left(\frac{2\pi^2 T_2
l_{11}^3\Lambda_1^6}{9P_{\theta}}\right)^{2/3} \left(l^3\Delta
r\right)^{1/2} F_D^{(2)}\left(1/2;-1,-1/2;3/2;-\frac{\Delta
r}{2l},-\frac{\Delta r}{4l}\right)
\\ \label{SCls2} &&=8\pi^{1/3}\left(\frac{2\pi^2 T_2 l_{11}^3\Lambda_1^6}{9P_{\theta}}\right)^{2/3}
\left(l^3\Delta r\right)^{1/2} \left(1+\frac{\Delta
r}{2l}\right)\left(1+\frac{\Delta r}{4l}\right)^{1/2}
\\ \nn &&\times F_D^{(2)}\left(1;-1,-1/2;3/2;\frac{1}{1+\frac{2l}{\Delta r}},
\frac{1}{1+\frac{4l}{\Delta r}}\right).\eea

{\bf Third type of membrane embedding \cite{9}}

Again, we want the membrane to move in the flat, four dimensional
part of the eleven dimensional background metric (\ref{11db}),
with constant energy $E$ and constant momenta $P_{I}$. On the
curved part of the metric, the membrane is extended along the
radial coordinate $r$, rotates in the plane given by the angle
$\psi_+= \psi+\tilde{\psi}$, and is wrapped along the angular
coordinate $\psi_-=\psi-\tilde{\psi}$. This membrane configuration
is given by \bea\nn &&X^0\equiv t=\Lambda_0^0\tau,\h
X^I=\Lambda_0^I\tau,\h X^4\equiv r(\s),
\\ \label{A3} &&\psi_+=\Lambda_0^+\tau,\h \psi_-=\Lambda_1^-\delta + \Lambda_2^-\sigma,
\h \psi_\pm = \psi\pm\tilde{\psi}.\eea In this case, the target
space metric seen by the membrane is \bea\nn &&g_{00}\equiv
g_{tt}=-l_{11}^{2},\h g_{IJ}=l_{11}^{2}\delta_{IJ}, \h
g_{44}\equiv g_{rr}=\frac{l_{11}^{2}}{C^2(r)},
\\ \label{b2} &&g_{++}=l_{11}^{2}\left(\frac{2l}{3}\right)^2 C^2(r),\h g_{--}=l_{11}^{2}D^2(r).\eea
Hence, in our notations, we have $\mu=(0,I,+,-)$, $a=4\equiv r$.
Now, the metric induced on the membrane worldvolume is \bea\nn
&&G_{00}=-l_{11}^{2}\left[(\Lambda_0^0)^2-\mathbf{\Lambda}_0^2
-(\Lambda_0^+)^2 \left(\frac{2l}{3}\right)^2 C^2\right],\\ \nn
&&G_{11}=l_{11}^{2}(\Lambda_1^-)^2 D^2,\h
G_{12}=l_{11}^{2}\Lambda_1^-\Lambda_2^- D^2,\h
G_{22}=l_{11}^{2}\left[(\Lambda_2^-)^2 D^2 +
\frac{r'^2}{C^2}\right].\eea The constraints are satisfied
identically, and $\mathcal{P}^2_\mu\equiv 0$. The Lagrangian takes
the form \bea\nn &&\mathcal{L}^{A}(\sigma)
=\frac{1}{4\lambda^0}\left(K_{rr}r'^2 - V\right),\h
K_{rr}=-(2\lambda^0 T_2 l_{11}^2\Lambda_1^-)^2\frac{D^2}{C^2},\\
\nn &&V=U= l_{11}^{2}\left[(\Lambda_0^0)^2-\mathbf{\Lambda}_0^2
-(\Lambda_0^+)^2 \left(\frac{2l}{3}\right)^2 C^2\right].\eea The
turning points read \bea\nn &&r_{min}=3l,\h
r_{max}=r_1=l\sqrt{1+\frac{8}{1-\frac{9v_0^2}{4l^2(\Lambda_0^+)^2}}}>3l,
\\ \nn &&r_2=-l\sqrt{1+\frac{8}{1-\frac{9v_0^2}{4l^2(\Lambda_0^+)^2}}}<0, \h v_0^2=(\Lambda_0^0)^2-\mathbf{\Lambda}_0^2.\eea

For the present embedding, we derive the following membrane
solution \bea\label{s3}
&&\sigma(r)=\int_{3l}^{r}\left[-\frac{K_{rr}(t)}{V(t)}\right]^{1/2}dt=
\frac{2\lambda^0 T_2 l_{11}\Lambda_1^-}{\left[(\Lambda_0^+)^2
\left(\frac{2l}{3}\right)^2-v_0^2\right]^{1/2}}\left[\frac{2^7
l^5\Delta r} {3\left(r_1-3l\right)
\left(3l-r_2\right)}\right]^{1/2}\times
\\ \nn &&F_D^{(6)}\left(1/2;-1,-1,-1,1/2,1/2,1/2;3/2;-\frac{\Delta r}{2l},-\frac{\Delta r}{3l},-\frac{\Delta r}{4l},-\frac{\Delta r}{6l},
-\frac{\Delta r}{3l-r_2},\frac{\Delta r}{r_1-3l}\right).\eea The
normalization condition leads to \bea\label{nc3} &&\frac{\lambda^0
T_2 l_{11}\Lambda_1^-}{\left[(\Lambda_0^+)^2
\left(\frac{2l}{3}\right)^2-v_0^2\right]^{1/2}}\left[\frac{2^7
l^5} {3\left(3l-r_2\right)}\right]^{1/2}\times
\\ \nn &&F_D^{(5)}\left(1/2;-1,-1,-1,1/2,1/2;1;-\frac{\Delta r_1}{2l},-\frac{\Delta r_1}{3l},-\frac{\Delta r_1}{4l},-\frac{\Delta r_1}{6l},
-\frac{\Delta r_1}{3l-r_2}\right)=
\\ \nn &&\frac{\lambda^0 T_2 l_{11}\Lambda_1^-}{\left[(\Lambda_0^+)^2 \left(\frac{2l}{3}\right)^2-v_0^2\right]^{1/2}}\left[\frac{2^7 l^5}
{3\left(3l-r_2\right)}\right]^{1/2}\times
\\ \nn &&\left(1+\frac{\Delta r_1}{2l}\right)\left(1+\frac{\Delta r_1}{3l}\right)
\left(1+\frac{\Delta r_1}{4l}\right)\left(1+\frac{\Delta
r_1}{6l}\right)^{-1/2} \left(1+\frac{\Delta
r_1}{3l-r_2}\right)^{-1/2}\times
\\ \nn &&F_D^{(5)}\left(1/2;-1,-1,-1,1/2,1/2,;1;\frac{1}{1+\frac{2l}{\Delta r_1}}, \frac{1}{1+\frac{3l}{\Delta r_1}},
\frac{1}{1+\frac{4l}{\Delta r_1}}, \frac{1}{1+\frac{6l}{\Delta
r_1}},\frac{1}{1+\frac{3l-r_2}{\Delta r_1}}\right)=1.\eea The
computation of the conserved momentum $P_+ \equiv P_{\psi_+}$
gives \bea\label{cmom3} &&P_+ = \frac{\pi^2 T_2
l_{11}^3\Lambda_0^+\Lambda_1^-}{\left[(\Lambda_0^+)^2
\left(\frac{2l}{3}\right)^2-v_0^2\right]^{1/2}}\left[\frac{2^5
l^7} {3^3\left(3l-r_2\right)}\right]^{1/2}\times
\\ \nn &&\Delta r_1 F_D^{(3)}\left(3/2;-1,-1/2,1/2;2;-\frac{\Delta r_1}{3l},-\frac{\Delta r_1}{6l},-\frac{\Delta r_1}{3l-r_2}\right)=
\\ \nn &&\frac{\pi^2 T_2 l_{11}^3\Lambda_0^+\Lambda_1^-}{\left[(\Lambda_0^+)^2 \left(\frac{2l}{3}\right)^2-v_0^2\right]^{1/2}}\left[\frac{2^5 l^7}
{3^3\left(3l-r_2\right)}\right]^{1/2}\times
\\ \nn &&\Delta r_1\left(1+\frac{\Delta r_1}{3l}\right)\left(1+\frac{\Delta r_1}{6l}\right)^{1/2}
\left(1+\frac{\Delta r_1}{3l-r_2}\right)^{-1/2}\times
\\ \nn &&F_D^{(3)}\left(1/2;-1,-1/2,1/2;2;\frac{1}{1+\frac{3l}{\Delta r_1}},
\frac{1}{1+\frac{6l}{\Delta r_1}},\frac{1}{1+\frac{3l-r_2}{\Delta
r_1}}\right).\eea Let us note that for the embedding (\ref{A3}),
the momentum $P_{\psi_-}$ is zero.

Going to the semiclassical limit $r_1\to\infty$, which in the case
under consideration leads to $9v_0^2/[4l^2(\Lambda_0^+)^2]\to
1_-$, one obtains that (\ref{nc3}) and (\ref{cmom3}) reduce to
\bea\nn
\Lambda_0^+\left[1-\frac{9v_0^2}{4l^2(\Lambda_0^+)^2}\right]^{3/2}=2\lambda^0
T_2 l_{11}\Lambda_1^-l, \h P_+ = \frac{2^{5/2}\pi^2T_2
l_{11}^3\Lambda_1^-l^3}{9\left[1-\frac{9v_0^2}{4l^2(\Lambda_0^+)^2}\right]^{3/2}}.\eea
These two equalities, together with (\ref{EP}), give the following
relation between the energy and the conserved momenta
\bea\label{EPc3} E^2= \mathbf{P}^2 + \frac{9}{2l^2}P_+^2 -
(6\pi^2T_2 l_{11}^3\Lambda_1^-)^{2/3}P_+^{4/3}.\eea In the
particular case when $\mathbf{P}=0$, (\ref{EPc3}) can be rewritten
as \bea\nn
E=\frac{3}{\sqrt{2}l}P_+\sqrt{1-\left(\frac{4\sqrt{2}\pi^2T_2
l_{11}^3\Lambda_1^-l^3}{9P_+}\right)^{2/3}}.\eea Expanding the
square root and neglecting the higher order terms, one derives
energy-charge relation of the type $E-K\sim K^{1/3}$, first found
for backgrounds of $G_2$-holonomy in \cite{27}.

Now, let us write down the semiclassical limit of our membrane
solution (\ref{s3}): \bea\label{s3scl} &&\sigma_{scl}(r)=
\frac{\pi^2T_2l_{11}^3\Lambda_1^-}{P_+}\left(\frac{2^7
l^5}{3^3}\right)^{1/2}\times \\ \nn &&\Delta r^{1/2}
F_D^{(4)}\left(1/2;-1,-1,-1,1/2;3/2;-\frac{\Delta
r}{2l},-\frac{\Delta r}{3l},-\frac{\Delta r}{4l},-\frac{\Delta
r}{6l}\right)=
\\ \nn &&\frac{\pi^2T_2l_{11}^3\Lambda_1^-}{P_+}\left(\frac{2^7 l^5}{3^3}\right)^{1/2}
\Delta r^{1/2}\left(1+\frac{\Delta
r}{2l}\right)\left(1+\frac{\Delta r}{3l}\right)
\left(1+\frac{\Delta r}{4l}\right)\left(1+\frac{\Delta
r}{6l}\right)^{-1/2}\times
\\ \nn &&F_D^{(4)}\left(1;-1,-1,-1,1/2;3/2;\frac{1}{1+\frac{2l}{\Delta r}}, \frac{1}{1+\frac{3l}{\Delta r}},
\frac{1}{1+\frac{4l}{\Delta r}}, \frac{1}{1+\frac{6l}{\Delta
r}}\right).\eea

{\bf Forth type of membrane embedding \cite{9}}

Let us consider membrane configuration given by the following
ansatz: \bea\nn &&X^0\equiv
t=\Lambda_0^0\tau+\frac{1}{\Lambda_0^0}\left[\left(\mathbf{\Lambda}_0.\mathbf{\Lambda}_1\right)\delta
+ \left(\mathbf{\Lambda}_0.\mathbf{\Lambda}_2\right)\sigma\right],
\h X^I=\Lambda_0^I\tau + \Lambda_1^I\delta + \Lambda_2^I\sigma,
\\ \label{A4} &&X^4\equiv r(\s),\h \psi_+=\Lambda_0^+\tau,\h \psi_-=\Lambda_0^-\tau ,
\h \psi_{\pm}=\psi\pm\tilde{\psi}.\eea It is analogous to
(\ref{A1}), but now the rotations are in the planes defined by the
angles $\psi_{\pm}=\psi\pm\tilde{\psi}$ instead of $\theta$ and
$\tilde{\theta}$.

The background felt by the membrane is as given in (\ref{b2}).
However, the metric induced on the membrane worldvolume is
different and it is the following \bea\nn
&&G_{00}=-l_{11}^{2}\left[(\Lambda_0^0)^2-\mathbf{\Lambda}_0^2
-(\Lambda_0^+)^2 \left(\frac{2l}{3}\right)^2 C^2 - (\Lambda_0^-)^2
D^2 \right],\\ \nn &&G_{11}=l_{11}^{2}M_{11},\h
G_{12}=l_{11}^{2}M_{12},\h G_{22}=l_{11}^{2}\left[M_{22} +
\frac{r'^2}{C^2}\right],\eea where $M_{ij}$ are defined in
(\ref{DM}). The constraints are identically satisfied, and the
constants of the motion $\mathcal{P}^2_\mu$ are given by
(\ref{cm1}). The membrane Lagrangian now takes the form \bea\nn
&&\mathcal{L}^{A}(\sigma) =\frac{1}{4\lambda^0}\left(K_{rr}r'^2 -
V\right),\h K_{rr}=-(2\lambda^0 T_2
l_{11}^2)^2\frac{M_{11}}{C^2},\\ \nn &&V=(2\lambda^0 T_2
l_{11}^2)^2 \det M_{ij} +
l_{11}^{2}\left[(\Lambda_0^0)^2-\mathbf{\Lambda}_0^2
-(\Lambda_0^+)^2 \left(\frac{2l}{3}\right)^2 C^2 - (\Lambda_0^-)^2
D^2 \right].\eea

Let us first consider the particular case when $\Lambda_0^-=0$,
i.e. $\psi=\tilde{\psi}$. The turning points now are \bea\nn
r_{min}=3l,\h
r_{max}=r_1=l\sqrt{1+\frac{8}{1-\frac{9u_0^2}{4l^2(\Lambda_0^+)^2}}}>3l,
\h
r_2=-l\sqrt{1+\frac{8}{1-\frac{9u_0^2}{4l^2(\Lambda_0^+)^2}}}<0,\eea
where $u_0^2$ is introduced in (\ref{u02}). Now one arrives at the
following membrane solution \bea\label{s4} &&\sigma(r)=
\frac{2\lambda^0 T_2 l_{11}}{\left[(\Lambda_0^+)^2
\left(\frac{2l}{3}\right)^2-u_0^2\right]^{1/2}}\left[\frac{2^7
l^3M_{11}\Delta r}{3\left(r_1-3l\right)
\left(3l-r_2\right)}\right]^{1/2}
\\ \nn &&\times F_D^{(5)}\left(1/2;-1,-1,1/2,1/2,1/2;3/2;-\frac{\Delta r}{2l},-\frac{\Delta r}{4l},-\frac{\Delta r}{6l},
-\frac{\Delta r}{3l-r_2},\frac{\Delta r}{r_1-3l}\right).\eea The
normalization condition gives \bea\label{nc4} &&\frac{\lambda^0
T_2 l_{11}}{\left[(\Lambda_0^+)^2
\left(\frac{2l}{3}\right)^2-u_0^2\right]^{1/2}}\left[\frac{2^7
l^3M_{11}}{3\left(3l-r_2\right)}\right]^{1/2}\times
\\ \nn &&F_D^{(4)}\left(1/2;-1,-1,1/2,1/2;1;-\frac{\Delta r_1}{2l},-\frac{\Delta r_1}{4l},-\frac{\Delta r_1}{6l},
-\frac{\Delta r_1}{3l-r_2}\right)=
\\ \nn &&\frac{\lambda^0 T_2 l_{11}}{\left[(\Lambda_0^+)^2 \left(\frac{2l}{3}\right)^2-u_0^2\right]^{1/2}}
\left[\frac{2^7 l^3M_{11}}{3\left(3l-r_2\right)}\right]^{1/2}\times
\\ \nn &&\left(1+\frac{\Delta r_1}{2l}\right)\left(1+\frac{\Delta r_1}{4l}\right)\left(1+\frac{\Delta r_1}{6l}\right)^{-1/2}
\left(1+\frac{\Delta r_1}{3l-r_2}\right)^{-1/2}\times
\\ \nn &&F_D^{(4)}\left(1/2;-1,-1,1/2,1/2,;1;\frac{1}{1+\frac{2l}{\Delta r_1}}, \frac{1}{1+\frac{4l}{\Delta r_1}},
\frac{1}{1+\frac{6l}{\Delta r_1}},\frac{1}{1+\frac{3l-r_2}{\Delta
r_1}}\right)=1.\eea We derive for the conserved momentum
$P_+\equiv P_{\psi_+}$ the following expression ($P_-\equiv
P_{\psi_-}=0$ as a consequence of $\Lambda_0^-=0$): \bea\nn &&P_+
=\frac{\pi^2 T_2 l_{11}^3\Lambda_0^+}{\left[(\Lambda_0^+)^2
\left(\frac{2l}{3}\right)^2-u_0^2\right]^{1/2}}\left[\frac{2^5 l^5
M_{11}}{3^3\left(3l-r_2\right)}\right]^{1/2}\times
\\ \nn &&\Delta r_1 F_D^{(2)}\left(3/2;-1/2,1/2;2;-\frac{\Delta r_1}{6l},-\frac{\Delta r_1}{3l-r_2}\right)=\\
\nn
&&\frac{\pi^2 T_2 l_{11}^3\Lambda_0^+}{\left[(\Lambda_0^+)^2
\left(\frac{2l}{3}\right)^2-u_0^2\right]^{1/2}}\left[\frac{2^5 l^5
M_{11}}{3^3\left(3l-r_2\right)}\right]^{1/2} \Delta r_1
\left(1+\frac{\Delta r_1}{6l}\right)^{1/2}\left(1+\frac{\Delta
r_1}{3l-r_2}\right)^{-1/2}\times
\\ \label{cmom4} &&F_D^{(2)}\left(1/2;-1/2,1/2;2;\frac{1}{1+\frac{6l}{\Delta r_1}},\frac{1}{1+\frac{3l-r_2}{\Delta r_1}}\right).\eea

In the semiclassical limit, (\ref{nc4}) and (\ref{cmom4}) simplify
to \bea\nn
\pi\Lambda_0^+\left[1-\frac{9u_0^2}{4l^2(\Lambda_0^+)^2}\right]=2^{3/2}3\lambda^0
T_2 l_{11}M_{11}^{1/2}, \h P_+ = \frac{2^{7/2}\pi T_2
l_{11}^3l^2M_{11}^{1/2}}{3\left[1-\frac{9u_0^2}{4l^2(\Lambda_0^+)^2}\right]}.\eea
Taking also into account (\ref{EP}), we obtain the following {\it
fourth} order algebraic equation for $E^2$ as a function of
$\mathbf{P}$ and $P_+$ \bea\nn
&&\left\{E^2\left[E^2-\mathbf{P}^2-(3/l)^2P_+^2\right] - (2\pi^2
T_2
l_{11}^3)^2\left\{\left(\mathbf{\Lambda}_1\times\mathbf{\Lambda}_2\right)^2
E^2 -
\left[\left(\mathbf{\Lambda}_1\times\mathbf{\Lambda}_2\right)\times\mathbf{P}\right]^2\right\}\right\}^2
\\ \label{Eg4} &&-2^7(3\pi T_2 l_{11}^3)^2E^2\left[\mathbf{\Lambda}_1^2 E^2
- \left(\mathbf{\Lambda}_1.\mathbf{P}\right)^2\right]P_{+}^2 =
0.\eea Let us consider a few simple cases. When $\Lambda_0^I=0$
and $\Lambda_2^I=c\Lambda_1^I$, (\ref{Eg4}) reduces to
\bea\label{p41} E^2=(3/l)^2P_+^2 + 2^{7/2}3\pi T_2
l_{11}^3\mid\mathbf{\Lambda}_1\mid P_{+},\eea or \bea\nn
E=\frac{3}{l}P_+\sqrt{1+ \frac{2^{7/2}\pi T_2
l_{11}^3l^2\mid\mathbf{\Lambda}_1\mid}{3P_+}}.\eea Expanding the
square root and neglecting the higher order terms, one derives
energy-charge relation of the type $E-K\sim const$. If we impose
only the conditions $\Lambda_0^I=0$, (\ref{Eg4}) gives
\bea\label{p42} E^2 =(2\pi^2 T_2
l_{11}^3)^2\left(\mathbf{\Lambda}_1\times\mathbf{\Lambda}_2\right)^2
+ (3/l)^2P_+^2+2^{7/2}3\pi T_2 l_{11}^3\mid\mathbf{\Lambda}_1\mid
P_{+}.\eea If we take $\Lambda_0^I\ne 0$,
$\Lambda_2^I=c\Lambda_1^I$, (\ref{Eg4}) simplifies to \bea\nn
E^2\left\{\left[E^2-\mathbf{P}^2-(3/l)^2P_+^2\right]^2 - 2^7(3\pi
T_2 l_{11}^3)^2 \mathbf{\Lambda}_1^2P_{+}^2\right\} +2^7(3\pi T_2
l_{11}^3)^2\left(\mathbf{\Lambda}_1.\mathbf{P}\right)^2P_{+}^2=0
,\eea which is {\it third} order algebraic equation for $E^2$.
Suppose that $\mathbf{\Lambda}_1$ and $\mathbf{P}$ are orthogonal
to each other, i.e.
$\left(\mathbf{\Lambda}_1.\mathbf{P}\right)=0$. Then, the above
relation becomes \bea\label{p43} E^2=\mathbf{P}^2 +
(3/l)^2P_+^2+2^{7/2}3\pi T_2 l_{11}^3\mid\mathbf{\Lambda}_1\mid
P_{+}.\eea

Finally, we give the semiclassical limit of the membrane solution
(\ref{s4}) \bea\nn &&\sigma_{scl}(r)=2\pi^2T_2
l_{11}^3\left(\frac{4l}{3}\right)^{3/2}\left[\mathbf{\Lambda}_1^2
-
\frac{1}{E^2}\left(\mathbf{\Lambda}_1.\mathbf{P}\right)^2\right]^{1/2}\frac{\Delta
r^{1/2}}{P_+}
\\ \nn &&\times F_D^{(3)}\left(1/2;-1,-1,1/2;3/2;-\frac{\Delta r}{2l},-\frac{\Delta r}{4l},
-\frac{\Delta r}{6l}\right)
\\ \nn &&=2\pi^2T_2 l_{11}^3\left(\frac{4l}{3}\right)^{3/2}\left[\mathbf{\Lambda}_1^2  - \frac{1}{E^2}
\left(\mathbf{\Lambda}_1.\mathbf{P}\right)^2\right]^{1/2}\frac{\Delta
r^{1/2}}{P_+} \left(1+\frac{\Delta
r}{2l}\right)\left(1+\frac{\Delta r}{4l}\right)
\left(1+\frac{\Delta r}{6l}\right)^{-1/2}\\ \nn &&\times
F_D^{(3)}\left(1;-1,-1,1/2;3/2;\frac{1}{1+\frac{2l}{\Delta r}},
\frac{1}{1+\frac{4l}{\Delta r}}, \frac{1}{1+\frac{6l}{\Delta
r}}\right).\eea

Now, we turn to the case $\Lambda_0^-\ne 0$, when the solutions of
the equation $r'=0$ are \bea\nn &&r_{min}=3l,\h
r_{max}=r_1=\frac{l}{\sqrt{2}}\sqrt{1+u^2-\Lambda^2}\sqrt{1+
\sqrt{1-\frac{4(u^2-9\Lambda^2)}{(1+u^2-\Lambda^2)^2}}},
\\ \nn &&r_2=\frac{l}{\sqrt{2}}\sqrt{1+u^2-\Lambda^2}\sqrt{1- \sqrt{1-\frac{4(u^2-9\Lambda^2)}{(1+u^2-\Lambda^2)^2}}},
\\ \nn &&r_3=-\frac{l}{\sqrt{2}}\sqrt{1+u^2-\Lambda^2}\sqrt{1+ \sqrt{1-\frac{4(u^2-9\Lambda^2)}{(1+u^2-\Lambda^2)^2}}},
\\ \nn &&r_4=-\frac{l}{\sqrt{2}}\sqrt{1+u^2-\Lambda^2}\sqrt{1- \sqrt{1-\frac{4(u^2-9\Lambda^2)}{(1+u^2-\Lambda^2)^2}}},
\\ \nn &&u^2=\left(\frac{3u_0}{l\Lambda_0^-}\right)^2,\h \Lambda^2=\left(2\frac{\Lambda_0^+}{\Lambda_0^-}\right)^2.\eea
Correspondingly, we obtain the following solution for $\sigma(r)$:
\bea\nn &&\sigma(r)= \frac{\lambda^0 T_2
l_{11}}{\Lambda_0^-}\left[\frac{2^9 3 l^3M_{11}\Delta r}
{\left(r_1-3l\right) \left(3l-r_2\right) \left(3l-r_3\right)
\left(3l-r_4\right)}\right]^{1/2}\times
\\ \label{s4g} &&F_D^{(7)}\left(1/2;-1,-1,1/2,1/2,1/2,1/2,1/2;3/2;\right.
\\ \nn &&-\left.\frac{\Delta r}{2l},-\frac{\Delta r}{4l},-\frac{\Delta r}{6l},-\frac{\Delta r}{3l-r_2},
-\frac{\Delta r}{3l-r_3},-\frac{\Delta r}{3l-r_4}, \frac{\Delta
r}{r_1-3l}\right).\eea For the normalization condition, we derive
the result \bea\label{nc4g} &&\frac{\lambda^0 T_2
l_{11}}{\Lambda_0^-}\left[\frac{2^7 3 l^3M_{11}}
{\left(3l-r_2\right) \left(3l-r_3\right)
\left(3l-r_4\right)}\right]^{1/2}\times
\\ \nn &&F_D^{(6)}\left(1/2;-1,-1,1/2,1/2,1/2,1/2;1;\right.
\\ \nn &&-\left.\frac{\Delta r_1}{2l},-\frac{\Delta r_1}{4l},-\frac{\Delta r_1}{6l},-\frac{\Delta r_1}{3l-r_2},
-\frac{\Delta r_1}{3l-r_3},-\frac{\Delta r_1}{3l-r_4}\right)=
\\ \nn &&\frac{\lambda^0 T_2 l_{11}}{\Lambda_0^-}\left[\frac{2^7 3 l^3M_{11}}
{\left(3l-r_2\right) \left(3l-r_3\right)
\left(3l-r_4\right)}\right]^{1/2} \left(1+ \frac{\Delta
r}{2l}\right)\left(1+ \frac{\Delta r}{4l}\right)\left(1+
\frac{\Delta r}{6l}\right)^{-1/2}\times
\\ \nn &&\left(1+ \frac{\Delta r}{3l-r_2}\right)^{-1/2} \left(1+ \frac{\Delta r}{3l-r_3}\right)^{-1/2}
\left(1+ \frac{\Delta r}{3l-r_4}\right)^{-1/2}\times
\\ \nn &&F_D^{(6)}\left(1/2;-1,-1,1/2,1/2,1/2,1/2;1;\right.
\\ \nn &&\left.\frac{1}{1+\frac{2l}{\Delta r_1}}, \frac{1}{1+\frac{4l}{\Delta r_1}}, \frac{1}{1+\frac{6l}{\Delta r_1}},
\frac{1}{1+\frac{3l-r_2}{\Delta r_1}}
,\frac{1}{1+\frac{3l-r_3}{\Delta r_1}}
,\frac{1}{1+\frac{3l-r_4}{\Delta r_1}}\right)=1.\eea The
computation of the conserved quantities $P_+$ and $P_-$ gives
\bea\label{cmom4g+} &&P_+=\pi^2T_2
l_{11}^3\frac{\Lambda_0^+}{\Lambda_0^-}\left[\frac{2^5 l^5 M_{11}}
{3\left(3l-r_2\right) \left(3l-r_3\right)
\left(3l-r_4\right)}\right]^{1/2}\Delta r_1\times
\\ \nn &&F_D^{(4)}\left(3/2;-1/2,1/2,1/2,1/2;2;-\frac{\Delta r_1}{6l},-\frac{\Delta r_1}{3l-r_2},
-\frac{\Delta r_1}{3l-r_3},-\frac{\Delta r_1}{3l-r_4}\right)=
\\ \nn &&\pi^2T_2 l_{11}^3\frac{\Lambda_0^+}{\Lambda_0^-}\left[\frac{2^5 l^5 M_{11}}
{3\left(3l-r_2\right) \left(3l-r_3\right)
\left(3l-r_4\right)}\right]^{1/2}\times
\\ \nn &&\Delta r_1\left(1+ \frac{\Delta r_1}{6l}\right)^{1/2}\left(1+ \frac{\Delta r_1}{3l-r_2}\right)^{-1/2}
 \left(1+ \frac{\Delta r_1}{3l-r_3}\right)^{-1/2} \left(1+ \frac{\Delta r_1}{3l-r_4}\right)^{-1/2}\times
\\ \nn &&F_D^{(4)}\left(1/2;-1/2,1/2,1/2,1/2;2; \frac{1}{1+\frac{6l}{\Delta r_1}},\frac{1}{1+\frac{3l-r_2}{\Delta r_1}} ,
\frac{1}{1+\frac{3l-r_3}{\Delta r_1}}
,\frac{1}{1+\frac{3l-r_4}{\Delta r_1}}\right),\eea

\bea\label{cmom4g-} &&P_-=\pi^2T_2 l_{11}^3\left[\frac{2^7 3 l^7
M_{11}} {\left(3l-r_2\right) \left(3l-r_3\right)
\left(3l-r_4\right)}\right]^{1/2}\times
\\ \nn &&F_D^{(7)}\left(1/2;-1,-2,-1,1/2,1/2,1/2,1/2;1;\right.
\\ \nn &&-\left.\frac{\Delta r_1}{2l},-\frac{\Delta r_1}{3l},-\frac{\Delta r_1}{4l},-\frac{\Delta r_1}{6l},-\frac{\Delta r_1}{3l-r_2},
-\frac{\Delta r_1}{3l-r_3},-\frac{\Delta r_1}{3l-r_4}\right)=
\\ \nn &&\pi^2T_2 l_{11}^3\left[\frac{2^7 3 l^7 M_{11}}
{\left(3l-r_2\right) \left(3l-r_3\right)
\left(3l-r_4\right)}\right]^{1/2}\times
\\ \nn &&\left(1+ \frac{\Delta r_1}{2l}\right)\left(1+ \frac{\Delta r_1}{3l}\right)^2\left(1+ \frac{\Delta r_1}{4l}\right)
\left(1+ \frac{\Delta r_1}{6l}\right)^{-1/2}\times
\\ \nn &&\left(1+ \frac{\Delta r_1}{3l-r_2}\right)^{-1/2} \left(1+ \frac{\Delta r_1}{3l-r_3}\right)^{-1/2}
\left(1+ \frac{\Delta r_1}{3l-r_4}\right)^{-1/2}\times
\\ \nn &&F_D^{(7)}\left(1/2;-1,-2,-1,1/2,1/2,1/2,1/2;1;\right.
\\ \nn &&\left.\frac{1}{1+\frac{2l}{\Delta r_1}},\frac{1}{1+\frac{3l}{\Delta r_1}},\frac{1}{1+\frac{4l}{\Delta r_1}},
\frac{1}{1+\frac{6l}{\Delta r_1}},\frac{1}{1+\frac{3l-r_2}{\Delta
r_1}} ,\frac{1}{1+\frac{3l-r_3}{\Delta r_1}},
\frac{1}{1+\frac{3l-r_4}{\Delta r_1}}\right).\eea

Let us now take the semiclassical limit $r_1\to\infty$.  In this
limit, (\ref{nc4g}), (\ref{cmom4g+}) and (\ref{cmom4g-}) reduce
correspondingly to \bea\nn &&\Lambda_0^- =
3\lambda^0T_2l_{11}M_{11}^{1/2},\h P_+ =
\frac{4}{3}\pi^2T_2l_{11}^3l^2M_{11}^{1/2}\frac{\Lambda_0^+}{\Lambda_0^-},\\
\nn &&P_- = \frac{1}{6}\pi^2T_2l_{11}^3l^2M_{11}^{1/2}
\left[\left(\frac{3u_0}{l\Lambda_0^-}\right)^2-
\left(2\frac{\Lambda_0^+}{\Lambda_0^-}\right)^2\right].\eea These
equalities, together with (\ref{EP}), lead to the following
relation between the energy $E$ and the conserved charges
$\mathbf{P}$, $P_+$ and $P_-$: \bea\nn
&&\left\{E^2\left[E^2-\mathbf{P}^2-(3/2l)^2 P_+^2\right] - (2\pi^2
T_2
l_{11}^3)^2\left\{\left(\mathbf{\Lambda}_1\times\mathbf{\Lambda}_2\right)^2
E^2 -
\left[\left(\mathbf{\Lambda}_1\times\mathbf{\Lambda}_2\right)\times\mathbf{P}\right]^2\right\}\right\}^2
\\ \label{Egg4} &&-(6\pi^2 T_2 l_{11}^3)^2E^2\left[\mathbf{\Lambda}_1^2 E^2
- \left(\mathbf{\Lambda}_1.\mathbf{P}\right)^2\right]P_{-}^2 =
0.\eea Let us point out that the above relation is only valid for
$P_-\ne 0$, whereas we can always set $\mathbf{P}$ or $P_+$ equal
to zero. Below, we give a few simple solutions of (\ref{Egg4}).

Choosing $\Lambda_0^I=0$ and $\Lambda_2^I=c\Lambda_1^I$, one
obtains \bea\label{p41g} E^2=(3/2l)^2P_+^2 + 6\pi^2 T_2
l_{11}^3\mid\mathbf{\Lambda}_1\mid P_{-},\eea which can be
rewritten as \bea\nn E=\frac{3}{2l}P_+\sqrt{1+ \frac{8\pi^2 T_2
l_{11}^3l^2\mid\mathbf{\Lambda}_1\mid P_{-}}{3P_+^2}}.\eea
Expanding the square root and neglecting the higher order terms,
one arrives at \bea\nn E= \frac{3}{2l}P_+ + 2\pi^2 T_2
l_{11}^3l\mid\mathbf{\Lambda}_1\mid \frac{P_{-}}{P_{+}}.\eea If
only the conditions $\Lambda_0^I=0$ are imposed, (\ref{Egg4})
gives \bea\label{p42g} E^2 =(2\pi^2 T_2
l_{11}^3)^2\left(\mathbf{\Lambda}_1\times\mathbf{\Lambda}_2\right)^2
+ (3/2l)^2P_+^2 + 6\pi^2 T_2 l_{11}^3\mid\mathbf{\Lambda}_1\mid
P_{-}.\eea If we choose $\Lambda_0^I\ne 0$,
$\Lambda_2^I=c\Lambda_1^I$, then (\ref{Egg4}) simplifies to a {\it
third} order algebraic equation for $E^2$ \bea\nn
E^2\left\{\left[E^2-\mathbf{P}^2-(3/2l)^2P_+^2\right]^2 - (6\pi^2
T_2 l_{11}^3)^2\mathbf{\Lambda}_1^2P_{-}^2\right\} +(6\pi^2 T_2
l_{11}^3)^2\left(\mathbf{\Lambda}_1.\mathbf{P}\right)^2P_{-}^2=0
.\eea If $\left(\mathbf{\Lambda}_1.\mathbf{P}\right)=0$, the above
relation reduces to \bea\label{p43g} E^2=\mathbf{P}^2 +
(3/2l)^2P_+^2+6\pi^2 T_2 l_{11}^3\mid\mathbf{\Lambda}_1\mid
P_{-}.\eea

Finally, let us write down the semiclassical limit of the membrane
solution (\ref{s4g}): \bea\nn &&\sigma_{scl}(r)=
\left(\frac{2^8\pi^2T_2
l_{11}^3l}{3^4P_-}\right)^{1/2}\left[\mathbf{\Lambda}_1^2  -
\frac{1}{E^2}\left(\mathbf{\Lambda}_1.\mathbf{P}\right)^2\right]^{1/4}\Delta
r^{1/2}
\\ \nn &&\times F_D^{(4)}\left(1/2;-1,1,-1,1/2;3/2;-\frac{\Delta r}{2l},-\frac{\Delta r}{3l},
-\frac{\Delta r}{4l},-\frac{\Delta r}{6l}\right)
\\ \nn &&=\left(\frac{2^8\pi^2T_2 l_{11}^3l}{3^4P_-}\right)^{1/2}\left[\mathbf{\Lambda}_1^2  - \frac{1}{E^2}\left(\mathbf{\Lambda}_1.
\mathbf{P}\right)^2\right]^{1/4}
\\ \nn &&\times \Delta r^{1/2}
\left(1+\frac{\Delta r}{2l}\right)\left(1+\frac{\Delta
r}{3l}\right)^{-1}\left(1+\frac{\Delta r}{4l}\right)
\left(1+\frac{\Delta r}{6l}\right)^{-1/2}\\ \nn &&\times
F_D^{(4)}\left(1;-1,1,-1,1/2;3/2;\frac{1}{1+\frac{2l}{\Delta r}},
\frac{1}{1+\frac{3l}{\Delta r}}, \frac{1}{1+\frac{4l}{\Delta r}},
\frac{1}{1+\frac{6l}{\Delta r}}\right).\eea

More rotating membrane solutions in this eleven dimensional
supergravity background can be found in Appendix B of \cite{9}.

{\bf Some mathematical results}

It is known that the Lauricella hypergeometric functions of $n$
variables $F_D^{(n)}$ \\ are defined as \cite{PBM-III}
\bea\label{Lhgfsd} &&F_D^{(n)}(a;b_1,\ldots,b_n;c;z_1,\ldots,z_n)=
\\ \nn
&&\sum_{k_1,\ldots,k_n=0}^{\infty}\frac{(a)_{k_1+\ldots+k_n}(b_1)_{k_1}\ldots(b_n)_{k_n}}
{(c)_{k_1+\ldots+k_n}}\frac{z_1^{k_1}\ldots z_n^{k_n}}{k_1!\ldots
k_n!},\h |z_j|<1,\h (a)_k = \frac{\Gamma(a+k)}{\Gamma(a)},\eea and
have the following integral representation \cite{PBM-III}
\bea\label{ir} &&F_D^{(n)}(a,b_1,\ldots ,b_n;c;z_1,\ldots ,z_n)=
\\ \nn &&
\frac{\Gamma(c)}{\Gamma (a)\Gamma(c-a)} \int_{0}^{1}
x^{a-1}(1-x)^{c-a-1}(1-z_1 x)^{-b_1}\ldots (1-z_n x)^{-b_n}dx,
\\ \nn && Re(a)>0,\h Re(c-a)>0.\eea

However, in order to perform our calculations, we need to know
more about the properties of these functions. That is why, we
proved in \cite{9} that the following equalities hold \bea\nn {\bf
1.}
&&F_D^{(n)}(a;b_1,\ldots,b_i,\ldots,b_j,\ldots,b_n;c;z_1,\ldots,z_i,\ldots,z_j,\ldots,z_n)=
\\ \nn &&F_D^{(n)}(a;b_1,\ldots,b_j,\ldots,b_i,\ldots,b_n;c;z_1,\ldots,z_j,\ldots,z_i,\ldots,z_n).
\\ \nn
{\bf 2.} &&F_D^{(n)}(a;b_1,\ldots,b_n;c;z_1,\ldots,z_n)=
\\ \nn &&\prod_{i=1}^{n}\left(1-z_i\right)^{-b_i}
F_D^{(n)}\left(c-a;b_1,\ldots,b_n;c;\frac{z_1}{z_1-1},\ldots,\frac{z_n}{z_n-1}\right).
\\ \nn
{\bf 3.}
&&F_D^{(n)}(a;b_1,\ldots,b_{i-1},b_i,b_{i+1},\ldots,b_n;c;z_1,\ldots,z_{i-1},1,z_{i+1},\ldots,z_n)=
\\ \nn &&\frac{\Gamma(c)\Gamma(c-a-b_i)}{\Gamma(c-a)\Gamma(c-b_i)}
F_D^{(n-1)}(a;b_1,\ldots,b_{i-1},b_{i+1},\ldots,b_n;c-b_i;z_1,\ldots,z_{i-1},z_{i+1},\ldots,z_n).
\\ \nn
{\bf 4.}
&&F_D^{(n)}(a;b_1,\ldots,b_{i-1},b_i,b_{i+1},\ldots,b_n;c;z_1,\ldots,z_{i-1},0,z_{i+1},\ldots,z_n)=
\\ \nn &&F_D^{(n-1)}(a;b_1,\ldots,b_{i-1},b_{i+1},\ldots,b_n;c;z_1,\ldots,z_{i-1},z_{i+1},\ldots,z_n).
\\ \nn
{\bf 5.}
&&F_D^{(n)}(a;b_1,\ldots,b_{i-1},0,b_{i+1},\ldots,b_n;c;z_1,\ldots,z_{i-1},z_i,z_{i+1},\ldots,z_n)=
\\ \nn &&F_D^{(n-1)}(a;b_1,\ldots,b_{i-1},b_{i+1},\ldots,b_n;c;z_1,\ldots,z_{i-1},z_{i+1},\ldots,z_n).
\\ \nn
{\bf 6.}
&&F_D^{(n)}(a;b_1,\ldots,b_i,\ldots,b_j,\ldots,b_n;c;z_1,\ldots,z_i,\ldots,z_i,\ldots,z_n)=
\\ \nn &&F_D^{(n-1)}(a;b_1,\ldots,b_i+b_j,\ldots,b_n;c;z_1,\ldots,z_i,\ldots,z_n).
\\ \nn
{\bf 7.}
&&F_D^{(2n+1)}(a;a-c+1,b_2,b_2,\ldots,b_{2n},b_{2n};c;-1,z_2,-z_2\ldots,z_{2n},-z_{2n})=
\\ \nn &&\frac{\Gamma(a/2)\Gamma(c)}{2\Gamma(a)\Gamma(c-a/2)}
F_D^{(n)}(a/2;b_2,\ldots,b_{2n};c-a/2;z_2^2,\ldots,z_{2n}^2).
\\ \nn
{\bf 8.}
&&F_D^{(2n+1)}\left(c-a;a-c+1,b_2,b_2,\ldots,b_{2n},b_{2n};c;\right.
\\ \nn
&&\left.1/2,-\frac{z_2}{1-z_2},\frac{z_2}{1+z_2},\ldots,-\frac{z_{2n}}{1-z_{2n}},\frac{z_{2n}}{1+z_{2n}}\right)=
\\ \nn &&\frac{\Gamma(a/2)\Gamma(c)}{2^{c-a}\Gamma(a)\Gamma(c-a/2)}
F_D^{(n)}\left(c-a;b_2,\ldots,b_{2n};c-a/2;-\frac{z_2^2}{1-z_2^2},\ldots,-\frac{z_{2n}^2}{1-z_{2n}^2}\right).
\eea

\subsubsection{Rotating D2-branes in type IIA reduction of
M-theory\\ on $G_2$ manifold and their semiclassical limits}
Here, we will consider D2-branes rotating in the background
(\ref{10db}).

We begin with the following D2-brane embedding in the target
space: \bea\label{DA1} &&X^0=
\Lambda_0^0\xi^0+\frac{\left(\mathbf{\Lambda}_0.\mathbf{\Lambda}_1\right)}{\Lambda_0^0}\left(\xi^1
+ c\xi^2\right), \h X^I=\Lambda_0^I\xi^0 + \Lambda_1^I\left(\xi^1
+ c\xi^2\right),
\\ \nn &&r=r(\xi^2),\h \theta_1=\Lambda_0^{\theta_1}\xi^0,\h \theta_2=\Lambda_0^{\theta_2}\xi^0;
\h
\left(\mathbf{\Lambda}_0.\mathbf{\Lambda}_1\right)=\delta_{IJ}\Lambda_0^I\Lambda_1^J,
\h c=constant.\eea It corresponds to D2-brane extended in the
radial direction $r$, and rotating in the planes given by the
angles $\theta_1$ and $\theta_2$ with constant angular momenta
$P_{\theta_1}$ and $P_{\theta_2}$. It is nontrivially spanned
along $x^0$ and $x^I$ and moves with constant energy $E$, and
constant momenta $P_{I}$.

The expression for the D2-brane solution found in \cite{10} for
this case is given by \bea\label{D2s1} &&\xi^2(r)=
\frac{16}{3}\lambda^0 T_{D2}\left[\frac{M l}
{\left(\Lambda_+^2+\Lambda_-^2\right)\left(3l-r_2\right)\Delta
r_1}\right]^{1/2}(2\Delta r)^{3/4}\times
\\ \nn &&F_D^{(5)}\left(3/4;-1/4,-1/4,1/4,1/2,1/2;7/4;
-\frac{\Delta r}{2l},-\frac{\Delta r}{4l},-\frac{\Delta r}{6l},
-\frac{\Delta r}{3l-r_2},\frac{\Delta r}{\Delta r_1}\right).\eea

Now, we compute the conserved momenta on the obtained solution:
\bea\label{DEP1} &&\frac{E}{\Lambda_0^0}=\frac{P_I}{\Lambda_0^I}=
8\pi^2T_{D2}\left[\frac{Ml}{\left(\Lambda_+^2+
\Lambda_-^2\right)(3l-r_2)}\right]^{1/2} \times
\\ \nn &&\left(1+\frac{\Delta
r_1}{2l}\right)^{1/2}\left(1+\frac{\Delta
r_1}{4l}\right)^{1/2}\left(1+\frac{\Delta
r_1}{6l}\right)^{-1/2}\left(1+\frac{\Delta
r_1}{3l-r_2}\right)^{-1/2}\times
\\ \nn && F_D^{(4)}\left(1/2;-1/2,-1/2,1/2,1/2;1;\frac{1}{1+\frac{2l}{\Delta
r_1}},\frac{1}{1+\frac{4l}{\Delta
r_1}},\frac{1}{1+\frac{6l}{\Delta
r_1}},\frac{1}{1+\frac{3l-r_2}{\Delta r_1}}\right),\eea
\bea\label{Dpt12}
&&P_{\theta_1}=\left(\Lambda_0^{\theta_1}-\Lambda_0^{\theta_2}\cos\psi_1^0\right)I_{A1}^D
+\left(\Lambda_0^{\theta_1}+\Lambda_0^{\theta_2}\cos\psi_1^0\right)I_{B1}^D,\\
\nn
&&P_{\theta_2}=\left(\Lambda_0^{\theta_2}-\Lambda_0^{\theta_1}\cos\psi_1^0\right)I_{A1}^D
+\left(\Lambda_0^{\theta_2}+\Lambda_0^{\theta_1}\cos\psi_1^0\right)I_{B1}^D,\eea
where \bea\label{IDA1} I_{A1}^D&=&
8\pi^2T_{D2}\left[\frac{Ml^5}{\left(\Lambda_+^2+
\Lambda_-^2\right)(3l-r_2)}\right]^{1/2} \times
\\ \nn &&\left(1+\frac{\Delta
r_1}{2l}\right)^{3/2}\left(1+\frac{\Delta
r_1}{4l}\right)^{1/2}\left(1+\frac{\Delta
r_1}{6l}\right)^{1/2}\left(1+\frac{\Delta
r_1}{3l-r_2}\right)^{-1/2}\times
\\ \nn && F_D^{(4)}\left(1/2;-3/2,-1/2,-1/2,1/2;1;\frac{1}{1+\frac{2l}{\Delta
r_1}},\frac{1}{1+\frac{4l}{\Delta
r_1}},\frac{1}{1+\frac{6l}{\Delta
r_1}},\frac{1}{1+\frac{3l-r_2}{\Delta r_1}}\right),\eea
\bea\label{IDB1} I_{B1}^D&=&
\frac{4}{3}\pi^2T_{D2}\left[\frac{Ml^3}{\left(\Lambda_+^2+
\Lambda_-^2\right)(3l-r_2)}\right]^{1/2} \times
\\ \nn &&\Delta r_1\left(1+\frac{\Delta
r_1}{2l}\right)^{1/2}\left(1+\frac{\Delta
r_1}{4l}\right)^{3/2}\left(1+\frac{\Delta
r_1}{6l}\right)^{-1/2}\left(1+\frac{\Delta
r_1}{3l-r_2}\right)^{-1/2}\times
\\ \nn && F_D^{(4)}\left(1/2;-1/2,-3/2,1/2,1/2;2;\frac{1}{1+\frac{2l}{\Delta
r_1}},\frac{1}{1+\frac{4l}{\Delta
r_1}},\frac{1}{1+\frac{6l}{\Delta
r_1}},\frac{1}{1+\frac{3l-r_2}{\Delta r_1}}\right).\eea

In the semiclassical limit, (\ref{DEP1}) - (\ref{IDB1}) simplify
to \bea\nn &&\frac{E}{\Lambda_0^0}=\frac{P_I}{\Lambda_0^I}=
\frac{2}{3}\pi^2T_{D2}\left(\frac{M}{\Lambda_+^2+\Lambda_-^2}\right)^{1/2},
\\ \nn  &&P_{\theta_1}=2\Lambda_0^{\theta_1}I_{A1}^D,\h
P_{\theta_2}=2\Lambda_0^{\theta_2}I_{A1}^D,\h
I_{A1}^D=I_{B1}^D=\frac{\sqrt{3}\pi^2T_{D2}M^{1/2}
}{\left(\Lambda_+^2+\Lambda_-^2\right)^{3/2}}v_0^2.\eea From here,
one obtains the following relation between the energy and the
conserved charges \bea\label{ED1}
E^2\left(E^2-\mathbf{P}^2\right)^2 -\frac{2^3}{3^5}(\pi^2
T_{D2})^2\left[\mathbf{\Lambda}_1^2 E^2 -
\left(\mathbf{\Lambda}_1.\mathbf{P}\right)^2\right]
\left(P^2_{\theta_1}+P^2_{\theta_2}\right) = 0,\eea which is third
order algebraic equation for $E^2$.

For $\left(\mathbf{\Lambda}_1.\mathbf{P}\right)=0$, (\ref{ED1})
reduces to \bea\nn E^2=\mathbf{P}^2+\frac{2^{3/2}}{3^{5/2}}\pi^2
T_{D2}\mid\mathbf{\Lambda}_1\mid
\left(P^2_{\theta_1}+P^2_{\theta_2}\right)^{1/2}.\eea This is the
same type energy-charge relation as the one obtained for the
string in (\ref{EG1}).

Let us now consider the other possible D2-brane embedding for the
same background metric. It is given by \bea\label{DA2}
&&X^0= \Lambda_0^0\xi^0, \h X^I=\Lambda_0^I\xi^0, \h r=r(\xi^2),\\
\nn
&&\theta_1=\Lambda_0^{\theta}\xi^0+\Lambda_1^{\theta}\xi^1+\Lambda_2^{\theta}\xi^2
,\h
\theta_2=\Lambda_0^{\theta}\xi^0-\Lambda_1^{\theta}\xi^1-\Lambda_2^{\theta}\xi^2
.\eea This ansatz describes D2-brane, which is extended along the
radial direction $r$ and rotates in the planes defined by the
angles $\theta_1$ and $\theta_2$, with equal angular momenta
$P_{\theta_1}=P_{\theta_2}=P_{\theta}$. Now we have nontrivial
wrapping along $\theta_1$ and $\theta_2$. In addition, the
D2-brane moves along $x^0$ and $x^I$ with constant energy $E$ and
constant momenta $P_{I}$ respectively.

Now, one finds the following D2-brane solution \cite{10}: \bea\nn
&&\xi^2(r)=\frac{8}{3}\lambda^0
T_{D2}\left[\frac{l\left(\Lambda_{1+}^2+\Lambda_{1-}^2\right)(3l-v_+)(3l-v_-)}
{3\left(\check{\Lambda}_+^2+\check{\Lambda}_-^2\right)\left(3l-r_2\right)\Delta
r_1}\right]^{1/2}(2\Delta r)^{3/4}\times
\\ \label{D2s2} &&F_D^{(7)}\left(3/4;-1/4,-1/4,1/4,-1/2,-1/2,1/2,1/2;7/4;\right.\\
\nn &&\left.-\frac{\Delta r}{2l},-\frac{\Delta
r}{4l},-\frac{\Delta r}{6l}, -\frac{\Delta
r}{3l-v_+},-\frac{\Delta r}{3l-v_-}, -\frac{\Delta
r}{3l-r_2},\frac{\Delta r}{\Delta r_1}\right),\eea where $v_{\pm}$
are the zeros of the polynomial \bea\nn
t^2-2l\frac{\Lambda_{1+}^2-\Lambda_{1-}^2}{\Lambda_{1+}^2+\Lambda_{1-}^2}t-3l^2=(t-v_+)(t-v_-).\eea

In the case under consideration, the conserved quantities are $E$,
${P_I}$ and $P_{\theta}$. We derive the following result for them
\cite{10} \bea\label{DEP2}
&&\frac{E}{\Lambda_0^0}=\frac{P_I}{\Lambda_0^I}=
4\pi^2T_{D2}\left[\frac{l\left(\Lambda_{1+}^2+\Lambda_{1-}^2\right)
(3l-v_+)(3l-v_-)}{3\left(\check{\Lambda}_{+}^2+\check{\Lambda}_{-}^2\right)
(3l-r_2)}\right]^{1/2}\times
\\ \nn &&\left(1+\frac{\Delta
r_1}{2l}\right)^{1/2}\left(1+\frac{\Delta
r_1}{4l}\right)^{1/2}\left(1+\frac{\Delta r_1}{6l}\right)^{-1/2}\times \\
\nn &&\left(1+\frac{\Delta r_1}{3l-v_+}\right)^{1/2}
\left(1+\frac{\Delta r_1}{3l-v_-}\right)^{1/2}\left(1+\frac{\Delta
r_1}{3l-r_2}\right)^{-1/2}\times
\\ \nn && F_D^{(6)}\left(1/2;-1/2,-1/2,1/2,-1/2,-1/2,1/2;1;\right.
\\ \nn &&\left.\frac{1}{1+\frac{2l}{\Delta
r_1}},\frac{1}{1+\frac{4l}{\Delta
r_1}},\frac{1}{1+\frac{6l}{\Delta
r_1}},\frac{1}{1+\frac{3l-v+}{\Delta
r_1}},\frac{1}{1+\frac{3l-v-}{\Delta
r_1}}\frac{1}{1+\frac{3l-r_2}{\Delta r_1}}\right),\eea \bea\nn
P_{\theta}=
\Lambda_0^{\theta}\left[\left(1-\cos\psi_1^0\right)I_{A2}^D +
\left(1+\cos\psi_1^0\right)I_{B2}^D\right],\eea where \bea\nn
&&I_{A2}^D=
4\pi^2T_{D2}\left[\frac{l^5\left(\Lambda_{1+}^2+\Lambda_{1-}^2\right)
(3l-v_+)(3l-v_-)}{3\left(\check{\Lambda}_{+}^2+\check{\Lambda}_{-}^2\right)
(3l-r_2)}\right]^{1/2}\times
\\ \nn &&\left(1+\frac{\Delta
r_1}{2l}\right)^{3/2}\left(1+\frac{\Delta
r_1}{4l}\right)^{1/2}\left(1+\frac{\Delta r_1}{6l}\right)^{1/2}\times \\
\nn &&\left(1+\frac{\Delta r_1}{3l-v_+}\right)^{1/2}
\left(1+\frac{\Delta r_1}{3l-v_-}\right)^{1/2}\left(1+\frac{\Delta
r_1}{3l-r_2}\right)^{-1/2}\times
\\ \nn && F_D^{(6)}\left(1/2;-3/2,-1/2,-1/2,-1/2,-1/2,1/2;1;\right.
\\ \nn &&\left.\frac{1}{1+\frac{2l}{\Delta
r_1}},\frac{1}{1+\frac{4l}{\Delta
r_1}},\frac{1}{1+\frac{6l}{\Delta
r_1}},\frac{1}{1+\frac{3l-v+}{\Delta
r_1}},\frac{1}{1+\frac{3l-v-}{\Delta
r_1}}\frac{1}{1+\frac{3l-r_2}{\Delta r_1}}\right),\eea \bea\nn
&&I_{B2}^D=
2\pi^2T_{D2}\left[\frac{l^3\left(\Lambda_{1+}^2+\Lambda_{1-}^2\right)
(3l-v_+)(3l-v_-)}{3^3\left(\check{\Lambda}_{+}^2+\check{\Lambda}_{-}^2\right)
(3l-r_2)}\right]^{1/2}\times
\\ \nn &&\Delta r_1\left(1+\frac{\Delta
r_1}{2l}\right)^{1/2}\left(1+\frac{\Delta
r_1}{4l}\right)^{3/2}\left(1+\frac{\Delta r_1}{6l}\right)^{-1/2}\times \\
\nn &&\left(1+\frac{\Delta r_1}{3l-v_+}\right)^{1/2}
\left(1+\frac{\Delta r_1}{3l-v_-}\right)^{1/2}\left(1+\frac{\Delta
r_1}{3l-r_2}\right)^{-1/2}\times
\\ \nn && F_D^{(6)}\left(1/2;-1/2,-3/2,1/2,-1/2,-1/2,1/2;2;\right.
\\ \nn &&\left.\frac{1}{1+\frac{2l}{\Delta
r_1}},\frac{1}{1+\frac{4l}{\Delta
r_1}},\frac{1}{1+\frac{6l}{\Delta
r_1}},\frac{1}{1+\frac{3l-v+}{\Delta
r_1}},\frac{1}{1+\frac{3l-v-}{\Delta
r_1}}\frac{1}{1+\frac{3l-r_2}{\Delta r_1}}\right).\eea

Taking the semiclassical limit in the above
expressions\footnote{In this limit $v_{\pm}$ remain finite.}, we
obtain the following dependence of the energy on $P_I$ and
$P_{\theta}$: \bea\label{ED2} E^2=\mathbf{P}^2+3^{5/3}(2\pi T_{D2}
\Lambda_1^{\theta})^{2/3}P_{\theta}^{4/3}.\eea

Now, we turn to the case of D2-brane embedded in the following way
\bea\label{DA3} &&X^0=
\Lambda_0^0\xi^0+\frac{\left(\mathbf{\Lambda}_0.\mathbf{\Lambda}_1\right)}{\Lambda_0^0}\left(\xi^1
+ c\xi^2\right), \h X^I=\Lambda_0^I\xi^0 + \Lambda_1^I\left(\xi^1
+ c\xi^2\right),
\\ \nn &&r=r(\xi^2),\h \theta_1=\Lambda_0^{\theta}\xi^0,\h
\phi_2=\Lambda_0^{\phi}\xi^0.\eea (\ref{DA3}) is analogous to
(\ref{DA1}), but now the rotations are in the planes defined by
the angles $\theta_1$ and $\phi_2$ instead of $\theta_1$ and
$\theta_2$.

The solution $\xi^2(r)$ can be obtained from (\ref{D2s1}) by the
replacement \bea\label{R1}\Lambda_+^2+\Lambda_-^2\to
\bar{\Lambda}_+^2+\bar{\Lambda}_-^2+4\Lambda_D^2/3.\eea The
explicit expressions for $E$ and $P_I$ can be obtained in the same
way from (\ref{DEP1}). The computation of the conserved angular
momenta $P_{\theta}$ and $P_{\phi}$ gives \bea\nn
P_{\theta}&=&\left(\Lambda_0^{\theta}-\Lambda_0^{\phi}\sin\psi_1^0\sin\theta_2^0\right)J_{A1}^D
+\left(\Lambda_0^{\theta}+\Lambda_0^{\phi}\sin\psi_1^0\sin\theta_2^0\right)J_{B1}^D,
\\ \nn
P_{\phi}&=&\left(\Lambda_0^{\phi}\sin\theta_2^0-\Lambda_0^{\theta}\sin\psi_1^0\right)\sin\theta_2^0J_{A1}^D
+\left(\Lambda_0^{\phi}\sin\theta_2^0+\Lambda_0^{\theta}\sin\psi_1^0\right)\sin\theta_2^0J_{B1}^D
\\ \nn &&+\Lambda_0^{\phi}\cos^2\theta_2^0 J_{D1}^D, \eea where one
obtains $J_{A1}^D$, $J_{B1}^D$ from (\ref{IDA1}), (\ref{IDB1}) by
the replacement (\ref{R1}), and \bea\nn J_{D1}^D&=&
8\pi^2T_{D2}\left[\frac{Ml^5}{\left(\bar{\Lambda}_+^2+
\bar{\Lambda}_-^2+4\Lambda_D^2/3\right)(3l-r_2)}\right]^{1/2}
\times
\\ \nn &&\left(1+\frac{\Delta
r_1}{2l}\right)^{1/2}\left(1+\frac{\Delta
r_1}{3l}\right)^{2}\left(1+\frac{\Delta
r_1}{4l}\right)^{1/2}\left(1+\frac{\Delta
r_1}{6l}\right)^{-1/2}\left(1+\frac{\Delta
r_1}{3l-r_2}\right)^{-1/2}\times
\\ \nn && F_D^{(5)}\left(1/2;-1/2,-2,-1/2,1/2,1/2;1;\right.
\\ \nn &&\left.\frac{1}{1+\frac{2l}{\Delta
r_1}},\frac{1}{1+\frac{3l}{\Delta
r_1}},\frac{1}{1+\frac{4l}{\Delta
r_1}},\frac{1}{1+\frac{6l}{\Delta
r_1}},\frac{1}{1+\frac{3l-r_2}{\Delta r_1}}\right).\eea

Taking $r_1\to\infty$ in the above expressions, one obtains that
in the semiclassical limit the following energy-charge relation
holds \bea\nn \frac{E^2\left(
E^2-\mathbf{P}^2\right)^2}{\mathbf{\Lambda}_1^2 E^2
-\left(\mathbf{\Lambda}_1.\mathbf{P}\right)^2}
=\frac{2^3}{3^5}(\pi^2 T_{D2})^2
\left(P^2_{\theta}+\frac{3P_{\phi}^2}{3-\cos^2\theta_2^0}\right).\eea
Obviously, this is a generalization of the relation (\ref{ED1})
and for $\theta_2^0=\pi/2$ has the same form.

Another possible ansatz for the D2-brane embedding is
\bea\label{DA4} X^0= \Lambda_0^0\xi^0, \h X^I=\Lambda_0^I\xi^0, \h
r=r(\xi^2),\h
\theta_1=\Lambda_1^{\theta}\xi^1+\Lambda_2^{\theta}\xi^2 ,\h
\phi_2=\Lambda_0^{\phi}\xi^0 ,\eea i.e., we have D2-brane extended
in the radial direction $r$, wrapped along the angular coordinate
$\theta_1$ and rotating in the plane given by the angle $\phi_2$.

Then one obtains the solution: \bea\nn
&&\xi^2(r)=\frac{8}{3}\lambda^0 T_{D2}\Lambda_1^{\theta}
\left[\frac{l(3l-w_+)(3l-w_-)}
{\left(3\Lambda^2+2\Lambda_D^2\right)\left(3l-r_2\right)\Delta
r_1}\right]^{1/2}(2\Delta r)^{3/4}\times
\\ \label{D2s4} &&F_D^{(7)}\left(3/4;-1/4,-1/4,1/4,-1/2,-1/2,1/2,1/2;7/4;\right.\\
\nn &&\left.-\frac{\Delta r}{2l},-\frac{\Delta
r}{4l},-\frac{\Delta r}{6l}, -\frac{\Delta
r}{3l-w_+},-\frac{\Delta r}{3l-w_-}, -\frac{\Delta
r}{3l-r_2},\frac{\Delta r}{\Delta r_1}\right),\h
w_{\pm}=\pm\sqrt{3} l.\eea The computation of the conserved
quantities $E$, $P_I$ and $P_{\phi_2}\equiv P_{\phi}$, gives
\bea\label{DEP4} &&\frac{E}{\Lambda_0^0}=\frac{P_I}{\Lambda_0^I}=
4\pi^2 T_{D2}\Lambda_1^{\theta} \left[\frac{l(3l-w_+)(3l-w_-)}
{\left(3\Lambda^2+2\Lambda_D^2\right)\left(3l-r_2\right)}\right]^{1/2}
\times\\ \nn &&\left(1+\frac{\Delta
r_1}{2l}\right)^{1/2}\left(1+\frac{\Delta
r_1}{4l}\right)^{1/2}\left(1+\frac{\Delta r_1}{6l}\right)^{-1/2}\times \\
\nn &&\left(1+\frac{\Delta r_1}{3l-w_+}\right)^{1/2}
\left(1+\frac{\Delta r_1}{3l-w_-}\right)^{1/2}\left(1+\frac{\Delta
r_1}{3l-r_2}\right)^{-1/2}\times
\\ \nn &&F_D^{(6)}\left(1/2;-1/2,-1/2,1/2,-1/2,-1/2,1/2;1;\right.\\
\nn &&\left.\frac{1}{1+\frac{2l}{\Delta
r_1}},\frac{1}{1+\frac{4l}{\Delta
r_1}},\frac{1}{1+\frac{6l}{\Delta
r_1}},\frac{1}{1+\frac{3l-w+}{\Delta
r_1}},\frac{1}{1+\frac{3l-w-}{\Delta
r_1}}\frac{1}{1+\frac{3l-r_2}{\Delta r_1}}\right),\eea \bea\nn
P_\phi=\sin^2\theta_2^0\left(J_{A2}^D+J_{B2}^D\right)
+\cos^2\theta_2^0 J_{D2}^D,\eea where \bea\nn && J_{A2}^D=4\pi^2
T_{D2}\Lambda_0^\phi \Lambda_1^{\theta}
\left[\frac{l^5(3l-w_+)(3l-w_-)}
{\left(3\Lambda^2+2\Lambda_D^2\right)\left(3l-r_2\right)}\right]^{1/2}
\times\\ \nn &&\left(1+\frac{\Delta
r_1}{2l}\right)^{3/2}\left(1+\frac{\Delta
r_1}{4l}\right)^{1/2}\left(1+\frac{\Delta r_1}{6l}\right)^{1/2}\times \\
\nn &&\left(1+\frac{\Delta r_1}{3l-w_+}\right)^{1/2}
\left(1+\frac{\Delta r_1}{3l-w_-}\right)^{1/2}\left(1+\frac{\Delta
r_1}{3l-r_2}\right)^{-1/2}\times
\\ \nn &&F_D^{(6)}\left(1/2;-3/2,-1/2,-1/2,-1/2,-1/2,1/2;1;\right.\\
\nn &&\left.\frac{1}{1+\frac{2l}{\Delta
r_1}},\frac{1}{1+\frac{4l}{\Delta
r_1}},\frac{1}{1+\frac{6l}{\Delta
r_1}},\frac{1}{1+\frac{3l-w+}{\Delta
r_1}},\frac{1}{1+\frac{3l-w-}{\Delta
r_1}}\frac{1}{1+\frac{3l-r_2}{\Delta r_1}}\right),\eea \bea\nn &&
J_{B2}^D=\frac{2}{3}\pi^2 T_{D2}\Lambda_0^\phi \Lambda_1^{\theta}
\left[\frac{l^3(3l-w_+)(3l-w_-)}
{\left(3\Lambda^2+2\Lambda_D^2\right)\left(3l-r_2\right)}\right]^{1/2}
\times\\ \nn &&\Delta r_1\left(1+\frac{\Delta
r_1}{2l}\right)^{1/2}\left(1+\frac{\Delta
r_1}{4l}\right)^{3/2}\left(1+\frac{\Delta r_1}{6l}\right)^{-1/2}\times \\
\nn &&\left(1+\frac{\Delta r_1}{3l-w_+}\right)^{1/2}
\left(1+\frac{\Delta r_1}{3l-w_-}\right)^{1/2}\left(1+\frac{\Delta
r_1}{3l-r_2}\right)^{-1/2}\times
\\ \nn &&F_D^{(6)}\left(1/2;-1/2,-3/2,1/2,-1/2,-1/2,1/2;2;\right.\\
\nn &&\left.\frac{1}{1+\frac{2l}{\Delta
r_1}},\frac{1}{1+\frac{4l}{\Delta
r_1}},\frac{1}{1+\frac{6l}{\Delta
r_1}},\frac{1}{1+\frac{3l-w+}{\Delta
r_1}},\frac{1}{1+\frac{3l-w-}{\Delta
r_1}}\frac{1}{1+\frac{3l-r_2}{\Delta r_1}}\right),\eea \bea\nn &&
J_{D2}^D=4\pi^2 T_{D2}\Lambda_0^\phi \Lambda_1^{\theta}
\left[\frac{l^5(3l-w_+)(3l-w_-)}
{\left(3\Lambda^2+2\Lambda_D^2\right)\left(3l-r_2\right)}\right]^{1/2}
\times\\ \nn &&\left(1+\frac{\Delta
r_1}{2l}\right)^{1/2}\left(1+\frac{\Delta
r_1}{3l}\right)^{2}\left(1+\frac{\Delta
r_1}{4l}\right)^{1/2}\left(1+\frac{\Delta r_1}{6l}\right)^{-1/2}\times \\
\nn &&\left(1+\frac{\Delta r_1}{3l-w_+}\right)^{1/2}
\left(1+\frac{\Delta r_1}{3l-w_-}\right)^{1/2}\left(1+\frac{\Delta
r_1}{3l-r_2}\right)^{-1/2}\times
\\ \nn &&F_D^{(7)}\left(1/2;-1/2,-2,-1/2,1/2,-1/2,-1/2,1/2;1;\right.\\
\nn &&\left.\frac{1}{1+\frac{2l}{\Delta
r_1}},\frac{1}{1+\frac{3l}{\Delta
r_1}},\frac{1}{1+\frac{4l}{\Delta
r_1}},\frac{1}{1+\frac{6l}{\Delta
r_1}},\frac{1}{1+\frac{3l-w+}{\Delta
r_1}},\frac{1}{1+\frac{3l-w-}{\Delta
r_1}}\frac{1}{1+\frac{3l-r_2}{\Delta r_1}}\right).\eea

Going to the semiclassical limit $r_1\to\infty$ in the above
expressions for the conserved quantities, one obtains the
following relation between them \bea\label{ED4}
E^2=\mathbf{P}^2+\frac{3^{7/3}}{2^{1/3}}\left(\frac{\pi T_{D2}
\Lambda_1^{\theta}}{3-\cos^2\theta_2^0}\right)^{2/3}P_{\phi}^{4/3}.\eea
This is a generalization of the energy-charge relation received in
(\ref{ED2}).

Another admissible embedding is \bea\label{DA5} &&X^0=
\Lambda_0^0\xi^0+\frac{\left(\mathbf{\Lambda}_0.\mathbf{\Lambda}_1\right)}{\Lambda_0^0}\left(\xi^1
+ c\xi^2\right), \h X^I=\Lambda_0^I\xi^0 + \Lambda_1^I\left(\xi^1
+ c\xi^2\right),
\\ \nn &&r=r(\xi^2),\h \phi_1=\Lambda_0^{\phi_1}\xi^0,\h
\phi_2=\Lambda_0^{\phi_2}\xi^0.\eea It is analogous to (\ref{DA1})
and (\ref{DA3}), but now the rotations are in the planes given by
the angles $\phi_1$ and $\phi_2$.

The solution $\xi^2(r)$, and the expressions for $E$, $P_I$, may
be obtained from the corresponding quantities for the embedding
(\ref{DA3}) by the replacements
$\bar{\Lambda}_{\mp}^2\to\tilde{\Lambda}_{\pm}^2$, $\Lambda_D^2\to
\tilde{\Lambda}_D^2$. For the conserved angular momenta
$P_{\phi_1}$ and $P_{\phi_2}$ one finds \bea\label{D5Pf1}
P_{\phi_1}&=&
\left(\Lambda_0^{\phi_1}\sin\theta_1^0+\Lambda_0^{\phi_2}\cos\psi_1^0\sin\theta_2^0\right)\sin\theta_1^0K_{A1}^D
\\ \nn
&+&\left(\Lambda_0^{\phi_1}\sin\theta_1^0-\Lambda_0^{\phi_2}\cos\psi_1^0\sin\theta_2^0\right)\sin\theta_1^0K_{B1}^D
\\ \nn &+&\left(\Lambda_0^{\phi_1}\cos\theta_1^0 +
\Lambda_0^{\phi_2}\cos\theta_2^0\right)\cos\theta_1^0K_{D1}^D,\eea
\bea\label{D5Pf2} P_{\phi_2}&=&
\left(\Lambda_0^{\phi_2}\sin\theta_2^0+\Lambda_0^{\phi_1}\cos\psi_1^0\sin\theta_1^0\right)\sin\theta_2^0K_{A1}^D
\\ \nn
&+&\left(\Lambda_0^{\phi_2}\sin\theta_2^0-\Lambda_0^{\phi_1}\cos\psi_1^0\sin\theta_1^0\right)\sin\theta_2^0K_{B1}^D
\\ \nn &+&\left(\Lambda_0^{\phi_1}\cos\theta_1^0 +
\Lambda_0^{\phi_2}\cos\theta_2^0\right)\cos\theta_2^0K_{D1}^D,\eea
where $K_{A1}^D$, $K_{B1}^D$ and $K_{D1}^D$ can be obtained from
$J_{A1}^D$, $J_{B1}^D$ and $J_{D1}^D$ through the above mentioned
replacements.

The calculations show that in the semiclassical limit, the
dependence of the energy on the conserved charges, for the present
case, is given by the equality: \bea\label{ED5} &&\frac{E^2\left(
E^2-\mathbf{P}^2\right)^2}{\mathbf{\Lambda}_1^2 E^2
-\left(\mathbf{\Lambda}_1.\mathbf{P}\right)^2} =\\ \nn
&&\frac{2^3}{3^5}(\pi^2
T_{D2})^2\frac{\left(3-\cos^2\theta_2^0\right)P_{\phi_1}^2+
\left(3-\cos^2\theta_1^0\right)P_{\phi_2}^2
-4P_{\phi_1}P_{\phi_2}\cos\theta_1^0\cos\theta_2^0}
{3-\cos^2\theta_1^0-
\cos^2\theta_2^0-\cos^2\theta_1^0\cos^2\theta_2^0}.\eea

Finally, let us consider the following possible D2-brane embedding
\bea\label{DA6}
&&X^0= \Lambda_0^0\xi^0, \h X^I=\Lambda_0^I\xi^0, \h r=r(\xi^2),\\
\nn
&&\phi_1=\Lambda_0^{\phi}\xi^0+\Lambda_1^{\phi}\xi^1+\Lambda_2^{\phi}\xi^2
,\h
\phi_2=\Lambda_0^{\phi}\xi^0-\Lambda_1^{\phi}\xi^1-\Lambda_2^{\phi}\xi^2
.\eea It describes D2-brane configuration, which is analogous to
the one in (\ref{DA2}), but now the rotations are in the planes
defined by the angles $\phi_1$ and $\phi_2$ instead of $\theta_1$
and $\theta_2$.

For this embedding, one obtains \bea\nn
&&\xi^2(r)=\frac{8}{3}\lambda^0
T_{D2}\left[\frac{l\left(\hat{\Lambda}_{1+}^2+\hat{\Lambda}_{1-}^2\right)(3l-u_+)(3l-u_-)}
{3\left(\hat{\Lambda}_+^2+\hat{\Lambda}_-^2+4\hat{\Lambda}_D^2/3\right)\left(3l-r_2\right)\Delta
r_1}\right]^{1/2}(2\Delta r)^{3/4}\times
\\ \nn &&F_D^{(7)}\left(3/4;-1/4,-1/4,1/4,-1/2,-1/2,1/2,1/2;7/4;\right.\\
\nn &&\left.-\frac{\Delta r}{2l},-\frac{\Delta
r}{4l},-\frac{\Delta r}{6l}, -\frac{\Delta
r}{3l-u_+},-\frac{\Delta r}{3l-u_-}, -\frac{\Delta
r}{3l-r_2},\frac{\Delta r}{\Delta r_1}\right),\eea where \bea\nn
u_\pm=l\left[\frac{\hat{\Lambda}_{1+}^2-\hat{\Lambda}_{1-}^2}
{\hat{\Lambda}_{1+}^2+\hat{\Lambda}_{1-}^2}\pm
\sqrt{3+\left(\frac{\hat{\Lambda}_{1+}^2-\hat{\Lambda}_{1-}^2}
{\hat{\Lambda}_{1+}^2+\hat{\Lambda}_{1-}^2}\right)^2}\right].\eea

The computation of the conserved charges results in
\bea\label{DEP6} &&\frac{E}{\Lambda_0^0}=\frac{P_I}{\Lambda_0^I}=
4\pi^2T_{D2}\left[\frac{l\left(\hat{\Lambda}_{1+}^2+\hat{\Lambda}_{1-}^2\right)(3l-u_+)(3l-u_-)}
{3\left(\hat{\Lambda}_+^2+\hat{\Lambda}_-^2+4\hat{\Lambda}_D^2/3\right)\left(3l-r_2\right)}\right]^{1/2}\times
\\ \nn &&\left(1+\frac{\Delta
r_1}{2l}\right)^{1/2}\left(1+\frac{\Delta
r_1}{4l}\right)^{1/2}\left(1+\frac{\Delta r_1}{6l}\right)^{-1/2}\times \\
\nn &&\left(1+\frac{\Delta r_1}{3l-u_+}\right)^{1/2}
\left(1+\frac{\Delta r_1}{3l-u_-}\right)^{1/2}\left(1+\frac{\Delta
r_1}{3l-r_2}\right)^{-1/2}\times
\\ \nn && F_D^{(6)}\left(1/2;-1/2,-1/2,1/2,-1/2,-1/2,1/2;1;\right.
\\ \nn &&\left.\frac{1}{1+\frac{2l}{\Delta
r_1}},\frac{1}{1+\frac{4l}{\Delta
r_1}},\frac{1}{1+\frac{6l}{\Delta
r_1}},\frac{1}{1+\frac{3l-u+}{\Delta
r_1}},\frac{1}{1+\frac{3l-u-}{\Delta
r_1}}\frac{1}{1+\frac{3l-r_2}{\Delta r_1}}\right),\eea \bea\nn
&&P_{\phi} \equiv P_{\phi_1}=P_{\phi_2}=\\
\nn
&&\Lambda_0^\phi\left\{\sin^2\theta^0\left[\left(1+\cos\psi_1^0\right)K^D_{A2}
+\left(1-\cos\psi_1^0\right)K^D_{B2}\right]+2\cos^2\theta^0K^D_{D2}\right\},\eea
where \bea\nn &&K^D_{A2}=
4\pi^2T_{D2}\left[\frac{l^5\left(\hat{\Lambda}_{1+}^2+\hat{\Lambda}_{1-}^2\right)(3l-u_+)(3l-u_-)}
{3\left(\hat{\Lambda}_+^2+\hat{\Lambda}_-^2+4\hat{\Lambda}_D^2/3\right)\left(3l-r_2\right)}\right]^{1/2}\times
\\ \nn &&\left(1+\frac{\Delta
r_1}{2l}\right)^{3/2}\left(1+\frac{\Delta
r_1}{4l}\right)^{1/2}\left(1+\frac{\Delta r_1}{6l}\right)^{1/2}\times \\
\nn &&\left(1+\frac{\Delta r_1}{3l-u_+}\right)^{1/2}
\left(1+\frac{\Delta r_1}{3l-u_-}\right)^{1/2}\left(1+\frac{\Delta
r_1}{3l-r_2}\right)^{-1/2}\times
\\ \nn && F_D^{(6)}\left(1/2;-3/2,-1/2,-1/2,-1/2,-1/2,1/2;1;\right.
\\ \nn &&\left.\frac{1}{1+\frac{2l}{\Delta
r_1}},\frac{1}{1+\frac{4l}{\Delta
r_1}},\frac{1}{1+\frac{6l}{\Delta
r_1}},\frac{1}{1+\frac{3l-u+}{\Delta
r_1}},\frac{1}{1+\frac{3l-u-}{\Delta
r_1}}\frac{1}{1+\frac{3l-r_2}{\Delta r_1}}\right),\eea
\bea\nn
&&K^D_{B2}=
2\pi^2T_{D2}\left[\frac{l^3\left(\hat{\Lambda}_{1+}^2+\hat{\Lambda}_{1-}^2\right)(3l-u_+)(3l-u_-)}
{3^3\left(\hat{\Lambda}_+^2+\hat{\Lambda}_-^2+4\hat{\Lambda}_D^2/3\right)\left(3l-r_2\right)}\right]^{1/2}\times
\\ \nn &&\Delta r_1\left(1+\frac{\Delta
r_1}{2l}\right)^{1/2}\left(1+\frac{\Delta
r_1}{4l}\right)^{3/2}\left(1+\frac{\Delta r_1}{6l}\right)^{-1/2}\times \\
\nn &&\left(1+\frac{\Delta r_1}{3l-u_+}\right)^{1/2}
\left(1+\frac{\Delta r_1}{3l-u_-}\right)^{1/2}\left(1+\frac{\Delta
r_1}{3l-r_2}\right)^{-1/2}\times
\\ \nn && F_D^{(6)}\left(1/2;-1/2,-3/2,1/2,-1/2,-1/2,1/2;2;\right.
\\ \nn &&\left.\frac{1}{1+\frac{2l}{\Delta
r_1}},\frac{1}{1+\frac{4l}{\Delta
r_1}},\frac{1}{1+\frac{6l}{\Delta
r_1}},\frac{1}{1+\frac{3l-u+}{\Delta
r_1}},\frac{1}{1+\frac{3l-u-}{\Delta
r_1}}\frac{1}{1+\frac{3l-r_2}{\Delta r_1}}\right),\eea
\bea\nn
&&K^D_{D2}=
4\pi^2T_{D2}\left[\frac{l^5\left(\hat{\Lambda}_{1+}^2+\hat{\Lambda}_{1-}^2\right)(3l-u_+)(3l-u_-)}
{3\left(\hat{\Lambda}_+^2+\hat{\Lambda}_-^2+4\hat{\Lambda}_D^2/3\right)\left(3l-r_2\right)}\right]^{1/2}\times
\\ \nn &&\left(1+\frac{\Delta
r_1}{2l}\right)^{1/2}\left(1+\frac{\Delta
r_1}{3l}\right)^{2}\left(1+\frac{\Delta
r_1}{4l}\right)^{1/2}\left(1+\frac{\Delta r_1}{6l}\right)^{-1/2}\times \\
\nn &&\left(1+\frac{\Delta r_1}{3l-u_+}\right)^{1/2}
\left(1+\frac{\Delta r_1}{3l-u_-}\right)^{1/2}\left(1+\frac{\Delta
r_1}{3l-r_2}\right)^{-1/2}\times
\\ \nn && F_D^{(7)}\left(1/2;-1/2,-2,-1/2,1/2,-1/2,-1/2,1/2;1;\right.
\\ \nn &&\left.\frac{1}{1+\frac{2l}{\Delta
r_1}},\frac{1}{1+\frac{3l}{\Delta
r_1}},\frac{1}{1+\frac{4l}{\Delta
r_1}},\frac{1}{1+\frac{6l}{\Delta
r_1}},\frac{1}{1+\frac{3l-u+}{\Delta
r_1}},\frac{1}{1+\frac{3l-u-}{\Delta
r_1}}\frac{1}{1+\frac{3l-r_2}{\Delta r_1}}\right).\eea

Taking the semiclassical limit in the above expressions for $E$,
$P_I$ and $P_\phi$, which in the case under consideration
corresponds to \bea\nn r_{1,2}\to \pm
2\sqrt{\frac{3v_0^2}{\hat{\Lambda}_+^2+\hat{\Lambda}_-^2+4\hat{\Lambda}_D^2/3}}\to\infty,\eea
we receive that the energy depends on $P_I$ and $P_\phi$ as
follows \bea\label{ED6} E^2=\mathbf{P}^2+3^{7/3}\left(\frac{2\pi
T_{D2}
\Lambda_1^{\phi}\sin\theta^0}{4-\sin^2\theta^0}\right)^{2/3}P_{\phi}^{4/3}.\eea
This is another generalization of the energy-charge relation given
in (\ref{ED2}).

\subsubsection{Two-spin magnon-like energy-charge relations
\\from M-theory viewpoint}

We showed in \cite{11} that for each M-theory background, having
subspaces with metrics of given type, there exist M2-brane
configurations, which in appropriate limit lead to two-spin
magnon-like energy-charge relations, established for strings on
$AdS_5\times S^5$, its $\beta$-deformation, and for membrane in
$AdS_4\times S^7$.

If we split the target space coordinates as $x^M=(x^\mu, x^a)$,
where $x^\mu$ are those on which the background does not depend,
the conserved charges are given by the expression \cite{4}
\bea\label{cc} Q_\mu =\frac{1}{2\lambda^0}\int d\xi^1 d\xi^2
g_{\mu N}\p_0 X^N.\eea

Now, let us turn to our particular tasks. Consider backgrounds of
the type \bea\label{mtb}
ds^2=c^2\left[-dt^2+c_1^2d\theta^2+c_2^2\cos^2\theta d\varphi_1^2
+c_3^2\sin^2\theta
d\varphi_2^2+c_4^2f(\theta)d\varphi_3^2\right],\eea where $c$,
$c_1$, $c_2$, $c_3$, $c_4$ are arbitrary constants, and
$f(\theta)$ takes two values: $f(\theta)=1$ and
$f(\theta)=\sin^2\theta$. We embed the membrane into (\ref{mtb})
in the following way \bea\label{ra} &&X^0(\xi^m)\equiv t(\xi^m)=
\La_0^0\xi^0, \h X^1(\xi^m)=\theta(\xi^2),
\\ \nn &&X^2(\xi^m)\equiv \varphi_1(\xi^m)=
\La_0^2\xi^0,
\\ \nn &&X^3(\xi^m)\equiv \varphi_2(\xi^m)=
\La_0^3\xi^0,
\\ \nn &&X^4(\xi^m)=\varphi_3(\xi^m)=
\La_i^4\xi^i,
\\ \nn &&\mu=0,2,3,4,\h a=1,\h
\La_0^0,\ldots,\La_i^4=constants.\eea This ansatz corresponds to
M2-brane extended in the $\theta$- direction, moving with constant
energy $E$ along the $t$-coordinate, rotating in the planes
defined by the angles $\varphi_1$, $\varphi_2$, with constant
angular momenta $J_1$, $J_2$, and {\it wrapped} along $\varphi_3$.

We begin with the case $f(\theta)=1$, when we have \cite{11}
\bea\label{ecp} &&K\theta'^2+V(\theta)=0,\\ \nn
&&K=-(2\lambda^0T_2c^2c_1c_4\La_1^4)^2,
\\ \nn &&V(\theta)=c^2\left\{(\Lambda_0^0)^2-(\Lambda_0^2c_2)^2
-\left[(\Lambda_0^3c_3)^2-(\Lambda_0^2c_2)^2\right]\sin^2\theta\right\}.\eea
From (\ref{ecp}) one obtains the turning point ($\theta'=0$) for
the effective one dimensional motion \bea\label{M^2}
M^2=\frac{(\Lambda_0^0)^2-(\Lambda_0^2c_2)^2}{(\Lambda_0^3c_3)^2-(\Lambda_0^2c_2)^2}.\eea
The solution of (\ref{ecp}) is \bea\label{s1} \xi^2(\theta)=
\frac{2\lambda^0T_2cc_1c_4\Lambda_1^4\sin\theta}{M\left[(\Lambda_0^3c_3)^2-(\Lambda_0^2c_2)^2\right]^{1/2}}
F_1(1/2,1/2,1/2;3/2;\sin^2\theta,\frac{\sin^2\theta}{M^2}).\eea On
this solution, the conserved charges (\ref{cc}) take the form
($Q_0\equiv -E$, $Q_2\equiv J_1$, $Q_3\equiv J_2$, $Q_4=0$)
\bea\label{1E} &&\frac{E}{\Lambda_0^0} =
\frac{2\pi^2T_2c^3c_1c_4\Lambda_1^4}{\left[(\Lambda_0^3c_3)^2-(\Lambda_0^2c_2)^2\right]^{1/2}}
\mbox{}_2F_1(1/2,1/2;1;M^2),
\\ \label{1J1} &&\frac{J_1}{\Lambda_0^2} =
\frac{2\pi^2T_2c^3c_1c_2^2c_4\Lambda_1^4}{\left[(\Lambda_0^3c_3)^2-(\Lambda_0^2c_2)^2\right]^{1/2}}
\mbox{}_2F_1(-1/2,1/2;1;M^2),
\\ \label{1J2} &&\frac{J_2}{\Lambda_0^3} =
\frac{2\pi^2T_2c^3c_1c_3^2c_4\Lambda_1^4}{\left[(\Lambda_0^3c_3)^2-(\Lambda_0^2c_2)^2\right]^{1/2}}
\left[\mbox{}_2F_1(1/2,1/2;1;M^2)-\mbox{}_2F_1(-1/2,1/2;1;M^2)
\right].\eea

Our next aim is to consider the limit, in which $M$ tends to its
maximum value: $M\to 1_-$. In this case, by using (\ref{M^2}) and
(\ref{1E})-(\ref{1J2}), one arrives at the energy-charge relation
\bea\label{2sgm} E-\frac{J_2}{c_3}=
\sqrt{\left(\frac{J_1}{c_2}\right)^2+\left(4\pi
T_2c^3c_1c_4\Lambda_1^4\right)^2} \eea for \bea\label{lim} E,
J_2/c_3\to\infty,\h E-J_2/c_3, J_1/c_2 - \mbox{finite}.\eea

Now, we are going to consider the case $f(\theta)=\sin^2\theta$
(see (\ref{mtb})), when we have \cite{11}
\bea\label{ecp1}
&&\tilde{K}\theta'^2+V(\theta)=0,\\ \nn
&&\tilde{K}=-(2\lambda^0T_2c^2c_1c_4\La_1^4)^2\sin^2\theta
=K\sin^2\theta, \eea where $V(\theta)$ and correspondingly $M^2$
are the same as in (\ref{ecp}) and (\ref{M^2}). The solution of
(\ref{ecp1}) is given by the equality \bea\label{s2}
\xi^2(\theta)= \frac{\lambda^0T_2cc_1c_4\Lambda_1^4\sin^2\theta}
{M\left[(\Lambda_0^3c_3)^2-(\Lambda_0^2c_2)^2\right]^{1/2}}
F_1(1,1/2,1/2;2;\sin^2\theta,\frac{\sin^2\theta}{M^2}),\eea and is
obviously different from the previously obtained one. The
computations show that on (\ref{s2}) the conserved charges
(\ref{cc}) are as follows \bea\label{2E} &&\frac{E}{\Lambda_0^0} =
\frac{2\pi
T_2c^3c_1c_4\Lambda_1^4}{\left[(\Lambda_0^3c_3)^2-(\Lambda_0^2c_2)^2\right]^{1/2}}
\ln\left(\frac{1+M}{1-M}\right),
\\ \label{2J1} &&\frac{J_1}{\Lambda_0^2} =
\frac{2\pi
T_2c^3c_1c_2^2c_4\Lambda_1^4}{\left[(\Lambda_0^3c_3)^2-(\Lambda_0^2c_2)^2\right]^{1/2}}
\left[\frac{1-M^2}{2}\ln\left(\frac{1+M}{1-M}\right)+M\right],
\\ \label{2J2} &&\frac{J_2}{\Lambda_0^3} =
\frac{2\pi
T_2c^3c_1c_3^2c_4\Lambda_1^4}{\left[(\Lambda_0^3c_3)^2-(\Lambda_0^2c_2)^2\right]^{1/2}}
\left[\frac{1+M^2}{2}\ln\left(\frac{1+M}{1-M}\right)-M
\right].\eea

Taking $M\to 1_-$, one sees that it corresponds again to the limit
(\ref{lim}), and the two-spin energy-charge relation is
\bea\label{2sgm'} E-\frac{J_2}{c_3}=
\sqrt{\left(\frac{J_1}{c_2}\right)^2+\left(2\pi
T_2c^3c_1c_4\Lambda_1^4\right)^2},\eea which differs from
(\ref{2sgm}) only by a factor of 4 in the second term on the right
hand side.

{\bf Discussion}

We have shown here that for each M-theory background, having
subspaces with metrics of the type (\ref{mtb}), there exist
M2-brane configurations given by (\ref{ra}), which in the limit
(\ref{lim}) lead to the two-spin, magnon-like, energy-charge
relations (\ref{2sgm}) and (\ref{2sgm'}).

Examples for target space metrics of the type (\ref{mtb}) are
several subspaces of $R\times S^7$, contained in the $AdS_4\times
S^7$ solution of M-theory.

More examples for target space metrics of the type (\ref{mtb}),
for which there exist the membrane configurations (\ref{ra})
giving rise to two-spin magnon-like energy-charge relations, can
be found for instance in different subspaces of the $AdS_7\times
S^4$ solution of M-theory and not only there.

\subsubsection{Integrable systems from membranes on $AdS_4 \times S^7$}

It is known that large class of classical string solutions in the
type IIB $AdS_5\times S^5$ background is related to the Neumann
and Neumann-Rosochatius integrable systems, including spiky
strings and giant magnons \cite{NR2}. It is also interesting if
these integrable systems can be associated with some membrane
configurations in M-theory. We explain here how this can be
achieved by considering membrane embedding in $AdS_4\times S^7$
solution of $M$-theory, with the desired properties \cite{11}.

On the other hand, we will show the existence of membrane
configurations in $AdS_4\times S^7$ \cite{12}, which correspond to
the continuous limit of the $SU(2)$ integrable spin chain, arising
in $\mathcal{N}=4$ SYM in four dimensions, dual to strings in
$AdS_5\times S^5$ \cite{K0311}.

Here we will work with the action (\ref{oa}) written for membranes
($p=2$) in diagonal worldvolume gauge $\lambda^{i}=0$, in which
the action and the constraints simplify to
\bea\label{omagf}
&&S_{M}=\int d^{3}\xi \mathcal{L}_{M}= \int
d^{3}\xi\left\{\frac{1}{4\lambda^0}\Bigl[G_{00}-\left(2\lambda^0T_2\right)^2\det
G_{ij}\Bigr] + T_2 C_{012}\right\},
\\ \label{00gf} &&G_{00}+\left(2\lambda^0T_2\right)^2\det G_{ij}=0,
\\ \label{0igf} &&G_{0i}=0.\eea

Searching for membrane configurations in $AdS_4\times S^7$, which
correspond to the Neumann or Neumann-Rosochatius integrable
systems, we should first eliminate the membrane interaction with
the background 3-form field on $AdS_4$, to ensure more close
analogy with the strings on $AdS_5\times S^5$. To make our choice,
let us write down the background. It can be parameterized as
follows \bea\nn &&ds^2=(2l_p\mathcal{R})^2\left[-\cosh^2\rho
dt^2+d\rho^2+\sinh^2\rho\left(d\alpha^2+\sin^2\alpha
d\beta^2\right)+ 4d\Omega_7^2\right],
\\ \nn &&d\Omega_7^2= d\psi_1^2+\cos^2\psi_1
d\varphi_1^2\\ \nn &&+\sin^2\psi_1\left[d\psi_2^2+\cos^2\psi_2
d\varphi_2^2+ \sin^2\psi_2\left(d\psi_3^2+\cos^2\psi_3
d\varphi_3^2+\sin^2\psi_3 d\varphi_4^2\right)\right],
\\ \nn &&c_{(3)}=(2l_p\mathcal{R})^3\sinh^3\rho\sin\alpha dt\wedge
d\alpha\wedge d\beta.\eea Since we want the membrane to have
nonzero conserved energy and spin on $AdS$, one possible choice,
for which the interaction with the $c_{(3)}$ field disappears, is
to fix the angle $\alpha$: $\alpha=\alpha_0=const.$ The metric of
the corresponding subspace of $AdS_4$ is \bea\nn
ds^2_{sub}=(2l_p\mathcal{R})^2\left[-\cosh^2\rho
dt^2+d\rho^2+\sinh^2\rho d(\beta\sin\alpha_0)^2\right].\eea The
appropriate membrane embedding into $ds^2_{sub}$ and $S^7$ is
\bea\nn
&&Z_{\mu}=2l_p\mathcal{R}\mbox{r}_\mu(\xi^m)e^{i\phi_\mu(\xi^m)},\h
\mu=(0,1),\h \phi_{\mu}=(\phi_0,\phi_1)=(t,\beta\sin\alpha_0), \\
\nn &&W_{a}=4l_p\mathcal{R}r_a(\xi^m)e^{i\varphi_a(\xi^m)},\h
a=(1,2,3,4),\eea where $\mbox{r}_\mu$ and $r_a$ are real functions
of $\xi^m$, while $\phi_\mu$ and $\varphi_a$ are the isometric
coordinates on which the background metric does not depend. The
six complex coordinates $Z_{\mu}$, $W_{a}$ are restricted by the
two real embedding constraints \bea\nn
\eta^{\mu\nu}Z_{\mu}\bar{Z}_{\nu}+\left(2l_p\mathcal{R}\right)^2=0,\h
\eta^{\mu\nu}=(-1,1),\h
\delta_{ab}W_{a}\bar{W}_{b}-\left(4l_p\mathcal{R}\right)^2=0,\eea
or equivalently \bea\nn \eta^{\mu\nu}
\mbox{r}_\mu\mbox{r}_\nu+1=0, \h \delta_{ab} r_ar_b-1=0.\eea The
coordinates $\mbox{r}_\mu$, $r_a$ are connected to the initial
coordinates, on which the background depends, through the
equalities \bea\nn &&\mbox{r}_0=\cosh\rho,\h \mbox{r}_1=\sinh\rho,
\\ \nn &&r_1= \cos\psi_1,\h r_2= \sin\psi_1\cos\psi_2,
\\ \nn &&r_3= \sin\psi_1\sin\psi_2\cos\psi_3,\h
r_4= \sin\psi_1\sin\psi_2\sin\psi_3.\eea

For the embedding described above, the induced metric is given by
\bea\label{im}
&&G_{mn}=\eta^{\mu\nu}\p_{(m}Z_\mu\p_{n)}\bar{Z_\nu} +
\delta_{ab}\p_{(m}W_a\p_{n)}\bar{W_b}= \\ \nn
&&(2l_p\mathcal{R})^2\left[\sum_{\mu,\nu=0}^{1}\eta^{\mu\nu}
\left(\p_m\mbox{r}_\mu\p_n\mbox{r}_\nu +
\mbox{r}_\mu^2\p_m\phi_\mu\p_n\phi_\nu\right)+
4\sum_{a=1}^{4}\left(\p_mr_a\p_nr_a +
r_a^2\p_m\varphi_a\p_n\varphi_a\right)\right].\eea
Correspondingly, the membrane Lagrangian becomes \bea\nn
\mathcal{L}=\mathcal{L}_{M}+\Lambda_A(\eta^{\mu\nu}
\mbox{r}_\mu\mbox{r}_\nu+1)+\Lambda_S(\delta_{ab} r_ar_b-1),\eea
where $\Lambda_A$ and $\Lambda_S$ are Lagrange multipliers.

{\bf Neumann and Neumann-Rosochatius integrable systems from
membranes}

Let us consider the following particular case of the above
membrane embedding \bea\label{nra}
&&Z_{0}=2l_p\mathcal{R}e^{i\kappa\tau}, Z_1=0,\h
W_{a}=4l_p\mathcal{R}r_a(\xi,\eta)e^{i\left[\omega_{a}\tau+\mu_a(\xi,\eta)\right]},
\\ \nn && \xi=\alpha\sigma_1+\beta\tau, \eta=\gamma\sigma_2+\delta\tau,\eea
which implies \bea\label{i} \mbox{r}_0=1,\h \mbox{r}_1=0,\h
\phi_0=t=\kappa\tau,\h
\varphi_a(\xi^m)=\varphi_a(\tau,\sigma_1,\sigma_2)=
\omega_{a}\tau+\mu_a(\xi,\eta).\eea Here $\kappa$, $\omega_{a}$,
$\alpha$, $\beta$, $\gamma$, $\delta$ are parameters, whereas
$r_a(\xi,\eta)$, $\mu_a(\xi,\eta)$ are arbitrary functions. As a
consequence, the embedding constraint $\eta^{\mu\nu}
\mbox{r}_\mu\mbox{r}_\nu+1=0$ is satisfied identically. For this
ansatz, the membrane Lagrangian takes the form ($\p_\xi=\p/\p\xi$,
$\p_\eta=\p/\p\eta$) \bea\nn
&&\mathcal{L}=-\frac{(4l_p\mathcal{R})^2}{4\lambda^0}
\left\{\left(8\lambda^0T_2l_p\mathcal{R}\alpha\gamma\right)^2
\sum_{a<b=1}^{4}\left[(\p_\xi r_a\p_\eta r_b-\p_\eta r_a\p_\xi r_b)^2\right. \right. \\
\nn &&+ \left. \left. (\p_\xi r_a\p_\eta\mu_b-\p_\eta
r_a\p_\xi\mu_b)^2r_b^2 + (\p_\xi\mu_a\p_\eta
r_b-\p_\eta\mu_a\p_\xi r_b)^2r_a^2\right.\right.
\\ \nn &&+\left.\left.(\p_\xi\mu_a\p_\eta\mu_b-\p_\eta\mu_a\p_\xi\mu_b)^2r_a^2r_b^2 \right]\right. \\
\nn &&+
\left.\sum_{a=1}^{4}\left[\left(8\lambda^0T_2l_p\mathcal{R}\alpha\gamma\right)^2
(\p_\xi r_a\p_\eta\mu_a-\p_\eta r_a\p_\xi\mu_a)^2-
\left(\beta\p_\xi\mu_a+\delta\p_\eta\mu_a+\omega_a\right)^2\right]r_a^2\right.
\\
\nn &&-\left.\sum_{a=1}^{4}\left(\beta\p_\xi r_a+\delta\p_\eta
r_a\right)^2+(\kappa/2)^2\right\}+\Lambda_S\left(\sum_{a=1}^{4}r_a^2-1\right).\eea
Now, we make the choice \bea\nn &&r_{1}=r_{1}(\xi),\h
r_{2}=r_{2}(\xi),\h \omega_3=\pm\omega_4=\omega,\\ \nn
&&r_3=r_3(\eta)=\epsilon\sin(b\eta+c),\h r_4=r_4(\eta)=\epsilon\cos(b\eta+c),\\
\nn &&\mu_1=\mu_1(\xi),\h \mu_2=\mu_2(\xi),\h
\mu_3,\mu_4=constants,\eea and receive (prime is used for
$d/d\xi$) \bea\nn
&&\mathcal{L}=-\frac{(4l_p\mathcal{R})^2}{4\lambda^0}
\left\{\sum_{a=1}^{2}\left[(A^2-\beta^2)r_a'^2\right.+
\left.(A^2-\beta^2)r_a^2\left(\mu'_a-\frac{\beta\omega_a}{A^2-\beta^2}\right)^2
- \frac{A^2}{A^2-\beta^2}\omega_a^2 r_a^2\right]\right. \\
\nn &&+\left.
(\kappa/2)^2-\epsilon^2(\omega^2+b^2\delta^2)\right\} +
\Lambda_S\left[\sum_{a=1}^{2}r_a^2-(1-\epsilon^2)\right],\eea
where $ A^2\equiv \left(8\lambda^0T_2l_p\mathcal{R}\epsilon
b\alpha\gamma\right)^2$. A single time integration of the
equations of motion for $\mu_a$ following from the above
Lagrangian gives \bea\nn
\mu'_a=\frac{1}{A^2-\beta^2}\left(\frac{C_a}{r_a^2}+\beta\omega_a\right),\eea
where $C_a$ are arbitrary constants. Taking this into account, one
obtains the following effective Lagrangian for the coordinates
$r_a(\xi)$ \bea\nn &&L=\frac{(4l_p\mathcal{R})^2}{4\lambda^0}
\sum_{a=1}^{2}\left[(A^2-\beta^2)r_a'^2 -
\frac{1}{A^2-\beta^2}\frac{C_a^2}{r_a^2} -
\frac{A^2}{A^2-\beta^2}\omega_a^2 r_a^2\right]\\ \nn
&&+\Lambda_S\left[\sum_{a=1}^{2}r_a^2-(1-\epsilon^2)\right].\eea
This Lagrangian in full analogy with the string considerations
corresponds to particular case of the $n$-dimensional {\it
Neumann-Rosochatius integrable system}. For $C_a=0$ one obtains
{\it Neumann integrable system}, which describes two-dimensional
harmonic oscillator, constrained to remain on a circle of radius
$\sqrt{1-\epsilon^2}$.

Let us write down the three constraints (\ref{00gf}), (\ref{0igf})
for the present case. To achieve more close correspondence with
the string on $AdS_5\times S^5$, we want the third one,
$G_{02}=0$, to be satisfied identically. To this end, since
$G_{02}\sim (ab)^2\gamma\delta,$ we set $\delta=0$, i.e.
$\eta=\gamma\sigma_2$. Then, the first two constraints give
\bea\nn &&\sum_{a=1}^{2}\left[(A^2-\beta^2)r_a'^2+
\frac{1}{A^2-\beta^2}\frac{C_a^2}{r_a^2}+
\frac{A^2}{A^2-\beta^2}\omega_a^2
r_a^2\right]=\frac{A^2+\beta^2}{A^2-\beta^2}\left[(\kappa/2)^2-(\epsilon\omega)^2\right],
\\ \label{effcs}&&\sum_{a=1}^{2}\omega_{a}C_a +
\beta\left[(\kappa/2)^2-(\epsilon\omega)^2\right]=0.\eea

Now, let us compute the energy and angular momenta for the
membrane configuration we are considering. Due to the background
isometries, there exist global conserved charges. In our case, the
background does not depend on $\phi_0=t$ and $\varphi_a$.
Therefore, the corresponding conserved quantities are the membrane
energy $E$ and four angular momenta $J_a$, given as spatial
integrals of the conjugated to these coordinates momentum
densities \bea\nn E=-\int d^2\sigma\frac{\p\mathcal{L}}{\p(\p_0
t)},\h J_a=\int
d^2\sigma\frac{\p\mathcal{L}}{\p(\p_0\varphi_a)},\h a=1,2,3,4.\eea
$E$ and $J_a$ can be computed by using the expression (\ref{im})
for the induced metric and the ansats (\ref{nra}), (\ref{i}).

In order to reproduce the string case, we can set $\omega=0$, and
thus $J_3=J_4=0$. The energy and the other two angular momenta are
given by \bea\nn
E=\frac{4\pi(l_p\mathcal{R})^2\kappa}{\lambda^0\alpha}\int d\xi,\h
J_a=\frac{\pi(4l_p\mathcal{R})^2}{\lambda^0\alpha(A^2-\beta^2)}\int
d\xi \left(\beta C_a + A^2\omega_a r_a^2\right),\h a=1,2.\eea From
here, by using the constraints (\ref{effcs}), one obtains the
energy-charge relation \bea\nn
\frac{4}{A^2-\beta^2}\left[A^2(1-\epsilon^2) +
\beta\sum_{a=1}^{2}\frac{C_a}{\omega_a}\right]\frac{E}{\kappa}
=\sum_{a=1}^{2}\frac{J_a}{\omega_a},\eea in full analogy with the
string case. Namely, for strings on $AdS_5\times S^5$, the result
in conformal gauge is \cite{NR2} \bea\nn
\frac{1}{\alpha^2-\beta^2}\left(\alpha^2+
\beta\sum_{a}\frac{C_a}{\omega_a}\right)\frac{E}{\kappa}
=\sum_{a}\frac{J_a}{\omega_a}.\eea

{\bf $SU(2)$ spin chain from membrane}

One of the predictions of AdS/CFT duality is that the string
theory on $AdS_5\times S^5$ should be dual to $\mathcal{N}=4$ SYM
theory in four dimensions. The spectrum of the string states and
of the operators in SYM should be the same. The first checks of
this conjecture {\it beyond} the supergravity approximation
revealed that there exist string configurations, whose energies in
the semiclassical limit are related to the anomalous dimensions of
certain gauge invariant operators in the planar SYM. On the field
theory side, it was found that the corresponding dilatation
operator is connected to the Hamiltonian of integrable Heisenberg
spin chain. On the other hand, it was established that there is
agreement at the level of actions between the continuous limit of
the $SU(2)$ spin chain arising in $\mathcal{N}=4$ SYM theory and a
certain limit of the string action in $AdS_5\times S^5$
background. Shortly after, it was shown that such equivalence also
holds for the $SU(3)$ and $SL(2)$ cases.

Here, we are interested in answering the question: is it possible
to reproduce this type of string/spin chain correspondence from
membranes on eleven dimensional curved backgrounds? It turns out
that the answer is positive at least for the case of M2-branes on
$AdS_4\times S^7$, as we will show below.

We will use our initial membrane embedding and fix \bea\nn
Z_{0}=2l_p\mathcal{R}e^{i\kappa\tau},\h Z_1=0,\eea which implies
$\mbox{r}_0=1$, $\mbox{r}_1=0$, $\phi_0=t=\kappa\tau$. Let us now
introduce new coordinates by setting \bea\nn
(\varphi_1,\varphi_2,\varphi_3,\varphi_4)=
\left(\frac{\kappa}{2}\tau+\alpha+\varphi,\frac{\kappa}{2}\tau+\alpha-\varphi,
\frac{\kappa}{2}\tau+\alpha+\phi,
\frac{\kappa}{2}\tau+\alpha-\tilde{\phi}\right)\eea and take the
limit $\kappa\to\infty$, $\p_0\to 0$, $\kappa\p_0$ - finite. In
this limit, we obtain the following expression for the membrane
Lagrangian
\bea\nn
\mathcal{L}&=&\frac{(2l_p\mathcal{R})^2}{\lambda^0}\kappa\left(\p_0\alpha
+
\sum_{k=1}^{3}\nu_k\p_0\rho_k\right)-\lambda^0T_2^2(4l_p\mathcal{R})^4
\left\{\sum_{a<b=1}^{4}(\p_1r_a\p_2r_b-\p_2r_a\p_1r_b)^2\right. \\
\nn &+&
\left.\sum_{a=1}^{4}\sum_{k=1}^{3}\mu_k(\p_1r_a\p_2\rho_k-\p_2r_a\p_1\rho_k)^2-
\sum_{a=1}^{4}\left(\p_1r_a\sum_{k=1}^{3}\nu_k\p_2\rho_k -
\p_2r_a\sum_{k=1}^{3}\nu_k\p_1\rho_k\right)^2\right.
\\ \nn &+&
\left.\sum_{k<n=1}^{3}\mu_k\mu_n(\p_1\rho_k\p_2\rho_n-\p_2\rho_k\p_1\rho_n)^2\right.
\\
\nn
&-&\left.\sum_{k=1}^{3}\mu_k\left(\p_1\rho_k\sum_{n=1}^{3}\nu_n\p_2\rho_n
- \p_2\rho_k\sum_{n=1}^{3}\nu_n\p_1\rho_n\right)^2\right\}+
\Lambda_S\left(\sum_{a=1}^{4}r_a^2-1\right),\eea where \bea\nn
(\mu_1,\mu_2,\mu_3)=(r_1^2+r_2^2,r_3^2,r_4^2),
(\nu_1,\nu_2,\nu_3)=(r_1^2-r_2^2,r_3^2,-r_4^2),
(\rho_1,\rho_2,\rho_3)=(\varphi,\phi,\tilde{\phi}) .\eea

Now, we are ready to face our main problem: how to reduce this
Lagrangian to the one corresponding to the thermodynamic limit of
spin chain, {\it without shrinking the membrane to string}? We
propose the following solution of this task: \bea\nn &&
\alpha=\alpha(\tau,\s_1),\h r_1=r_1(\tau,\s_1),\h
r_2=r_2(\tau,\s_1),
\\ \nn
&&r_3=r_3(\tau,\s_2)=\epsilon\sin[b\s_2+c(\tau)],\h
r_4=r_4(\tau,\s_2)=\epsilon\cos[b\s_2+c(\tau)],
\\ \nn
&&\varphi=\varphi(\tau,\s_1),\h
\epsilon,b,\phi,\tilde{\phi}=constants,\h \epsilon^2<1. \eea These
restrictions lead to
\bea\nn \mathcal{L}
&=&\frac{(2l_p\mathcal{R})^2}{\lambda^0}\kappa\left[\p_0\alpha +
(r_1^2-r_2^2)\p_0\varphi\right]-\lambda^0(\epsilon
bT_2)^2(4l_p\mathcal{R})^4
\left\{\sum_{a=1}^{2}(\p_1r_a)^2\right.
\\ \nn &+&
\left.\left[(r_1^2+r_2^2)-(r_1^2-r_2^2)^2\right](\p_1\varphi)^2\right\}
+\Lambda_S\left[\sum_{a=1}^{2}r_a^2-(1-\epsilon^2)\right].\eea If
we introduce the parametrization \bea\nn
r_1=(1-\epsilon^2)^{1/2}\cos\psi,\h
r_2=(1-\epsilon^2)^{1/2}\sin\psi,\eea the new variable
$\tilde{\alpha}$=$\alpha/(1-\epsilon^2)$, and take the limit
$\epsilon^2\to 0$ neglecting the terms of order higher than
$\epsilon^2$, we will receive \bea\nn
\frac{\mathcal{L}}{1-\epsilon^2}=\frac{(2l_p\mathcal{R})^2}{\lambda^0}\kappa\left[\p_0\tilde{\alpha}
+\cos(2\psi)\p_0\varphi\right] -\lambda^0(\epsilon
bT_2)^2(4l_p\mathcal{R})^4\left[(\p_1\psi)^2
+\sin^2(2\psi)(\p_1\varphi)^2\right].\eea As for the membrane
action corresponding to the above Lagrangian, it can be
represented in the form \bea\nn S_M=\frac{\mathcal{J}}{2\pi}\int
dt d\s\left[\p_t\tilde{\alpha} +
\cos(2\psi)\p_t\varphi\right]-\frac{\tilde{\lambda}}{4\pi\mathcal{J}}\int
dt d\sigma
\left[\left(\p_\s\psi\right)^2+\sin^2(2\psi)(\p_\s\varphi)^2\right],\eea
where $\mathcal{J}$ is the angular momentum conjugated to
$\tilde{\alpha}$, $t=\kappa\tau$ and \bea\nn
\tilde{\lambda}=2^{15}(\pi^2\epsilon
bT_2)^2(l_p\mathcal{R})^6.\eea This action corresponds to the
thermodynamic limit of $SU(2)$ {\it integrable spin chain}
\cite{K0311}.

\subsubsection{M2-brane perspective on $\mathcal{N}=6$
super Chern-Simons-matter theory at level k}

In 2008, O. Aharony, O. Bergman, D. L. Jafferis and J. Maldacena
(ABJM) proposed three-dimensional super Chern-Simons-matter
theory, which at level $k$ is supposed to describe the low energy
limit of $N$ M2-branes \cite{ABJM}. For large $N$ and $k$, but
fixed 't Hooft coupling $\lambda=N/k$, it is dual to type IIA
string theory on $AdS_4\times \mathbb{CP}^3$. For large $N$ but
{\it finite} $k$, it is dual to M theory on $AdS_4\times S^7/Z_k$.
Here, relying on the second duality, we find exact giant magnon
and single spike solutions of membrane configurations on
$AdS_4\times S^7/Z_k$ by reducing the system to the
Neumann-Rosochatius integrable model. We derive the dispersion
relations and their finite-size corrections with explicit
dependence on the level $k$ \cite{16}.

Let us introduce the following complex coordinates on the
$S^7/Z_k$ subspace \bea\nn &&z_1=\cos\psi\cos\frac{\theta_1}{2}
e^{i\left[\frac{\varphi}{k}+\frac{1}{2}\left(\phi_1+\phi_3\right)\right]},
\h z_2=\cos\psi\sin\frac{\theta_1}{2}
e^{i\left[\frac{\varphi}{k}-\frac{1}{2}\left(\phi_1-\phi_3\right)\right]},
\\ \nn &&z_3=\sin\psi\cos\frac{\theta_2}{2}
e^{i\left[\frac{\varphi}{k}+\frac{1}{2}\left(\phi_2-\phi_3\right)\right]},\h
z_4=\sin\psi\sin\frac{\theta_2}{2}
e^{i\left[\frac{\varphi}{k}-\frac{1}{2}\left(\phi_2+\phi_3\right)\right]}.
\eea Obviously, they satisfy the relation \bea\nn
\sum_{a=1}^{4}z_a\bar{z}_a\equiv 1.\eea Next, we compute the
metric \bea\nn ds^2_{S^7/Z_k}=\sum_{a=1}^{4}dz_a
d\bar{z}_a=\frac{1}{k^2}\left(d\varphi+kA_1\right)^2+ds^2_{\mathbb
{CP}^3},\eea where \bea\nn
&&A_1=\frac{1}{2}\left[\cos^2\psi\cos\theta_1
d\phi_1+\sin^2\psi\cos\theta_2 d\phi_2 +
\left(\cos^2\psi-\sin^2\psi\right)d\phi_3\right], \\ \nn
&&ds^2_{\mathbb {CP}^3}=
d\psi^2+\sin^2\psi\cos^2\psi\left(\frac{1}{2}\cos\theta_1
d\phi_1-\frac{1}{2}\cos\theta_2 d\phi_2+d\phi_3\right)^2 \\
\label{cp3m}
&&+\frac{1}{4}\cos^2\psi\left(d\theta_1^2+\sin^2\theta_1
d\phi_1^2\right)+\frac{1}{4}\sin^2\psi\left(d\theta_2^2+\sin^2\theta_2
d\phi_2^2\right).\eea

The membrane embedding into $R_t\times S^7/Z_k$, appropriate for
our purposes, is \bea\nn X_{0}=\frac{R}{2}t(\xi^m), \h W_{a}=R
r_a(\xi^m)e^{i\varphi_a(\xi^m)},\h a=(1,2,3,4),\eea where $t$ is
the $AdS$ time, $r_a$ are real functions of $\xi^m$, while
$\varphi_a$ are the isometric coordinates on which the background
metric does not depend. The four complex coordinates $W_{a}$ are
restricted by the real embedding condition \bea\nn
\sum_{a=1}^{4}W_{a}\bar{W}_{a}=R^2,\h \mbox{or}\h
\sum_{a=1}^{4}r_a^2=1.\eea The coordinates $r_a$ are connected to
the initial coordinates, on which the background depends, in an
obvious way.

For the embedding described above, the metric induced on the
M2-brane worldvolume is given by \bea\label{pim}
G_{mn}=\frac{R^2}{4}\left[-\p_mt\p_nt+
4\sum_{a=1}^{4}\left(\p_mr_a\p_nr_a +
r_a^2\p_m\varphi_a\p_n\varphi_a\right)\right].\eea
Correspondingly, the membrane Lagrangian becomes \bea\nn
\mathcal{L}=\mathcal{L}_{M}+\Lambda\left(\sum_{a=1}^{4}r_a^2-1\right),\eea
where $\Lambda$ is a Lagrange multiplier.

{\bf NR integrable system for M2-branes on $R_t\times S^7/Z_k$}

Let us consider the following particular case of the above
membrane embedding \cite{11} \bea\label{pnra}
&&X_{0}=\frac{R}{2}\kappa\tau,\h
W_{a}=Rr_a(\xi,\eta)e^{i\left[\omega_{a}\tau+\mu_a(\xi,\eta)\right]},
\\ \nn && \xi=\alpha\sigma_1+\beta\tau,\h \eta=\gamma\sigma_2+\delta\tau,\eea
which implies \bea\label{pi} t=\kappa\tau,\h
\varphi_a(\xi^m)=\varphi_a(\tau,\sigma_1,\sigma_2)=
\omega_{a}\tau+\mu_a(\xi,\eta).\eea Here $\kappa$, $\omega_{a}$,
$\alpha$, $\beta$, $\gamma$, $\delta$ are parameters. For this
ansatz, the membrane Lagrangian takes the form ($\p_\xi=\p/\p\xi$,
$\p_\eta=\p/\p\eta$) \bea\nn &&\mathcal{L}=-\frac{R^2}{4\lambda^0}
\left\{\left(2\lambda^0T_2R\alpha\gamma\right)^2
\sum_{a<b=1}^{4}\left[(\p_\xi r_a\p_\eta r_b-\p_\eta r_a\p_\xi r_b)^2\right. \right. \\
\nn &&+ \left. \left. (\p_\xi r_a\p_\eta\mu_b-\p_\eta
r_a\p_\xi\mu_b)^2r_b^2 + (\p_\xi\mu_a\p_\eta
r_b-\p_\eta\mu_a\p_\xi r_b)^2r_a^2\right.\right.
\\ \nn &&+\left.\left.(\p_\xi\mu_a\p_\eta\mu_b-\p_\eta\mu_a\p_\xi\mu_b)^2r_a^2r_b^2 \right]\right. \\
\nn &&+
\left.\sum_{a=1}^{4}\left[\left(2\lambda^0T_2R\alpha\gamma\right)^2
(\p_\xi r_a\p_\eta\mu_a-\p_\eta r_a\p_\xi\mu_a)^2-
\left(\beta\p_\xi\mu_a+\delta\p_\eta\mu_a+\omega_a\right)^2\right]r_a^2\right.
\\
\nn &&-\left.\sum_{a=1}^{4}\left(\beta\p_\xi r_a+\delta\p_\eta
r_a\right)^2+(\kappa/2)^2\right\}+\Lambda\left(\sum_{a=1}^{4}r_a^2-1\right).\eea

We have found a set of sufficient conditions, which reduce the
above Lagrangian to the NR one. First of all, two of the angles
$\varphi_a$ should be set to zero. The corresponding $r_a$
coordinates must depend only on $\eta$ in a specific way. The
remaining variables $r_a$ and $\mu_a$ can depend only on
$\xi$\footnote{Of course, the roles of $\xi$ and $\eta$ can be
interchanged in this context.}. In principle, there are six such
possibilities. How they are realized for the $R_t\times S^7/Z_k$
background, we will discuss in the next section. Here, we will
work out the following example \bea\nn &&r_{1}=r_{1}(\xi),\h
r_{2}=r_{2}(\xi),\h \mu_1=\mu_1(\xi),\h \mu_2=\mu_2(\xi),\\
\label{pNRA}
&&r_3=r_3(\eta)=r_0\sin\eta,\h r_4=r_4(\eta)=r_0\cos\eta,\h r_0<1,\\
\nn &&\varphi_3=\varphi_4=0.\eea For this choice, we receive
(prime is used for $d/d\xi$) \bea\nn
&&\mathcal{L}=-\frac{R^2}{4\lambda^0}
\left\{\sum_{a=1}^{2}\left[(\tilde{A}^2-\beta^2)r_a'^2\right.+
\left.(\tilde{A}^2-\beta^2)r_a^2\left(\mu'_a-\frac{\beta\omega_a}{\tilde{A}^2-\beta^2}\right)^2
- \frac{\tilde{A}^2}{\tilde{A}^2-\beta^2}\omega_a^2 r_a^2\right]\right. \\
\nn &&+\left. (\kappa/2)^2-r_0^2\delta^2\right\} +
\Lambda\left[\sum_{a=1}^{2}r_a^2-(1-r_0^2)\right],\eea where $
\tilde{A}^2\equiv \left(2\lambda^0T_2R\alpha\gamma r_0\right)^2$.
Now we can integrate once the equations of motion for $\mu_a$
following from the above Lagrangian to get \bea\label{mus}
\mu'_a=\frac{1}{\tilde{A}^2-\beta^2}\left(\frac{C_a}{r_a^2}+\beta\omega_a\right),\eea
where $C_a$ are arbitrary constants. By using (\ref{mus}) in the
equations of motion for $r_a(\xi)$, one finds that they can be
obtained from the effective Lagrangian \bea\nn L_{NR}=
\sum_{a=1}^{2}\left[(\tilde{A}^2-\beta^2)r_a'^2 -
\frac{1}{\tilde{A}^2-\beta^2}\left(\frac{C_a^2}{r_a^2} +
\tilde{A}^2\omega_a^2
r_a^2\right)\right]+\Lambda_M\left[\sum_{a=1}^{2}r_a^2-(1-r_0^2)\right].\eea
This Lagrangian, in full analogy with the string considerations,
corresponds to particular case of the NR integrable system. For
$C_a=0$ one obtains the Neumann integrable model, which in the
case at hand describes two-dimensional harmonic oscillator,
constrained to a circle of radius $\sqrt{1-r_0^2}$.

Let us consider the three constraints (\ref{00gf}), (\ref{0igf})
for the present case. For more close correspondence with the
string case, we want the third one, $G_{02}=0$, to be identically
satisfied. To this end, since $G_{02}\sim r_0^2\gamma\delta,$ we
set $\delta=0$, i.e. $\eta=\gamma\sigma_2$. Then, the first two
constraints give the conserved Hamiltonian $H_{NR}$ and a relation
between the parameters involved: \bea\nn
&&H_{NR}=\sum_{a=1}^{2}\left[(\tilde{A}^2-\beta^2)r_a'^2+
\frac{1}{\tilde{A}^2-\beta^2}\left(\frac{C_a^2}{r_a^2} +
\tilde{A}^2\omega_a^2
r_a^2\right)\right]=\frac{\tilde{A}^2+\beta^2}{\tilde{A}^2-\beta^2}
(\kappa/2)^2,
\\ \label{peffcs}&&\sum_{a=1}^{2}\omega_{a}C_a +
\beta(\kappa/2)^2=0.\eea

For closed membranes, $r_a$ and $\mu_a$ must satisfy the following
periodicity conditions \bea
r_a(\xi+2\pi\alpha,\eta+2\pi\gamma)=r_a(\xi,\eta),\h
\mu_a(\xi+2\pi\alpha,\eta+2\pi\gamma)=\mu_a(\xi,\eta)+2\pi
n_a,\label{ppbc} \eea where $n_a$ are integer winding numbers. In
particular, $\gamma$ is a non-zero integer.

Since the background metric does not depend on $t$ and
$\varphi_a$, the corresponding conserved quantities are the
membrane energy $E$ and four angular momenta $J_a$, defined by
\bea\nn E=-\int d^2\sigma\frac{\p\mathcal{L}}{\p(\p_0 t)},\h
J_a=\int d^2\sigma\frac{\p\mathcal{L}}{\p(\p_0\varphi_a)},\h
a=1,2,3,4.\eea For our ansatz (\ref{pNRA}) $J_3=J_4=0$. The energy
and the other two angular momenta are given by \bea\label{pcqs}
E=\frac{\pi R^2\kappa}{4\lambda^0\alpha}\int d\xi,\h J_a=\frac{\pi
R^2}{\lambda^0\alpha(\tilde{A}^2-\beta^2)}\int d\xi \left(\beta
C_a + \tilde{A}^2\omega_a r_a^2\right),\h a=1,2.\eea From here, by
using the constraints (\ref{peffcs}), one obtains the
energy-charge relation \bea\nn
\frac{4}{\tilde{A}^2-\beta^2}\left[\tilde{A}^2(1-r_0^2) +
\beta\sum_{a=1}^{2}\frac{C_a}{\omega_a}\right]\frac{E}{\kappa}
=\sum_{a=1}^{2}\frac{J_a}{\omega_a}.\eea As usual, it is linear
with respect to $E$ and $J_a$ before taking the semiclassical
limit.

To identically satisfy the embedding condition \bea\nn
\sum_{a=1}^{2}r_a^2-(1-r_0^2)=0,\eea we set \bea\nn
r_1(\xi)=\sqrt{1-r_0^2}\sin\theta(\xi),\h
r_2(\xi)=\sqrt{1-r_0^2}\cos\theta(\xi).\eea Then from the
conservation of the NR Hamiltonian (\ref{peffcs}) one finds
\bea\label{mtsol} &&\theta'=\frac{\pm 1}{\tilde{A}^2-\beta^2}
\left[(\tilde{A}^2+\beta^2)\tilde{\kappa}^2 -
\frac{\tilde{C}_1^2}{\sin^2{\theta}} -
\frac{\tilde{C}_2^2}{\cos^2{\theta}} -
\tilde{A}^2\left(\omega_1^2\sin^2{\theta}
+\omega_2^2\cos^2{\theta}\right)\right]^{1/2},\\ \nn
&&\sum_{a=1}^{2}\omega_{a}\tilde{C}_a + \beta\tilde{\kappa}^2=0,\h
\tilde{\kappa}^2=\frac{(\kappa/2)^2}{1-r_0^2},\h
\tilde{C}_a^2=\frac{C_a^2}{(1-r_0^2)^2}.\eea By replacing the
solution for $\theta(\xi)$ received from (\ref{mtsol}) into
(\ref{mus}), one obtains the solutions for $\mu_a$:
\bea\label{mu12s}
\mu_1=\frac{1}{\tilde{A}^2-\beta^2}\left(\tilde{C}_1\int\frac{d\xi}{\sin^2{\theta}}
+ \beta\omega_1\xi\right),\h
\mu_2=\frac{1}{\tilde{A}^2-\beta^2}\left(\tilde{C}_2\int\frac{d\xi}{\cos^2{\theta}}
+ \beta\omega_2\xi\right).\eea

The above analysis shows that the NR integrable models for
membranes on $R_t\times S^7$ and $R_t\times S^7/Z_k$ are the same
\cite{15}. Therefore, we can use the results obtained in \cite{15}
for the present case. For convenience, the corresponding solutions
and dispersion relations are given in Appendix.

{\bf M2-brane solutions on $R_t\times S^7/Z_k$ and dispersion
relations}

For our membrane embedding in $R_t\times S^7/Z_k$, the angular
variables $\varphi_a$ are related to the corresponding background
coordinates as follows \bea\nn
&&\varphi_1=\frac{\varphi}{k}+\frac{1}{2}\left(\phi_1+\phi_3\right),
\h\varphi_2=\frac{\varphi}{k}-\frac{1}{2}\left(\phi_1-\phi_3\right),
\\ \nn &&\varphi_3=\frac{\varphi}{k}+\frac{1}{2}\left(\phi_2-\phi_3\right),
\h\varphi_4=\frac{\varphi}{k}-\frac{1}{2}\left(\phi_2+\phi_3\right).\eea
As a consequence, for the angular momenta we have \bea\nn
&&J_{\varphi_1}=\frac{J_{\varphi}}{k}+\frac{1}{2}\left(J_{\phi_1}+J_{\phi_3}\right),
\h
J_{\varphi_2}=\frac{J_{\varphi}}{k}-\frac{1}{2}\left(J_{\phi_1}-J_{\phi_3}\right),
\\ \nn && J_{\varphi_3}=\frac{J_{\varphi}}{k}+\frac{1}{2}\left(J_{\phi_2}-J_{\phi_3}\right),
\h
J_{\varphi_4}=\frac{J_{\varphi}}{k}-\frac{1}{2}\left(J_{\phi_2}+J_{\phi_3}\right).\eea
$\varphi_a$ and $J_{\varphi_a}$ satisfy the equalities \bea\nn
\sum_{a=1}^{4}\varphi_a=\frac{4}{k}\varphi,\h
\sum_{a=1}^{4}J_{\varphi_a}=\frac{4}{k}J_{\varphi}.\eea

One of the conditions for the existence of NR description of the
M2-brane dynamics is that two of the angles $\varphi_a$ must be
zero, which means that two of the four angular momenta
$J_{\varphi_a}$ vanish. The six possible cases are \bea\nn
&&\bullet\
\varphi_1=\phi_3+\frac{\phi_1}{2}=\frac{2}{k}\varphi+\frac{\phi_1}{2},\h
\varphi_2=\phi_3-\frac{\phi_1}{2}=\frac{2}{k}\varphi-\frac{\phi_1}{2},
\h \varphi_3=0,\h \varphi_4=0; \\ \nn &&\bullet\
\varphi_1=\phi_1=\frac{2}{k}\varphi+\phi_3,\h
\varphi_3=\phi_2=\frac{2}{k}\varphi-\phi_3,\h \varphi_2=0,\h
\varphi_4=0; \\ \label{spc} &&\bullet\
\varphi_1=\phi_1=\frac{2}{k}\varphi+\phi_3,\h
\varphi_4=-\phi_2=\frac{2}{k}\varphi-\phi_3,\h \varphi_2=0,\h
\varphi_3=0;
\\ \nn &&\bullet\ \varphi_2=-\phi_1=\frac{2}{k}\varphi+\phi_3,\h \varphi_3=\phi_2=\frac{2}{k}\varphi-\phi_3,\h
\varphi_1=0,\h \varphi_4=0;
\\ \nn &&\bullet\ \varphi_2=-\phi_1=\frac{2}{k}\varphi+\phi_3,\h \varphi_4=-\phi_2=\frac{2}{k}\varphi-\phi_3,\h
\varphi_1=0,\h \varphi_3=0; \\ \nn &&\bullet\
\varphi_3=-\phi_3+\frac{\phi_2}{2}=\frac{2}{k}\varphi+\frac{\phi_2}{2},\h
\varphi_4=-\phi_3-\frac{\phi_2}{2}=\frac{2}{k}\varphi-\frac{\phi_2}{2},
\h \varphi_1=0,\h \varphi_2=0 .\eea Here, $\phi_1$ and $\phi_2$
are the isometry angles on the two two-spheres inside $\mathbb
{CP}^3$, while $\phi_3$ is isometry angle on the $U(1)$ fiber over
$S^2\times S^2$, as can be seen from (\ref{cp3m}).

From (\ref{spc}) it is clear that we have two alternative
descriptions for $\varphi_a$. One is only in terms of the isometry
angles on $\mathbb {CP}^3$, and the other includes the eleventh
coordinate $\varphi$. This is a consequence of our restriction to
M2-brane configurations, which can be described by the NR
integrable system.

The six cases above can be divided into two classes. The first one
contains the first and last possibilities, and the other one - the
remaining ones. The cases belonging to the first class are related
to each other by the exchange of $\phi_1$ and $\phi_2$. This
corresponds to exchanging the two $S^2$ inside $\mathbb {CP}^3$.
Since these spheres enter symmetrically, the two cases are
equivalent. In terms of ($\varphi$,$\phi_3$), the four cases from
the second class are actually identical. That is why, all of them
can be described simultaneously by choosing one representative
from the class.

Let us first give the M2-brane solutions for cases in the first
class. Since they correspond to our example in the previous
section, the membrane configuration reads \bea\nn
&&W_1=Rr_1(\xi)\exp\left\{i\varphi_1(\tau,\xi)
\right\}=R\sqrt{1-r_0^2}\sin\theta(\xi)\exp\left\{i\left[\frac{2}{k}\varphi(\tau,\xi)
+ \frac{\phi(\tau,\xi)}{2}\right]\right\},\\ \nn
&&W_2=Rr_2(\xi)\exp\left\{i\varphi_2(\tau,\xi)
\right\}=R\sqrt{1-r_0^2}\cos\theta(\xi)\exp\left\{i\left[\frac{2}{k}\varphi(\tau,\xi)
-\frac{\phi(\tau,\xi)}{2}\right]\right\},\\ \nn
&&W_3=Rr_0\sin(\gamma\sigma_2),\h
W_4=Rr_0\cos(\gamma\sigma_2),\eea where $\phi$ is equal to
$\phi_1$ or $\phi_2$.

From the NR system viewpoint, the membrane solutions for the
second class configurations differ from the ones just given by the
exchange of $W_2$, $W_3$, and by the replacement
$\phi/2\to\phi_3$. In other words, we have \bea\nn
&&W_1=Rr_1(\xi)\exp\left\{i\varphi_1(\tau,\xi)
\right\}=R\sqrt{1-r_0^2}\sin\theta(\xi)\exp\left\{i\left[\frac{2}{k}\varphi(\tau,\xi)
+ \phi_3(\tau,\xi)\right]\right\},\\ \nn
&&W_2=Rr_0\sin(\gamma\sigma_2),\\ \nn
&&W_3=Rr_3(\xi)\exp\left\{i\varphi_3(\tau,\xi)
\right\}=R\sqrt{1-r_0^2}\cos\theta(\xi)\exp\left\{i\left[\frac{2}{k}\varphi(\tau,\xi)
-\phi_3(\tau,\xi)\right]\right\},\\ \nn
&&W_4=Rr_0\cos(\gamma\sigma_2),\eea

The explicit solutions for $\theta(\xi)$ and
$\varphi_{1,2,3}(\tau,\xi)$, of the M2-brane GM and SS, along with
the energy-charge relations for the infinite and finite sizes are
given in the Appendix E (see \cite{16}). Here, we will present
them in terms of $\varphi$ and $\phi_{1,2,3}$.

In accordance with (\ref{EJGM0}), we have for the M2-brane GM with
two angular momenta the following dispersion relation
\bea\label{gm0}
\sqrt{1-r_0^2}E-\frac{1}{2}\left(\frac{2}{k}J_{\varphi}+J_{\phi}\right)
=\sqrt{\frac{1}{4}\left(\frac{2}{k}J_{\varphi}-J_{\phi}\right)^2 +
8\lambda
k^2\left[r_0(1-r_0^2)\gamma\right]^2\sin^2\frac{p}{2}},\eea where
$J_{\phi}$ can be equal to $J_{\phi_1}/2$, $J_{\phi_2}/2$ or
$J_{\phi_3}$. In writing (\ref{gm0}), we have used that \bea\nn
R=l_p\left(2^5\pi^2kN\right)^{1/6},\h
T_2=\frac{1}{(2\pi)^2l_p^3},\eea and the 't Hooft coupling is
defined by $\lambda=N/k$.

If we introduce the notations \bea\label{mnot}
\mathcal{E}=a\sqrt{1-r_0^2}E,\h
\mathcal{J}_{\varphi}=a\frac{J_{\varphi}}{2},\h
\mathcal{J}_{\phi}=a\frac{J_{\phi}}{2},\h
a=\frac{1}{\sqrt{2\lambda}kr_0(1-r_0^2)\gamma},\eea the above
energy-charge relation takes the form \bea\nn
\mathcal{E}-\mathcal{J}_1(k) =\sqrt{\mathcal{J}^2_2(k) +
4\sin^2\frac{p}{2}},\eea where \bea\label{Jnot}
\mathcal{J}_1(k)=\frac{2}{k}\mathcal{J}_{\varphi}+\mathcal{J}_{\phi},\h
\mathcal{J}_2(k)=\frac{2}{k}\mathcal{J}_{\varphi}-\mathcal{J}_{\phi}.\eea

By using (\ref{mnot}), (\ref{Jnot}) and (\ref{pIEJ1}), we can
write down the dispersion relation for the dyonic GM, including
the leading finite-size correction as \bea\nn
&&\mathcal{E}-\mathcal{J}_1(k) =\sqrt{\mathcal{J}^2_2(k) +
4\sin^2\frac{p}{2}} - \frac{16 \sin^4\frac{p}{2}}
{\sqrt{\mathcal{J}^2_2(k)+4\sin^2\frac{p}{2}}}\\
\nn &&\exp\left[-\frac{2\sin^2\frac{p}{2}\left(\mathcal{J}_1(k) +
\sqrt{\mathcal{J}^2_2(k)+4\sin^2\frac{p}{2}}\right)
\sqrt{\mathcal{J}^2_2(k) +4\sin^2\frac{p}{2}}}{\mathcal{J}^2_2(k)
+4\sin^4\frac{p}{2}}\right].\eea

The reason to introduce $\mathcal{E}$, $\mathcal{J}_{\varphi}$ and
$\mathcal{J}_{\phi}$ namely in this way is the following. For GM
on the $R_t\times S^3$ subspace of $AdS_5\times S^5$, in terms of
\bea\nn \mathcal{E}=\frac{2\pi}{\sqrt{\lambda}}E,\h
\mathcal{J}_{1}=\frac{2\pi}{\sqrt{\lambda}}J_1,\h
\mathcal{J}_{2}=\frac{2\pi}{\sqrt{\lambda}}J_2,\eea we have
\bea\nn \mathcal{E}-\mathcal{J}_{1} =\sqrt{\mathcal{J}_{2}^2 +
4\sin^2\frac{p}{2}}.\eea The same result can be obtained for the
GM on the $R_t\times \mathbb {CP}^3$ subspace of $AdS_4\times
\mathbb {CP}^3$, if we use the identification \cite{ABR} \bea\nn
\mathcal{E}=\frac{E}{\sqrt{2\lambda}},\h
\mathcal{J}_{1}=\frac{J_1}{\sqrt{2\lambda}},\h
\mathcal{J}_{2}=\frac{J_2}{\sqrt{2\lambda}}.\eea In the all three
cases, the second term under the square root is the same. In this
description it is universal - for different backgrounds and for
different extended objects.

Analogously, for the SS case one can find (see (\ref{pssS3c}))
\bea\nn &&\mathcal{E}-\Delta\varphi_1=
p+8\sin^2\frac{p}{2}\tan\frac{p}{2}
\exp\left(-\frac{(\Delta\varphi_1+ p)\tan\frac{p}{2}}
{\mathcal{J}^2_2(k) \csc^2p+\tan^2\frac{p}{2}}\right) \\ \nn
&&\mathcal{J}_1(k)= \sqrt{\mathcal{J}^2_2(k)
+4\sin^2\frac{p}{2}}.\eea

Let us point out that for $k=1$ the above dispersion relations
coincide with the ones obtained earlier in \cite{15}. We can also
reproduce the energy-charge relations for dyonic GM and SS strings
on $R_t\times \mathbb {CP}^3$ by taking an appropriate limit. To
show this, let us consider the second case in (\ref{spc}), for
which \bea\nn J_{\phi_1}=\frac{2}{k}J_{\varphi}+J_{\phi_3},\h
J_{\phi_2}=\frac{2}{k}J_{\varphi}-J_{\phi_3}.\eea In accordance
with our membrane embedding, the following identification should
be made \bea\nn J_1^{str}=\frac{J_{\phi_1}}{2},\h
J_2^{str}=\frac{J_{\phi_2}}{2}.\eea Then in the limit $k\to
\infty$, $r_0\to 0$, such that $kr_0\gamma=1$, we obtain from
(\ref{gm0}) \bea\nn E-J_1^{str}=\sqrt{(J_2^{str})^2 + 8\lambda
\sin^2\frac{p}{2}}.\eea This is exactly what we have derived in
\cite{ABR} for dyonic GM strings on $R_t\times \mathbb {CP}^3$.
Obviously, this also applies for the leading finite-size
correction. In the same way, the SS string dispersion relation for
$R_t\times \mathbb {CP}^3$ background can be reproduced.

\newpage

\setcounter{equation}{0}
\section{Three-point correlation functions}

The  AdS/CFT conjecture \cite{AdS/CFT} implies   that the
correlation functions in the dual   (boundary)   quantum field
theory  can be computed alternatively  in   string theory,  i.e.,
essentially by the methods of  a  two - dimensional   theory. The
first computations however  were  mostly   performed  in the
supergravity approximation, representing the correlators in terms
of integrals over the target (bulk) coordinates  \cite{FMMR,LMRS}.

On the string side one may start with a  semiclassical approach,
when the string path integral for the correlation functions   is
evaluated in the saddle-point approximation with  large  't Hooft
coupling $\lambda >>1$. In this calculation one has to identify
the correct vertex operators \cite{P,AT} and to find the
corresponding classical solutions, which provide the appropriate
saddle-point approximation.

It is known that the correlation functions of any conformal field
theory can be determined  in principle in terms of the basic
conformal data $\{\Delta_i,C_{ijk}\}$, where $\Delta_i$ are the
conformal dimensions defined by the two-point correlation
functions
\begin{equation}\nn
\left\langle{\cal O}^{\dagger}_i(x_1){\cal O}_j(x_2)\right\rangle=
\frac{C_{12}\delta_{ij}}{|x_1-x_2|^{2\Delta_i}}
\end{equation}
and $C_{ijk}$ are the structure constants in the operator product
expansion
\begin{equation}\nn
\left\langle{\cal O}_i(x_1){\cal O}_j(x_2){\cal
O}_k(x_3)\right\rangle=
\frac{C_{ijk}}{|x_1-x_2|^{\Delta_1+\Delta_2-\Delta_3}
|x_1-x_3|^{\Delta_1+\Delta_3-\Delta_2}|x_2-x_3|^{\Delta_2+\Delta_3-\Delta_1}}.
\end{equation}
Therefore, the determination of the initial conformal data for a
given conformal field theory is the most important step in the
conformal bootstrap approach.

The three-point functions of two ``heavy'' operators and a
``light'' operator can be approximated by a supergravity vertex
operator evaluated at the ``heavy'' classical string configuration
\cite{rt10,Hernandez2}: \bea \nn \langle
V_{H}(x_1)V_{H}(x_2)V_{L}(x_3)\rangle=V_L(x_3)_{\rm classical}.
\eea For $\vert x_1\vert=\vert x_2\vert=1$, $x_3=0$, the
correlation function reduces to \bea \nn \langle
V_{H}(x_1)V_{H}(x_2)V_{L}(0)\rangle=\frac{C_{123}}{\vert
x_1-x_2\vert^{2\Delta_{H}}}. \eea Then, the normalized structure
constants \bea \nn \mathcal{C}=\frac{C_{123}}{C_{12}} \eea can be
found from \bea \label{nsc} \mathcal{C}=c_{\Delta}V_L(0)_{\rm
classical}, \eea were $c_{\Delta}$ is the normalized constant of
the corresponding ``light'' vertex operator.

\subsection{Semiclassical three-point correlation functions
in $AdS_5 \times S^5$}

In \cite{19} we computed holographic three-point correlation
functions or structure constants of a zero-momentum dilaton
operator and two (dyonic) giant magnon string states with a
finite-size length in the semiclassical approximation. We show
that the semiclassical structure constants match exactly with the
three-point functions between two $su(2)$ magnon single trace
operators with finite size and the Lagrangian in the large 't
Hooft coupling constant limit. A special limit
$J\gg\sqrt{\lambda}$ of our result is compared with the relevant
result based on the L\"uscher corrections.

\cite{19} is the first paper where the finite-size effects on the
semiclassical three-point correlation functions have been taken
into account.

In \cite{22,23,25} we extended the results found in \cite{19} to
the following three cases:
\begin{enumerate}
\item{Dilaton operator with non-zero momentum: $V_L=V^d_j$}
\item{Primary scalar operators: $V_L=V^{pr}_j$} \item{Singlet
scalar operators on higher string levels: $V_L= V^q$}
\end{enumerate}

\subsubsection{Two GM states and dilaton with zero momentum}

Let us start with the case of two GM and the zero-momentum dilaton
operator, namely the Lagrangian whose vertex operator is given by
\bea \label{dv}
V^d=\left(Y_4+Y_5\right)^{-4}\left[z^{-2}\left(\p_+x_{m}\p_-x^{m}+\p_+z\p_-z\right)
+\p_+X_{k}\p_-X_{k}\right], \eea where \bea \nn
Y_4=\frac{1}{2z}\left(x^mx_m+z^2-1\right), \h
Y_5=\frac{1}{2z}\left(x^mx_m+z^2+1\right), \eea and $x_m$, $z$ are
coordinates on $AdS_5$, while $X_k$ are the coordinates on $S^5$.

{\bf Giant magnons with finite size}

The finite-size giant magnon solution can be represented as
($i\tau=\tau_e$) \bea\nn &&x_{0e}=\tanh(\kappa\tau_e),\h x_i=0,\h
z=\frac{1}{\cosh(\kappa\tau_e)},
\\ \label{fsgm} &&\cos\theta=\sqrt{1-v^2\kappa^2}\
\mathbf{dn}\left(\frac{\sqrt{1-v^2\kappa^2}}{1-v^2}(\sigma-v\tau)\Big\vert
1-\epsilon\right) ,
\\ \nn &&\phi= \frac{\tau-v\sigma}{1-v^2}+\frac{1}{v\sqrt{1-v^2\kappa^2}}\times
\\ \nn &&
\Pi\left(am\left(\frac{\sqrt{1-v^2\kappa^2}}{1-v^2}(\sigma-v\tau)\right),
-\frac{1-v^2\kappa^2}{v^2\kappa^2}\left(1-\epsilon\right),
\Big\vert 1-\epsilon\right),\eea where
\bea\nn
\epsilon=\frac{1-\kappa^2}{1-v^2\kappa^2}.\eea

To find the finite-size effect on the three-point correlator, we
will use (\ref{nsc}) and (\ref{dv}), which computed on
(\ref{fsgm}) gives \bea\label{c3d}
\mathcal{C}^d=c_{\Delta}^{d}\int_{-\infty}^{\infty}\frac{d\tau_e}{\cosh^4(\kappa\tau_e)}
\int_{-L}^{L}d\sigma\left[\kappa^2 +\p_+X_{k}\p_-X_{k}\right],\eea
where\bea\nn \p_+X_{k}\p_-X_{k}&=&-\frac{1}{(1-v^2)\sin^2\theta}
\left\{2-(1+v^2)\kappa^2\right.
\\ \nn &&-\left.\cos^2\theta\left[4-(1+v^2)\kappa^2
-2\cos^2\theta\right]\right\} .\eea

Performing the integrations in (\ref{c3d}), one finds
\bea\label{exact1} \mathcal{C}^d=\frac{16}{3}c_{\Delta}^{d}
\sqrt{\frac{1-v^2}{1-\epsilon}}
\left[\mathbf{E}(1-\epsilon)-\epsilon \
\mathbf{K}(1-\epsilon)\right].\eea Let us point out that the
parameter $L$ in (\ref{exact1}) is given by \bea\nn
L=\frac{1-v^2}{\sqrt{1-v^2\kappa^2}}\mathbf{K}(1-\epsilon).\eea
This is our {\it exact} result for the normalized coefficient
$\mathcal{C}^d$ in the semiclassical three-point correlation
function, corresponding to the case when the "heavy" vertex
operators are {\it finite-size} giant magnons, and the light
vertex is taken to be the zero-momentum dilaton operator.

For the case of this dilaton operator, the three-point function of
the SYM can be easily related to the conformal dimension of the
heavy operators. This corresponds to shift `t Hooft coupling
constant which is the overall coefficient of the Lagrangian
\cite{Costa}. This gives an important relation between the
structure constant and the conformal dimension as follows:
\bea\label{rel1}
\mathcal{C}_3^d=\frac{32\pi}{3}c_{\Delta}^{d}\sqrt{\lambda}\p_\lambda\Delta.
\eea We want to show that this relation is correct for the case of
the giant magnons with arbitrary finite size.

In the context of the AdS/CFT correspondence, it is now
well-established that the conformal dimension of a single trace
operator with one magnon state is the same as $E-J$ in the strong
coupling limit. For an exact relation from the gauge theory side,
one should solve the thermodynamic Bethe equations. It has been
shown that finite-size corrections to the conformal dimensions of
the SYM (dyonic) giant magnon operators computed by the L\"uscher
formula for $J\gg\sqrt{\lambda}$ match exactly with $E-J$ of
corresponding string state configurations. Based on these results,
we can assume that the conformal dimensions $\Delta$ of the SYM
operators are the same as $E-J$ of corresponding string states.

The {\it exact} classical expression for finite-size giant magnon
energy-charge relation is given by  \bea\label{eEJ} E-J\equiv
\Delta= \frac{\sqrt{\lambda}}{\pi}
\sqrt{\frac{1-v^2}{1-v^2\epsilon}}
\left[\mathbf{E}(1-\epsilon)-\left(1-\sqrt{(1-v^2\epsilon)
(1-\epsilon)}\right)\mathbf{K}(1-\epsilon)\right]. \eea The
corresponding expressions for $J$ and $p$ are \bea\nn
&&J=\frac{\sqrt{\lambda}}{\pi}\sqrt{\frac{1-v^2}{1-v^2\epsilon}}
\left[\mathbf{K}(1-\epsilon)-\mathbf{E}(1-\epsilon)\right],\\ \nn
&&p=2v\sqrt{\frac{1-v^2\epsilon}{1-v^2}}
\left[\frac{1}{v^2}\mathbf{\Pi}\left(1-\frac{1}{v^2}\Big\vert
1-\epsilon\right)-\mathbf{K}(1-\epsilon)\right],\eea where $J$ is
the angular momentum of the string, and $p$ is the magnon
momentum. One can obtain $E-J$ in terms of $J$ and $p$ by
eliminating $v, \epsilon$ from these expressions.

To take $\lambda$-derivative on $\Delta$, we need know $\lambda$
dependence of $v$ and $\epsilon$. Our strategy is to find
$v'(\lambda)$ and $\epsilon'(\lambda)$ from the conditions that
$J$ and $p$ are independent variables of $\lambda$, namely,
\bea\label{eqs} \frac{dJ}{d\lambda}=\frac{dp}{d\lambda}=0.\eea
Solving these conditions, we find the derivatives of the functions
$v(\lambda)$ and $\epsilon(\lambda)$ \bea\label{spd}
&&\frac{dv}{d\lambda}= -\frac{v
(1-v^2)\epsilon\left[\mathbf{E}(1-\epsilon)-\mathbf{K}(1-\epsilon)\right]^{2}}
{2\lambda(1-\epsilon)\left[\mathbf{E}(1-\epsilon)^2
-v^2\epsilon\mathbf{K}(1-\epsilon)^2\right]}, \\ \nn
&&\frac{d\epsilon}{d\lambda}=-
\frac{\epsilon\left[\mathbf{E}(1-\epsilon)-\mathbf{K}(1-\epsilon)\right]
\left[\mathbf{E}(1-\epsilon)-v^2\epsilon\mathbf{K}(1-\epsilon)\right]}
{\lambda\left[\mathbf{E}(1-\epsilon)^2
-v^2\epsilon\mathbf{K}(1-\epsilon)^2\right]}.\eea Replacing
(\ref{spd}) into the derivative of (\ref{eEJ}), one finds
\bea\label{ddr}
\lambda\p_\lambda\Delta=\frac{\sqrt{\lambda}}{2\pi}
\sqrt{\frac{1-v^2}{1-\epsilon}}
\left[\mathbf{E}(1-\epsilon)-\epsilon\mathbf{K}(1-\epsilon)\right].\eea
Comparing (\ref{exact1}) and (\ref{ddr}), we conclude that the
equality (\ref{rel1}) holds.

Next, we would like to compare (\ref{exact1}) with the known
leading finite-size correction to the giant magnon dispersion
relation \cite{AFZ06}. To this end, we have to consider the limit
$\epsilon\to 0$ in (\ref{exact1}). Taking into account the
behavior of the elliptic integrals in the $\epsilon\to 0$ limit,
we can use the ansatz \bea\label{vexp} v(\epsilon)=v_0+v_1\epsilon
+ v_2\epsilon\log(\epsilon).\eea Actually, all parameters in
(\ref{vexp}) are already known and are given by  \bea\label{vsol}
&&v_0=\cos(p/2),\h
v_1=\frac{1}{4}\sin^2(p/2)\cos(p/2)(1-\log(16)),\\ \nn &&
v_2=\frac{1}{4}\sin^2(p/2)\cos(p/2),\h\epsilon=16
\exp{\left(-\frac{2\pi J}{\sqrt{\lambda}\sin(p/2)}-2\right)}.\eea
Expanding (\ref{exact1}) in $\epsilon$ and using (\ref{vexp}),
(\ref{vsol}), we obtain \bea\label{Cexp}
\mathcal{C}^d=\frac{16}{3}c_{\Delta}^{d}\sin(p/2)
\left[1-4\sin(p/2)\left(\sin(p/2)+ \frac{2\pi
J}{\sqrt{\lambda}}\right)\exp{\left(-\frac{2\pi
J}{\sqrt{\lambda}\sin(p/2)}-2\right)}\right].\eea

On the other hand, from the giant magnon dispersion relation,
including the leading finite-size effect,  \bea\nn
\Delta=\frac{\sqrt{\lambda}}{\pi}\sin(p/2)\left[1-4\sin^2(p/2)
\exp{\left(-\frac{2\pi
J}{\sqrt{\lambda}\sin(p/2)}-2\right)}\right],\eea one finds
\bea\label{der1}
\lambda\p_\lambda\Delta=\frac{\sqrt{\lambda}}{2\pi} \sin(p/2)
\left[1-4\sin(p/2)\left(\sin(p/2)+ \frac{2\pi
J}{\sqrt{\lambda}}\right)\exp{\left(-\frac{2\pi
J}{\sqrt{\lambda}\sin(p/2)}-2\right)}\right].\eea This confirms
explicitly that the relation (\ref{rel1}) holds in the small
$\epsilon$ i.e. $J\gg\sqrt{\lambda}$ limit.

{\bf Dyonic giant magnons with finite size}

The dyonic finite-size giant magnon solution is given by \bea\nn
&&x_{0e}=\tanh(\kappa\tau_e),\h x_i=0,\h
z=\frac{1}{\cosh(\kappa\tau_e)},
\\ \label{fsgm2} &&\cos\theta=z_+
\mathbf{dn}\left(\frac{\sqrt{1-u^2}}{1-v^2}z_+(\sigma-v\tau)\Big\vert
1-\epsilon\right) ,
\\ \nn &&\phi_1= \frac{\tau-v\sigma}{1-v^2}+\frac{vW}{\sqrt{1-u^2}z_+(1-z_+^2)}\times
\\ \nn &&
\Pi\left(am\left(\frac{\sqrt{1-u^2}}{1-v^2}z_+(\sigma-v\tau)\right),
-\frac{z_+^2}{1-z_+^2}\left(1-\epsilon\right),
\Big\vert 1-\epsilon\right) \\ \nn &&\phi_2=
u\frac{\tau-v\sigma}{1-v^2},\eea where \bea\nn
\epsilon=\frac{z_-^2}{z_+^2},\h W=\kappa^2.\eea $z_\pm^2$ can be
written as \bea\nn &&z^2_\pm=\frac{1}{2(1-u^2)} \left\{q_1+q_2-u^2
\pm\sqrt{(q_1-q_2)^2-\left[2\left(q_1+q_2-2q_1
q_2\right)-u^2\right] u^2}\right\}, \\ \nn &&q_1=1-W,\h q_2=1-v^2W
.\eea

Now, we have to replace into (\ref{c3d}) the following expression
obtained from the above solution \bea\nn
\p_+X_{k}\p_-X_{k}&=&-\frac{1}{(1-v^2)\sin^2\theta}
\left\{1-v^2W^2+(1-u^2)z_+^4\epsilon +2(1-u^2)\cos^4\theta\right. \\
\nn &&-
\left.\cos^2\theta\left[2+z_+^2(1+\epsilon)-u^2\left(1+z_+^2(1+\epsilon)\right)\right]\right\}
.\eea Computing the integrals in (\ref{c3d}), we find
\bea\label{exact2} &&\mathcal{C}^d=\frac{8}{3}c_{\Delta}^{d}
\frac{1}{\sqrt{(1-u^2)W\chi_p}(1-\chi_p)}
\Bigg\{(1-\chi_p)\left[2(1-u^2)\chi_p\mathbf{E}(1-\epsilon)\right.
\\ \nn &&
-\left.\left(u^2-\left(1-v^2\right)W+(1-u^2)(1+\epsilon)\chi_p\right)\mathbf{K}(1-\epsilon)\right]
\\ \nn &&
-\left(1-v^2W^2-\chi_p-(1-\chi_p)\left(\epsilon\chi_p
+u^2(1-\epsilon\chi_p)\right)\right)\times
\\ \nn &&
\mathbf{\Pi}\left(-\frac{\chi_p}{1-\chi_p}\left(1-\epsilon\right)\Big\vert
1-\epsilon\right)\Bigg\},\eea where we introduced the notations
\bea\nn \chi_p=z_+^2,\h \chi_m=z_-^2,\h \Rightarrow
\epsilon=\frac{\chi_m}{\chi_p}.\eea This is our {\it exact} result
for the normalized coefficient $\mathcal{C}^d$ in the three-point
correlation function, corresponding to the case when the "heavy"
vertex operators are {\it finite-size} dyonic giant magnons.

To check the relation (\ref{rel1}), we need to know $\Delta$. As
GM case, we claim that this is given by $E-J_1$. The explicit
results are given by \cite{15} \bea\nn &&\mathcal{E}
=\frac{2\sqrt{W}(1-v^2)} {\sqrt{1-u^2}\sqrt{\chi_p}}\mathbf{K}
\left(1-\epsilon\right), \\ \label{cqsGMd} &&\mathcal{J}_1=
\frac{2\sqrt{\chi_p}} {\sqrt{1-u^2}}\left[
\frac{1-v^2W}{\chi_p}\mathbf{K} \left(1-\epsilon\right)-\mathbf{E}
\left(1-\epsilon\right)\right], \\ \nn &&\mathcal{J}_2=
\frac{2u\sqrt{\chi_p}} {\sqrt{1-u^2}}\mathbf{E}
\left(1-\epsilon\right)\\
&&p=\frac{2v} {\sqrt{1-u^2}\sqrt{\chi_p}}
\left[\frac{W}{1-\chi_p}\mathbf{\Pi}\left(-\frac{\chi_p}{1-\chi_p}(1-\epsilon)\bigg\vert
1-\epsilon\right) -\mathbf{K} \left(1-\epsilon\right)\right], \eea
and \bea\nn  \mathcal{E}=\frac{2\pi
E}{\sqrt{\lambda}},\qquad\mathcal{J}_{1,2}=\frac{2\pi
J_{1,2}}{\sqrt{\lambda}}.\eea

In this case, we need to obtain $v'(\lambda), \epsilon'(\lambda),
u'(\lambda)$ from the condition that $J_1, J_2, p$ be independent
of $\lambda$. It turns out that the exact calculations for these
are too complicated. Instead, we will just focus on the
$\epsilon\to 0$ limit of (\ref{exact2}) and $\lambda$ derivative
of $\Delta$ from the L\"uscher formula to check (\ref{rel1}). To
this end, we will use the expansions \bea\nn
&&\chi_p=\chi_{p0}+\left(\chi_{p1}+\chi_{p2}\log(\epsilon)\right)\epsilon,
\h\chi_m=\chi_{m1}\epsilon,
\\
\label{Dparsd} &&v=v_0+\left(v_1+v_2\log(\epsilon)\right)\epsilon,
\h u=u_0+\left(u_1+u_2\log(\epsilon)\right)\epsilon,
\\ \nn &&W=1+W_{1}\epsilon
.\eea

First note that $\chi_p$ and $\chi_m$ satisfy the following
relations \bea\label{chirel} &&\chi_p+\chi_m= \frac{2-(1+v^2)W-u^2}{1-u^2}\\
\nn && \chi_p\chi_m=\frac{1-(1+v^2)W-v^2W^2}{1-u^2}.\eea Expanding
(\ref{chirel}) and using the definition of $\epsilon$, we arrive
at
\bea\label{chid} &&\chi_{p0}=1-\frac{v_0^2}{1-u_0^2}, \\
\nn &&\chi_{p1}=
\frac{v_0}{\left(1-v_0^2\right)\left(1-u_0^2\right)^2}
\Big\{v_0\left[(1-v_0^2)^2-3(1-v_0^2)u_0^2+2u_0^4-2(1-v_0^2)u_0u_1\right]
\\ \nn
 &&\h\h -2\left(1-v_0^2\right)\left(1-u_0^2\right)v_1\Big\},\\ \nn
&&\chi_{p2}=
-2v_0\frac{v_2+(v_0u_2-u_0v_2)u_0}{\left(1-u_0^2\right)^2} \\
\nn &&\chi_{m1}=1-\frac{v_0^2}{1-u_0^2},
\\ \nn &&W_1=-\frac{(1-u_0^2-v_0^2)^2}
{(1-u_0^2)(1-v_0^2)}.\eea

The coefficients in the expansions of $v$ and $u$, we take from
\cite{PB10}, where for the case under consideration we have to set
$K_1=\chi_{n1}=0$, or equivalently $\Phi=0$. This gives
\bea\label{zmsd}
&&v_0=\frac{\sin(p)}{\sqrt{\mathcal{J}_2^2+4\sin^2(p/2)}},\h
u_0=\frac{\mathcal{J}_2}{\sqrt{\mathcal{J}_2^2+4\sin^2(p/2)}}
\\ \nn &&v_1=\frac{v_0(1-v_0^2-u_0^2)}{4(1-u_0^2)(1-v_0^2)} \left[(1-v_0^2)(1-\log(16))
-u_0^2\left(5-v_0^2(1+\log(16))-\log(4096)\right)\right]
\\ \nn &&v_2=\frac{v_0(1-v_0^2-u_0^2)}{4(1-u_0^2)(1-v_0^2)} \left[1-v_0^2-u_0^2(3+v_0^2)\right]
\\ \nn &&u_1=\frac{u_0(1-v_0^2-u_0^2)}{4(1-v_0^2)} \left[1-\log(16)-v_0^2(1+\log(16))\right]
\\ \nn &&u_2= \frac{u_0(1-v_0^2-u_0^2)}{4(1-v_0^2)} (1+v_0^2).\eea

We need also the expression for $\epsilon$. To the leading order,
it can be written as \cite{PB10} \bea\label{epsd}
\epsilon=16\exp\left(-\frac{2-\frac{2
v_0^2}{1-u_0^2}+\mathcal{J}_1\sqrt{1-v_0^2-u_0^2}}{1-v_0^2}\right).\eea

By using (\ref{chid}), (\ref{zmsd}) and (\ref{epsd}) in the
$\epsilon$-expansion of (\ref{exact2}), we derive \bea\label{exp2}
&&\mathcal{C}^d=\frac{16}{3}c_{\Delta}^{d}
\Bigg\{\frac{\mathcal{J}_2^2+4\sin^2(p/2)-16\sin^4(p/2)\exp(f)}{2\sqrt{\mathcal{J}_2^2+4\sin^2(p/2)}}
\\ \nn &&
+\frac{1}{\left(\mathcal{J}_2^2+4\sin^2(p/2)\right)\left(\mathcal{J}_2^2+4\sin^4(p/2)\right)^2}
\left[32\exp(f)\left(2\mathcal{J}_2^2\sqrt{\mathcal{J}_2^2+4\sin^2(p/2)}-3\mathcal{J}_1\right.\right.
\\ \nn &&
+2\left.\left.\left(\mathcal{J}_1\left(2+\mathcal{J}_2^2\right)
+\mathcal{J}_2^2
\sqrt{\mathcal{J}_2^2+4\sin^2(p/2)}\right)\cos(p)-\mathcal{J}_1\cos(2p)\right)\sin^8(p/2)\right]
\\ \nn &&-\frac{\mathcal{J}_2^2}{2\sqrt{\mathcal{J}_2^2+4\sin^2(p/2)}}
-\frac{8\mathcal{J}_2^2\sin^4(p/2)}{\left(\mathcal{J}_2^2+4\sin^2(p/2)\right)^{3/2}}\exp(f)\Bigg\},\eea
where \bea\nn f= -\frac{2\left(\mathcal{J}_1 +
\sqrt{\mathcal{J}_2^2+4\sin^2(p/2)}\right)
\sqrt{\mathcal{J}_2^2+4\sin^2(p/2)}\sin^2(p/2)}{\mathcal{J}_2^2+4\sin^4(p/2)}
.\eea

On the other hand, from the dyonic giant magnon dispersion
relation, including the leading finite-size correction,
\bea\label{dfsdr}
\Delta_{dyonic}=\frac{\sqrt{\lambda}}{2\pi}\left[
\sqrt{\mathcal{J}_2^2+4\sin^2(p/2)} - \frac{16 \sin^4(p/2)}
{\sqrt{\mathcal{J}_2^2+4\sin^2(p/2)}}\exp(f)\right],\eea one
obtains \bea\label{der2} &&\lambda\p_\lambda\Delta_{dyonic}=
\frac{\sqrt{\lambda}}{2\pi}\Bigg\{\frac{\mathcal{J}_2^2+4\sin^2(p/2)-16\sin^4(p/2)\exp(f)}{2\sqrt{\mathcal{J}_2^2+4\sin^2(p/2)}}
\\ \nn &&
+\frac{1}{\left(\mathcal{J}_2^2+4\sin^2(p/2)\right)\left(\mathcal{J}_2^2+4\sin^4(p/2)\right)^2}
\left[32\exp(f)\left(2\mathcal{J}_2^2\sqrt{\mathcal{J}_2^2+4\sin^2(p/2)}-3\mathcal{J}_1\right.\right.
\\ \nn &&
+2\left.\left.\left(\mathcal{J}_1\left(2+\mathcal{J}_2^2\right)
+\mathcal{J}_2^2
\sqrt{\mathcal{J}_2^2+4\sin^2(p/2)}\right)\cos(p)-\mathcal{J}_1\cos(2p)\right)\sin^8(p/2)\right]
\\ \nn &&-\frac{\mathcal{J}_2^2}{2\sqrt{\mathcal{J}_2^2+4\sin^2(p/2)}}
-\frac{8\mathcal{J}_2^2\sin^4(p/2)}{\left(\mathcal{J}_2^2+4\sin^2(p/2)\right)^{3/2}}\exp(f)\Bigg\}.\eea
Comparing (\ref{exp2}) and (\ref{der2}), we see that the relation
(\ref{rel1}) is also valid for finite-size dyonic giant magnons,
as it should be.

\subsubsection{Two GM states and dilaton with non-zero momentum}

For the dilaton vertex we have \cite{rt10} \bea \label{dvj}
V^d=\left(Y_4+Y_5\right)^{-\Delta_d}
\left(X_1+iX_2\right)^j\left[z^{-2}\left(\p_+x_{m}\p_-x^{m}+\p_+z\p_-z\right)
+\p_+X_{k}\p_-X_{k}\right], \eea where the scaling dimension
$\Delta_d=4+j$ to the leading order in the large $\sqrt{\lambda}$
expansion. The corresponding operator in the dual gauge theory is
proportional to $Tr\left(F_{\mu\nu}^2 \ Z^j+\ldots\right)$, or for
$j=0$, just to the SYM Lagrangian.

The normalized structure constant (\ref{nsc}) can be computed by
using (\ref{dvj}), applied for the case of giant magnons, to be
\cite{22} \bea\label{c3dfn} &&
\mathcal{C}_j^d=2\pi^{3/2}c_{\Delta}^{d}
\frac{\Gamma\left(\frac{4+j}{2}\right)}{\Gamma\left(\frac{5+j}{2}\right)}
\frac{\chi_p^{\frac{j-1}{2}}}{\sqrt{(1-u^2)W}}
\\ \nn &&\left[(1-u^2)\chi_p
\
{}_2F_1\left(\frac{1}{2},-\frac{1}{2}-\frac{j}{2};1;1-\frac{\chi_m}{\chi_p}\right)\right.
\\ \nn && -\left.\left(1-W\right)
\
{}_2F_1\left(\frac{1}{2},\frac{1}{2}-\frac{j}{2};1;1-\frac{\chi_m}{\chi_p}\right)\right].\eea

A few comments are in order. The structure constant in
(\ref{c3dfn}) corresponds to finite-size dyonic giant magnons,
i.e. with two angular momenta. The case of finite-size giant
magnons with one angular momentum nonzero can be obtained by
setting $u=0$ ($\chi_p$, $\chi_m$ also depend on $u$ according to
(\ref{chirel})). The infinite size case \cite{Hernandez2} is
reproduced for $W=1$, $\chi_m=0$. For $j=0$, (\ref{c3dfn}) reduces
to the result of \cite{19}.

{\bf Leading finite-size effect}

Expanding (\ref{c3dfn}) in $\epsilon$ one finds \cite{25}

$j=1$: \bea\nn &&\mathcal{C}_1^d\approx\frac{3}{4}\pi^2
c_{5}^{d}\sin^3(p/2)
\Bigg\{\frac{1}{\sqrt{\mathcal{J}_2^2+4\sin^2(p/2)}}
\\ \nn
&&-\frac{1}{128\left(\mathcal{J}_2^2+4\sin^2(p/2)\right)^{3/2}
\left(\mathcal{J}_2^2+4\sin^4(p/2)\right)^{2}}\Big[\Big(840
+826\mathcal{J}_2^2+258\mathcal{J}_2^4-24\mathcal{J}_2^6 \\ \nn
&&-2\left(744+707\mathcal{J}_2^2+244\mathcal{J}_2^4+72\mathcal{J}_2^6\right)\cos(p)
\\ \nn &&+4\left(255+218\mathcal{J}_2^2+62\mathcal{J}_2^4-6\mathcal{J}_2^6\right)\cos(2p)
-\left(520+367\mathcal{J}_2^2+24\mathcal{J}_2^4\right)\cos(3p)
\\ \nn &&+2\left(92+47\mathcal{J}_2^2+3\mathcal{J}_2^4\right)\cos(4p)
-\left(40+11\mathcal{J}_2^2\right)\cos(5p)+4\cos(6p)\Big)
\\ \nn &&+8\mathcal{J}_1\sin^2(p/2)\sqrt{\mathcal{J}_2^2+4\sin^2(p/2)}\Big(
\left(8+19\mathcal{J}_2^2+12\mathcal{J}_2^4\right)\cos(p)+
\left(8-16\mathcal{J}_2^2\right)\cos(2p)
\\ \nn &&-\left(8+3\mathcal{J}_2^2\right)\cos(3p)
-2\left(5+5\mathcal{J}_2^2-2\mathcal{J}_2^4-\cos(4p)\right)\big)
\Big]\epsilon\Bigg\},\eea

$j=2$: \bea\nn &&\mathcal{C}_2^d\approx \frac{2^8}{3^2 5}
c_{6}^{d}\sin^4(p/2)
\Bigg\{\frac{1}{\sqrt{\mathcal{J}_2^2+4\sin^2(p/2)}}
\\ \nn
&&-\frac{1}{128\left(\mathcal{J}_2^2+4\sin^2(p/2)\right)^{3/2}
\left(\mathcal{J}_2^2+4\sin^4(p/2)\right)^{2}}\Big[\Big(210
+8\mathcal{J}_2^2\left(6-\mathcal{J}_2^2\right)\left(7+4\mathcal{J}_2^2\right)
\\ \nn
&&-8\left(63+84\mathcal{J}_2^2+38\mathcal{J}_2^4+16\mathcal{J}_2^6\right)\cos(p)
\\ \nn &&+\left(585+576\mathcal{J}_2^2+176\mathcal{J}_2^4-32\mathcal{J}_2^6\right)\cos(2p)
-4\left(115+84\mathcal{J}_2^2+4\mathcal{J}_2^4\right)\cos(3p)
\\ \nn &&+2\left(111+56\mathcal{J}_2^2+4\mathcal{J}_2^4\right)\cos(4p)
-4\left(15+4\mathcal{J}_2^2\right)\cos(5p)+7\cos(6p)\Big)
\\ \nn &&-8\mathcal{J}_1\sin^2(p/2)\sqrt{\mathcal{J}_2^2+4\sin^2(p/2)}\Big(
15+8\mathcal{J}_2^2-8\mathcal{J}_2^4-4\left(3+5\mathcal{J}_2^2
+4\mathcal{J}_2^4\right)\cos(p)
\\ \nn &&-
\left(12-8\mathcal{J}_2^2\right)\cos(2p)+4\left(3+\mathcal{J}_2^2\right)\cos(3p)
-3\cos(4p)\Big)\Big]\epsilon\Bigg\},\eea

$j=3$: \bea\nn &&\mathcal{C}_3^d\approx \frac{3. 5}{2^{5}}\pi^2
c_{7}^{d}\sin^5(p/2)
\Bigg\{\frac{1}{\sqrt{\mathcal{J}_2^2+4\sin^2(p/2)}}
\\ \nn
&&+\frac{1}{960\left(\mathcal{J}_2^2+4\sin^2(p/2)\right)^{3/2}
\left(\mathcal{J}_2^2+4\sin^4(p/2)\right)^{2}}
\Big[20\Big(256\left(13+15\cos(p)\right)\sin^{10}(p/2)
\\ \nn &&+288\mathcal{J}_2^2\left(5+7\cos(p)\right)\sin^{8}(p/2)
+\mathcal{J}_2^4\left(54+241\cos(p)+10\cos(2p)\right.
\\ \nn &&+15\left.\cos(3p)\right)\sin^{2}(p/2)+ 10\mathcal{J}_2^6\cos(p)
\left(5+3\cos(p)\right)\Big)
\\ \nn && +60\mathcal{J}_1\sin^2(p/2)\sqrt{\mathcal{J}_2^2+4\sin^2(p/2)}
\Big(20+6\mathcal{J}_2^2-12\mathcal{J}_2^4
\\ \nn
&&-\left(16+21\mathcal{J}_2^2+20\mathcal{J}_2^4\right)\cos(p)
-2\left(8-5\mathcal{J}_2^2\right)\cos(2p)
\\ \nn &&+\left(16+5\mathcal{J}_2^2\right)\cos(3p)
-4\cos(4p)\Big )\Big]\epsilon\Bigg\},\eea

$j=4$: \bea\nn &&\mathcal{C}_4^d\approx \frac{2^{11}}{3.5^2.7}
c_{8}^{d}\sin^6(p/2)
\Bigg\{\frac{1}{\sqrt{\mathcal{J}_2^2+4\sin^2(p/2)}}
\\ \nn
&&+\frac{1}{8192\left(\mathcal{J}_2^2+4\sin^2(p/2)\right)^{3/2}
\left(\mathcal{J}_2^2+4\sin^4(p/2)\right)^{2}} \Big[64\Big(294
+14\mathcal{J}_2^2-60\mathcal{J}_2^4+48\mathcal{J}_2^6 \\ \nn
&&-4\left(51-49\mathcal{J}_2^2-53\mathcal{J}_2^4-36\mathcal{J}_2^6\right)\cos(p)
\\ \nn &&-\left(435+8\mathcal{J}_2^2\left(61+19\mathcal{J}_2^2-6\mathcal{J}_2^4\right)\right)\cos(2p)
+2\left(305+209\mathcal{J}_2^2+6\mathcal{J}_2^4\right)\cos(3p)
\\ \nn &&-2\left(179+83\mathcal{J}_2^2+6\mathcal{J}_2^4\right)\cos(4p)
+2\left(53+13\mathcal{J}_2^2\right)\cos(5p)-13\cos(6p)\Big)
\\ \nn &&+512\mathcal{J}_1\sin^2(p/2)\sqrt{\mathcal{J}_2^2+4\sin^2(p/2)}\Big(
25+4\mathcal{J}_2^2-16\mathcal{J}_2^4-\left(20+22\mathcal{J}_2^2+24\mathcal{J}_2^4\right)\cos(p)
\\ \nn &&-4\left(5-3\mathcal{J}_2^2\right)\cos(2p)+2\left(10+3\mathcal{J}_2^2\right)\cos(3p)
-5\cos(4p)\Big)\Big]\epsilon\Bigg\}.\eea In the four formulas
above $\epsilon$ is given by \bea\label{epsilone} &&\epsilon=16
\exp\left[-\frac{2\left(\mathcal{J}_1 +
\sqrt{\mathcal{J}_2^2+4\sin^2(p/2)}\right)
\sqrt{\mathcal{J}_2^2+4\sin^2(p/2)}\sin^2(p/2)}{\mathcal{J}_2^2+4\sin^4(p/2)}
\right].\eea

Actually, we computed the normalized coefficients in the
three-point correlators up to $j=10$. However, since the
expressions for them are too complicated, we give here only the
results for the first two odd and two even values of $j$. Knowing
these expressions, the conclusion is that they have the same
structure for any $j$ in the small $\epsilon$ limit\footnote{The
only difference in that sense is that for $j$ odd an additional
overall factor of $\pi^2$ appears, as can be seen from the
formulas above.}. Namely

\bea\nn &&\mathcal{C}_j^d\approx A_j
c_{j+4}^{d}\sin^{j+2}\left(\frac{p}{2}\right)
\Bigg\{\frac{1}{\sqrt{\mathcal{J}_2^2+4\sin^2\left(\frac{p}{2}\right)}}
+\frac{a_j}{\left(\mathcal{J}_2^2+4\sin^2\left(\frac{p}{2}\right)\right)^{3/2}
\left(\mathcal{J}_2^2+4\sin^4\left(\frac{p}{2}\right)\right)^{2}}
\\ \label{av} &&
\Big[P_j^3(\mathcal{J}_2^2)+\mathcal{J}_1\sin^2\left(\frac{p}{2}\right)
\sqrt{\mathcal{J}_2^2+4\sin^2\left(\frac{p}{2}\right)} \
Q_j^2(\mathcal{J}_2^2)\Big]\epsilon\Bigg\}.\eea where $\epsilon$
is given in (\ref{epsilone}), $A_j$ and $a_j$ are numerical
coefficients, while $P_j^3(\mathcal{J}_2^2)$ and
$Q_j^2(\mathcal{J}_2^2)$ are polynomials of third and second order
respectively, with coefficients depending on $p$ in a
trigonometric way.

Now, let us restrict ourselves to the simpler case when
$\mathcal{J}_2=0$, i.e. giant magnon string states with one
(large) angular momentum $\mathcal{J}_1\neq 0$. Knowing the above
results for $1\le j \le 10$, one can conclude that the normalized
structure constants in the three-point correlators for any $j\ge
1$ in the small $\epsilon$ limit look
like\footnote{$\mathcal{C}_{j0}^d$ is used for $\mathcal{C}_{j}^d$
computed for $\mathcal{J}_2=0$ case.} \bea\label{av0}
&&\mathcal{C}_{j0}^d\approx \frac{A_j}{2} c_{j+4}^{d}
\sin^{j}\left(\frac{p}{2}\right)\left[\sin\left(\frac{p}{2}\right)\right.
\\ \nn
&&+\left.\left(B_{j0}\sin\left(\frac{p}{2}\right)+C_{j0}\sin\left(\frac{3p}{2}\right)
+D_{j0}\left(1+\cos(p)\right)\mathcal{J}_1\right)
e^{-2-\frac{\mathcal{J}_1}{\sin\frac{p}{2}}}\right],\eea where
\bea\nn &&B_{j0}=(-2^2,3,\frac{2.11}{3},11,\frac{2^3
3^2}{5},\frac{53}{3},\frac{2.73}{7},2^3 3,\ldots)\h \mbox{for}\h
j=(1,\ldots,8,\ldots),
\\ \nn &&C_{j0}=1+3j,\h D_{j0}=2(j+1).\eea

\subsubsection{Two GM states and primary scalar operators}

The primary scalar vertex is \cite{BCFM98,Zarembo,rt10}
\bea
\label{prv} V^{pr}=\left(Y_4+Y_5\right)^{-\Delta_{pr}}
\left(X_1+iX_2\right)^j\left[z^{-2}\left(\p_+x_{m}\p_-x^{m}-\p_+z\p_-z\right)
-\p_+X_{k}\p_-X_{k}\right],\eea where now the scaling dimension is
$\Delta_{pr}=j$. The corresponding operator in the dual gauge
theory is $Tr\left( Z^j\right)$.

For giant magnons we have \cite{Hernandez2} \bea\nn
z^{-2}\left(\p_+x_{m}\p_-x^{m}-\p_+z\p_-z\right)=
\kappa^2\left(\frac{2}{\cosh^2(\kappa\tau_e)}-1\right).\eea Then
the light vertex operator becomes \bea\nn  V^{pr}=
\frac{\cos^j\theta}{\cosh^j(\kappa\tau_e)}
\left[\kappa^2\left(\frac{2}{\cosh^2(\kappa\tau_e)}-1\right)-\mathcal{L}_{S^3}^{gm}\right],\eea
where the infinite-size case was considered in \cite{Hernandez2},
while for the finite-size giant magnons $\mathcal{L}_{S^3}^{gm}$
should be taken from \bea\label{Lgm} &&\mathcal{L}_{S^3}^{gm}=
-\frac{1}{1-v^2}\left[2-(1+v^2)W
-2\left(1-u^2\right)\chi\right].\eea Let us also note that the
first integral for $\chi$ is given by \bea\label{fichi} \chi'=
\frac{2\sqrt{1-u^{2}}}{1-v^2}
\sqrt{\chi(\chi_{p}-\chi)(\chi-\chi_{m})}.\eea As a consequence,
the normalized structure constant in the corresponding three-point
function, for the case under consideration, takes the form:
\bea\label{c3pr} \mathcal{C}_j^{pr}&=&c_{\Delta}^{pr}
\left[\int_{-\infty}^{\infty}d\tau_e
\frac{W}{\cosh^{j}(\sqrt{W}\tau_e)}
\left(\frac{2}{\cosh^2(\sqrt{W}\tau_e)}-1\right)
\int_{-L}^{L}d\sigma\chi^{\frac{j}{2}} \right.
\\ \nn &&-\left.\int_{-\infty}^{\infty} \frac{d\tau_e}{\cosh^{j}(\sqrt{W}\tau_e)}
\int_{-L}^{L}d\sigma\chi^{\frac{j}{2}}\mathcal{L}_{S^3}^{gm}\right].\eea

Performing the integrations in (\ref{c3pr}) one finally finds
\cite{22} \bea\label{c3prf}
&&\mathcal{C}_j^{pr}=\pi^{3/2}c_{\Delta}^{pr}
\frac{\Gamma\left(\frac{j}{2}\right)}{\Gamma\left(\frac{3+j}{2}\right)}
\frac{\chi_p^{\frac{j-1}{2}}}{\sqrt{(1-u^2)W}}
\\ \nn &&\left[\left(1-W+j(1-v^2W)\right)
\
{}_2F_1\left(\frac{1}{2},\frac{1}{2}-\frac{j}{2};1;1-\epsilon\right)\right.
\\ \nn &&-\left.\left(1+j\right)\left(1-u^2\right)\chi_p
\
{}_2F_1\left(\frac{1}{2},-\frac{1}{2}-\frac{j}{2};1;1-\epsilon\right)\right].\eea

{\bf Leading finite-size effect}


Let us start with the simpler case when $J_2=0$, or equivalently
$u=0$. Expanding (\ref{c3prf}) in $\epsilon$ one finds \cite{25}
\bea\label{Cpr0} &&\mathcal{C}_{10}^{pr}\approx 0,\h
\mathcal{C}_{20}^{pr}\approx
\frac{4}{3}c_{2}^{pr}\mathcal{J}_1\sin^2(p/2)\ \epsilon,
\\ \nn
&&\mathcal{C}_{j0}^{pr}\approx c_{j}^{pr}a_{j}\sin(p/2)^{j+1}\
\epsilon,\h j = 3, ..., 10,\eea where \bea\label{e0} \epsilon= 16
\exp[-2 - \mathcal{J}_1\csc(p/2)],\eea for the case under
consideration. The numerical coefficients $a_{j}$ are given by
\bea\nn a_{j}=\left(\frac{1}{4}\pi^{2},\frac{2^{4}}{3.
5},\frac{1}{16}\pi^{2},\frac{2^7}{3^2.5.7},\frac{3.5}{2^{9}}\pi^{2},
\frac{2^{10}}{3^3.5^2.7},\frac{5.7}{2^{11}}\pi^2,\frac{2^{14}}{3^2.5^2.7^2.11}\right).\eea

A few comments are in order. From (\ref{Cpr0}) one can conclude
that the $\mathcal{C}_{10}^{pr}$ and $\mathcal{C}_{20}^{pr}$ cases
are exceptional, while $\mathcal{C}_{j0}^{pr}$ have the same
structure for $j\ge 3$. $\mathcal{C}_{10}^{pr}\approx 0$ means
that the small $\epsilon$ - contribution to the three point
correlator is zero to the leading order in $\epsilon$.
$\mathcal{C}_{20}^{pr}$ is the only one normalized structure
constant of this type proportional to $\mathcal{J}_1$. It is still
exponentially suppressed by $\epsilon$. The common feature of
$\mathcal{C}_{j0}^{pr}$ in (\ref{Cpr0}) is that they all vanish in
the {\it infinite size} case, i.e., for $\epsilon =0$. This
property was established in \cite{Hernandez2}. Here, we obtained
the leading finite-size corrections to it.

Now, let us turn to the dyonic case, i.e. $J_2\ne 0$. Working in
the same way, but with $u\ne 0$, we derive

$j=1$: \bea\label{c1pre} &&\mathcal{C}_{1}^{pr}\approx
c_{1}^{pr}\frac{\pi^2}{16}\frac{\mathcal{J}_2^2 \csc(p/2)}{
[\mathcal{J}_2^2+4
\sin^2(p/2)]^{3/2}[\mathcal{J}_2^2+4\sin^4(p/2)]}\times
\\ \nn
&& \left\{8[\mathcal{J}_2^2+4
\sin^2(p/2)][\mathcal{J}_2^2+4\sin^4(p/2)]\right.
\\ \nn  &&+\left.\sin^2(p/2)\left[40+17\mathcal{J}_2^2+2\mathcal{J}_2^4
-20(3+\mathcal{J}_2^2)\cos(p)\right.\right. \\ \nn
&&+\left.\left.3(8+\mathcal{J}_2^2)\cos(2p)-4\cos(3p)-4\frac{\mathcal{J}_2^2+8
\sin^2(p/2)}{\mathcal{J}_2^2+4\sin^4(p/2)}\times\right.\right.
\\ \nn && \left.\left.\left(\mathcal{J}_1\sqrt{\mathcal{J}_2^2+4
\sin^2(p/2)}+\mathcal{J}_2^2+4
\sin^2(p/2)\right)\times\right.\right.
\\ \nn &&\left.\left.
\left(\mathcal{J}_2^2+4\sin^4(p/2)+2\sin^2(p)\right)\sin^2(p/2)\right]\epsilon\right\},\eea

$j=2$: \bea\label{c2pre} &&\mathcal{C}_{2}^{pr}\approx
\frac{4}{3}c_{2}^{pr} \frac{1}{[\mathcal{J}_2^2+4
\sin^2(p/2)]^{3/2}[\mathcal{J}_2^2+4\sin^4(p/2)]}\times
\\ \nn
&&\left\{2\mathcal{J}_2^2[\mathcal{J}_2^2+4 \sin^2(p/2)]
\left[\mathcal{J}_2^2+4\sin^4(p/2)\right]-\sin^4(p/2)\times\right.
\\ \nn
&&\left.\left[20+3\mathcal{J}_2^2-2\mathcal{J}_2^4-2(15+2\mathcal{J}_2^2)\cos(p)
+(12+\mathcal{J}_2^2)\cos(2p)-2\cos(3p)\right.\right.
\\ \nn &&+
\left.\left.\frac{8}{\mathcal{J}_2^2+4\sin^4(p/2)}
\left(\mathcal{J}_1\sqrt{\mathcal{J}_2^2+4
\sin^2(p/2)}+\mathcal{J}_2^2+4
\sin^2(p/2)\right)\times\right.\right.
\\ \nn &&\left.\left.\left(-3+2(2+\mathcal{J}_2^2)\cos(p)
-\cos(2p)\right)\sin^4(p/2)\right]\epsilon\right\},\eea

$j=3$: \bea\label{c3pre} &&\mathcal{C}_{3}^{pr}\approx c_{3}^{pr}
\frac{\pi^2}{256}\csc(p/2) \frac{[\mathcal{J}_2^2+4
\sin^2(p/2)]^{5/2}}{\mathcal{J}_2^2+4 \sin^4(p/2)}\times \\
\nn && \left\{48\mathcal{J}_2^2 \sin^2(p/2)
\frac{\mathcal{J}_2^2+4 \sin^4(p/2)}{[\mathcal{J}_2^2+4
\sin^2(p/2)]^{3}} -
\left[\frac{25\mathcal{J}_2^4}{[\mathcal{J}_2^2+4
\sin^2(p/2)]^{2}}\right.\right.\\ \nn
&&\left.\left.-\mathcal{J}_2^2 \frac{\mathcal{J}_2^2+4
\sin^4(p/2)}{[\mathcal{J}_2^2+4
\sin^2(p/2)]^{3}}\left(21-16\cos(p)-5\cos(2p)+8\mathcal{J}_2^2\right)
\right.\right.
\\ \nn &&\left.\left.
-\frac{3}{2}\mathcal{J}_2^6
\frac{11-12\cos(p)+\cos(2p)+6\mathcal{J}_2^2}{[\mathcal{J}_2^2+4
\sin^2(p/2)]^{4}} \right.\right. \\ \nn && \left.\left.
+\left(3\mathcal{J}_2^2 \left(\mathcal{J}_2^2+4
\sin^2(p/2)+\mathcal{J}_1\sqrt{\mathcal{J}_2^2+4
\sin^2(p/2)}\right)\times\right.\right.\right.
\\ \nn &&\left.\left.\left.\left(80+42\mathcal{J}_2^2+12\mathcal{J}_2^4
-\left(120+47\mathcal{J}_2^2-4\mathcal{J}_2^4\right)\cos(p)\right.\right.\right.\right.
\\ \nn &&\left.\left.\left.\left.+\left(8+\mathcal{J}_2^2\right)\left(6\cos(2p)-\cos(3p)\right)\right)\sin^4(p/2)
\right)\frac{1}{[\mathcal{J}_2^2+4
\sin^2(p/2)]^4[\mathcal{J}_2^2+4 \sin^4(p/2)]}\right.\right.
\\ \nn &&\left.\left.-\frac{20\mathcal{J}_2^4\sin^2(p)}{[\mathcal{J}_2^2+4
\sin^2(p/2)]^3}+\frac{3\mathcal{J}_2^4\sin^4(p)}{[\mathcal{J}_2^2+4
\sin^2(p/2)]^4}-8\left(\frac{\mathcal{J}_2^2+4
\sin^4(p/2)}{\mathcal{J}_2^2+4
\sin^2(p/2)}\right)^2\right]\epsilon\right\},\eea

$j=4$: \bea\label{c4pre} &&\mathcal{C}_{4}^{pr}\approx
\frac{2}{45}c_{4}^{pr} \frac{[\mathcal{J}_2^2+4
\sin^2(p/2)]^{5/2}}{\mathcal{J}_2^2+4 \sin^4(p/2)} \\ \nn
&&\left\{\frac{32\mathcal{J}_2^2[\mathcal{J}_2^2+4
\sin^4(p/2)]\sin^2(p/2)}{[\mathcal{J}_2^2+4
\sin^2(p/2)]^3}-\left[\frac{17\mathcal{J}_2^4}{[\mathcal{J}_2^2+4
\sin^2(p/2)]^2}\right.\right. \\ \nn && \left.\left.
-\frac{1}{2}\mathcal{J}_2^2 \frac{\mathcal{J}_2^2+4
\sin^4(p/2)}{\mathcal{J}_2^2+4 \sin^2(p/2)]^3}
\left(39-32\cos(p)-7\cos(2p)+16\mathcal{J}_2^2\right)\right.\right.
\\ \nn &&\left.\left. -\mathcal{J}_2^6
\frac{11-12\cos(p)+\cos(2p)+6\mathcal{J}_2^2}{[\mathcal{J}_2^2+4
\sin^2(p/2)]^4}\right.\right. \\ \nn
&&\left.\left.+\left(2\mathcal{J}_2^2 \left(\mathcal{J}_2^2+4
\sin^2(p/2)+\mathcal{J}_1\sqrt{\mathcal{J}_2^2+4
\sin^2(p/2)}\right)\times\right.\right.\right.
\\ \nn &&\left.\left.\left.
\left(75+44\mathcal{J}_2^2+16\mathcal{J}_2^4
-2\left(58+23\mathcal{J}_2^2-4\mathcal{J}_2^4\right)\cos(p)\right.\right.\right.\right.
\\ \nn &&\left.\left.\left.\left.+ 4\left(13+\mathcal{J}_2^2\right)\cos(2p) -
2\left(6+\mathcal{J}_2^2\right)\cos(3p)+\cos(4p)\right)\sin^4(p/2)
\right)\times\right.\right.
\\ \nn &&\left.\left.\frac{1}{[\mathcal{J}_2^2+4
\sin^2(p/2)]^4[\mathcal{J}_2^2+4\sin^4(p/2)]}\right.\right.
\\ \nn &&\left.\left.-\frac{13\mathcal{J}_2^4\sin^2(p)}{[\mathcal{J}_2^2+4
\sin^2(p/2)]^3}+\frac{2\mathcal{J}_2^4\sin^4(p)}{[\mathcal{J}_2^2+4
\sin^2(p/2)]^4}-3\left(\frac{\mathcal{J}_2^2+4
\sin^4(p/2)}{\mathcal{J}_2^2+4
\sin^2(p/2)}\right)^2\right]\epsilon\right\}.\eea

In the four formulas above $\epsilon$ is given by
(\ref{epsilone}).

\subsubsection{Two GM states and singlet
scalar operators on higher string levels}

As explained in \cite{rt10}, there exist special massive string
states vertex operators with finite  quantum numbers  for which
the leading-order bosonic part is known explicitly and thus they
can be used as candidates for ``light'' vertex operators in the
semiclassical computation of the correlation functions. These are
singlet operators which do not mix  with other operators to
leading nontrivial order in $\frac{1}{\sqrt{\lambda}}$
\cite{AT,rt9}. An example of such scalar operator carrying no
spins is \cite{rt10}
 \bea\label{Vq}   V^q= (Y_4+Y_5)^{- \Delta} \Big[(\p X_k \bar{\p} X_k)^q+ ... \Big]
.\eea
The marginality condition for  this operator is \cite{rt10}:
 $2(1-q)   +  \frac{1}{2\sqrt{\lambda}}\Big[\Delta (\Delta-4) +  2q(q-1)\Big]
\\ \nn  +  {\frac{1}{(\sqrt{\lambda})^2} } \big[ \frac{2}{3} q (q-1) (q-
\frac{7}{2})  +
 4q \big]  +\mathcal{O}\left(\frac{1}{(\sqrt{\lambda})^3}\right)=0$.
This  operator  corresponds to a scalar string state at level
$n=q-1$ so that the fermionic contributions   should make the
$q=1$ state massless (BPS), with $\Delta=4$  following from the
marginality condition. The $q=2$ choice   corresponds to a scalar
state on the first excited string level. In that case, we have
\cite{rt9} \bea\nn
\Delta(\Delta-4)=4(\sqrt{\lambda}-1)+\mathcal{O}\left(\frac{1}{\sqrt{\lambda}}\right),\eea
with solution \bea\nn
\Delta=2(\lambda^{1/4}+1)+\frac{0}{\lambda^{1/4}}+\mathcal{O}\left(\frac{1}{\lambda^{3/4}}\right).\eea
However, the subleading terms here should not be trusted as far as
the fermions are expected to change the $\Delta$-independent terms
in the 1-loop anomalous dimension. For arbitrary string level $n$,
the solution of the marginality condition with respect to
$\Delta$, to leading order in $\frac{1}{\sqrt{\lambda}}$, is given
by \bea\label{mc}
\Delta_q=2\left(\sqrt{(q-1)\sqrt{\lambda}+1-\frac{1}{2}q(q-1)}+1\right).\eea

Let us also point out that the number $q$ of $\p X_k \bar{\p} X_k$
factors in an operator never increases due to
renormalization~\cite{AT}. That is why, it can be used as a
quantum number to characterize the leading term in the
corresponding operator \cite{Wegner90}.

The normalized structure constant can be computed by using
(\ref{Vq}), applied for the case of giant magnons, to be \cite{23}

\bea\label{c3df} &&\mathcal{C}^q=c_{\Delta_q}\pi^{3/2}
\frac{\Gamma\left(\frac{\Delta_q}{2}\right)}{\Gamma\left(\frac{\Delta_q+1}{2}\right)}
\frac{(-1)^q\left[2-(1+v^2)W\right]^q}{(1-v^2)^{q-1}\sqrt{(1-u^2)W\chi_p}}
\\ \nn &&\sum_{k=0}^{q}\frac{q!}{k!(q-k)!}\left[-\frac{1-u^2}{1-\frac{1}{2}
(1+v^2)W}\right]^{k}\chi_p^{k}\
{}_2F_1\left(\frac{1}{2},\frac{1}{2}-k;1;1-\frac{\chi_m}{\chi_p}\right),\eea
where  \bea\nn &&\chi_p=\frac{1}{2(1-u^2)} \left\{q_1+q_2-u^2
+\sqrt{(q_1-q_2)^2-\left[2\left(q_1+q_2-2q_1 q_2\right)-u^2\right]
u^2}\right\},
\\ \nn &&\chi_m=\frac{1}{2(1-u^2)} \left\{q_1+q_2-u^2
-\sqrt{(q_1-q_2)^2-\left[2\left(q_1+q_2-2q_1 q_2\right)-u^2\right]
u^2}\right\},
\\ \label{roots} &&q_1=1-W,\h q_2=1-v^2W .\eea

This is our general result corresponding to finite-size giant
magnons with two angular momenta and to arbitrary string level
$n=q-1=0,1,2,...$. Now, let us give some particular examples
contained in (\ref{c3df}).

{\bf Giant magnons with one angular momentum}

The case of finite-size giant magnons with one angular momentum
$J_1\ne 0$ corresponds to $u=0$. This can be seen from the
explicit expression for the second angular momentum $J_2$: \bea\nn
\mathcal{J}_2\equiv\frac{2\pi J_2}{\sqrt{\lambda}} =\frac{2u
\sqrt{\chi_{p}}}{\sqrt{1-u^2}}\
\mathbf{E}\left(1-\frac{\chi_m}{\chi_p}\right).\eea Then from
(\ref{roots}) one obtains the following simplified expressions for
$\chi_p$, $\chi_m$: \bea\nn \chi_p=1-v^2W,\h \chi_m=1-W.\eea
Taking this into account, and using (\ref{c3df}), one can find
that the normalized structure constants for the first three string
levels, for the case at hand, are given by:

$q=1$ (level $n=0$) \bea\nn &&\mathcal{C}^1=2c_{\Delta_1}\pi^{1/2}
\frac{\Gamma\left(\frac{\Delta_1}{2}\right)}{\Gamma\left(\frac{\Delta_1+1}{2}\right)}
\frac{1}{\sqrt{W(1-v^2W)}}
\\ \nn &&\left[2(1-v^2W) \
\mathbf{E}\left(1-\frac{1-W}{1-v^2W}\right)
-\left(2-(1+v^2)W\right)\mathbf{K}\left(1-\frac{1-W}{1-v^2W}\right)\right].\eea

$q=2$ (level $n=1$) \bea\nn &&\mathcal{C}^2=2c_{\Delta_2}\pi^{1/2}
\frac{\Gamma\left(\frac{\Delta_2}{2}\right)}{\Gamma\left(\frac{\Delta_2+1}{2}\right)}
\frac{1}{(1-v^2)\sqrt{W(1-v^2W)}}
\\ \nn
&&\left[\left(2-(1+v^2)W\right)^2\mathbf{K}\left(1-\frac{1-W}{1-v^2W}\right)\right.
\\ \nn &&-4\left.\left(2-(1+v^2)W\right)(1-v^2W) \
\mathbf{E}\left(1-\frac{1-W}{1-v^2W}\right)\right.
\\ \nn &&+\left. 2\pi(1-v^2W)^2 \ {}_2F_1\left(\frac{1}{2},-\frac{3}{2};1;1-\frac{1-W}{1-v^2W}\right)\right].\eea

$q=3$ (level $n=2$) \bea\nn
&&\mathcal{C}^3=-2c_{\Delta_3}\pi^{1/2}
\frac{\Gamma\left(\frac{\Delta_3}{2}\right)}{\Gamma\left(\frac{\Delta_3+1}{2}\right)}
\frac{\left(2-(1+v^2)W\right)^3}{(1-v^2)^2\sqrt{W(1-v^2W)}}
\\ \nn &&\left[\mathbf{K}\left(1-\frac{1-W}{1-v^2W}\right)
-\frac{6(1-v^2W)}{2-(1+v^2)W} \
\mathbf{E}\left(1-\frac{1-W}{1-v^2W}\right)\right.
\\ \nn &&+\left.\frac{6\pi(1-v^2W)^2}{\left(2-(1+v^2)W\right)^2} \
{}_2F_1\left(\frac{1}{2},-\frac{3}{2};1;1-\frac{1-W}{1-v^2W}\right)\right.
\\ \nn &&-\left.\frac{4\pi(1-v^2W)^3}{\left(2-(1+v^2)W\right)^3} \
{}_2F_1\left(\frac{1}{2},-\frac{5}{2};1;1-\frac{1-W}{1-v^2W}\right)
\right].\eea

{\bf Giant magnons with two angular momenta}

$q=1$ (level $n=0$): \bea\nn
&&\mathcal{C}^1=2c_{\Delta_1}\pi^{1/2}
\frac{\Gamma\left(\frac{\Delta_1}{2}\right)}{\Gamma\left(\frac{\Delta_1+1}{2}\right)}
\frac{1}{\sqrt{(1-u^2)W\chi_p}}
\\ \nn &&\left[2(1-u^2)\chi_p \
\mathbf{E}\left(1-\frac{\chi_m}{\chi_p}\right)
-\left(2-(1+v^2)W\right)\mathbf{K}\left(1-\frac{\chi_m}{\chi_p}\right)\right].\eea

$q=2$ (level $n=1$): \bea\nn
&&\mathcal{C}^2=2c_{\Delta_2}\pi^{1/2}
\frac{\Gamma\left(\frac{\Delta_2}{2}\right)}{\Gamma\left(\frac{\Delta_2+1}{2}\right)}
\frac{1}{(1-v^2)\sqrt{(1-u^2)W\chi_p}}
\\ \nn &&\left[\left(2-(1+v^2)W\right)^2\mathbf{K}\left(1-\frac{\chi_m}{\chi_p}\right)
-4(1-u^2)\left(2-(1+v^2)W\right)\chi_p \
\mathbf{E}\left(1-\frac{\chi_m}{\chi_p}\right)\right.
\\ \nn &&+\left. 2\pi(1-u^2)^2\chi_p^2 \ {}_2F_1\left(\frac{1}{2},-\frac{3}{2};1;1-\frac{\chi_m}{\chi_p}\right)\right].\eea

$q=3$ (level $n=2$): \bea\nn
&&\mathcal{C}^3=-2c_{\Delta_3}\pi^{1/2}
\frac{\Gamma\left(\frac{\Delta_3}{2}\right)}{\Gamma\left(\frac{\Delta_3+1}{2}\right)}
\frac{\left(2-(1+v^2)W\right)^3}{(1-v^2)^2\sqrt{(1-u^2)W\chi_p}}
\\ \nn &&\left[\mathbf{K}\left(1-\frac{\chi_m}{\chi_p}\right)
-\frac{6(1-u^2)\chi_p}{2-(1+v^2)W} \
\mathbf{E}\left(1-\frac{\chi_m}{\chi_p}\right)\right.
\\ \nn &&+\left.\frac{6\pi(1-u^2)^2\chi_p^2}{\left(2-(1+v^2)W\right)^2} \
{}_2F_1\left(\frac{1}{2},-\frac{3}{2};1;1-\frac{\chi_m}{\chi_p}\right)\right.
\\ \nn &&-\left.\frac{4\pi(1-u^2)^3\chi_p^3}{\left(2-(1+v^2)W\right)^3} \
{}_2F_1\left(\frac{1}{2},-\frac{5}{2};1;1-\frac{\chi_m}{\chi_p}\right)
\right].\eea

\subsection{Semiclassical three-point correlation functions in\\
TsT-deformed $AdS_5 \times S^5$}

\subsubsection{Two GM states and dilaton with zero momentum}

Working as in the undeformed case and taking into account the
deformation of the sphere, one finds that the normalized structure
constant is given by \cite{20} \bea\label{exact}
&&\mathcal{C}_{\tilde{\gamma}}^{d}=\frac{16}{3}c_{\Delta}^{d}
\frac{1}{\sqrt{(1-u^2)W(\chi_p-\chi_n)}}\times \\ \nn
&&\left[\left((1-u^2)(1-\tilde{\gamma}K)-\tilde{\gamma}uvW\right)
\sqrt{\chi_p-\chi_n}\mathbf{E}(1-\epsilon)\right.
\\ \nn &&+\left.\left(\left(W(1-\tilde{\gamma}uv\chi_n)-(1-\tilde{\gamma}K)
\left(1-(1-u^2)\chi_n\right)\right) \mathbf{K}(1-\epsilon)\right)
\right],\eea where \bea\label{eg}
\epsilon=\frac{\chi_{m}-\chi_{n}}{\chi_{p}-\chi_{n}},\eea and
\bea\nn &&\chi_p+\chi_m+\chi_n=\frac{2-(1+v^2)W-u^2}{1
-u^2},\\
\label{eqsg} &&\chi_p \chi_m+\chi_p \chi_n+\chi_m
\chi_n=\frac{1-(1+v^2)W+(v W-u K)^2-K^2}{1 -u^2},\\ \nn && \chi_p
\chi_m \chi_n=- \frac{K^2}{1 -u^2}.\eea
 The case of dyonic finite-size giant magnons we are interested in, corresponds to
\bea\nn 0<u<1,\h 0<v<1,\h 0<W<1,\h 0<\chi_{m}<\chi< \chi_{p}<1,\h
\chi_{n}<0.\eea

This is our {\it exact} semiclassical result for the normalized
coefficient $\mathcal{C}_{\tilde{\gamma}}^{d}$ in the three-point
correlation function, corresponding to the case when the heavy
vertex operators are {\it finite-size} dyonic giant magnons living
on the $\gamma$-deformed three-sphere.

{\bf Leading finite-size effect}

For the case of the dilaton operator, the three-point function of
the SYM can be easily related to the conformal dimension of the
heavy operators. This corresponds to shift `t Hooft coupling
constant which is the overall coefficient of the Lagrangian
\cite{Costa}. This gives an important relation between the
structure constant and the conformal dimension as follows:
\bea\label{rel1g}
\mathcal{C}_{\tilde{\gamma}}^{d}=\frac{32\pi}{3}c_{\Delta}^{d}\sqrt{\lambda}\p_\lambda\Delta.
\eea We want to show here that this relation holds for the case of
finite-size giant magnons ($J_2=0$), assuming that $\Delta=E-J_1$,
and considering the limit $\epsilon\to 0$. To this end, we
introduce the expansions \bea\nn
&&\chi_p=\chi_{p0}+\left(\chi_{p1}+\chi_{p2}\log(\epsilon)\right)\epsilon,
\\ \nn &&\chi_m=\chi_{m0}+\left(\chi_{m1}+\chi_{m2}\log(\epsilon)\right)\epsilon,
\\ \nn &&\chi_n=\chi_{n0}+\left(\chi_{n1}+\chi_{n2}\log(\epsilon)\right)\epsilon,
\\
\label{Dparsg} &&v=v_0+\left(v_1+v_2\log(\epsilon)\right)\epsilon, \\
\nn &&u=u_0+\left(u_1+u_2\log(\epsilon)\right)\epsilon,
\\ \nn &&W=W_0+\left(W_1+W_2\log(\epsilon)\right)\epsilon,
\\ \nn &&K=K_0+\left(K_1+K_2\log(\epsilon)\right)\epsilon .\eea
A few comments are in order. To be able to reproduce the
dispersion relation for the infinite-size giant magnons, we set
\bea\label{isg} \chi_{m0}=\chi_{n0}=K_0=0,\h W_0=1.\eea In
addition, one can check that if we keep the coefficients
$\chi_{m2}$, $\chi_{n2}$, $W_2$ and $K_2$ nonzero, the known
leading correction to the giant magnon energy-charge relation
\cite{HS08} will be modified by a term proportional to
$\mathcal{J}_1^2$. That is why we choose \bea\label{k2g}
\chi_{m2}= \chi_{n2}=W_2=K_2=0.\eea Finally, since we are
considering for simplicity giant magnons with one angular
momentum, we also set \bea\label{u0g} u_0=0,\eea because the
leading term in the $\epsilon$-expansion of $\mathcal{J}_2$ is
proportional to $u_0$.

By replacing (\ref{Dparsg}) in (\ref{eg}) and (\ref{eqsg}), and
taking into account (\ref{isg}), (\ref{k2g}), (\ref{u0g}), we
obtain
\bea\label{chig} &&\chi_{p0}=1-v_0^2, \\
\nn &&\chi_{p1}= \frac{v_0}{1-v_0^2}
\Big[v_0\sqrt{(1-v_0^2)^4-4K_1^2(1-v_0^2)}-2(1-v_0^2)v_1 \Big ],\\
\nn &&\chi_{p2}=
-2v_0v_2 ,\\
\nn &&\chi_{m1}=
\frac{(1-v_0^2)^2+\sqrt{(1-v_0^2)^4-4K_1^2(1-v_0^2)}}
{2(1-v_0^2)}, \\
\nn &&\chi_{n1}=
-\frac{(1-v_0^2)^2-\sqrt{(1-v_0^2)^4-4K_1^2(1-v_0^2)}}
{2(1-v_0^2)},
\\ \nn &&W_1=-\frac{\sqrt{(1-v_0^2)^4-4K_1^2(1-v_0^2)}}
{1-v_0^2}.\eea

The other parameters in (\ref{Dparsg}) and (\ref{chig}) can be
found in the following way. First, we impose the conditions
$J_2=0$ and $p_1$ to be independent of $\epsilon$. This leads to
four equations with solution \bea\label{uv}
&&v_1=\frac{v_0\sqrt{(1-v_0^2)^4-4K_1^2(1-v_0^2)} \left(1-\log
16\right)}{4(1-v_0^2)}, \\ \nn &&v_2=
\frac{v_0\sqrt{(1-v_0^2)^4-4K_1^2(1-v_0^2)}}{4(1-v_0^2)}, \\ \nn
&&u_1=\frac{K_1v_0\log 4}{1-v_0^2}, \\ \nn &&u_2=
-\frac{K_1v_0}{2(1-v_0^2)},\eea where \bea\label{v0g}
v_0=\cos\frac{p_1}{2}.\eea Next, to the leading order, the
expansions for $\mathcal{J}_1$ and $p_2=2\pi n_2$ $(n_2\in
\mathbb{Z})$ give \bea\label{ekf}
\epsilon=16\exp\left(-2-\frac{\mathcal{J}_1}{\sin\frac{p_1}{2}}\right),
\h K_1=\frac{1}{2}\sin^3\frac{p_1}{2}\sin\Phi, \h
\Phi=2\pi\left(n_2-\frac{\tilde{\gamma}}{\sqrt{\lambda}}J_1\right).\eea

Now, we consider the limit $\epsilon\to 0$ in the expression
(\ref{exact}) for the structure constant in the 3-point
correlation function, by using (\ref{Dparsg}), (\ref{isg}),
(\ref{k2g}), (\ref{u0g}), (\ref{chig}), (\ref{uv}), and obtain
\bea\label{C3se} \mathcal{C}_{\tilde{\gamma}}^{d}&\approx
&\frac{4}{3}c_{\Delta}^{d}
\frac{1}{(1-v_0^2)^{3/2}}\Bigg[4+4v_0^4\Big(1-\tilde{\gamma}K_1(1-\log
4)\ \epsilon\Big)\\ \nn
&&-v_0^2\left(8+\left(\sqrt{(1-v_0^2)^4-4K_1^2(1-v_0^2)}\ (1-\log
16)-8\tilde{\gamma}K_1(1-\log 4)\right)\epsilon\right)
\\ \nn
&&-\left(4\tilde{\gamma}K_1(1-\log
4)-\sqrt{(1-v_0^2)^4-4K_1^2(1-v_0^2)}\ (1-\log 256)\right)\epsilon
\\ \nn
&&-\left(v_0^2\sqrt{(1-v_0^2)^4-4K_1^2(1-v_0^2)}+2\tilde{\gamma}K_1(1-v_0^2)^2\right)
\epsilon\log\epsilon
\\ \nn
&&+\sqrt{(1-v_0^2)^4-4K_1^2(1-v_0^2)}\ \epsilon\log(16\
\epsilon)\Bigg] .\eea According to (\ref{v0g}), (\ref{ekf}), the
above expression for $\mathcal{C}_{\tilde{\gamma}}^{d}$ can be
rewritten in terms of $p_1$, $\mathcal{J}_1$, as \bea\label{pJ}
\mathcal{C}_{\tilde{\gamma}}^{d}&\approx &
\frac{16}{3}c_{\Delta}^{d}
\sin\frac{p_1}{2}\left[1-4\sin^2\frac{p_1}{2}\left(\cos\Phi
+\mathcal{J}_1\csc\frac{p_1}{2}\cos\Phi
-\tilde{\gamma}\mathcal{J}_1\sin\Phi\right)
e^{-2-\frac{\mathcal{J}_1}{\sin\frac{p_1}{2}}}\right].\eea

In order to check if the equality (\ref{rel1g}) holds for the
present case, let us now consider the dispersion relation of giant
magnons on $TsT$-transformed $AdS_5\times S^5$, including the
leading finite-size correction, which is known to be
\cite{BF08,18} \bea\label{dr}
E-J_1=\frac{\sqrt{\lambda}}{\pi}\sin(p/2)\left[1-4\sin^2(p/2)\cos\Phi
\exp{\left(-2-\frac{2\pi
J_1}{\sqrt{\lambda}\sin(p/2)}\right)}\right].\eea Taking the
$\lambda$ derivative of (\ref{dr}), one finds \bea\label{der1g}
\lambda\p_\lambda\Delta=\frac{\sqrt{\lambda}}{2\pi}
\sin\frac{p}{2} \left[1-4\sin^2\frac{p}{2}\left(\cos\Phi
+\mathcal{J}_1\csc\frac{p}{2}\cos\Phi
-\tilde{\gamma}\mathcal{J}_1\sin\Phi\right)
e^{-2-\frac{\mathcal{J}_1}{\sin\frac{p}{2}}}\right].\eea
Identifying $p\equiv p_1$, and comparing (\ref{pJ}) with
(\ref{der1g}), we see that the equality (\ref{rel1g}) is also
valid for the $\gamma$-deformed case.

\subsubsection{Two GM states and dilaton with non-zero momentum}

The normalized structure constant for the case at hand is given by
\cite{22} \bea\label{cdjg} &&\mathcal{C}_{j\tilde{\gamma}}^d=
2\pi^{3/2}c_{\Delta}^{d}
\frac{\Gamma\left(\frac{4+j}{2}\right)}{\Gamma\left(\frac{5+j}{2}\right)}
\frac{\chi_p^{j/2}}{\sqrt{(1-u^2)W\left(\chi_p-\chi_n\right)}}
\\ \nn &&
\left\{\left[1-\tilde{\gamma}K-u\left(u+\tilde{\gamma}(vW-uK)\right)\right]\chi_p
F_1\left(1/2,1/2,-1-j/2;1;1-\epsilon,1-\frac{\chi_m}{\chi_p}\right)\right.
\\ \nn
&&-\left.\left(1-W-\tilde{\gamma}K\right)
F_1\left(1/2,1/2,-j/2;1;1-\epsilon,1-\frac{\chi_m}{\chi_p}\right)\right\}.\eea

The small $\epsilon$ limit corresponds to considering the leading
finite-size effect, while $\epsilon=0$, $\chi_m=0$, $\chi_n=0$,
$K=0$, $W=1$, describes the infinite-size case.

{\bf Leading finite-size effect}

Here, we restrict ourselves to the case $\mathcal{J}_2=0$,
$\mathcal{J}_{1}\equiv \mathcal{J}$ large but finite, i.e.
$J_1\equiv J\gg \sqrt{\lambda}$.

Expanding (\ref{cdjg}) for this case to the leading order in
$\epsilon$, one finds \cite{26} ($j\ge 1$) \bea\nn
&&\mathcal{C}_{j\tilde{\gamma}}^{d}\approx c_{4+j}^{d}
\frac{\sqrt{\pi}}{2}
\frac{\Gamma\left(\frac{4+j}{2}\right)}{\Gamma\left(\frac{5+j}{2}\right)}
\chi_{p0}^{\frac{1}{2}(j-1)}\Big\{\epsilon\left[4W_1+(2+j)(2\chi_{m1}-\chi_{p0})
+4\tilde{\gamma}K_1\right]
\\ \nn
&& \times\log\frac{16}{\epsilon}\
{}_1F_0\left(-\frac{j}{2},1\right)
\\ \label{d1}
&&+\sqrt{\pi}\frac{\Gamma\left(\frac{j}{2}\right)}{\Gamma\left(\frac{3+j}{2}\right)}
\Big[2 j
\chi_{p0}+\left(2\chi_{m1}-\chi_{p0}+W_1\left(2+j(2-\chi_{p0})\right)\right.
\\ \nn &&+\left.j(\chi_{m1}+\chi_{n1}+(1+j)\chi_{p1})
+2\tilde{\gamma}(K_1+j(K_1-(K_1+v_0 u_1) \chi_{p0}))\right)\
\epsilon
\\ \nn
&&+(j(1+j)\chi_{p2}-2\tilde{\gamma}jv_0u_2\chi_{p0})\
\epsilon\log\epsilon\Big]\Big\}.\eea
This can be rewritten as
\bea\nn &&\mathcal{C}_{j\tilde{\gamma}}^{d}\approx
c_{4+j}^{d}\pi\frac{\Gamma\left(\frac{j}{2}\right)\Gamma\left(\frac{4+j}{2}\right)}
{\Gamma\left(\frac{3+j}{2}\right)\Gamma\left(\frac{5+j}{2}\right)}
\sin^{1+j}(p/2) \Big\{j-\frac{1}{8}\Big[\left(4-j(1+3j)(1+\cos
p)\right.
\\ \nn &&-\left. j(1+j)(1+\cos p)\csc(p/2)\right)\mathcal{J}\cos\Phi
\\ \nn &&-\tilde{\gamma}\left(4\sin(p/2)- j(1+\cos p)\mathcal{J}\right)
\sin\Phi\Big]\epsilon\Big\}.\eea

\subsubsection{Two GM states and primary scalar operators}

According to \cite{22} the normalized structure constant for this
case is given by \bea\label{c3prgf}
&&\mathcal{C}_{j\tilde{\gamma}}^{pr}= \pi^{3/2}c_{\Delta}^{pr}
\frac{\Gamma\left(\frac{j}{2}\right)}{\Gamma\left(\frac{1+j}{2}\right)}
\frac{(1-v^2)\chi_p^{j/2}}{\sqrt{(1-u^2)\left(\chi_p-\chi_n\right)}}
\\ \nn &&
\left\{\left[\sqrt{W}\frac{j-1}{j+1} +\frac{1}{\sqrt{W}(1-v^2)}
\left(2-(1+v^2)W-2\tilde{\gamma}K\right)\right]\right.
\\ \nn && \left.\times
F_1\left(1/2,1/2,-j/2;1;1-\epsilon,1-\frac{\chi_m}{\chi_p}\right)\right.
\\ \nn
&&- \left.\frac{2}{\sqrt{W}(1-v^2)}\left[1-\tilde{\gamma}K
-u\left(u-\tilde{\gamma}u
K+\tilde{\gamma}vW\right)\right]\chi_p\right.
\\ \nn && \left. \times
F_1\left(1/2,1/2,-1-j/2;1;1-\epsilon,1-\frac{\chi_m}{\chi_p}\right)\right\}.\eea

It can be shown that (\ref{c3prgf}) reduces to the undeformed case
if we fix \bea\nn \tilde{\gamma}=K=\chi_n=0\h \Rightarrow
\epsilon=\frac{\chi_m}{\chi_p}.\eea This can be done by using the
following property of the hypergeometric function $F_1$ \bea\nn
F_1(a,b_1,b_2;c;z,z)={}_2F_1\left(a,b_1+b_2;c;z\right).\eea

For the infinite-size case, (\ref{c3prgf}) gives \bea\label{cginf}
&&\mathcal{C}_{j\tilde{\gamma}\infty}^{pr}= \pi c_{\Delta}^{pr}
\frac{\Gamma\left(\frac{j}{2}\right)\Gamma\left(1+\frac{j}{2}\right)}
{\Gamma\left(\frac{3+j}{2}\right)\Gamma\left(\frac{1+j}{2}\right)}
\mathcal{J}_2\frac{\sin^{j-2}(p/2)}{\sqrt{\mathcal{J}_2^2+4\sin^2(p/2)}}
\left[\mathcal{J}_2+\tilde{\gamma}\sin^2(p/2)\sin(p)\right].\eea

{\bf Leading finite-size effect}

Again, we will consider here the particular case
$\mathcal{J}_2=0$, $\mathcal{J}_{1}\equiv \mathcal{J}$ large but
finite, i.e. $J_1\equiv J\gg \sqrt{\lambda}$.

As was pointed out in \cite{25}, where the undeformed case has
been considered, $j=1$ and $j=2$ are special values. That is why
we will start with these two cases first.

{\it The case $j=1$}

Expanding the coefficients in
$\mathcal{C}_{1\tilde{\gamma}}^{pr}$, one can rewrite it in the
following form \bea\nn &&\mathcal{C}_{1\tilde{\gamma}}^{pr}\approx
c_1^{pr}\frac{\pi^2}{2} \Bigg\{\frac{1}{\chi_{p0}}
F_1\left(1/2,1/2,-1/2;1;1-\epsilon,1-\frac{\chi_{m1}}{\chi_{p0}}\
\epsilon\right) \Big[\left(1-v_0^2\right)\chi_{n1}\ \epsilon
\\ \nn
&&+\left(2-(4v_0v_1+3W_1+4\tilde{\gamma}K_1)\ \epsilon
-v_0^2(2+W_1\ \epsilon)\right){\chi_{p0}}-4v_0v_2{\chi_{p0}}\
\epsilon\log(\epsilon)\Big]
\\ \label{pr1} &&- F_1\left(1/2,1/2,-3/2;1;1-\epsilon,1-\frac{\chi_{m1}}{\chi_{p0}}\
\epsilon\right)
\\ \nn &&\times
\Big[4\chi_{p0}+2\left(\chi_{n1}-(W_1+2\tilde{\gamma}(K_1+v_0u_1))
\chi_{p0}+2\chi_{p1}\right)\ \epsilon
\\ \nn
&&+4(\chi_{p2}-\tilde{\gamma}v_0u_2\chi_{p0})\ \epsilon
\log(\epsilon)\Big]\Bigg\}.\eea
$\mathcal{C}_{1\tilde{\gamma}}^{pr}$ can be represented as a
function of $\mathcal{J}$, $p$ and $\Phi$ in the following way
\cite{26} \bea\nn &&\mathcal{C}_{1\tilde{\gamma}}^{pr}\approx
-c_1^{pr} \frac{\pi^2}{4} \sin^2(p/2) \Bigg\{8
F_1\left(1/2,1/2,-3/2;1;1-\epsilon,1-\frac{1}{2}\left(1+\cos\Phi\right)\
\epsilon\right)
\\ \nn &&-4 F_1\left(1/2,1/2,-1/2;1;1-\epsilon,1-\frac{1}{2}\left(1+\cos\Phi\right)\
\epsilon\right)
\\ \label{pr11} &&+\Bigg[F_1\left(1/2,1/2,-1/2;1;1-\epsilon,1-\frac{1}{2}\left(1+\cos\Phi\right)\
\epsilon\right)
\\ \nn &&\times\left(1-\cos\Phi
\left(9+2 \cos p + \mathcal{J}(1+\cos p)\csc(p/2)\right)+4
\tilde{\gamma}\sin(p/2)\sin\Phi\right) \\ \nn
&&-F_1\left(1/2,1/2,-3/2;1;1-\epsilon,1-\frac{1}{2}\left(1+\cos\Phi\right)\
\epsilon\right)
\\ \nn &&\times\left(2-2\cos\Phi\left(5+2\cos p+
\mathcal{J}(1+\cos p)\csc(p/2)\right)\right.
\\ \nn
&&+\left.\tilde{\gamma}\left(\mathcal{J}(1+\cos
p)+4\sin(p/2)\right)\sin\Phi\right)\Bigg]\epsilon\Bigg\}.\eea

For the undeformed case, when $\tilde{\gamma}=0$, $\Phi=0$,
(\ref{pr11}) simplifies to \bea\label{pr1u}
\mathcal{C}_{1}^{pr}\approx -c_1^{pr} \frac{\pi^2}{4} \sin(p/2)
\left[3\sin(p/2)+\sin(3p/2)+\mathcal{J}(1+\cos
p)\right]\epsilon^2.\eea This is in accordance with the result
$\mathcal{C}_{1}^{pr}\approx 0$ found in \cite{25}, where only the
leading order in $\epsilon$ was taken into account.

{\it The case $j=2$}

Now we have \bea\label{pr2} &&\mathcal{C}_{2\tilde{\gamma}}^{pr}=
- \frac{8}{3}c_2^{pr}
\frac{1}{(1-\epsilon)^2\sqrt{(1-u^2)W(\chi_p-\chi_n)}}
\Big\{\Big[3-(1+2v^2)W-3\tilde{\gamma}K\Big](1-\epsilon)
\\ \nn &&\times \Big[\left(\chi_{m}-\chi_{p}\right) \mathbf{E}(1-\epsilon)
-(\chi_{m}-\chi_{p}\ \epsilon)
\mathbf{K}(1-\epsilon)+\left(1-u(u-\tilde{\gamma}(Ku-v W))
-\tilde{\gamma}K\right)
\\ \nn &&\times
(2(\chi_{p}-\chi_{m})((2-\epsilon)\chi_{m}+(1-2\epsilon)\chi_{p})
\mathbf{E}(1-\epsilon)+((3-\epsilon)\chi_{m}^{2}-4\chi_{m}\chi_{p}\
\epsilon
\\ \nn
&&-\chi_{p}^{2}(1-3\epsilon)\
\epsilon)\mathbf{K}(1-\epsilon)\Big]\Big\}.\eea

Expanding (\ref{pr2}) in $\epsilon$, one finds \bea\label{pr2f}
&&\mathcal{C}_{2\tilde{\gamma}}^{pr}\approx \frac{2}{3}c_2^{pr}
\sin^2(p/2) \left[2\mathcal{J}\cos\Phi-\tilde{\gamma}
\left(2\sin(p/2)-\mathcal{J}(1+\cos
p)\right)\sin(p/2)\sin\Phi\right]\epsilon.\eea Obviously, the
result for the undeformed case is properly reproduced by the above
formula.

Now, we will deal with $j\ge 3$, when we can use the following
representation of $F_1(a,b_1,b_2;c;z_1,z_2)$ \cite{w}:
\bea\label{F1exp2F1} F_1(a,b_1,b_2;c;z_1,z_2)= \sum_{k=
0}^{\infty} \frac{(a)_{k}(b_2)_{k}}{(c)_{k}}\
{}_2F_1\left(a+k,b_1;c+k;z_1\right)\frac{z_2^k}{k!} .\eea

Then, expending
${}_2F_1\left(\frac{1}{2}+k,\frac{1}{2};1+k;1-\epsilon\right)(1-\chi_m/\chi_p)^k$
around $\epsilon=0$, one finds \bea\nn
&&{}_2F_1\left(\frac{1}{2}+k,\frac{1}{2};1+k;1-\epsilon\right)\left(1-\frac{\chi_m}{\chi_p}\right)^k
\approx \frac{\Gamma(1+k)}{\sqrt{\pi}\ \Gamma(\frac{1}{2}+k)}
\left\{\log(4)-H_{k-\frac{1}{2}}\right. \\ \nn &&-\left.
\frac{1}{4\chi_{p0}} \left[2\chi_{p0}+\left(4k
\chi_{m1}-(1+2k)\chi_{p0}\right)\left(\log(4)-H_{k-\frac{1}{2}}\right)\right]\epsilon
-\log(\epsilon)\right.
\\ \label{expF}
&&-\left.\frac{\chi_{p0}+2k(\chi_{p0}-2\chi_{m1})}{4\chi_{p0}}\
\epsilon\log(\epsilon)\right\},\eea where $H_{z}$ is defined as
\cite{w}\bea\nn H_{z}=\psi(z+1)+\gamma.\eea

The replacement of (\ref{expF}) in (\ref{F1exp2F1}), taking into
account that \bea\nn a=\frac{1}{2},\h b_1=\frac{1}{2},\h c=1,\h
z_1=1-\epsilon,\h z_2=1-\frac{\chi_m}{\chi_p},\eea gives
\bea\label{F1exp}
F_1\left(\frac{1}{2},\frac{1}{2},b_2;1;1-\epsilon,1-\frac{\chi_m}{\chi_p}\right)\approx
C_0 + C_1\ \epsilon + C_2\ \epsilon \log(\epsilon) + C_3
\log(\epsilon),\eea where \bea\label{Ccoeff} &&C_0=
\frac{\Gamma(-b_2)}{\sqrt{\pi}\
\Gamma\left(\frac{1}{2}-b_2\right)} + \frac{\log(16)}{\pi}\
{}_1F_0(b_2,1),\\
\nn &&C_1= \frac{1}{4\pi}\Bigg
\{\frac{1}{\chi_{p0}}\Bigg[-\frac{\sqrt{\pi}\
\Gamma(-1-b_2)}{\Gamma\left(\frac{1}{2}-b_2\right)}\ (\chi_{p0}+2
b_2 \chi_{m1})
\\ \nn &&+8 \log(2)\ b_2 (\chi_{p0}-2 \chi_{m1})\
{}_1F_0(1+b_2,1)\Bigg ]-2(1-\log(4))\ {}_1F_0(b_2,1)\Bigg \},\\
\nn &&C_2=-\frac{1}{4\pi\chi_{p0}} \Big[\chi_{p0}\
{}_1F_0(b_2,1)+2 b_2(\chi_{p0}-2\chi_{m1})\ {}_1F_0(1+b_2,1)\Big],\\
\nn &&C_3= -\frac{1}{\pi}\ {}_1F_0(b_2,1).\eea

In the normalized structure constants (\ref{c3prgf}), there are
two hypergeometric functions\\
$F_1\left(\frac{1}{2},\frac{1}{2},b_2;1;1-\epsilon,1-\frac{\chi_m}{\chi_p}\right)$
with $b_2=-j/2$ and $b_2=-1-j/2$.

By using (\ref{F1exp}), (\ref{Ccoeff}) in (\ref{c3prgf}) and
expanding it about $\epsilon=0$, we can write down the following
approximate equality for $j\ge 3$ \bea\label{cprjgg2}
&&\mathcal{C}_{j\tilde{\gamma}}^{pr}\approx A_0+A_1 \epsilon+A_2
\epsilon\log(\epsilon),\eea where the coefficients are given by
\bea\label{Acoeffs} &&A_0= c_j^{pr}\pi
\frac{\Gamma(\frac{j}{2})^2}{\Gamma(\frac{1+j}{2})\Gamma(\frac{3+j}{2})}
\ j \ \chi_{p0}^{\frac{1}{2}(j-1)} (1-v_0^2-\chi_{p0}),\\ \nn
&&A_1= c_j^{pr}\frac{\pi}{4}
\frac{\Gamma(\frac{j}{2})\Gamma(\frac{j}{2}-1)}{\Gamma(\frac{1+j}{2})\Gamma(\frac{3+j}{2})}
\ \chi_{p0}^{\frac{1}{2}(j-3)}
\Big\{4(W_1+\chi_{m1})\chi_{p0}-2\chi_{p0}^2
\\ \nn
&&-\left[2\chi_{n1}(1-v_0^2-\chi_{p0})+\chi_{p0}(1-v_0(v_0+8v_1+2v_0W_1)\right.
\\ \nn
&&-\left.\chi_{p0}(1-2W_1))-2(1-v_0^2+\chi_{p0})\chi_{p1}\right]j
\\ \nn
&&+\left[\chi_{n1}-4v_0v_1\chi_{p0}+\chi_{m1}(1-v_0^2-\chi_{p0})
-v_0^2(\chi_{n1}+W_1\chi_{p0}-3\chi_{p1})-3\chi_{p1}\right.
\\ \nn
&&+\left.\chi_{p0}\left(-\chi_{n1}+W_1(-1+\chi_{p0})+\chi_{p1}\right)\right]j^2
\\ \nn
&&+(1-v_0^2-\chi_{p0})\chi_{p1}\ j^3
\\ \nn
&&+\tilde{\gamma}\left[4K_1\chi_{p0}+\left(2\chi_{p0}\left(K_1-2(K_1+v_0u_1)\chi_{p0}\right)\right)j
+\left(2\chi_{p0}\left(v_0u_1\chi_{p0}-K_1(1-\chi_{p0})\right)\right)j^2\right]\Big\},\\
\nn &&A_2= -c_j^{pr}\frac{\pi}{2}
\frac{\Gamma(\frac{j}{2})^2}{\Gamma(\frac{1+j}{2})\Gamma(\frac{3+j}{2})}
\ j \
\chi_{p0}^{\frac{1}{2}(j-3)}\left[4v_0v_2\chi_{p0}+(1-v_0^2+\chi_{p0})\chi_{p2}\right.
\\ \nn
&&-\left.(1-v_0^2-\chi_{p0})\chi_{p2}\
j-2\tilde{\gamma}v_0u_2\chi_{p0}^2\right].\eea

Now, our goal is to express (\ref{cprjgg2}) in terms of
$\mathcal{J}$, $p$, and $\Phi$. The result is given by \cite{26}
\bea\label{Cprf} &&\mathcal{C}_{j\tilde{\gamma}}^{pr}\approx
c_j^{pr}\frac{\pi}{8}
\frac{\Gamma\left(\frac{j}{2}\right)}{\Gamma\left(\frac{1+j}{2}\right)\Gamma\left(\frac{3+j}{2}\right)}
\sin\left(\frac{p}{2}\right)^{1+j}\left[4(j-1)\Gamma\left(\frac{j}{2}-1\right)\cos(\Phi)\right.
\\ \nn
&&-\left.\tilde{\gamma}\Gamma\left(\frac{j}{2}\right)
\left(4\sin\left(\frac{p}{2}\right)-j\left(1+\cos\left(p\right)\right)\mathcal{J}\right)
\sin(\Phi)\right]\epsilon.\eea Let us point out that (\ref{Cprf})
reduces exactly to the result found for the undeformed case in
\cite{25}, when $\tilde{\gamma}=0$, $\Phi=0$. Moreover, it
generalizes it for any $j\ge 3$.

\subsubsection{Two GM states and singlet
scalar operators on higher string levels}

According to \cite{23}, the normalized structure constant for the
present case is given by \bea\label{c3dgf}
&&\mathcal{C}_{\tilde{\gamma}}^{q}= c_{\Delta_q}\pi^{3/2}
\frac{\Gamma\left(\frac{\Delta_q}{2}\right)}{\Gamma\left(\frac{\Delta_q+1}{2}\right)}
\frac{(-2A)^q}{(1-v^2)^{q-1}\sqrt{(1-u^2)W(\chi_p-\chi_n)}}
\\ \nn &&\sum_{k=0}^{q}\frac{q!}{k!(q-k)!}\left(-\frac{B}{A}\right)^{k}\chi_p^{k}\
F_1\left(\frac{1}{2},\frac{1}{2},-k;1;1-\epsilon,1-\frac{\chi_m}{\chi_p}\right),\eea
where \bea\label{deg} &&A=1-\frac{1}{2}(1+v^2)W-\tilde{\gamma}K
,\h
B=1-\tilde{\gamma}K -u\left[u-\tilde{\gamma}(K u-v W)\right],\\
\nn &&\epsilon=\frac{\chi_{m}-\chi_{n}}{\chi_{p}-\chi_{n}},\eea

Now, let us write down what the general formula (\ref{c3dgf}) for
the normalized structure constant in the $\gamma$-deformed case
gives for the first two string levels.

$q=1$ (level $n=0$): \bea\nn
&&\mathcal{C}_{\tilde{\gamma}}^1=2c_{\Delta_1}\pi^{3/2}
\frac{\Gamma\left(\frac{\Delta_1}{2}\right)}{\Gamma\left(\frac{\Delta_1+1}{2}\right)}
\frac{1-\frac{1}{2}(1+v^2)W-\tilde{\gamma}K}
{\sqrt{(1-u^2)W(\chi_p-\chi_n)}}
\\ \nn &&\Bigg[\frac{1-u^2-\tilde{\gamma}\left(uvW+(1-u^2)K\right)}{1-\frac{1}{2}(1+v^2)W-\tilde{\gamma}K}
\chi_p \ F_1\left(1/2,1/2,-1;1;1-\epsilon,1-\chi_m/\chi_p\right)
\\ \nn
&&-\frac{2}{\pi}\mathbf{K}\left(1-\epsilon\right)\Bigg].\eea

$q=2$ (level $n=1$): \bea\nn
&&\mathcal{C}_{\tilde{\gamma}}^2=4c_{\Delta_2}\pi^{3/2}
\frac{\Gamma\left(\frac{\Delta_2}{2}\right)}{\Gamma\left(\frac{\Delta_2+1}{2}\right)}
\frac{\left(1-\frac{1}{2}(1+v^2)W-\tilde{\gamma}K\right)^2}
{(1-v^2)\sqrt{(1-u^2)W(\chi_p-\chi_n)}}
\\ \nn &&\Bigg[\frac{2}{\pi}\mathbf{K}\left(1-\epsilon\right) -
2\frac{1-u^2-\tilde{\gamma}\left(uvW+(1-u^2)K\right)}{1-\frac{1}{2}(1+v^2)W-\tilde{\gamma}K}
\chi_p \ F_1\left(1/2,1/2,-1;1;1-\epsilon,1-\chi_m/\chi_p\right)
\\ \nn
&&+\left(\frac{1-u^2-\tilde{\gamma}\left(uvW+(1-u^2)K\right)}{1-\frac{1}{2}(1+v^2)W-\tilde{\gamma}K}\right)^2\chi_p^2
\
F_1\left(1/2,1/2,-2;1;1-\epsilon,1-\chi_m/\chi_p\right)\Bigg].\eea

{\bf Leading finite-size effect}

Here, we restrict ourselves to the case $\mathcal{J}_2=0$,
$\mathcal{J}_{1}=\mathcal{J}$ large but finite, i.e. $J_1\gg
\sqrt{\lambda}$.

For this case, we were not able to obtain a general formula for
the leading finite-size corrections to the three-point correlation
functions in terms of $\mathcal{J}$, $p$, and $\Phi$, for any
$q\ge 1$. That is why, we are going to present here the results
for $q=1,...,5$ (string levels $n=0,1,2,3,4$).

Here, we are interested in the case of small $\epsilon$ (or,
equivalently, large $\mathcal{J}$) limit. So, we will expand
everything in $\epsilon$. Since the computations are similar to
the previously considered cases, we will write down the final
results only. They are given by the following approximate
equalities: \bea\nn &&\mathcal{C}_{\tilde{\gamma}}^{1}\approx
c_{\Delta_1}\frac{\sqrt{\pi}}{8}\frac{\Gamma\left(\frac{\Delta_1}{2}\right)}
{\Gamma\left(\frac{1+\Delta_1}{2}\right)}\sin(p/2)
\Big\{16-8\mathcal{J}\csc(p/2)+\left[4-\left(2\left(1-\cos p
+\mathcal{J}^2\cot^2(p/2)\right)\right.\right.
\\ \nn
&&+\left.\left.\mathcal{J}(5-\cos
p)\csc(p/2)\right)\cos\Phi+8\tilde{\gamma}\mathcal{J}\sin^2(p/2)
\sin\Phi\right]\epsilon\Big\},\eea

\bea\nn &&\mathcal{C}_{\tilde{\gamma}}^{2}\approx
-c_{\Delta_2}\frac{\sqrt{\pi}}{24}\frac{\Gamma\left(\frac{\Delta_2}{2}\right)}
{\Gamma\left(\frac{1+\Delta_2}{2}\right)}
\Big\{8(2\sin(p/2)-3\mathcal{J})+\Big[12\sin(p/2) \\ \nn &&+
\left(2(27+5\cos p)\sin(p/2)-\mathcal{J}(31+13\cos
p+3\mathcal{J}(1+\cos p)\csc (p/2))\right)\cos\Phi
\\ \nn &&-8\tilde{\gamma}\sin(p/2)(8\sin(p/2)-\mathcal{J}(7+\cos p))\sin\Phi\Big]\epsilon\Big\},\eea

\bea\nn &&\mathcal{C}_{\tilde{\gamma}}^{3}\approx c_{\Delta_3}
\frac{\sqrt{\pi}}{120}\frac{\Gamma\left(\frac{\Delta_3}{2}\right)}
{\Gamma\left(\frac{1+\Delta_3}{2}\right)}
\Big\{8(38\sin(p/2)-15\mathcal{J})+\Big[60\sin(p/2) \\ \nn &&+
\left(18(13+19\cos p)\sin(p/2)-\mathcal{J}(187+97\cos
p+15\mathcal{J}(1+\cos p)\csc (p/2))\right)\cos\Phi
\\ \nn &&-12\tilde{\gamma}\sin(p/2)(48\sin(p/2)-\mathcal{J}(23-7\cos p))\sin\Phi\Big]\epsilon\Big\},\eea

\bea\nn &&\mathcal{C}_{\tilde{\gamma}}^{4}\approx -c_{\Delta_4}
\frac{\sqrt{\pi}}{840}\frac{\Gamma\left(\frac{\Delta_4}{2}\right)}
{\Gamma\left(\frac{1+\Delta_4}{2}\right)} \Big\{1264\sin(p/2)-
840\mathcal{J}+\Big[\sin(p/2)\left(420 \right.
\\ \nn
&&+\left.\left(4730+2054\cos p-\mathcal{J}(1837+1207\cos
p)\csc(p/2)-210\mathcal{J}^2\cot^2(p/2)\right)\cos\Phi \right.
\\ \nn &&-16\left.\tilde{\gamma} \left(424\sin(p/2)-3\mathcal{J}
(79+9\cos p)\right) \sin\Phi\right)\Big]\epsilon\Big\},\eea

\bea\nn &&\mathcal{C}_{\tilde{\gamma}}^{5}\approx c_{\Delta_5}
\frac{\sqrt{\pi}}{2520}\frac{\Gamma\left(\frac{\Delta_5}{2}\right)}
{\Gamma\left(\frac{1+\Delta_5}{2}\right)}
\Big\{8(902\sin(p/2)-315\mathcal{J})+\Big[1260\sin(p/2)
\\ \nn
&&+\left(2(6093+7667\cos
p)\sin(p/2)-\mathcal{J}\left(6343+4453\cos p\right.\right.
\\ \nn
&&+\left.\left.315\mathcal{J}(1+\cos
p)\csc(p/2)\right)\right)\cos\Phi
\\ \nn
&&-20\tilde{\gamma}\sin(p/2)(1376\sin(p/2)-\mathcal{J}(523-107\cos
p))\sin\Phi\Big]\epsilon\Big\}.\eea

\subsection{Semiclassical three-point correlation functions in
$\eta$-deformed $AdS_5 \times S^5$}

\subsubsection{Two GM states and dilaton with zero momentum}

We derive the 3-point correlation function between two giant
magnons heavy string states and the light dilaton operator with
zero momentum in the $\eta$-deformed $AdS_5\times S^5$ valid for
any $J_1$ and $\eta$ in the semiclassical limit. We show that this
result satisfies a consistency relation between the 3-point
correlation function and the conformal dimension of the giant
magnon. We also provide a leading finite $J_1$ correction
explicitly \cite{29}.

The normalized structure constant in the 3-point correlation
function for the case under consideration can be written as
follows \bea\label{C1} C_{\tilde{\eta}}^d=
\frac{16c_\Delta^d}{3\tilde{\eta}}\frac{\chi_m}{\sqrt{\chi_p(1-\chi_m)
(\chi_\eta-\chi_m)}} \left[
\mathbf{\Pi}\left(1-\frac{\chi_m}{\chi_p},1-\epsilon\right)
-\mathbf{K}\left(1-\epsilon\right) \right], \eea where \bea
\label{epsil}
\epsilon=\frac{\chi_{m}(\chi_\eta-\chi_p)}{\chi_{p}(\chi_\eta-\chi_m)}.
\eea

Eq.(\ref{C1}) is the main result of this paper, which is an {\it
exact} semiclassical result for the normalized structure constant
$C_{\tilde{\eta}}^d$ valid for any value of $\tilde\eta$ and
$J_1$. Here, $\chi_p$ and $\chi_m$ are determined by the angular
momentum $J_1$ and world-sheet momentum $p$ from the following
equations: \footnote{We express $J_1$ and $p$ in terms of
different but equivalent combinations of elliptic functions
compared with Eqs. (3.23) and (3.25) in \cite{28}.} \bea
J_1&=&\frac{2T}{\tilde\eta}
\frac{1}{\sqrt{\chi_p(\chi_\eta-\chi_m)}}
\left[\chi_p\mathbf{K}\left(1-\epsilon\right)
-\chi_m\mathbf{\Pi}\left(1-\frac{\chi_m}{\chi_p},
1-\epsilon\right) \right],\label{J1e}
\\
p&=&\frac{2\chi_m}{\tilde\eta}
\sqrt{\frac{1-\chi_p}{\chi_p(1-\chi_m)(\chi_\eta-\chi_m)}}
\left[\mathbf{K}\left(1-\epsilon\right)
-\mathbf{\Pi}\left(\frac{\chi_p-\chi_m}{\chi_p(1-\chi_m)},
1-\epsilon\right)\right]. \eea The world-sheet energy of the giant
magnon is given by \bea E=\frac{2T}{\tilde\eta}
\frac{\chi_p-\chi_m}{\sqrt{\chi_p(1-\chi_m)(\chi_\eta-\chi_m)}}\
\mathbf{K} \left(1-\epsilon\right).\label{E1} \eea

One of nontrivial check is that the $g$ derivative of
$\Delta=E-J_1$ should be proportional to the normalized structure
constant $C_{\tilde{\eta}}^d$ since the $g$ derivative of the
two-point function inserts the dilaton (Lagrangian) operator into
the two-point function of the heavy operators \cite{Costa}. This
can be expressed by \bea C_{\tilde{\eta}}^d=\frac{8 c_\Delta^d}{3
\sqrt{1+\tilde{\eta}^2}} \frac{\p\Delta}{\p g}.\label{check} \eea
To check that Eqs.(\ref{C1}), (\ref{J1e})-(\ref{E1}) satisfy
Eq.(\ref{check}), we use the fact that \bea \frac{\p J_1}{\p
g}=\frac{\p p}{\p g}=0 \eea as noticed in \cite{19} for the case
of undeformed giant magnon. From these, we can obtain the
expressions for $\p\chi_p/\p g$ and $\p\chi_m/\p g$ which can be
inserted to $\p\Delta/\p g$. The $\eta$-deformed case involves
much more complicated expressions which can be dealt with the
Mathematica. In the Appendix of \cite{29}, we provided our
Mathematica code which confirms that the structure constant
$C_{\tilde{\eta}}^d$ in Eq.(\ref{C1}) do sastisfy the consistency
condition (\ref{check}) exactly.

In the limit $\tilde{\eta}\to 0$ with $\tilde{\eta}^2\chi_\eta\to
1$, Eq.(\ref{C1}) becomes \bea\label{C10}C_3^{0}=
\frac{16c_\Delta^d}{3}\sqrt{\frac{\chi_p}{1-\chi_m}} \ \left[
\mathbf{E}\left(1-\epsilon\right)
-\epsilon\mathbf{K}\left(1-\epsilon\right) \right], \h
\epsilon=\frac{\chi_m}{\chi_p} \eea where we used the identity
$(1-a)\mathbf{\Pi}(a,a)=\mathbf{E}(a)$. This is the structure
constant of the undeformed theory derived in \cite{19}.

{\bf Leading finite-size effect}

It is straightforward to compute the leading finite-size effect on
$C_{\tilde{\eta}}^d$ for $J_1\gg g$ by taking the limit
$\epsilon\to 0$ in (\ref{C1}).

First we expand the parameters $\chi_p$, $W$ and $v$ for small
$\epsilon$ as follows: \bea\label{chie} &&\chi_{p}= \chi_{p
0}+(\chi_{p 1}+\chi_{p 2}\log\epsilon)\epsilon,
\\ \nn &&W= 1+W_1 \epsilon,
\\ \nn &&v= v_0+(v_1+v_2\log\epsilon)\epsilon.\eea
Inserting into Eq.(\ref{C1}), we obtain \bea\label{CLO1}
&&C_{\tilde{\eta}}^d\approx
\frac{16c_{\Delta}^{d}}{3\tilde{\eta}^2
\sqrt{\left(1+\frac{1}{\tilde{\eta}^2}\right)\chi_{p0}}}
\Bigg\{\sqrt{(1+\tilde{\eta}^2)\chi_{p0}}\
\mbox{arctanh}\frac{\tilde{\eta}\sqrt{\chi_{p0}}}{\sqrt{1+\tilde{\eta}^2}}
\\ \nn &&-\Bigg[\frac{W_1}{2} \sqrt{(1+\tilde{\eta}^2)\chi_{p0}}\ \mbox{arctanh}\frac{\tilde{\eta}\sqrt{\chi_{p0}}}{\sqrt{1+\tilde{\eta}^2}}
+\frac{\tilde{\eta}}{4\left(1+\tilde{\eta}^2(1-\chi_{p0})\right)}\times
\\ \nn &&\left((1+\tilde{\eta}^2)(\chi_{p0}-2\chi_{p1})-
4\left((1+\tilde{\eta}^2)\chi_{p0}+2W_1\left(1+\tilde{\eta}^2(1-\chi_{p0})\right)\right)
\log 2\right)\Bigg]\epsilon
\\ \nn &&-\frac{\tilde{\eta}}{4\left(1+\tilde{\eta}^2(1-\chi_{p0})\right)}
\left(\left((1+\tilde{\eta}^2)(\chi_{p0}-2\chi_{p2})
+2W_1\left(1+\tilde{\eta}^2(1-\chi_{p0})\right)\right)\right)\epsilon\log\epsilon\Bigg\}
.\eea

In view of the equations \bea \chi_{m}=1-W,\h \chi_{p}=1-v^2 W,\h
\chi_{\eta}=1+\frac{1}{{\tilde\eta}^2} \label{param} \eea and
(\ref{epsil}), we can express all the auxiliary parameters in
terms of $v$ (or its coefficients $v_0$, $v_1$, and $v_2$):
\bea\label{chisole} &&\chi_{p 0}=1-v_0^2,\h \chi_{p
1}=1-v_0^2-2v_0 v_1- \frac{(1-v_0^2)^2}{1+\tilde{\eta}^2 v_0^2},\h
\chi_{p 2}=-2v_0 v_2,
\\ \nn &&W_1=-\frac{(1+\tilde{\eta}^2)(1-v_0^2)}{1+\tilde{\eta}^2 v_0^2}.\eea
This leads to \bea\label{CLO2}  &&C_{\tilde{\eta}}^d\approx
\frac{16 c_\Delta^d}{3 \tilde{\eta}}
\Bigg\{\mbox{arctanh}\frac{\tilde{\eta}\sqrt{1-v_0^2}}{\sqrt{1+\tilde{\eta}^2}}
+\frac{1}{4\sqrt{(1+\tilde{\eta}^2)(1-v_0^2)}{\left(1+\tilde{\eta}^2
v_0^2\right)^2}}\times
\\ \nn &&\Bigg[(1+\tilde{\eta}^2)\left((1-v_0^2)\left(1+\tilde{\eta}^2 v_0^2\right)
\left(2\sqrt{\left(1+\tilde{\eta}^2\right)((1-v_0^2)}\mbox{arctanh}\frac{\tilde{\eta}\sqrt{1-v_0^2}}{\sqrt{1+\tilde{\eta}^2}}
-\tilde{\eta}\log 16\right)\right.
\\ \nn &&\left. -\tilde{\eta} \left(1-v_0(3v_0-2v_0^3-4v_1+v_0(1-v_0^2-4v_0 v_1)\tilde{\eta}^2)\right)\right)\Bigg]\epsilon
\\ \nn &&+\frac{\tilde{\eta}(1+\tilde{\eta}^2)(1-v_0^2-4v_0 v_2)}
{4\sqrt{(1+\tilde{\eta}^2)(1-v_0^2)}(1+\tilde{\eta}^2 v_0^2)}\
\epsilon\log\epsilon\Bigg\}.\eea

To fix $v_0$, $v_1$, and $v_2$, one can use the small $\epsilon$
expansion of the angular difference \bea\nn
\Delta\phi_1=\phi_1(\tau,L)-\phi_1(\tau,-L)\equiv p,\eea where we
identified the angular difference $\Delta\phi_1$ with the magnon
momentum $p$ on the dual spin chain. The result is \cite{28}
\bea\label{v0sole} v_0=\frac{\cot\frac{p}{2}}
{\sqrt{\tilde{\eta}^2+\csc^2\frac{p}{2}}},\eea and
\bea\label{v1v2sol} v_1=\frac{v_0(1-v_0^2)\left[1-\log 16
+\tilde{\eta}^2 \left(2-v_0^2(1+\log 16
)\right)\right]}{4(1+\tilde{\eta}^2 v_0^2)}, \h
v_2=\frac{1}{4}v_0(1-v_0^2).\eea By using (\ref{v0sole}),
(\ref{v1v2sol}) in (\ref{CLO2}), one finds \bea\label{C3}
&&C_3^{\tilde{\eta}}\approx \frac{16 c_\Delta^d}{3 \tilde{\eta}}
\Bigg\{\mbox{arcsinh}\left( \tilde{\eta} \sin\frac{p}{2}\right)+
\frac{(1+\tilde{\eta}^2)\sin^2\frac{p}{2}}{4\sqrt{\tilde{\eta}^2+\csc^2
\frac{p}{2}}}\times
\\ \nn &&\Bigg[\Bigg(2 \sqrt{\tilde{\eta}^2+\csc^2
\frac{p}{2}}\mbox{arcsinh}\left( \tilde{\eta}
\sin\frac{p}{2}\right) -\tilde{\eta}(1+\log 16)\Bigg)\epsilon
+\tilde{\eta}\epsilon\log\epsilon\Bigg] \Bigg\}.\eea

The expansion parameter $\epsilon$ in the leading order is given
by \cite{28} \bea\label{eps1} \epsilon =16\
\exp\left[-\left(\frac{J_1}{g}
+\frac{2\sqrt{1+\tilde{\eta}^2}}{\tilde{\eta}}\mbox{arcsinh}\left(\tilde{\eta}
\sin\frac{p}{2}\right)
\right)\sqrt{\frac{1+\tilde{\eta}^2\sin^2\frac{p}{2}}{\left(1+\tilde{\eta}^2\right)\sin^2\frac{p}{2}}}
\right].\eea Here we used Eq.({\ref{T}) for the string tension
$T$.

The final expression for the normalized structure costant is given
by \bea\label{C3f} &&C_{\tilde{\eta}}^d\approx \frac{16
c_\Delta^d}{3 \tilde{\eta}} \Bigg\{
\mbox{arcsinh}\left(\tilde{\eta}\sin\frac {p}{2}\right)- 4
\frac{\tilde{\eta}(1+\tilde{\eta}^2)\sin^3
\frac{p}{2}}{\sqrt{1+\tilde{\eta}^2
\sin^2\frac{p}{2}}}\Bigg[1+\frac{J_1}{g}
\sqrt{\frac{\tilde{\eta}^2+\csc^2\frac{p}{2}}{1+\tilde{\eta}^2}}\Bigg]
\\ \nn &&\times \exp\left[-\left(\frac{J_1}{g}
+\frac{2\sqrt{1+\tilde{\eta}^2}}{\tilde{\eta}}\mbox{arcsinh}\left(\tilde{\eta}
\sin\frac{p}{2}\right)
\right)\sqrt{\frac{1+\tilde{\eta}^2\sin^2\frac{p}{2}}{\left(1+\tilde{\eta}^2\right)\sin^2\frac{p}{2}}}
\right]\Bigg\}.\eea

Let us point out that in the limit $\tilde{\eta} \to 0$,
(\ref{C3f}) reduces to \bea\nn C_3\approx \frac{16}{3} c_\Delta^d
\sin\frac{p}{2}
\left[1-4\sin\frac{p}{2}\left(\sin\frac{p}{2}+\frac{J_1}{g}\right)
\exp\left(-\frac{J_1}{g \sin\frac{p}{2}}-2\right)\right],\eea
which reproduces the result for the undeformed case found in
\cite{19}. Another check is that this satisfies Eq.(\ref{check})
with $\Delta$ computed in \cite{28} \bea\label{fre} &&\Delta\equiv
E-J_1\approx 2 g \sqrt{1+\tilde{\eta}^2}
\Bigg\{\frac{1}{\tilde{\eta}} \mbox{arcsinh}\left(\tilde{\eta}
\sin\frac{p}{2}\right)-4\frac{(1+\tilde{\eta}^2)
\sin^3\frac{p}{2}}{\sqrt{1+\tilde{\eta}^2
\sin^2\frac{p}{2}}}\times
\\ \nn && \exp\left[-\left(\frac{J_1}{g}
+\frac{2\sqrt{1+\tilde{\eta}^2}}{\tilde{\eta}}\mbox{arcsinh}\left(\tilde{\eta}
\sin\frac{p}{2}\right)
\right)\sqrt{\frac{1+\tilde{\eta}^2\sin^2\frac{p}{2}}{\left(1+\tilde{\eta}^2\right)\sin^2\frac{p}{2}}}
\right] \Bigg\}.\eea

\subsubsection{Two GM states and dilaton with non-zero momentum}

Here we will be interested in the case when the dilaton momentum
$j>0$. According to \cite{30} the semiclassical normalized
structure constants for the case under consideration is given by
\bea\label{etadj} \mathcal{C}_{\tilde{\eta}}^{d,j}&=& \frac{2
\pi^{\frac{3}{2}}c_\Delta^{d,j}
\Gamma\left(2+\frac{j}{2}\right)(1-v^2 \kappa^2)^{\frac{j-1}{2}}}
{\Gamma\left(\frac{5+j}{2}\right)\sqrt{\kappa^2(1+\tilde{\eta}^2
\kappa^2)}}\times
\\ \nn
&&\Bigg[(1-v^2 \kappa^2)F_1
\left(\frac{1}{2},\frac{2+j}{2},-\frac{1+j}{2};1;\frac{\tilde{\eta}^2(1-v^2)\kappa^2}
{1+\tilde{\eta}^2\kappa^2},\frac{(1+\tilde{\eta}^2)(1-v^2)\kappa^2}
{(1+\tilde{\eta}^2\kappa^2)(1-v^2\kappa^2)}\right)
\\ \nn &&-(1-\kappa^2)F_1 \left(\frac{1}{2},\frac{j}{2},\frac{1-j}{2};1;\frac{\tilde{\eta}^2(1-v^2)\kappa^2}
{1+\tilde{\eta}^2\kappa^2},\frac{(1+\tilde{\eta}^2)(1-v^2)\kappa^2}
{(1+\tilde{\eta}^2\kappa^2)(1-v^2\kappa^2)}\right)\Bigg].\eea

Now we take the limit $\tilde{\eta}\to 0$ in (\ref{etadj}) and
obtain \bea\label{dud} \mathcal{C}^{d,j}&=&\frac{2
\pi^{\frac{3}{2}}c_\Delta^{d,j}
\Gamma\left(2+\frac{j}{2}\right)(1-v^2
\kappa^{2})^{\frac{j-1}{2}}}{\kappa
\Gamma\left(\frac{5+j}{2}\right)} \times
\\ \nn
&&\Bigg[(1-v^2
\kappa^2){}_2F_1\left(\frac{1}{2},-\frac{1+j}{2};1;\frac{(1-v^2)\kappa^2)}{1-v^2
\kappa^2}\right)
\\ \nn
&&-(1-\kappa^2){}_2F_1\left(\frac{1}{2},\frac{1-j}{2};1;\frac{(1-v^2)\kappa^2)}{1-v^2
\kappa^2}\right)\Bigg] .\eea This is exactly what was found in
\cite{22} for $u=0$, as it should be.

Let us also say that in the particular case when $j=1$,
(\ref{etadj}) simplifies to \bea\nn
\mathcal{C}_{\eta}^{d,1}=\frac{3\pi
c_\Delta^{d,1}\sqrt{\kappa^2(1+\tilde{\eta}^2 \kappa^2)}} {2
\tilde{\eta}^2 \kappa^2}\Bigg[\mathbf{K}\left(\tilde{\eta}^2
\frac{(1-v^2)\kappa^2}{1+\tilde{\eta}^2\kappa^2}\right)-\mathbf{E}\left(\tilde{\eta}^2
\frac{(1-v^2)\kappa^2}{1+\tilde{\eta}^2\kappa^2}\right)\Bigg]
.\eea In the limit $\tilde{\eta}\to 0$, $\mathcal{C}_{\eta}^{d,1}$
becomes \bea\nn\mathcal{C}^{d,1}=\frac{3}{8}\pi^2
c_\Delta^{d,1}\kappa (1-v^2).\eea

\subsubsection{Two GM states and primary scalar operators}

It was proven in \cite{30} that the normalized structure constant
for this case can be represented as \bea\label{c3prs}
\mathcal{C}_{\eta}^{pr,j}&=&\frac{2
c_{\Delta}^{pr,j}\sqrt{\pi}}{\tilde{\eta}\kappa} \frac{\Gamma
(\frac{j}{2})}{\Gamma (\frac{1+j}{2})}
 \left[\frac{1-\kappa^{2}+j(1-v^2\kappa^{2})}{1+j} J_j-J_{jp}\right], \eea
 where
\bea\label{Jj} J_j=\int_{\chi_m}^{\chi_p}
\frac{\chi^{\frac{j}{2}}}
{\sqrt{(\chi_\eta-\chi)(\chi_p-\chi)(\chi-\chi_m)\chi}}\
d\chi,\eea \bea\label{Jjp} J_{jp}=\int_{\chi_m}^{\chi_p}
\frac{\chi^{\frac{j}{2}+1}}
{\sqrt{(\chi_\eta-\chi)(\chi_p-\chi)(\chi-\chi_m)\chi}}\ d\chi
.\eea

To compute the above two integrals, we introduce the variable
\bea\nn x=\frac{\chi-\chi_m}{\chi_p-\chi_m} \in (0,1).\eea Then
$J_j$ becomes \bea\label{Jj1}
J_j=\chi_m^{\frac{j-1}{2}}(\chi_\eta-\chi_m)^{-\frac{1}{2}}
\int_{0}^{1}
x^{-\frac{1}{2}}(1-x)^{-\frac{1}{2}}\left(1-\frac{\chi_p-\chi_m}
{\chi_\eta-\chi_m}x\right)^{-\frac{1}{2}}
\left(1+\frac{\chi_p-\chi_m} {\chi_m}x\right)^{\frac{j-1}{2}}
dx.\eea Comparing the above expression with the integral
representation for the hypergeometric function of two variables
$F_1(a,b_1,b_2;c;z_1,z_2)$ \cite{PBM-III} \bea\nn
F_1(a,b_1,b_2;c;z_1,z_2)= \frac{\Gamma(c)}{\Gamma (a)\Gamma(c-a)}
\int_{0}^{1} x^{a-1}(1-x)^{c-a-1}(1-z_1 x)^{-b_1}(1-z_2 x)^{-b_2},
\\ \nn Re(a)>0,\h Re(c-a)>0,\eea
one finds \bea\label{Jj2}
J_j=\pi\chi_m^{\frac{j-1}{2}}(\chi_\eta-\chi_m)^{-\frac{1}{2}} F_1
\left(\frac{1}{2},\frac{1}{2},-\frac{j-1}{2};1;\frac{\chi_p-\chi_m}
{\chi_\eta-\chi_m},-\frac{\chi_p-\chi_m} {\chi_m}\right).\eea

In order to compute $J_{jp}$, we have to replace $j$ with $j+2$.
Doing this, we obtain \bea\label{Jjp2}
J_{jp}=\pi\chi_m^{\frac{j+1}{2}}(\chi_\eta-\chi_m)^{-\frac{1}{2}}
F_1
\left(\frac{1}{2},\frac{1}{2},-\frac{j+1}{2};1;\frac{\chi_p-\chi_m}
{\chi_\eta-\chi_m},-\frac{\chi_p-\chi_m} {\chi_m}\right).\eea

The replacement of (\ref{Jj2}) and (\ref{Jjp2}) into (\ref{c3prs})
gives \bea\label{c3prsr}
\mathcal{C}_{\eta}^{pr,j}&=&\frac{2\pi^{\frac{3}{2}}c_{\Delta}^{pr,j}}{\tilde{\eta}\kappa}
\frac{\Gamma (\frac{j}{2})}{\Gamma (\frac{1+j}{2})}
\frac{\chi_m^{\frac{j-1}{2}}}{\sqrt{\chi_\eta-\chi_m}}
\Bigg\{\left[1-\frac{(1+jv^2)\kappa^{2}}{1+j}\right]\times
\\ \nn &&F_1 \left(\frac{1}{2},\frac{1}{2},\frac{1-j}{2};1;\frac{\chi_p-\chi_m}
{\chi_\eta-\chi_m},-\frac{\chi_p-\chi_m} {\chi_m}\right)
\\ \nn &&-\chi_m F_1 \left(\frac{1}{2},\frac{1}{2},-\frac{1+j}{2};1;\frac{\chi_p-\chi_m}
{\chi_\eta-\chi_m},-\frac{\chi_p-\chi_m} {\chi_m}\right)\Bigg\}.
\eea

Knowing that \bea\nn \chi_\eta=1+\frac{1}{\tilde{\eta}^{2}}\h
\chi_p=1-v^2 \kappa^2,\h \chi_m=1-\kappa^2,\eea and using the
relation \cite{PBM-III} \bea\nn
F_1(a,b_1,b_2;c;z_1,z_2)=(1-z_1)^{c-a-b_1}(1-z_2)^{-b_2} F_1
\left(c-a,c-b_1-b_2,b_2;c;z_1,\frac{z_1-z_2}{1-z_2}\right),\eea we
can rewrite (\ref{c3prsr}) in the following form
\bea\label{c3prsf}
\mathcal{C}_{\tilde{\eta}}^{pr,j}&=&\frac{2\pi^{\frac{3}{2}}c_{\Delta}^{pr,j}\Gamma
(\frac{j}{2}) (1-v^2 \kappa^2)^{\frac{j-1}{2}}} {\Gamma
(\frac{1+j}{2})\sqrt{\kappa^2(1+\tilde{\eta}^2 \kappa^2)}}
\Bigg\{\left[1-\frac{(1+jv^2)\kappa^{2}}{1+j}\right]\times
\\ \nn &&F_1 \left(\frac{1}{2},\frac{j}{2},\frac{1-j}{2};1;\frac{\tilde{\eta}^2(1-v^2)\kappa^2}
{1+\tilde{\eta}^2\kappa^2},\frac{(1+\tilde{\eta}^2)(1-v^2)\kappa^2}
{(1+\tilde{\eta}^2\kappa^2)(1-v^2\kappa^2)}\right)
\\ \nn &&-(1-v^2\kappa^2) F_1 \left(\frac{1}{2},\frac{2+j}{2},-\frac{1+j}{2};1;\frac{\tilde{\eta}^2(1-v^2)\kappa^2}
{1+\tilde{\eta}^2\kappa^2},\frac{(1+\tilde{\eta}^2)(1-v^2)\kappa^2}
{(1+\tilde{\eta}^2\kappa^2)(1-v^2\kappa^2)}\right)\Bigg\} .\eea
This is our final exact semiclassical result for this type of
three-point correlation functions.

Next, we would like to compare (\ref{c3prsf}) with the known
expression for the undeformed case \cite{22}. To this end, we take
the limit $\tilde{\eta}\to 0$ and by using that \cite{PBM-III}
\bea\nn F_1 \left(a,b_1,b_2;c;0,z_2\right)=\
{}_2F_1(a,b_2;c;z_2),\eea  we find \bea\nn \mathcal{C}^{pr,j}&=&
\frac{2\pi^{\frac{3}{2}}c_{\Delta}^{pr,j}\Gamma\left(2+\frac{j}{2}\right)(1-v^2
\kappa^2)^{\frac{j-1}{2}}} {\kappa
\Gamma\left(\frac{5+j}{2}\right)} \Bigg[(1-v^2
\kappa^2){}_2F_1\left(\frac{1}{2},-\frac{1+j}{2};1;,\frac{(1-v^2)\kappa^2}
{1-v^2\kappa^2}\right)
\\ \nn &&-(1-\kappa^2){}_2F_1\left(\frac{1}{2},\frac{1-j}{2};1;\frac{(1-v^2)\kappa^2)}{1-v^2 \kappa^2}\right)\Bigg].\eea
This is exactly the same result found in \cite{22} for $u=0$
(finite-size giant magnons with one nonzero angular momentum) as
it should be.

Let us also give an example for the simplest case when $j=1$. In
that case (\ref{c3prsf}) reduces to
\bea\nn\mathcal{C}_{\eta}^{pr,1} &=&\frac{2 \pi c_{\Delta}^{pr,1}}
{\tilde{\eta}^2\sqrt{\kappa^2(1+\tilde{\eta}^2\kappa^2)}}
\Bigg[2(1+\tilde{\eta}^2\kappa^2)\mathbf{E}\left(\tilde{\eta}^2
\frac{(1-v^2)\kappa^2}{1+\tilde{\eta}^2\kappa^2}\right)
\\ \nn &&-(2+(1+v^2)\tilde{\eta}^2\kappa^2)\mathbf{K}\left(\tilde{\eta}^2
\frac{(1-v^2)\kappa^2}{1+\tilde{\eta}^2\kappa^2}\right)\Bigg].\eea

In the limit $\tilde{\eta}\to 0$, $\mathcal{C}_{\eta}^{pr,1}\to
0$.

\subsubsection{Two GM states and singlet
scalar operators on higher string levels}

It was found in \cite{30} that the normalized structure constants
for the case at hand are given by \bea\label{qf}
\mathcal{C}^q_{\tilde{\eta}} &=&
c_{\Delta}^{q}\frac{\sqrt{\pi}}{\kappa}
\frac{\Gamma\left(\frac{\Delta_q^\eta}{2}\right)}
{\Gamma\left(\frac{\Delta_q^\eta+1}{2}\right)}\frac{(-1)^q}{\tilde{\eta}(1-v^2)^{q-1}}
\int_{\chi_m}^{\chi_p}d \chi
\frac{\left[2-(1+v^2)\kappa^2-2\chi\right]^{q}}
{\sqrt{(\chi_\eta-\chi)(\chi_p-\chi)(\chi-\chi_m)}\chi}
\\ \nn &=& c_{\Delta}^{q}\frac{\pi^{\frac{3}{2}}}{\kappa} \frac{\Gamma\left(\frac{\Delta_q^\eta}{2}\right)}
{\Gamma\left(\frac{\Delta_q^\eta+1}{2}\right)}\frac{(-1)^q\left[2-(1+v^2)\kappa^2\right]^{q}}
{\tilde{\eta}(1-v^2)^{q-1}\sqrt{\chi_\eta-\chi_m}}\times
\\ \nn && \sum_{k=0}^{q}\frac{q!}{k!(q-k)!}\left[-\frac{1}{1-\frac{1}{2}(1+v^2)\kappa^2}\right]^k \chi_m^{k-\frac{1}{2}}
F_1
\left(\frac{1}{2},\frac{1}{2},\frac{1}{2}-k;1;\frac{\chi_p-\chi_m}{\chi_\eta-\chi_m},
-\frac{\chi_p-\chi_m}{\chi_m}\right)
\\ \nn &=&c_{\Delta}^{q}\pi^{\frac{3}{2}}\frac{\Gamma\left(\frac{\Delta_q^\eta}{2}\right)}
{\Gamma\left(\frac{\Delta_q^\eta+1}{2}\right)}\frac{(-1)^q\left[2-(1+v^2)\kappa^2\right]^{q}}
{(1-v^2)^{q-1}\sqrt{\kappa^2(1+\tilde{\eta}^2 \kappa^2)(1-v^2
\kappa^2)}}\sum_{k=0}^{q}\frac{q!}{k!(q-k)!}
\\ \nn &&  \times \left[-\frac{1-v^2\kappa^2}{1-\frac{1}{2}(1+v^2)\kappa^2}\right]^k
F_1
\left(\frac{1}{2},k,\frac{1}{2}-k;1;\frac{\tilde{\eta}^2(1-v^2)\kappa^2}
{1+\tilde{\eta}^2\kappa^2},\frac{(1+\tilde{\eta}^2)(1-v^2)\kappa^2}
{(1+\tilde{\eta}^2\kappa^2)(1-v^2\kappa^2)}\right),\eea where
\bea\label{dqe} \Delta_q^\eta=2\left(1+\sqrt{2\pi
g\sqrt{1+\tilde{\eta}^2}(q-1)+1-\frac{1}{2}q(q-1)}\right).\eea

In order to compare with the undeformed case, we take the limit
$\tilde{\eta}\to 0$ in (\ref{qf}) and obtain \bea\nn \mathcal{C}^q
&=&
c_{\Delta}^{q}\pi^{\frac{3}{2}}\frac{\Gamma\left(\frac{\Delta_q}{2}\right)}
{\Gamma\left(\frac{\Delta_q+1}{2}\right)}\frac{(-1)^q\left[2-(1+v^2)\kappa^2\right]^{q}}
{(1-v^2)^{q-1}\sqrt{\kappa^2(1-v^2
\kappa^2)}}\sum_{k=0}^{q}\frac{q!}{k!(q-k)!}
\\ \nn &&  \times \left[-\frac{1-v^2\kappa^2}{1-\frac{1}{2}(1+v^2)\kappa^2}\right]^k
{}_2 F_1 \left(\frac{1}{2},\frac{1}{2}-k;1;\frac{(1-v^2)\kappa^2}
{1-v^2\kappa^2}\right).\eea This is exactly what was found in
\cite{23} for finite-size giant magnons with one nonzero angular
momentum.

Let us consider two particular cases. From (\ref{qf}) it follows
that the normalized structure constants for the first two string
levels, for the case at hand, are given by

$q=1$ (level $n=0$)

\bea\nn C_{\tilde{\eta}}^1&=& 2
c_{\Delta}^{1}\pi^{\frac{1}{2}}\frac{\Gamma\left(\frac{\Delta_1^\eta}{2}\right)}
{\Gamma\left(\frac{\Delta_1^\eta+1}{2}\right)}\frac{1}
{\sqrt{\kappa^2(1+\tilde{\eta}^2 \kappa^2)(1-v^2 \kappa^2)}}\times
\\ \nn && \Bigg[\pi(1-v^2 \kappa^2) F_1 \left(\frac{1}{2},1,-\frac{1}{2};1;\frac{\tilde{\eta}^2(1-v^2)\kappa^2}
{1+\tilde{\eta}^2\kappa^2},\frac{(1+\tilde{\eta}^2)(1-v^2)\kappa^2}
{(1+\tilde{\eta}^2\kappa^2)(1-v^2\kappa^2)}\right)
\\ \nn &&-\left(2-(1+v^2)\kappa^2\right) \mathbf{K}\left(\frac{(1+\tilde{\eta}^2)(1-v^2)\kappa^2}
{(1+\tilde{\eta}^2\kappa^2)(1-v^2\kappa^2)}\right)\Bigg].\eea

$q=2$ (level $n=1$)

\bea\nn \mathcal{C}^2_{\tilde{\eta}}&=& 2
c_{\Delta}^{2}\pi^{\frac{3}{2}}\frac{\Gamma\left(\frac{\Delta_2^\eta}{2}\right)}
{\Gamma\left(\frac{\Delta_2^\eta+1}{2}\right)}\frac{\left(2-(1+v^2)\kappa^2\right)^2}
{(1-v^2)\sqrt{\kappa^2(1+\tilde{\eta}^2 \kappa^2)(1-v^2
\kappa^2)}}\times
\\ \nn &&\Bigg\{ \frac{1}{\pi} \mathbf{K}\left(\frac{(1+\tilde{\eta}^2)(1-v^2)\kappa^2}
{(1+\tilde{\eta}^2\kappa^2)(1-v^2\kappa^2)}\right)-\frac{2(1-v^2
\kappa^2)}{\left(2-(1+v^2)\kappa^2\right)^2} \times
\\ \nn &&\Bigg[\left(2-(1+v^2)\kappa^2\right) F_1 \left(\frac{1}{2},1,-\frac{1}{2};1;\frac{\tilde{\eta}^2(1-v^2)\kappa^2}
{1+\tilde{\eta}^2\kappa^2},\frac{(1+\tilde{\eta}^2)(1-v^2)\kappa^2}
{(1+\tilde{\eta}^2\kappa^2)(1-v^2\kappa^2)}\right)
\\ \nn &&-(1-v^2\kappa^2)F_1 \left(\frac{1}{2},2,-\frac{3}{2};1;\frac{\tilde{\eta}^2(1-v^2)\kappa^2}
{1+\tilde{\eta}^2\kappa^2},\frac{(1+\tilde{\eta}^2)(1-v^2)\kappa^2}
{(1+\tilde{\eta}^2\kappa^2)(1-v^2\kappa^2)}\right)\Bigg] \Bigg\}
.\eea

In the limit $\tilde{\eta}\to 0$, the above two expressions
simplify to \bea\nn \mathcal{C}^1 &=& 2
c_{\Delta}^{1}\pi^{\frac{1}{2}}\frac{\Gamma\left(\frac{\Delta_1}{2}\right)}
{\Gamma\left(\frac{\Delta_1+1}{2}\right)}\frac{1}
{\sqrt{\kappa^2(1-v^2 \kappa^2)}}\times
\\ \nn && \Bigg[2(1-v^2 \kappa^2) \mathbf{E}\left(\frac{(1-v^2)\kappa^2}
{1-v^2\kappa^2}\right)
\\ \nn
&&-\left(2-(1+v^2)\kappa^2\right)
\mathbf{K}\left(\frac{(1-v^2)\kappa^2}
{(1-v^2\kappa^2)}\right)\Bigg] ,\eea and \bea\nn \mathcal{C}^2 &=&
2
c_{\Delta}^{2}\pi^{\frac{1}{2}}\frac{\Gamma\left(\frac{\Delta_2}{2}\right)}
{\Gamma\left(\frac{\Delta_2+1}{2}\right)}\frac{1}
{(1-v^2)\sqrt{\kappa^2(1-v^2 \kappa^2)}}\times
\\ \nn &&\Bigg[\left(2-(1+v^2)\kappa^2\right)^2 \mathbf{K}\left(\frac{(1-v^2)\kappa^2}
{(1-v^2\kappa^2)}\right)
\\ \nn && -4\left(2-(1+v^2)\kappa^2\right)(1-v^2\kappa^2) \mathbf{E}\left(\frac{(1-v^2)\kappa^2}
{1-v^2\kappa^2}\right)
\\ \nn && +2\pi (1-v^2\kappa^2)^2 {}_2 F_1\left(\frac{1}{2},-\frac{3}{2};1;\frac{(1-v^2)\kappa^2}
{1-v^2\kappa^2}\right)\Bigg] \eea respectively.

\section{Contributions}

\begin{enumerate}

\item{In \cite{1} we consider null bosonic p-branes moving in
curved space-times. Some exact solutions of the classical
equations of motion and of the constraints for the null string and
the null membrane in Demianski-Newman background are found.}

\item{In \cite{2} we consider null bosonic p-branes moving in
curved space-times and develop a method for solving their
equations of motion and constraints, which is suitable for string
theory backgrounds. As an application, we give an exact solution
for such background in ten dimensions.}

\item{In \cite{3} we show how the classical string dynamics in
D-dimensional gravity background can be reduced to the dynamics of
a massless particle constrained on a certain surface whenever
there exists at least one Killing vector for the background
metric. We obtain a number of sufficient conditions, which ensure
the existence of exact solutions to the equations of motion and
constraints. These results are extended to include the Kalb-Ramond
background. The D1-brane dynamics is also analyzed and exact
solutions are found. Finally, we illustrate our considerations
with several examples in different dimensions. All this also
applies to the tensionless strings.}

\item{In \cite{4} we consider probe p-branes and Dp-branes
dynamics in D-dimensional string theory backgrounds of general
type. Unified description for the tensile and tensionless branes
is used. We obtain exact solutions of their equations of motion
and constraints in static gauge as well as in more general gauges.
Their dynamics in the whole space-time is also analyzed and exact
solutions are found.}

\item{In \cite{5} we classify almost all classical string
configurations, considered in the framework of the semi-classical
limit of the string/gauge theory duality. Then, we describe a
procedure for obtaining the conserved quantities and the exact
classical string solutions in general string theory backgrounds,
when the string embedding coordinates depend non-linearly on the
worldsheet time parameter.}

\item{In \cite{6}, based on the recently considered classical
string configurations, in the framework of the semi-classical
limit of the string/gauge theory correspondence, we describe a
procedure for obtaining exact classical string solutions in
general string theory backgrounds, when the string embedding
coordinates depend non-linearly on the worldsheet spatial
parameter. The tensionless limit, corresponding to small t'Hooft
coupling on the field theory side, is also considered. Applying
the developed approach, we find new string solutions - with two
spins in $AdS_5 \times S^5$ and in $AdS_5$-black hole background.}

\item{In \cite{7} we consider different M2-brane configurations in
the M-theory $AdS_7\times S^4$ background, with field theory dual
$A_{N-1}(2,0)$ SCFT. New membrane solutions are found and compared
with the recently obtained ones.}

\item{In \cite{8} motivated by the recent achievements in the
framework of the semiclassical limit of the M-theory/field theory
correspondence, we propose an approach for obtaining exact
membrane solutions in general enough M-theory backgrounds, having
field theory dual description. As an application of the derived
general results, we obtain several types of membrane solutions in
$AdS_4\times S^7$ M-theory background.}

\item{In \cite{9} we obtain exact rotating membrane solutions and
explicit expressions for the conserved charges on a manifold with
exactly known metric of $G_2$ holonomy in M-theory, with four
dimensional N=1 field theory dual. After that, we investigate
their semiclassical limits and derive different relations between
the energy and the other conserved quantities, which is a step
towards M-theory lift of the semiclassical string/gauge theory
correspondence for N=1 field theories.}

\item{In \cite{10} we consider rotating strings and D2-branes on
type IIA background, which arises as dimensional reduction of
M-theory on manifold of $G_2$ holonomy, dual to N=1 gauge theory
in four dimensions. We obtain exact solutions and explicit
expressions for the conserved charges. By taking the semiclassical
limit, we show that the rotating strings can reproduce only one
type of semiclassical behavior, exhibited by rotating M2-branes on
$G_2$ manifolds. Our further investigation leads to the conclusion
that the rotating D2-branes reproduce two types of the
semiclassical energy-charge relations known for membranes in
eleven dimensions.}

\item{In \cite{11} we show that for each M-theory background,
having subspaces with metrics of given type, there exist M2-brane
configurations, which in appropriate limit lead to two-spin
magnon-like energy-charge relations, established for strings on
$AdS_5\times S^5$, its $\beta$-deformation, and for membrane in
$AdS_4\times S^7$.}

\item{It is known that large class of classical string solutions
in the type IIB $AdS_5\times S^5$ background is related to the
Neumann and Neumann-Rosochatius integrable systems, including
spiky strings and giant magnons. It is also interesting if these
integrable systems can be associated with some membrane
configurations in M-theory. We show in \cite{12} that this is
indeed the case by presenting explicitly several types of membrane
embedding in $AdS_4\times S^7$ with the searched properties.}

\item{In \cite{13} we find membrane configurations in $AdS_4\times
S^7$, which correspond to the continuous limit of the SU(2)
integrable spin chain, considered as a limit of the SU(3) spin
chain, arising in N=4 SYM in four dimensions, dual to strings in
$AdS_5\times S^5$. We also discuss the relationship with the
Neumann-Rosochatius integrable system at the level of Lagrangians,
comparing the string and membrane cases.}

\item{In \cite{14} we describe how Neumann and Neumann-Rosochatius
type integrable systems, as well as the continuous limit of the
SU(2) integrable spin chain, can be obtained from M2-branes in the
framework of AdS/CFT correspondence.}

\item{In \cite{15} we use the reduction of the string dynamics on
$R_t\times S^3$ to the Neumann-Rosochatius integrable system to
map all string solutions described by this dynamical system onto
solutions of the complex sine-Gordon integrable model. This
mapping relates the parameters in the solutions on both sides of
the correspondence. In the framework of this approach, we find
finite-size string solutions, their images in the (complex)
sine-Gordon system, and the leading finite-size effects of the
single spike "$E-\Delta\phi$" relation for both $R_t\times S^2$
and $R_t\times S^3$ cases.}

\item{In \cite{16} we consider semi-classical solution of
membranes on the $AdS_4\times S^7$. This is supposed to be dual to
the $\mathcal{N}=6$ super Chern-Simons theory with level $k=1$ in
a planar limit recently proposed by Aharony, Bergmann, Jafferis,
and Maldacena (ABJM). We have identified giant magnon and single
spike states on the membrane by reducing them to the Neumamm -
Rosochatius integrable system. We also connect these to the
complex sine-Gordon integrable model. Based on this approach, we
find finite-size membrane solutions and obtain their images in the
complex sine-Gordon system along with the leading finite-size
corrections to the energy-charge relations.}

\item{Recently, O. Aharony, O. Bergman, D. L. Jafferis and J.
Maldacena (ABJM) proposed three-dimensional super
Chern-Simons-matter theory, which at level $k$ is supposed to
describe the low energy limit of $N$ M2-branes. For large $N$ and
$k$, but fixed 't Hooft coupling $\lambda=N/k$, it is dual to type
IIA string theory on $AdS_4\times CP^3$. For large $N$ but finite
$k$, it is dual to M-theory on $AdS_4\times S^7/Z_k$. In \cite{17}
, relying on the second duality, we find exact giant magnon and
single spike solutions of membrane configurations on $AdS_4\times
S^7/Z_k$ by reducing the system to the Neumann-Rosochatius
integrable model. We derive the dispersion relations and their
finite-size corrections with explicit dependence on the level
$k$.}

\item{In \cite{18} we consider finite-size effects for the dyonic
giant magnon of the type IIA string theory on $AdS_4\times CP^3$
by applying Luscher $\mu$-term formula which is derived from a
recently proposed S-matrix for the $\mathcal{N}=6$ super
Chern-Simons theory. We compute explicitly the effect for the case
of a symmetric configuration where the two external bound states,
each of A and B particles, have the same momentum $p$ and spin
$J_2$. We compare this with the classical string theory result
which we computed by reducing it to the Neumann-Rosochatius
system. The two results match perfectly.}

\item{In \cite{19} we investigate dyonic giant magnons propagating
on $\gamma$-deformed $AdS_5\times S^5$ by Neumann-Rosochatius
reduction method with a twisted boundary condition. We compute
finite-size effect of the dispersion relations of dyonic giant
magnons which generalizes the previously known case of the giant
magnons with one angular momentum found by Bykov and Frolov.}

\item{In \cite{20} we compute holographic three-point correlation
functions or structure constants of a zero-momentum dilaton
operator and two (dyonic) giant magnon string states with a
finite-size length in the semiclassical approximation. We show
that the semiclassical structure constants match exactly with the
three-point functions between two $su(2)$ magnon single trace
operators with finite size and the Lagrangian in the large 't
Hooft coupling constant limit. A special limit
$J\gg\sqrt{\lambda}$ of our result is compared with the relevant
result based on the L\"uscher corrections.}

\item{In \cite{21} we compute semiclassical three-point
correlation function, or structure constant, of two finite-size
(dyonic) giant magnon string states and a light dilaton mode in
the Lunin-Maldacena background, which is the $\gamma$-deformed, or
$TsT$-transformed $AdS_5\times S_{\gamma}^5$, dual to $\mathcal{N}
= 1$ super Yang-Mills theory. We also prove that an important
relation between the structure constant and the conformal
dimension, checked for the $\mathcal{N} = 4$ super Yang-Mills
case, still holds for the $\gamma$-deformed string background.}

\item{In \cite{22} we investigate finite-size giant magnons
propagating on $\gamma$-deformed  $AdS_4 \times CP^3_{\gamma}$
type IIA string theory background, dual to one parameter
deformation of the $\mathcal{N}=6$ super Chern-Simoms-matter
theory. Analyzing the finite-size effect on the dispersion
relation, we find that it is modified compared to the undeformed
case, acquiring $\gamma$ dependence.}

\item{In \cite{23}, in the framework of the semiclassical
approach, we compute the normalized structure constants in
three-point correlation functions, when two of the vertex
operators correspond to heavy string states, while the third
vertex corresponds to a light state. This is done for the case
when the heavy string states are {\it finite-size} giant magnons
with one or two angular momenta, and for two different choices of
the light state, corresponding to dilaton operator and primary
scalar operator. The relevant operators in the dual gauge theory
are $Tr\left(F_{\mu\nu}^2 \ Z^j+\ldots\right)$ and $Tr\left(
Z^j\right)$. We first consider the case of $AdS_5\times S^5$ and
$\mathcal{N} = 4$ super Yang-Mills. Then we extend the obtained
results to the $\gamma$-deformed $AdS_5\times S^5_\gamma$, dual to
$\mathcal{N} = 1$ super Yang-Mills theory, arising as an exactly
marginal deformation of $\mathcal{N} = 4$ super Yang-Mills.}

\item{In \cite{24}, in the framework of the semiclassical
approach, we compute the normalized structure constants in
three-point correlation functions, when two of the vertex
operators correspond to "heavy" string states, while the third
vertex corresponds to a "light" state. This is done for the case
when the "heavy" string states are {\it finite-size} giant
magnons, carrying one or two angular momenta. The "light" states
are taken to be {\it singlet scalar operators on higher string
levels}. We consider two cases: string theory on $AdS_5\times S^5$
and its $\gamma$-deformation.}

\item{In \cite{25} in the framework of the semiclassical approach,
we find the leading finite-size effects on the normalized
structure constants in some three-point correlation functions in
$AdS_5\times S^5$, expressed in terms of the conserved string
angular momenta $J_1$, $J_2$, and the worldsheet momentum $p_w$,
identified with the momentum $p$ of the magnon excitations in the
dual spin-chain arising in $\mathcal{N}=4$ SYM in four
dimensions.}

\item{In \cite{26} we compute the leading finite-size effects on
the normalized structure constants in semiclassical three-point
correlation functions of two finite-size giant magnon string
states and three different types of "light" states - primary
scalar operators, dilaton operator with nonzero momentum and
singlet scalar operators on higher string levels. This is done for
the case of $TsT$-transformed, or $\gamma$-deformed, $AdS_5\times
S^5$ string theory background.}

\item{In \cite{27} we develop an approach for solving the string
equations of motion and Virasoro constraints in any background
which has some (unfixed) number of commuting Killing vector
fields. It is based on a specific ansatz for the string embedding.
We apply the above mentioned approach for strings moving in
$AdS_3\times S^3\times T^4$ with 2-form NS-NS B-field. We
succeeded to find solutions for a large class of string
configurations on this background. In particular, we derive dyonic
giant magnon solutions in the $R_t \times S^3$ subspace, and
obtain the leading finite-size correction to the dispersion
relation.}

\item{In \cite{28} we consider strings moving in the $R_t\times
S^3_\eta$ subspace of the $\eta$-deformed $AdS_5\times S^5$ and
obtain a class of solutions depending on several parameters. They
are characterized by the string energy and two angular momenta.
Finite-size dyonic giant magnon belongs to this class of
solutions. Further on, we restrict ourselves to the case of giant
magnon with one nonzero angular momentum, and obtain the leading
finite-size correction to the dispersion relation.}

\item{In \cite{29} we derive the 3-point correlation function
between two giant magnons heavy string states and the light
dilaton operator with zero momentum in the $\eta$-deformed
$AdS_5\times S^5$ valid for any $J_1$ and $\eta$ in the
semiclassical limit. We show that this result satisfies a
consistency relation between the 3-point correlation function and
the conformal dimension of the giant magnon. We also provide a
leading finite $J_1$ correction explicitly.}

\item{In \cite{30} we compute some normalized structure constants
in the $\eta$-deformed $AdS_5\times S^5$ in the framework of the
semiclassical approach. This is done for the cases when the
``heavy'' string states are finite-size giant magnons carrying one
angular momentum and for three different choices of the ``light''
state: primary scalar operators, dilaton operator with nonzero
momentum, singlet scalar operators on higher string levels.}

\end{enumerate}

The above contributions can be described as investigations in the
following three areas:

\begin{enumerate}

\item{$P$-branes and $Dp$-branes dynamics in general
string/M-theory backgrounds. As applications of the proposed
approach, some particular cases have been considered
\cite{1}-\cite{4}}.

\item{AdS/CFT: string and membrane results:
\cite{5,6,10,15,18,19,22,27,28} and
\cite{7,8,9,10,11,12,13,14,16,17} correspondingly}.

\item{Semiclassical three-point correlation functions in which the
finite-size effect on the ``heavy'' giant magnon string states are
taken into account: \cite{20,21,23,24,25,26,29,30}}.

\end{enumerate}

\section*{Acknowledgements}
This work is partially supported by the NSFB grant DFNI T02/6.

\section*{Appendices}
\def\theequation{A.\arabic{equation}}
\setcounter{equation}{0}
\begin{appendix}

\section{Notations for the special functions}

\h Euler gamma function - $\Gamma(z)$

Pochhammer symbol - $(a)_n$

Jacobi elliptic functions - $\mathbf{sn}(z\vert m)$,
$\mathbf{cn}(z\vert m)$, $\mathbf{dn}(z\vert m)$

Jacobi amplitude - am(z)

Incomplete elliptic integrals of the first, second and third kind
- $F\left(z\vert m\right)$, $E\left(z\vert m\right)$,
$\Pi\left(n,z\vert m\right)$

Complete elliptic integrals of the first, second and third kind -
$\mathbf{K}\left(m\right)$, $\mathbf{E}\left(m\right)$,
$\mathbf{\Pi}\left(n\vert m\right)$

Hypergeometric functions of one variable - ${}_1F_0(a;z)$,
${}_2F_1(a,b;c;z)$

Hypergeometric functions of two variables -
$F_1(a,b_1,b_2;c;z_1,z_2)$, $F_2(a,b_1,b_2;c_1,c_2;z_1,z_2)$

Hypergeometric functions of $n$ variables -
$F_{D}^{(n)}(a,b_1,...,b_n;c;z_1,...,z_n)$

\section{Relation between the NR system and CSG}

\subsection{Explicit Relations between the Parameters}

In the general case, the relation between the parameters in the
solutions of the NR and CSG integrable systems is given by \bea\nn
K^2=R^2M^2,\h C_\phi= \frac{2}{\alpha^2-\beta^2}\left\{3M^2-
2\left[\kappa^2
+\frac{(\kappa^2-\omega_1^2)-\omega_2^2}{1-\beta^2/\alpha^2}\right]\right\},\eea
\bea\nn &&\frac{1}{4}M^4(\alpha^2-\beta^2)\frac{A^2}{\beta^2}=
M^4\left(M^2-\kappa^2+\frac{\omega_2^2}{1-\beta^2/\alpha^2}\right)
\\ \nn &&-\left(\frac{\kappa^2-\omega_1^2}{1-\beta^2/\alpha^2}\right)
\left\{M^4 +
\left[M^2-\left(\frac{\kappa^2-\omega_1^2}{1-\beta^2/\alpha^2}\right)\right]
\left(\frac{\kappa^2-\omega_1^2}{1-\beta^2/\alpha^2}\right)\right.\\
\label{equiv2}
&&-\left.\left[2M^2-\left(\frac{\kappa^2-\omega_1^2}{1-\beta^2/\alpha^2}\right)\right]
\left(\kappa^2-\frac{\omega_2^2}{1-\beta^2/\alpha^2}\right)\right\}
\\ \nn &&-\frac{(\omega_1^2-\omega_2^2)}{\omega_1^2\left(1-\beta^2/\alpha^2\right)^3}
\biggl\{\left[M^2\left(1-\beta^2/\alpha^2\right) - \kappa^2\right]
(\omega_1^2-\omega_2^2)\check{C}_2^2\\ \nn
&&-\left[M^2\left(1-\beta^2/\alpha^2\right) -
(\kappa^2-\omega_1^2)\right]
\left[2\frac{\beta}{\alpha}\omega_2\kappa^2\check{C}_2 +
\left(\kappa^2-\omega_1^2\right)
\left(\frac{\beta^2}{\alpha^2}\kappa^2-\omega_1^2\right)\right]\biggr\},\eea
\bea \nn &&\frac{1}{4}M^4(\alpha^2-\beta^2)C_\chi^2=
-\left(\frac{\kappa^2-\omega_1^2}{1-\beta^2/\alpha^2}\right)^2
\left[\kappa^2
-\frac{(\kappa^2-\omega_1^2)+\omega_2^2}{1-\beta^2/\alpha^2}\right]
\\ \nn &&+\frac{(\omega_1^2-\omega_2^2)}{\omega_1^2\left(1-\beta^2/\alpha^2\right)^3}
\left\{\kappa^2
(\omega_1^2-\omega_2^2)\check{C}_2^2-(\kappa^2-\omega_1^2)
\left[2\frac{\beta}{\alpha}\omega_2\kappa^2\check{C}_2 +
\left(\kappa^2-\omega_1^2\right)
\left(\frac{\beta^2}{\alpha^2}\kappa^2-\omega_1^2\right)\right]\right\}
,\eea where $\check{C}_2=C_2/\alpha$. Thus, we have expressed the
CSG parameters $C_\phi$, $A$ and $C_\chi$ through the NR
parameters $\alpha$, $\beta$, $\kappa$, $\omega_1$, $\omega_2$,
$C_2$. The mass parameter $M$ remains free.

Let us consider several examples, which illustrate the established
NR - CSG correspondence. We are interested in the GM and SS
configurations on $R_t\times S^2$ and $R_t\times S^3$. From the
NR-system viewpoint, we have to set $C_2=0$ in (\ref{tsol}),
(\ref{f1s}) and (\ref{f2s}) for the GM and SS string solutions.
This condition is to require one of the turning points, where
$\theta'=0$, to lay on the equator of the sphere, i.e.
$\theta=\pi/2$ \cite{KRT06}.

\subsection{On $R_t\times S^2$}
We begin with the $R_t\times S^2$ case, when $C_2=\omega_2=0$ and
$\theta'$ in (\ref{tsol}) takes the form \bea\label{tps2}
\theta'=\frac{\pm\alpha\omega_1}{(\alpha^2-\beta^2)\sin\theta}
\sqrt{\left(\frac{\beta^2\kappa^2}{\alpha^2\omega_1^2}-\sin^2\theta\right)
\left(\sin^2\theta-\frac{\kappa^2}{\omega_1^2}\right)}.\eea

\subsubsection{The Giant Magnon}

The GM solution corresponds to $\kappa^2=\omega_1^2$ with
$\alpha^2>\beta^2$, which is given by \bea\nn \cos\theta=
\frac{\sqrt{1-\beta^2/\alpha^2}}{\cosh\left(\omega_1\frac{\sigma+\tau\beta/\alpha}
{\sqrt{1-\beta^2/\alpha^2}}\right)}. \label{gmsol}\eea From
\ref{f1s}), one finds $f_2=0$ and \bea\nn
f_1=\arctan\left[\frac{\alpha}{\beta}\sqrt{1-\beta^2/\alpha^2}
\tanh\left(\omega_1\frac{\sigma+\tau\beta/\alpha}
{\sqrt{1-\beta^2/\alpha^2}}\right)\right].\eea For
$R^2=M^2=\omega_1^2=1$, $\beta/\alpha=-\sin\theta_0$, this string
solution coincides with the Hofman-Maldacena solution \cite{HM06},
and is equivalent to the solution in \cite{CDO06} for $R_t\times
S^2$ after the identification $W_1=Z_1\exp(i\pi/2)$, $W_2=Z_2$.
Now, the parameters in (\ref{GMp}) take the values \bea\nn
&&K^2=R^2\omega_1^2=1,\h M^2=\omega_1^2=1,\\ \nn
&&C_\phi=\frac{2\omega_1^2}{\alpha^2-\beta^2}
=\frac{2}{\alpha^2\cos^2\theta_0},\h A= C_{\chi}=0,\eea and from
Eqs.(\ref{ffi}),(\ref{equiv}) and (\ref{equiv1})) the
corresponding SG solution becomes \bea\nn
\sin(\phi/2)=\frac{1}{\cosh\left(\frac{\sigma-\tau\sin\theta_0}
{\cos\theta_0}-\eta_0\right)}.\eea This can be also obtained from
(\ref{S3SG}) by setting $\omega_2=0$.

However, for $M^2>\omega_1^2=\kappa^2$, we have \bea\nn
A=2\beta\sqrt{\frac{M^2-\omega_1^2}{\alpha^2-\beta^2}}\ne 0.\eea
This case is related to the CSG system instead of the SG one. It
is interesting to find the CSG solution associated with it. Using
(\ref{ffi}) again, we find \bea\nn
\sin(\phi/2)=\frac{\omega_1}{M\cosh\left(\omega_1\frac{\sigma+\tau\beta/\alpha}
{\sqrt{1-\beta^2/\alpha^2}}-\eta_0\right)},\h \chi=2
\sqrt{\frac{M^2-\omega_1^2}{1-\beta^2/\alpha^2}}
\left(\frac{\beta}{\alpha}\sigma+\tau\right).\eea

\subsubsection{The Single Spike}

The SS solution corresponds to
$\beta^2\kappa^2=\alpha^2\omega_1^2$. In this case, the
expressions for $\theta$ and $f_1$ are \bea\nn \cos\theta=
\frac{\sqrt{1-\alpha^2/\beta^2}}{\cosh\left(C\xi\right)},\quad
f_1=-\omega_1(\sigma\alpha/\beta+\tau)+\arctan\left[\frac{\beta}{\alpha}\sqrt{1-\alpha^2/\beta^2}
\tanh\left(C\xi\right)\right],\eea and the corresponding string
solution is \bea\nn
W_1&=&R\sqrt{1-\frac{1-\alpha^2/\beta^2}{\cosh^2\left(C\xi\right)}}
\exp{\left\{-i\omega_1\sigma\alpha/\beta +
i\arctan\left[\frac{\beta}{\alpha}\sqrt{1-\alpha^2/\beta^2}
\tanh\left(C\xi\right)\right]\right\}},\\
\nn W_2&=&
\frac{R\sqrt{1-\alpha^2/\beta^2}}{\cosh\left(C\xi\right)},\qquad
Z_0=R\exp\left(i\frac{\alpha}{\beta}\omega_1\tau\right),\eea where
we used a short notation \bea\nn
C\xi\equiv\omega_1\frac{\alpha}{\beta}\frac{\sigma\alpha/\beta+\tau}
{\sqrt{1-\alpha^2/\beta^2}}. \eea

The ``dual'' SG solution can be obtained from (\ref{phis}) by
setting $\omega_2=0$. If we choose $R=1$,
$\alpha/\beta=\sin\theta_1$, $\omega_1=-\cot\theta_1$, $\beta=1$,
the SS solution on $R_t\times S^2$ in \cite{IK07} is reproduced.

\subsection{On $R_t\times S^3$}

\subsubsection{The Giant Magnon}

Let us continue with the $R_t\times S^3$ case, when $C_2=0$,
$\omega_2\ne 0$. First, we would like to establish the
correspondence between the dyonic GM string solution \cite{KRT06}
$(\kappa^2=\omega_1^2)$ to those found in \cite{CDO06} \bea\nn
&&Z_1=\frac{1}{\sqrt{1+k^2}} \left\{\tanh\left[\cos\alpha^D
\left(\sigma\sqrt{1+k^2\cos^2\alpha^D}
-k\tau\cos\alpha^D\right)\right]-ik\right\}\exp(i\tau),\\
\nn &&Z_2=\frac{1}{\sqrt{1+k^2}}\frac{\exp\left[i\sin\alpha^D
\left(\tau\sqrt{1+k^2\cos^2\alpha^D}-k\sigma\cos\alpha^D\right)\right]}
{\cosh\left[\cos\alpha^D \left(\sigma\sqrt{1+k^2\cos^2\alpha^D}
-k\tau\cos\alpha^D\right)\right]},\eea where the parameter $k$ is
related to the soliton rapidity $\hat{\theta}$ through the
equality \bea\nn k=\frac{\sinh\hat{\theta}}{\cos\alpha^D},\eea and
$\alpha^D$ determines the $U(1)$ charge carried by the CSG soliton
\cite{CDO06}.

The solutions of Eqs.(\ref{tsol}), (\ref{f1s}) and (\ref{f2s}) are
given by \bea\nn
&&\cos\theta=\frac{\cos\theta_0}{\cosh\left(C\xi\right)},\h
f_1=\arctan\left[\cot\theta_0\tanh(C\xi)\right], \h
f_2=\frac{\beta\omega_2}{\alpha^2-\beta^2}\xi,
\\ \nn && \sin^2\theta_0\equiv\frac{\beta^2\omega_1^2}{\alpha^2(\omega_1^2-\omega_2^2)},\h
C\equiv\frac{\alpha\sqrt{\omega_1^2-\omega_2^2}}
{\alpha^2-\beta^2}\cos\theta_0.\eea Then, the comparison shows
that the two solutions are equivalent if \bea\nn
&&Z_1\exp(i\pi/2)=W_1=R\sin\theta\exp\left[i\left(\omega_1\tau+f_1\right)\right],
\\ \nn
&&Z_2=W_2=R\cos\theta\exp\left[i\left(\omega_2\tau+f_2\right)\right],
\\ \nn &&R=\kappa=\omega_1=1,\h
\alpha=\cos\alpha^D\sqrt{1+k^2\cos^2\alpha^D},\\ \nn
&&\beta=-k\cos^2\alpha^D,\h\omega_2=
\frac{\sin\alpha^D}{\sqrt{1+k^2\cos^2\alpha^D}}.\eea As a
consequence, the CSG parameters in (\ref{GMp}) reduce to \bea\nn
C_\phi= \frac{2}{\cos^2\alpha^D}\left(1+
2\sin^2\alpha^D\right)],\h A=k\sin(2\alpha^D), \h C_{\chi}=0,\h
K^2=1.\eea

\subsubsection{The Single Spike}

Now, let us turn to the SS solutions on $R_t\times S^3$ as
described by the NR integrable system \cite{BobR07}. By using the
SS-condition $\beta^2\kappa^2=\alpha^2\omega_1^2$ in (\ref{tsol})
one derives \bea\nn
\theta'=\frac{\alpha\sqrt{\omega_1^2-\omega_2^2}}{\alpha^2-\beta^2}
\frac{\cos\theta}{\sin\theta}
\sqrt{\sin^2\theta-\frac{\alpha^2\omega_1^2}{\beta^2(\omega_1^2-\omega_2^2)}},\eea
whose solution is given by \bea\nn \cos\theta =
\frac{\sqrt{\left(1-\alpha^2/\beta^2\right)\omega_1^2-
\omega_2^2}} {\sqrt{\omega_1^2-\omega_2^2}\cosh(C\xi)},\quad
C\xi\equiv \sqrt{\omega_1^2-\frac{\omega_2^2}{1-\alpha^2/\beta^2}}
\frac{\alpha(\sigma\alpha/\beta+\tau)}
{\sqrt{\beta^2-\alpha^2}}.\eea By using (\ref{f1s}), (\ref{f2s}),
one finds the following expressions for the string embedding
coordinates $\varphi_j=\omega_j\tau+f_j$ \bea\nn
\varphi_1&=&-\omega_1\sigma\alpha/\beta +
\arctan\left\{\frac{\beta}{\alpha\omega_1}\sqrt{\left(1-\frac{\alpha^2}{\beta^2}\right)
\left(\omega_1^2-\frac{\omega_2^2}{1-\alpha^2/\beta^2}\right)}
\tanh(C\xi)\right\}\\
\nn \varphi_2&=&-\omega_2\frac{\alpha(\sigma
+\tau\alpha/\beta)}{\beta(1-\alpha^2/\beta^2)}.\eea Comparing the
above results with the SS string solution given in (4.1) - (4.7)
of \cite{IKSV07}, we see that the two solutions coincide for
\bea\label{chp} R=1,\h
\sin\theta_1=-\frac{1}{\sqrt{\omega_1^2-\omega_2^2}}, \h
\sin\gamma_1=\frac{\omega_2}{\omega_1},\h
\omega_1=-\frac{\beta}{\alpha}.\eea From (\ref{SSp}), the CSG
parameters are \bea\nn &&C_\phi=
\frac{2}{\beta^2\left(1-\sin^2\theta_1\cos^2\gamma_1\right)}
\left[4-3M^2 +
\frac{2\cos^4\gamma_1}{\sin^2\gamma_1\left(1-\sin^2\theta_1\cos^2\gamma_1\right)}\right],
\h K^2=M^2,\\ \nn
&&A=\frac{M^2-1}{M^2\sqrt{1-\sin^2\theta_1\cos^2\gamma_1}}
\sqrt{\frac{\cos^4\gamma_1}{\sin^2\gamma_1\left(1-\sin^2\theta_1\cos^2\gamma_1\right)}
-M^2},\\ \nn &&C_\chi=
-\frac{2\sin\gamma_1}{M^2\beta\left(1-\sin^2\theta_1\cos^2\gamma_1\right)}.\eea
Comparing (\ref{chp}) with (\ref{asszo}), one sees that the
solution found in \cite{IKSV07} corresponds actually to $M^2=1$
which leads to $A_{SS}=0$. Hence, the ``dual'' CSG solution is of
the type (\ref{CSGsol}).

\section{Explicit exact solutions in $AdS_3\times S^3\times T^4$ with
NS-NS B-field}

Let start with the solutions for the string coordinates in $AdS_3$
subspace. By using (\ref{SP}),
 (\ref{idc}) and (\ref{bfs}), one can find that the scalar potential $U_r$ in  (\ref{Eqa}) is given by
\bea\label{ur}
&&U_r(r)=\frac{1}{2(\alpha^2-\beta^2)}\Bigg[\left(\alpha
\Lambda^\phi\right)^2  r^2  - \left(\alpha
\Lambda^t\right)^2(1+r^2)
\\ \nn &&+ \frac{\left(C_\phi +q\alpha \Lambda^t r^2\right)^2}{r^2}-
 \frac{\left(C_t- q\alpha \Lambda^\phi r^2\right)^2}{1+r^2} \Bigg] .\eea

After introducing the variable \bea\label{yr} y=r^2,\eea and
replacing (\ref{ur}) into (\ref{xi}) one can rewrite it in the
following form \bea\label{dxidy}
d\xi=\frac{\alpha^2-\beta^2}{2\alpha
\sqrt{(1-q^2)\left[\left(\Lambda^\phi\right)^2-\left(\Lambda^t\right)^2\right]}}
\frac{dy}{\sqrt{(y_p-y)(y-y_m)(y-y_n)}} ,\eea where \bea\nn 0\leq
y_m<y<y_p,\h y_n<0,\eea and $y_p$, $y_m$, $y_n$ satisfy the
relations \bea\nn
&&y_p+y_m+y_n=\frac{1}{\alpha^{2}(1-q^2)\left[\left(\Lambda^\phi\right)^2-\left(\Lambda^t\right)^2\right]}
\\ \nn
&&\left[C_r(\alpha^2-\beta^2)-\alpha\left(\alpha\left(\Lambda^\phi\right)^2-2\alpha\left(\Lambda^t\right)^2\right)
+2q\left(C_\phi\Lambda^t+C_t\Lambda^\phi\right)+q^2\alpha\left(\Lambda^t\right)^2\right],
\\ \label{ypm} && y_p y_m+y_p y_n+y_m y_n = -\frac{1}{\alpha^{2}(1-q^2)\left[\left(\Lambda^\phi\right)^2-\left(\Lambda^t\right)^2\right]}
\\ \nn &&\left[C_r(\alpha^2-\beta^2)+C_t^2-C_\phi^2+\alpha^2\left(\Lambda^t\right)^2-2q\alpha C_\phi\Lambda^t\right],
\\ \nn && y_p y_m y_n=- \frac{C_\phi^2}{\alpha^{2}(1-q^2)\left[\left(\Lambda^\phi\right)^2-\left(\Lambda^t\right)^2\right]}.\eea

Integrating (\ref{dxidy}) and inverting \bea\nn \xi(y)=
\frac{\alpha^2-\beta^2}{\alpha
\sqrt{(1-q^2)\left[\left(\Lambda^\phi\right)^2-\left(\Lambda^t\right)^2\right](y_p-y_n)}}\
F\left(\arcsin\sqrt{\frac{y_p-y}{y_p-y_m}},\frac{y_p-y_m}{y_p-y_n}\right)\eea
to $y(\xi)$, one finds the following solution \bea\label{yxi}
y(\xi)=(y_p-y_n)\ {\mathbf{dn}}^2\left[\frac{\alpha
\sqrt{(1-q^2)\left[\left(\Lambda^\phi\right)^2-\left(\Lambda^t\right)^2\right](y_p-y_n)}}
{\alpha^2-\beta^2}\ \xi,\frac{y_p-y_m}{y_p-y_n}\right]+y_n.\eea

Next, we will compute $\tilde{X}^t(\xi)$ and $\tilde{X}^\phi(\xi)$
entering (\ref{Xmu}). Integrating \bea\nn
&&\frac{d\tilde{X}^t}{d\xi}= \frac{1}{\alpha^2-\beta^2}
\left[\beta\Lambda^t +q\alpha\Lambda^\phi
-\left(C_t+q\alpha\Lambda^\phi\right)\frac{1}{1+y}\right],
\\ \nn &&\frac{d\tilde{X}^\phi}{d\xi}= \frac{1}{\alpha^2-\beta^2}
\left(\beta\Lambda^\phi+q\alpha\Lambda^t+\frac{C_\phi}{y}\right),\eea
and using (\ref{yxi}), we obtain the following solutions for the
string coordinates $t$, $\phi$, in accordance with our ansatz
\bea\label{sols}
&&t(\tau,\sigma)=\Lambda^t\tau+\frac{1}{\alpha\sqrt{(1-q^2)\left[\left(\Lambda^\phi\right)^2
-\left(\Lambda^t\right)^2\right](y_p-y_n)}}
\\ \nn &&\left[\left(\beta \Lambda^t+q \alpha\Lambda^\phi\right)\
F\left(\arcsin\sqrt{\frac{y_p-y}{y_p-y_m}},\frac{y_p-y_m}{y_p-y_n}\right)\right.
\\ \nn &&-\left. \frac{C_t+q \alpha\Lambda^\phi}{1+y_p}
\ \Pi\left(\arcsin\sqrt{\frac{y_p-y}{y_p-y_m}},
\frac{y_p-y_m}{1+y_p},\frac{y_p-y_m}{y_p-y_n}\right)\right]
\\ \label{Ff} &&\phi(\tau,\sigma) =\Lambda^\phi\tau + \frac{1}{\alpha\sqrt{(1-q^2)\left[\left(\Lambda^\phi\right)^2
-\left(\Lambda^t\right)^2\right](y_p-y_n)}}
\\ \nn &&\left[\left(\beta \Lambda^\phi+q\alpha\Lambda^t\right)\
F\left(\arcsin\sqrt{\frac{y_p-y}{y_p-y_m}},\frac{y_p-y_m}{y_p-y_n}\right)\right.
\\ \nn &&+\left. \frac{C_\phi}{y_p}
\ \Pi\left(\arcsin\sqrt{\frac{y_p-y}{y_p-y_m}},
\frac{y_p-y_m}{y_p},\frac{y_p-y_m}{y_p-y_n}\right)\right].\eea

Let us compute now the string energy and spin on the solutions
found. Starting from (\ref{EsS}), (\ref{Sf}), we obtain
\bea\label{Ess} &&E_s= \frac{2
T}{\sqrt{(1-q^2)\left[\left(\Lambda^\phi\right)^2
-\left(\Lambda^t\right)^2\right](y_p-y_n)}}
\\ \nn &&\left[\left(\Lambda^t-\frac{\beta}{\alpha^2}C_t-q\frac{C_\phi}{\alpha}\right)
\mathbf{K}\left(1-\frac{y_m-y_n}{y_p-y_n}\right) +\right.
\\ \nn &&\left. (1-q^2)\Lambda^t \left(y_n \ \mathbf{K}\left(1-\frac{y_m-y_n}{y_p-y_n}\right)
+(y_p-y_n)\
\mathbf{E}\left(1-\frac{y_m-y_n}{y_p-y_n}\right)\right)\right],
\\ \label{Ss} &&S= \frac{2 T}{\sqrt{(1-q^2)\left[\left(\Lambda^\phi\right)^2
-\left(\Lambda^t\right)^2\right](y_p-y_n)}}
\\ \nn &&\left[\left(\frac{\beta}{\alpha^2}C_\phi+q\frac{C_t}{\alpha}+\Lambda^\phi q^2\right)
\mathbf{K}\left(1-\frac{y_m-y_n}{y_p-y_n}\right) +\right.
\\ \nn &&\left. (1-q^2)\Lambda^\phi \left(y_n \ \mathbf{K}\left(1-\frac{y_m-y_n}{y_p-y_n}\right)
+(y_p-y_n)\
\mathbf{E}\left(1-\frac{y_m-y_n}{y_p-y_n}\right)\right)\right.
\\ \nn &&\left. -\frac{q\frac{C_t}{\alpha}+q^2\Lambda^\phi}{1+y_p}\
\mathbf{\Pi}\left(\frac{y_p-y_m}{1+y_p},1-\frac{y_m-y_n}{y_p-y_n}\right)\right].\eea

Now we turn to the $S^3$ subspace. By using (\ref{SP}),
(\ref{idc}) and (\ref{bfs}), one can show that the scalar
potential $U_\theta$ in (\ref{xi}) can be written as
\bea\label{Ux} &&U_\theta(\theta) = \frac{1}{2(\alpha^2-\beta^2)}
\left[\frac{\left(C_{\phi_2}-q\alpha\Lambda^{\phi_1}\
\chi\right)^2}{\chi}
+\frac{\left(C_{\phi_1}+q\alpha\Lambda^{\phi_2}\
\chi\right)^2}{1-\chi}\right.
\\ \nn &&\left. +\alpha^2 \left(\Lambda^{\phi_2}\right)^2 \chi+
\alpha^2\left(\Lambda^{\phi_1}\right)^2(1-\chi)\right],\eea where
we introduced the notation \bea\label{chidef} \chi\equiv
\cos^2\theta.\eea Replacing (\ref{Ux}) in (\ref{xi}), one can see
that it can be written in the form \bea\label{xiS3} d\xi
=\frac{\alpha^2-\beta^2}{2\alpha\sqrt{(1-q^2)
\left(\left(\Lambda^{\phi_1}\right)^2-\left(\Lambda^{\phi_2}\right)^2\right)}}
\frac{d\chi}{\sqrt{(\chi_p-\chi)(\chi-\chi_m)(\chi-\chi_n)}},\eea
where \bea\nn 0\leq \chi_m<\chi<\chi_p\leq 1,\h \chi_n\leq 0,\eea
and \bea\nn &&\chi_p+\chi_m+\chi_n =  \frac{1}{\alpha^2(1-q^2)
\left(\left(\Lambda^{\phi_1}\right)^2-\left(\Lambda^{\phi_2}\right)^2\right)}
\\ \nn &&
\left[-C_\theta(\alpha^2-\beta^2)
-\left(\alpha\Lambda^{\phi_2}\right)^2+
(2-q^2)\left(\alpha\Lambda^{\phi_1}\right)^2 \right.
\\ \nn &&\left.- 2q\alpha \left(C_{\phi_2}\Lambda^{\phi_1}
+C_{\phi_1}\Lambda^{\phi_2}\right)\right],\eea

\bea\nn &&\chi_p\chi_m+\chi_p\chi_n+\chi_m\chi_n=
 \frac{1}{\alpha^2(1-q^2)
\left(\left(\Lambda^{\phi_1}\right)^2-\left(\Lambda^{\phi_2}\right)^2\right)}
\\ \nn &&
\left[\left(\alpha\Lambda^{\phi_1}\right)^2+C_{\phi_1}^2
-C_{\phi_2}^2-C_\theta (\alpha^2-\beta^2) - 2q\alpha
C_{\phi_2}\Lambda^{\phi_1}\right],\eea \bea\nn
\chi_p\chi_m\chi_n=-\frac{\left(C_{\phi_2}\right)^2}{\alpha^2(1-q^2)
\left(\left(\Lambda^{\phi_1}\right)^2-\left(\Lambda^{\phi_2}\right)^2\right)}.\eea

Integrating (\ref{xiS3}), one finds the following solution for
$\chi$ \bea\label{chisol} \chi(\xi)=(\chi_p-\chi_n) \
\mathbf{dn}^2\left[\frac{\alpha\sqrt{(1-q^2)
\left(\left(\Lambda^{\phi_1}\right)^2-\left(\Lambda^{\phi_2}\right)^2\right)(\chi_p-\chi_n)}}
{\alpha^2-\beta^2}\ \xi,
\frac{\chi_p-\chi_m}{\chi_p-\chi_n}\right]+\chi_n.\eea

Now we are ready to find the ``orbit'' $r=r(x)$. Written in terms
of $y$ and $\chi$, it is given by \bea\label{orbit} &&y=
(y_p-y_n)\ \mathbf{dn}^2\left[\frac{\sqrt{
\left(\left(\Lambda^\phi\right)^2-\left(\Lambda^t\right)^2\right)(y_p-y_n)}}
{\sqrt{\left(\left(\Lambda^{\phi_1}\right)^2-\left(\Lambda^{\phi_2}\right)^2\right)(\chi_p-\chi_n)}}\right.
\\ \nn &&\left.\times F\left(\arcsin\sqrt{\frac{\chi_p-\chi}{\chi_p-\chi_m}},
\frac{\chi_p-\chi_m}{\chi_p-\chi_n}\right),
\frac{y_p-y_m}{y_p-y_n}\right]+y_n .\eea

Next, we compute $\tilde{X}^{\phi_1}(\xi)$ and
$\tilde{X}^{\phi_2}(\xi)$. Replacing the results in our ansatz, we
derive the following solutions for the isometric coordinates on
$S^3$

\bea\label{thetatsol} &&\phi_1 = \Lambda^{\phi_1}\tau+
\frac{1}{\alpha\sqrt{(1-q^2)
\left(\left(\Lambda^{\phi_1}\right)^2-\left(\Lambda^{\phi_2}\right)^2\right)(\chi_p-\chi_n)}}
\\ \nn &&\left[\left(\beta\Lambda^{\phi_1}
-q\alpha\Lambda^{\phi_2}\right)\
F\left(\arcsin\sqrt{\frac{\chi_p-\chi}{\chi_p-\chi_m}},\frac{\chi_p-\chi_m}{\chi_p-\chi_n}\right)\right.
\\ \nn &&\left. +\frac{C_{\phi_1}+q\alpha\Lambda^{\phi_2}}{1-\chi_p}
\
\Pi\left(\arcsin\sqrt{\frac{\chi_p-\chi}{\chi_p-\chi_m}},-\frac{\chi_p-\chi_m}{1-\chi_p},
\frac{\chi_p-\chi_m}{\chi_p-\chi_n}\right)\right].\eea

\bea\label{thetasol} &&\phi_2 =\Lambda^{\phi_2}\tau+
\frac{1}{\alpha\sqrt{(1-q^2)
\left(\left(\Lambda^{\phi_1}\right)^2-\left(\Lambda^{\phi_2}\right)^2\right)(\chi_p-\chi_n)}}
\\ \nn &&\left[\left(\beta\Lambda^{\phi_2}
-q\alpha\Lambda^{\phi_1}\right)\
F\left(\arcsin\sqrt{\frac{\chi_p-\chi}{\chi_p-\chi_m}},\frac{\chi_p-\chi_m}{\chi_p-\chi_n}\right)\right.
\\ \nn &&\left. +\frac{C_{\phi_1}}{\chi_p}
\
\Pi\left(\arcsin\sqrt{\frac{\chi_p-\chi}{\chi_p-\chi_m}},1-\frac{\chi_m}{\chi_p},
\frac{\chi_p-\chi_m}{\chi_p-\chi_n}\right)\right],\eea

Based on (\ref{Qt}) and the solutions for the string coordinates
on $S^3$ we found, we can write down the explicit expressions for
the conserved angular momenta $J_1$ and $J_2$ computed on the
solutions. The result is

\bea\label{Jtsol} &&J_1 = \frac{2T}{\sqrt{(1-q^2)
\left(\left(\Lambda^{\phi_1}\right)^2-\left(\Lambda^{\phi_2}\right)^2\right)(\chi_p-\chi_n)}}
\\ \nn &&\left[\left(\frac{\beta}{\alpha^2}C_{\phi_1}+\Lambda^{\phi_1}
-q\frac{C_{\phi_2}}{\alpha}\right)
\mathbf{K}\left(1-\frac{\chi_m-\chi_n}{\chi_p-\chi_n}\right)\right.
\\ \nn &&\left.
-(1-q^2)\Lambda^{\phi_1} \left(\chi_n\
\mathbf{K}\left(1-\frac{\chi_m-\chi_n}{\chi_p-\chi_n}\right)
+(\chi_p-\chi_n)\
\mathbf{E}\left(1-\frac{\chi_m-\chi_n}{\chi_p-\chi_n}\right)\right)\right].\eea

\bea\label{Jsol} &&J_2= \frac{2T}{\sqrt{(1-q^2)
\left(\left(\Lambda^{\phi_1}\right)^2-\left(\Lambda^{\phi_2}\right)^2\right)(\chi_p-\chi_n)}}
\\ \nn &&\left[\left(\frac{\beta}{\alpha^2}C_{\phi_2}-q\left(\frac{C_{{\phi_1}}}{\alpha}
+q\Lambda^{\phi_2}\right)\right)
\mathbf{K}\left(1-\frac{\chi_m-\chi_n}{\chi_p-\chi_n}\right)\right.
\\ \nn &&\left.
+(1-q^2)\Lambda^{\phi_2} \left(\chi_n\
\mathbf{K}\left(1-\frac{\chi_m-\chi_n}{\chi_p-\chi_n}\right)
+(\chi_p-\chi_n)\
\mathbf{E}\left(1-\frac{\chi_m-\chi_n}{\chi_p-\chi_n}\right)\right)\right.
\\ \nn &&\left.
+\frac{q\left(\frac{C_{\phi_1}}{\alpha}+q\Lambda^{\phi_2}\right)}{1-\chi_p}
\ \mathbf{\Pi}\left(-\frac{\chi_p-\chi_m}{1-\chi_p},
1-\frac{\chi_m-\chi_n}{\chi_p-\chi_n}\right)\right],\eea

Now, let us go to the $T^4$ subspace. Since in terms of
$\varphi^i$ coordinates the metric is flat and there is no
$B$-field, the solutions for the string coordinates are simple and
given by \bea\label{fi} \varphi^i(\tau,\sigma)= \Lambda^i\tau
+\frac{1}{\alpha^2-\beta^2} \left(C_i
+\beta\Lambda^i\right)\xi.\eea The conserved charges (\ref{Ji})
can be computed to be \bea\label{Jisol} J_i^T=\frac{2\pi\alpha
T}{\alpha^2-\beta^2}
\left(\frac{\beta}{\alpha}C_i+\alpha\Lambda^i\right).\eea

If we impose the periodicity conditions \bea\nn
\varphi^i(\tau,\sigma)= \varphi^i(\tau,\sigma+2L)+2\pi n_i,\h
n_i\in \mathbb{Z}_i,\eea the integration constants $C_i$ are fixed
in terms of the embedding parameters. Namely, \bea\label{Ci} C_i=
\frac{\pi n_i}{L\alpha}(\alpha^2-\beta^2) -\beta\Lambda^i.\eea
Replacing (\ref{Ci}) into (\ref{fi}) and (\ref{Jisol}), one
finally finds \bea\label{T4solJi} &&\varphi^i
=\left(\Lambda^i+\frac{\beta}{\alpha} \frac{\pi n_i}{L}\right)\tau
+\frac{\pi n_i}{L}\sigma,
\\ \nn &&J_i^T =2\pi T \left(\Lambda^i+\frac{\beta}{\alpha}\frac{\pi n_i}{L}\right).\eea

Let us finally point out that the Virasoro constraints impose the
following two conditions on the embedding parameters and
integrations constants in the solutions found \bea\label{Vc}
&&C_r+C_\theta=0,
\\ \nn &&\Lambda^t C_t+\Lambda^\phi C_\phi+\Lambda^{\phi_1} C_{\phi_1}+\Lambda^{\phi_2} C_{\phi_2}
-\Lambda^i\left(\beta\Lambda^i-(\alpha^2-\beta^2)\frac{\pi
n_i}{\alpha L}\right)=0.\eea

\section{M2-brane GM and SS}

For the GM-like case by using that  $\tilde{C}_2=0$,
$\tilde{\kappa}^2=\omega_1^2$ in (\ref{mtsol}), (\ref{mu12s}), one
finds \bea\nn
&&\cos\theta(\xi)=\frac{\cos\tilde{\theta}_0}{\cosh\left(D_0\xi\right)},\h
\sin^2\tilde{\theta}_0=\frac{\beta^2\omega_1^2}{\tilde{A}^2(\omega_1^2-\omega_2^2)},\h
D_0=\frac{\tilde{A}\sqrt{\omega_1^2-\omega_2^2}}
{\tilde{A}^2-\beta^2}\cos\tilde{\theta}_0,
\\ \nn &&\varphi_1(\tau,\xi)=\omega_1\tau +\arctan\left[\cot\tilde{\theta}_0\tanh(D_0\xi)\right],
\h\varphi_2(\tau,\xi)=\omega_2\left(\tau +
\frac{\beta}{\tilde{A}^2-\beta^2}\xi\right).\eea

For the SS-like solutions when $\tilde{C}_2=0$,
$\tilde{\kappa}^2=\omega_1^2\tilde{A}^2/\beta^2$, by solving the
equations (\ref{mtsol}), (\ref{mu12s}), one arrives at \bea\nn
&&\cos\theta(\xi)=\frac{\cos\tilde{\theta}_1}{\cosh\left(D_1\xi\right)},\h
\sin^2\tilde{\theta}_1=\frac{\tilde{A}^2\omega_1^2}{\beta^2(\omega_1^2-\omega_2^2)},\h
D_1=\frac{\tilde{A}\sqrt{\omega_1^2-\omega_2^2}}
{\tilde{A}^2-\beta^2}\cos\tilde{\theta}_1,
\\ \nn &&\varphi_1(\tau,\xi)=\omega_1\left(\tau
-\frac{\xi}{\beta}\right)
-\arctan\left[\cot\tilde{\theta}_1\tanh(D_1\xi)\right],
\h\varphi_2(\tau,\xi)=\omega_2\left(\tau +
\frac{\beta}{\tilde{A}^2-\beta^2}\xi\right) .\eea

The energy-charge relations computed on the above membrane
solutions were found in \cite{BR07}, and in our notations read
\bea\label{EJGM0} \sqrt{1-r_0^2}E-\frac{J_1}{2}
=\sqrt{\left(\frac{J_2}{2}\right)^2
+\frac{\tilde{\lambda}}{\pi^2}\sin^2\frac{p}{2}},\h
\frac{p}{2}=\frac{\pi}{2}-\tilde{\theta}_0, \eea for the GM-like
case, and \bea\label{EJSS0}
\sqrt{1-r_0^2}E-\frac{\sqrt{\tilde{\lambda}}}{2\pi}\Delta\varphi_1
=\frac{\sqrt{\tilde{\lambda}}}{\pi}\frac{p}{2}, \h \frac{J_1}{2}
=\sqrt{\left(\frac{J_2}{2}\right)^2
+\frac{\tilde{\lambda}}{\pi^2}\sin^2\frac{p}{2}},\h
\frac{p}{2}=\frac{\pi}{2}-\tilde{\theta}_1, \eea for the SS-like
solution, where \bea\label{tl}
\tilde{\lambda}=\left[2\pi^2T_2R^3r_0
(1-r_0^2)\gamma\right]^2.\eea

\subsection{Finite-Size Effects}

For $\tilde{C}_2=0$, Eq.(\ref{mtsol}) can be written as
\bea\label{ptS3eq}
(\cos\theta)'=\mp\frac{\tilde{A}\sqrt{\omega_1^2-\omega_2^2}}{\tilde{A}^2-\beta^2}
\sqrt{(z_+^2-\cos^2\theta)(\cos^2\theta-z_-^2)},\eea where \bea\nn
&&z^2_\pm=\frac{1}{2(1-\frac{\omega_2^2}{\omega_1^2})}
\left\{q_1+q_2-\frac{\omega_2^2}{\omega_1^2}
\pm\sqrt{(q_1-q_2)^2-\left[2\left(q_1+q_2-2q_1
q_2\right)-\frac{\omega_2^2}{\omega_1^2}\right]
\frac{\omega_2^2}{\omega_1^2}}\right\}, \\ \nn
&&q_1=1-\tilde{\kappa}^2/\omega_1^2,\h
q_2=1-\beta^2\tilde{\kappa}^2/\tilde{A}^2\omega_1^2 .\eea The
solution of (\ref{ptS3eq}) is \bea\label{ptS3sol} \cos\theta=z_+
\mathbf{dn}\left(C\xi|m\right),\h
C=\mp\frac{\tilde{A}\sqrt{\omega_1^2-\omega_2^2}}{\tilde{A}^2-\beta^2}
z_+,\h m\equiv 1-z^2_-/z^2_+ .\eea The solutions of
Eqs.(\ref{mu12s}) now read \bea\nn
&&\mu_1=\frac{2\beta/\tilde{A}}{z_+\sqrt{1-\omega_2^2/\omega_1^2}}
\left[C\xi -
\frac{\tilde{\kappa}^2/\omega_1^2}{1-z^2_+}\Pi\left(am(C\xi),-\frac{z^2_+
-z^2_-}{1-z^2_+}\bigg \vert m\right)\right],\\ \nn
&&\mu_2=\frac{2\beta\omega_2/\tilde{A}\omega_1}{z_+\sqrt{1-\omega_2^2/\omega_1^2}}
C\xi.\eea

Our next task is to find out what kind of energy-charge relations
can appear for the M2-brane solution in the limit when the energy
$E\to \infty$. It turns out that the semiclassical behavior
depends crucially on the sign of the difference
$\tilde{A}^2-\beta^2$.

\subsubsection{The M2-brane GM}

We begin with the M2-brane GM, i.e. $\tilde{A}^2>\beta^2$. In this
case, one obtains from (\ref{pcqs}) the following expressions for
the conserved energy $E$ and the angular momenta $J_1$, $J_2$
\bea\nn &&\mathcal{E}
=\frac{2\tilde{\kappa}(1-\beta^2/\tilde{A}^2)} {\omega_1
z_+\sqrt{1-\omega_2^2/\omega_1^2}}\mathbf{K}
\left(1-z^2_-/z^2_+\right), \\ \label{pcqsGM} &&\mathcal{J}_1=
\frac{2 z_+}{\sqrt{1-\omega_2^2/\omega_1^2}} \left[
\frac{1-\beta^2\tilde{\kappa}^2/\tilde{A}^2\omega_1^2}{z^2_+}\mathbf{K}
\left(1-z^2_-/z^2_+\right)-\mathbf{E}
\left(1-z^2_-/z^2_+\right)\right], \\ \nn &&\mathcal{J}_2= \frac{2
z_+ \omega_2/\omega_1 }{\sqrt{1-\omega_2^2/\omega_1^2}}\mathbf{E}
\left(1-z^2_-/z^2_+\right).\eea Here, we have used the notations
\bea\label{pnot} \mathcal{E}=\frac{2\pi}{\sqrt{\tilde{\lambda}}}
\sqrt{1-r_0^2} E ,\h
\mathcal{J}_1=\frac{2\pi}{\sqrt{\tilde{\lambda}}}\frac{J_1}{2}, \h
\mathcal{J}_2=\frac{2\pi}{\sqrt{\tilde{\lambda}}}\frac{J_2}{2},\eea
where $\tilde{\lambda}$ is defined in (\ref{tl}). The computation
of $\Delta\varphi_1$ gives \bea\label{ppws} p\equiv\Delta\varphi_1
&=& 2\int_{\theta_{min}}^{\theta_{max}}\frac{d
\theta}{\theta'}\mu'_1=
\\ \nn &-&\frac{2\beta/\tilde{A}}{z_+\sqrt{1-\omega_2^2/\omega_1^2}}
\left[\frac{\tilde{\kappa}^2/\omega_1^2}{1-z^2_+}\mathbf{\Pi}\left(-\frac{z^2_+
- z^2_-}{1-z^2_+}\bigg\vert 1-z^2_-/z^2_+\right) -\mathbf{K}
\left(1-z^2_-/z^2_+\right)\right].\eea

Expanding the elliptic integrals about $z_-^2=0$, one arrives at
\bea\label{pIEJ1} &&\mathcal{E}-\mathcal{J}_1 =
\sqrt{\mathcal{J}_2^2+4\sin^2(p/2)} - \frac{16 \sin^4(p/2)}
{\sqrt{\mathcal{J}_2^2+4\sin^2(p/2)}}\\ \nn
&&\exp\left[-\frac{2\left(\mathcal{J}_1 +
\sqrt{\mathcal{J}_2^2+4\sin^2(p/2)}\right)
\sqrt{\mathcal{J}_2^2+4\sin^2(p/2)}\sin^2(p/2)}{\mathcal{J}_2^2+4\sin^4(p/2)}
\right].\eea It is easy to check that the energy-charge relation
(\ref{pIEJ1}) coincides with the one found in \cite{HS08},
describing the finite-size effects for dyonic GM. The difference
is that in the string case the relations between $\mathcal{E}$,
$\mathcal{J}_1$, $\mathcal{J}_2$ and $E$, $J_1$, $J_2$ are given
by \bea\nn \mathcal{E}=\frac{2\pi}{\sqrt{\lambda}}E ,\h
\mathcal{J}_1=\frac{2\pi}{\sqrt{\lambda}}J_1, \h
\mathcal{J}_2=\frac{2\pi}{\sqrt{\lambda}}J_2,\eea while for the
M2-brane they are written in (\ref{pnot}).

\subsubsection{The M2-brane SS}

Let us turn our attention to the M2-brane SS, when
$\tilde{A}^2<\beta^2$. The computation of the conserved quantities
(\ref{pcqs}) and $\Delta\varphi_1$ now gives \bea\nn &&\mathcal{E}
=\frac{2\tilde{\kappa}(\beta^2/\tilde{A}^2-1)}
{\omega_1\sqrt{1-\omega_2^2/\omega_1^2}z_+}\mathbf{K}
\left(1-z^2_-/z^2_+\right), \\ \nn &&\mathcal{J}_1= \frac{2
z_+}{\sqrt{1-\omega_2^2/\omega_1^2}} \left[\mathbf{E}
\left(1-z^2_-/z^2_+\right)
-\frac{1-\beta^2\tilde{\kappa}^2/\tilde{A}^2\omega_1^2}{z^2_+}\mathbf{K}
\left(1-z^2_-/z^2_+\right)\right], \\ \nn &&\mathcal{J}_2=
-\frac{2 z_+ \omega_2/\omega_1
}{\sqrt{1-\omega_2^2/\omega_1^2}}\mathbf{E}
\left(1-z^2_-/z^2_+\right), \\ \nn &&\Delta\varphi_1=
-\frac{2\beta/\tilde{A}}{\sqrt{1-\omega_2^2/\omega_1^2}z_+}
\left[\frac{\tilde{\kappa}^2/\omega_1^2}{1-z^2_+}\mathbf{\Pi}\left(-\frac{z^2_+
- z^2_-}{1-z^2_+}|1-z^2_-/z^2_+\right) -\mathbf{K}
\left(1-z^2_-/z^2_+\right)\right].\eea

$\mathcal{E}-\Delta\varphi_1$ can be derived as
\bea\label{pDiffJ12}\mathcal{E}-\Delta\varphi_1&=& \arcsin
N(\mathcal{J}_1,\mathcal{J}_2) +
2\left(\mathcal{J}_1^2-\mathcal{J}_2^2\right)
\sqrt{\frac{4}{\left[4-\left(\mathcal{J}_1^2-\mathcal{J}_2^2\right)\right]}-1}\\
\nn
&\times&\exp\left[-\frac{2\left(\mathcal{J}_1^2-\mathcal{J}_2^2\right)
N(\mathcal{J}_1,\mathcal{J}_2)}
{\left(\mathcal{J}_1^2-\mathcal{J}_2^2\right)^2 +
4\mathcal{J}_2^2}\left[\Delta\varphi +\arcsin
N(\mathcal{J}_1,\mathcal{J}_2)\right]\right],\\
\nn N(\mathcal{J}_1,\mathcal{J}_2)&\equiv&
\frac{1}{2}\left[4-\left(\mathcal{J}_1^2-\mathcal{J}_2^2\right)\right]
\sqrt{\frac{4}{\left[4-\left(\mathcal{J}_1^2-\mathcal{J}_2^2\right)\right]}-1}.\eea
Finally, by using the SS relation between the angular momenta
\bea\nn \mathcal{J}_1=\sqrt{\mathcal{J}_2^2+4\sin^2(p/2)},\eea we
obtain \bea\label{pssS3c} \mathcal{E}-\Delta\varphi_1=
p+8\sin^2\frac{p}{2}\tan\frac{p}{2}
\exp\left(-\frac{\tan\frac{p}{2}(\Delta\varphi_1 + p)}
{\tan^2\frac{p}{2} + \mathcal{J}_2^2 \csc^2p}\right).\eea This
result coincides with the string result found in \cite{14}. As in
the GM case, the difference is in the identification (\ref{pnot}).

\end{appendix}

\end{document}